\documentclass[a4paper,11pt]{article}
\pdfoutput=1 % if your are submitting a pdflatex (i.e. if you have
\usepackage{jheppub} % for details on the use of the package, please
\usepackage[T1]{fontenc} % if needed
\usepackage{bm}% bold math
\usepackage{url}
\DeclareMathAlphabet{\pazocal}{OMS}{zplm}{m}{n}
\usepackage{braket}
\usepackage{amsmath}
\usepackage{amssymb}
\usepackage{leftidx}
\usepackage[colorinlistoftodos]{todonotes}
\usepackage{bm}
\usepackage{slashed}
\usepackage{calrsfs}
\usepackage{cancel}
\usepackage{dirtytalk}
\usepackage{upgreek}
\usepackage{comment}
\usepackage{xcolor}
\usepackage{ulem}
\usepackage{tensor}

\renewcommand{\emph}{\textit}

\usepackage{hyperref}

\title{\boldmath Unitary Rigid Supersymmetry for the Chiral Graviton and Chiral Gravitino in de~Sitter Spacetime}

\author[a]{Atsushi Higuchi}
\author[b]{and Vasileios A. Letsios}

\affiliation[a]{Department of Mathematics, University of York\\Ian Wand Building, Deramore Lane, York, YO10 5GH, United Kingdom }
\affiliation[b]{ Physique de l’Univers, Champs et Gravitation, Université de Mons – UMONS \\
Place du Parc 20, 7000 Mons, Belgium}

\emailAdd{atsushi.higuchi@york.ac.uk}
\emailAdd{vasileios.letsios@umons.ac.be}

\abstract{It is commonly believed that a unitary supersymmetric quantum field theory (QFT) involving graviton and gravitino fields on fixed 4-dimensional de Sitter spacetime ($dS_{4}$) cannot exist due to known challenges associated with supersymmetry (SUSY) in $dS_{4}$. In this paper, we contradict this expectation by presenting a new unitary supersymmetric QFT on $dS_{4}$: the free supersymmetric theory of the chiral graviton
and chiral gravitino fields. By chiral we mean that the corresponding field strengths are anti-self-dual, and the gauge potentials are complex, each carrying a single complex propagating degree of freedom. The global SUSY transformations are generated by the standard Dirac Killing spinors of $dS_{4}$. The theory overcomes the known obstacles to unitary global SUSY on $dS_{4}$ by closing the commutator between two SUSY transformations on $so(4,2) \oplus u(1)$ rather than the de Sitter algebra $so(4,1)$. Crucially, the $so(4,2)$ symmetry is realised through unconventional conformal-like transformations. This  free theory cannot become interacting while preserving SUSY  in  a way that makes the spin-2 sector the true graviton sector of General Relativity, as  the three-graviton coupling cannot be $u(1)$-invariant.

We establish the unitarity of the free supersymmetric theory in two complementary ways. First, by studying the action of the superalgebra generators on the space of physical gravitino and graviton mode solutions. In particular, we introduce positive-definite, invariant inner products and demonstrate that the SUSY representation is unitary, forming a direct sum of two unitary irreducible representations — one with negative-helicity modes and the other with positive-helicity modes. Second, by quantising the fields and explicitly constructing the complex quantum supercharges $Q_{A}$ and $Q^{A\dagger}$, we show that the trace $\sum_{A} \{ Q_{A}, Q^{A \dagger} \}$ is positive-definite.

Before constructing the supersymmetric theory, we examine the free graviton and gravitino fields on $dS_{4}$, where the gravitino is known to have an imaginary mass parameter. We introduce a Hermitian, gauge-invariant, and local Lagrangian for the free gravitino field and explain why the requirement of unitarity forces the field to be chiral, removing half of the propagating helicity states. }

\begin{document} 
\maketitle
\flushbottom

%\begin{comment}

\section{Introduction} \label{Introduction}

Apart from its significance in inflationary cosmology \cite{Baumann}, de Sitter (dS) spacetime is also relevant to the physics of our present Universe, as suggested by recent observational data supporting an accelerated spatial expansion \cite{PlanckCollab, Suzuki, SDSS, Cole}. Both these eras require a quantum understanding \cite{Witten, Anninos_musings}. It is thus important to develop tools for a deeper theoretical understanding of quantum de Sitter spacetime.

In recent times, the attempts towards a deeper theoretical understanding of dS spacetime have manifested themselves in (at least) two main approaches. The first approach concerns the study of lower-dimensional models in order to develop a more complete quantum understanding \cite{Dio_charm, Anninos_SUGRA, Letsios_Alan, Epstein1, Epstein2, Lasse, Hinter, Anninos_2sphere, Chiara, Eleanor, Silverstein2, Susskind, Anninos_centaur, Anninos_liquids, Maldacenaa, Maloney, Anninos_Schwinger, Beatrix_microscopic, Frob_again}. The second approach concerns the study of a large class of quantum fields in four-dimensional (or higher-dimensional) dS spacetime \cite{Higuchi_forbidden, STSHS, Higuchi_Instab, Higuchi_Instab2, Higuchi_Marolf, Letsios, Letsios_announce, Letsios_announce_II, Letsios_unconventional, Anninos_higher-spin, Anninos_dS/CFT, Anninos_path, Zhenya, Deser_Waldron_ArbitrarySR, Mixed_Symmetry_dS, Law, Sengor_scalar, bootstrap1, bootstrap2, bootstrap_twist, Hertog,Prof_Alan, Sleight, Penedones_bootstrap, Giataganas, Neiman, Neiman2, Neiman3}. In this approach, the Unitary Irreducible Representations of the de Sitter algebra $so(D,1)$ \cite{Ottoson, Schwarz, Hirai1, Hirai2} play a central role because they are identified with elementary particles on $D$-dimensional dS spacetime, as a generalisation of  Wigner's classification for Minkowski spacetime. In this paper, we take the latter approach, and we uncover new features of supersymmetric {quantum field theory} on four-dimensional de Sitter spacetime, placing special emphasis on group-theoretic aspects.  

 Four-dimensional dS spacetime ($dS_{4}$), is the maximally symmetric solution of the vacuum Einstein equations with positive cosmological constant 
 \cite{hawking},
 \begin{equation}
    R_{\mu \nu}-\frac{1}{2} g_{\mu \nu} R + \Lambda  g_{\mu \nu}=0,
\end{equation}
where $\Lambda = 3 \,\mathcal{R}_{dS}^{-2}$ is the cosmological constant, $\mathcal{R}_{dS}$ is the dS radius, $g_{\mu \nu}$ is the metric tensor,  $R_{\mu \nu} = 3  \mathcal{R}_{dS}^{-2}\,g_{\mu \nu}$ is the Ricci tensor and $R$ is the Ricci scalar. We will work in units where $\mathcal{R}_{dS}=1$.
 Unlike in anti-de Sitter and Minkowski spacetimes, formulating supersymmetric theories in de Sitter spacetime presents fundamental challenges. The main obstacles to the existence of unitary, unbroken de Sitter supersymmetry (SUSY), which differ depending on whether one considers global or local SUSY, can be summarised as follows {(see for instance  the discussion in Section 4 of \cite{Anninos_SUGRA})}:
\begin{itemize}
    \item \textbf{Problems concerning the unitarity of global SUSY on a fixed de Sitter background.}  This can be understood already at the level of abstract representation theory \cite{Lukierski, Nieuwenhuizen}. It is possible to supersymmetrise the dS algebra $so(4,1)$ by introducing spinorial supercharges ${Q}^{(i)}_{A}$, i.e.\ odd generators, which we take to be Dirac spinors for the sake of the discussion. The index $A$ is a spinor index referring to the fundamental spinor representation of $so(4,1)$, and $i$ is an extended SUSY index keeping track of the number of supercharges. Alternatively, one can double the number of supercharges and introduce a symplectic Majorana reality condition, as in \cite{Nieuwenhuizen}, but this will lead to the same representation-theoretic results. As shown in \cite{Lukierski, Nieuwenhuizen},  from the structure of the algebra it follows that $\sum_{A,i} \{ {Q}^{(i)}_{A}, {Q}^{(i)A \, \dagger}   \} = 0$, and thus,  all non-trivial representations of the dS superalgebra on a Hilbert space must be non-unitary (i.e.\ positive-norm and negative-norm states must exist). On the other hand, requiring that negative-norm states do not appear implies that all the ${Q}^{(i)}_{A}$'s, as well as all the dS generators, must annihilate all states in the Hilbert space, i.e.\ only the trivial representation is possible.\footnote{De Sitter supersymmetry has also been studied in the ambient space formalism in Ref.~\cite{Takook}. However, since the anti-commutator between two supercharges closes on $so(4,1)$, the theory is most likely non-unitary.} Super-extensions of $so(2,1)$, $so(3,1)$ and $so(5,1)$ also exist but unitary representations are allowed only in the case of $so(2,1)$ \cite{Lukierski}.

    \item \textbf{Problems concerning the unitarity of $dS_{4}$ Supergravity.} The explicit construction of the $N=2$ Supergravity action with a positive cosmological constant was carried out in \cite{Nieuwenhuizen}, involving a real vierbein, a real photon, and two symplectic Majorana gravitini. The number of gravitini was doubled in order to apply the symplectic Majorana condition  because the conventional Majorana condition for the gravitino on $dS_{4}$  \textbf{cannot} be used - see also Section \ref{Sec_gravitino}. This difficulty is related to  the fact that the mass parameter of the gravitino on $dS_{4}$ is imaginary. Similarly, the conventional Majorana condition is  \textbf{not} consistent with the Killing spinor equation on $dS_{4}$, but the sympelctic Majorana condition is. According to Ref.~\cite{Nieuwenhuizen}, although the $N=2$ dS Supergravity action is invariant under local SUSY, the photon kinetic term has the wrong sign, i.e.\ it is a ghost.
    
    On the other hand, by relaxing the requirement of unbroken SUSY, certain solutions are known. For example, an explicit $dS_{4}$ Supergravity action invariant under spontaneously broken local $N=1$ SUSY  was given in \cite{dS_SUGRA_broken}\footnote{See also \cite{Bansal, Skenderis}.}. This model includes a massive gravitino (this has a real mass parameter), and is  also consistent with unitarity. Another interesting example of stable dS vacuua corresponds to the matter-coupled Supergravity theories with  $N=2$ SUSY  as described in 
    Ref.~\cite{stable_dS_vacuua}. The  main ingredients of the construction  include non-Abelian non-compact gaugings, de Roo–Wagemans rotation angles and Fayet–Iliopoulos terms. However, the question of whether these vacuua can be lifted to string theory remains open\footnote{This is not an easy task, as the no-go theorem of Ref.~\cite{Maldacena} presents serious obstacles for obtaining $dS_{4}$ vacuua from smooth, classical compactifications of  higher-dimensional Supergravity. One possible way to circumvent the  no-go theorem  of \cite{Maldacena} is to include orientifolds in the construction. De Sitter solutions of 10-dimensional supergravity have been obtained in this way; see 
    e.g., \cite{Andriot1} for a review and \cite{Andriot2} for a recent example. Another approach is to consider time-dependent compactifications - see \cite{Andriot3} for recent examples.}.\footnote{Another important question concerns the  non-perturbative existence of dS vacua in string theory - see, e.g., \cite{Linde, Silverstein}.}
     Interestingly, as anticipated by \cite{Lukierski}, and shown in \cite{Anninos_SUGRA}, the problems related to unitary $dS_{4}$ Supergravity with unbroken SUSY can be bypassed in two dimensions.
\end{itemize}
  \textbf{A way out for global SUSY in $dS_{4}$.} In the case of global SUSY, it is possible to have unitary representations by enlarging the even symmetry algebra to the conformal algebra $so(4,2) \supset$ $so(4,1)$. Now the anti-commutator of two supercharges closes on $so(4,2)$ instead of $so(4,1)$, and the trace $\sum_{A,i} \{ {Q}^{(i)}_{A}, \ {Q}^{(i)A\,\dagger}  \}$ does \textbf{not} have to vanish. Thus, unitary representations exist \cite{Dobrev}. Such unitary representations are realised in the case of superconformal field theories on a fixed $dS_{4}$ background spacetime, like the ones constructed in \cite{Anous}.
  %%%%%%%%%%%%%%%%%%%%%%%%%%%%%%%%%%%%%%%%%%%%%%%%%%%%%%%%%%%%%%%%%%5
\subsection{New results}

In this paper, we  present a new unitary supersymmetric quantum field theory (QFT) on a fixed $dS_{4}$ background that includes (a version of) the fields of the supergravity multiplet. In particular, we present:
$$\text{The free \textit{supersymmetric theory of the  chiral graviton and chiral gravitino fields}.}$$ 
By `chiral' we mean that the corresponding 
field strengths are self-dual or anti-self-dual, and thus complex. The corresponding chiral graviton and chiral gravitino gauge potentials are complex and carry one complex propagating degree of freedom each. In this paper, we choose to work with the anti-self-dual case, without loss of generality. The global SUSY transformations of the chiral graviton and chiral gravitino are generated by the standard complex (Dirac) Killing spinors of $dS_{4}$.\footnote{The idea of dropping the reality conditions as an attempt to construct supersymmetric theories on $dS_{4}$ was first mentioned as a speculation by Deser and Waldron in 
Ref.~\cite{Deser_Waldron_ArbitrarySR}.} However, the theory is not associated with a local action functional as the splitting of helicities needed for the theory to become chiral can be achieved only on-shell.

We show that the new supersymmetric theory of the chiral graviton and chiral gravitino avoids the obstacles \cite{Lukierski, Nieuwenhuizen}  to unitary global SUSY on $dS_{4}$ {mentioned above} because the commutator between two SUSY variations  closes on the even algebra $so(4,2) \bigoplus u(1)$. 
Interestingly, unlike in the superconformal {theory} of Ref.~\cite{Anous}, the $so(4,2)$ symmetry in our theory is realised in an unconventional way that does not correspond to standard infinitesimal conformal transformations \cite{Vasiliev, Letsios_conformal-like}.\footnote{Such conformal-like symmetries were first known to exist in the case of strictly massless gauge potentials of any spin on $AdS_{4}$ \cite{Vasiliev}. Recently, conformal-like symmetries - the ones used in the present paper - were found for strictly massless fermionic gauge potentials on $dS_{4}$ \cite{Letsios_conformal-like}. Moreover, it was shown that the strictly massless tensor-spinor mode solutions that form the fermionic discrete series UIRs of $so(4,1)$, also form UIRs of $so(4,2)$. This result is generalised to the case of graviton modes on $dS_{4}$ in the present paper, and plays a central role in the final unitary supersymmetric theory of the chiral graviton and chiral gravitino.}

 Another point worth emphasising is that, although the non-closure of the superalgebra on $so(4,1)$  is a necessary condition for unitarity, it is  \textbf{not} sufficient.  This is demonstrated with the following example. As we discuss in detail, the theory of a standard (i.e.\ non-chiral) complex graviton and a standard Dirac gravitino also carries a representation of the same superalgebra as in the case of the chiral supermultiplet, but the representation is \textbf{non-unitary}, despite the closure of the commutator between two SUSY transformations on  $so(4,2) \bigoplus u(1)$. Interestingly, the unitarity of the theory is achieved by imposing the anti-self-duality constraint on the field strengths which removes all negative-norm states from the Hilbert space, i.e.\ it is the supermultiplet of the chiral graviton and chiral gravitino that carries a unitary representation of SUSY. The appearance of negative-norm states for helicity degrees of freedom that one would expect to be physical according to Minkowskian intuition \cite{Letsios_announce, Letsios_announce_II, Letsios_conformal-like}, and the necessity for the anti-self-duality (or self-duality) constraint on the field strength to remove the negative-norm states, appears already in the quantum theory of the free   Dirac gravitino field on  $dS_{4}$, and we will discuss it in detail. 
 
 The unitarity of the chiral graviton-chiral gravitino supermultiplet is demonstrated in  detail in two different ways:
 \begin{itemize}
     \item \textbf{Unitary SUSY on the space of mode solutions.} We study the action of our superalgebra on the space of standard gravitino~\cite{Letsios_announce, Letsios_announce_II, Letsios_conformal-like} and graviton \cite{HiguchiLinearised} physical mode solutions on global $dS_{4}$, which furnish discrete series UIRs of the dS algebra $so(4,1)$. The  Minkowskian short-distance behaviour of the modes allows us to distinguish between generalised positive-frequency and negative-frequency solutions, as is customary for field theories on global $dS_{4}$ ~\cite{Letsios_announce, Letsios_announce_II, Letsios_conformal-like, HiguchiLinearised}.  We also recall that the  discrete series UIRs {of $so(4,1)$} formed by gravitino modes extend to UIRs of  $so(4,2)$ \cite{Letsios_conformal-like} with the help of the conformal-like transformations \cite{Vasiliev, Letsios_conformal-like}. In addition, we show, for the first time, that the same happens for the graviton modes, i.e.\ the graviton modes furnishing discrete series UIRs of {$so(4,1)$} also furnish $so(4,2)$ UIRs.    Once we clarify how the spaces of fixed-helicity graviton and gravitino modes furnish  UIRs of $so(4,2)(\bigoplus u(1))$, we show that SUSY is represented irreducibly on these spaces. That is, there is a direct sum of two irreducible SUSY representations: a negative-helicity representation with helicities  $(-2, -{3}/{2})$, and a positive-helicity representation with helicities $(+2, +{3}/{2})$. We show that each of these irreducible SUSY representations is a UIR according to the group-theoretic definition of unitarity: we introduce positive-definite scalar products that are invariant under  even generators $\in so(4,2) \bigoplus u(1)$, as well as under SUSY transformations.  
     
     Each of  the two {afore-mentioned} SUSY UIRs can be formed by either positive-frequency or negative-frequency modes. {However, the unitary supersymmetric QFT of the chiral graviton and chiral gravitino, discussed in Section \ref{Sec_SUSY},  includes: a positive-frequency single-particle Hilbert space furnishing only the SUSY UIR  with helicities $(-2,-3/2)$, and a negative-frequency single-particle Hilbert space furnishing only the SUSY UIR with helicities $(+2,+3/2)$. Although allowed at the abstract representation theory level, the SUSY  UIR with helicities $(+2,+3/2)$ is omitted from the positive-frequency sector, and the SUSY UIR with helicities $(-2,-3/2)$ is omitted from the negative-frequency sector. These states, which are removed from the physical state space with the help of the anti-self-duality constraint, have negative norms because of the curious features of the quantum gravitino field on $dS_{4}$. This phenomenon is discussed in detail in Subsection \ref{Subsec_gravitino quantisation}. }

     \item  \textbf{Unitary SUSY on the QFT Fock space.} We  quantise the chiral graviton and chiral gravitino fields by fully fixing the gauge, and then, we construct the quantum operators corresponding to the four complex SUSY Noether charges $Q[\epsilon]= \overline{\eta}^{A} Q_{A} $ - one for each Dirac Killing spinor $\epsilon$ (\ref{dS Killing spinors explicit_ e=S(t,x)n}) of $dS_{4}$. Unitarity is demonstrated by showing that these quantum charges generate the {afore-mentioned} SUSY UIRs by acting on single-particle states: a negative-helicity UIR in the positive-frequency sector, and a positive-helicity UIR in the negative-frequency sector. We also demonstrate the desired positivity of the anti-commutator of spinorial supercharges $\sum^{4}_{A=1}\{ Q_{A}, Q^{A\, \dagger}\}$.
 \end{itemize}
 
%%%%%%%%%%%%%%%%%%%%%%%%%%%%%%%%%%%%%%%%%%%%%%%%%%%%%%%%%%%%5

\subsection{Key ingredients, new results as by-products, and outline}

Before presenting the new unitary supersymmetric theory, we will discuss its key ingredients in detail:   the free graviton and  gravitino fields on global $dS_{4}$, their $so(4,1)$ and $so(4,2)$ representation-theoretic properties, their quantisation, and the properties of the (Dirac)  Killing spinors on $dS_{4}$.  In the process of discussing these ingredients, we will present various new results as by-products which will play a significant role in our {unitary supersymmetric theory. Let us give the outline of the paper with emphasis on the new results that appear as by-products:
\\
\\
$\bullet$ \underline{In Section \ref{sec_back material and conventions}}, we review the basics about the geometry of global $dS_{4}$, and we give  our notation and conventions.
 \\
 \\
  $\bullet$ \underline{In Section \ref{Sec_gravitino}}, we study the  gravitino field on $dS_{4}$. We introduce an alternative local action functional (\ref{gravitino action}) for the Dirac gravitino that is \textbf{hermitian}, unlike the naive conventional Rarita-Schwinger action which is non-hermitian because of the imaginary mass parameter.
    In Subsection \ref{Subsec_gravitino modes so(4,1)}, we review how the gravitino modes with helicities $-3/2$ and $+3/2$ on global $dS_{4}$ form a direct sum of two discrete series UIRs of $so(4,1)$ \cite{Letsios_announce, Letsios_announce_II}.
    Then, in Subsection \ref{Subsec_gravitino modes so(4,2)}, we review how the gravitino modes with helicities $-3/2$ and $+3/2$ on global $dS_{4}$ form a direct sum of two UIRs of the conformal-like algebra $so(4,2)$ \cite{Letsios_conformal-like}.
    In Subsection \ref{Subsec_gravitino conf-like off-shell}, we show, \textit{for the first time}, that the hermitian action (\ref{gravitino action}) is not only dS-invariant but also invariant under conformal-like transformations.
    In Subsection \ref{Subsec_gravitino quantisation}, we study the quantisation of the Dirac gravitino on \textit{global} $dS_{4}$, \textit{for the first time} - a preliminary study of this question was initiated in \cite{Letsios_thesis}. We explain why unitarity requires the quantum gravitino field to be chiral. In particular, we show that the gravitino QFT associated with the hermitian local action functional (\ref{gravitino action}) has a curious feature: half of the propagating helicities have negative norm and the other half have positive norm, as was already suggested by the mode analysis in Refs.~\cite{Letsios_announce_II, Letsios_conformal-like}. We thus introduce the anti-self-duality constraint on the gravitino field strength, rendering the gravitino chiral, and this restricts the theory to its positive-norm sector.\footnote{This comes in contrast with the gravitino in Minkowski and $AdS$ spacetimes, where choosing a chiral gravitino field is optional rather than necessary.} 
    %We show explicitly that the mode expansion of the chiral gravitino  is consistent with the anti-self-duality constraint on the gravitino field strength.  We also show that the anti-self-duality constraint is consistent with the following:  dS-invariance, (conformal-like) $so(4,2)$-invariance, and the microcausality of the theory. Quantum dS charges and quantum conformal-like charges are also constructed. The way in which they generate $so(4,1)$ and $so(4,2)$ UIRs in the QFT Fock space is also clarified, indicating the invariance of the chiral gravitino vacuum under $so(4,2)$.
%%%%%%%%%%%%%%%%%%%%%%%%%%%%%%%%%%%%%%%%%%%%%%%%%%%%%%%%%%%%%%%
\\
\\
$\bullet$ \underline{In Section \ref{Sec_graviton}}, we study the graviton field on $dS_{4}$, with special emphasis on {the} {chiral} {graviton, as this is the superpartner of the chiral gravitino needed for our unitary supersymmetric theory}. 
In Subsection \ref{Subsec_graviton modes so(4,1)}, we review how the standard graviton modes with helicities $-2$ and $+2$ on global $dS_{4}$ form a direct sum of two discrete series UIRs of $so(4,1)$ \cite{HiguchiLinearised}.
In Subsection \ref{Subsec_graviton modes so(4,2)}, we discuss the conformal-like symmetry of the graviton on $dS_{4}$.
In particular, in Subsection \ref{subsubsec_real graviton and so(5,1)}, we present the expressions for the conformal-like transformations of the real graviton field on $dS_{4}$ generated by the five non-Killing conformal Killing vectors. We show that these are symmetries of the field equations. We also show that the symmetry algebra closes on $so(5,1)$ up to  gauge transformations. Interestingly, the conformal-like transformations preserve neither the linearised Einstein-Hilbert action nor the Klein-Gordon inner product.
In Subsection \ref{subsubsec_cmplx graviton and so(4,2)}, we discuss the conformal-like symmetry of the complex graviton (complex strictly massless spin-2 field). Redefining the conformal-like transformations of the real graviton by introducing a factor of $i = \sqrt{-1}$, we show that the symmetry algebra for the complex graviton closes on $so(4,2)$ up to  gauge transformations. The complex graviton  field equations are shown to  be invariant under the conformal-like symmetries. We also show, for the first time, that the hermitian action functional for the complex graviton (\ref{complex graviton action}) is invariant under the conformal-like symmetries, and so is the Klein-Gordon inner product. 
In Subsection \ref{subsubsec_cmplx graviton and so(4,2) ON MODES}, we show, for the first time, that the graviton modes of helicity $-2$ and $+2$ on global $dS_{4}$ furnish a direct sum of two $so(4,2)$ UIRs.
In Subsection \ref{Subsec_graviton quantisation}, we quantise the chiral graviton field.  
\\
\\
$\bullet$ \underline{In Section \ref{Subsec_Killing spinors dS4}}, we review the basics about Dirac Killing spinors, and their bilinears, on $dS_{4}$. We explain how explicit expressions for Killing spinors on $dS_{4}$ can be obtained by analytically continuing Killing spinors on $S^{4}$.  We also explain, for the first time, how the conformal-like $so(4,2)$ symmetry acts on dS Killing spinors.
\\
\\
$\bullet$ \underline{Section \ref{Sec_SUSY}} focuses on our main result: the supersymmetric QFT of the chiral graviton and chiral gravitino on $dS_{4}$ is unitary. This is discussed in Subsection \ref{Subsec_CHIRAL theory SUSY}. 
However, before presenting the unitary theory, in Subsection \ref{Subsec_non-chiral theory SUSY} we begin by discussing the non-chiral supersymmetric theory of a complex graviton and a complex gravitino on $dS_{4}$, each with two complex propagating degrees of freedom. Although we show that this theory is non-unitary, many of its features will be inherited by its unitary chiral counterpart. 
Therefore, in Subsection \ref{Subsec_non-chiral theory SUSY}, we begin by presenting the global SUSY transformations for the non-chiral theory.  We show that the field equations are SUSY-invariant, and so is the hermitian action functional of the theory. Then, we find the Noether charges and currents associated with SUSY invariance. We also calculate the commutator of two SUSY transformations, and we show that the SUSY algebra closes on $so(4,2) \bigoplus u(1)$. {We also} find the SUSY  transformations of the gauge-invariant field strengths. We show that duality transformations commute with SUSY transformations.  Then,  the non-unitarity of the non-chiral theory is discussed.
{Finally}, in Subsection \ref{Subsec_CHIRAL theory SUSY}, we present our unitary supersymmetric theory of the chiral graviton and chiral gravitino, and we clarify which features are inherited from the non-chiral theory of Subsection \ref{Subsec_non-chiral theory SUSY}. The unitarity of our chiral supersymmetric theory is demonstrated explicitly  at the level of mode solutions in Subsection \ref{subsubsec_unitary SUSY modes}. The unitarity of the theory in the supersymmetric QFT Fock space is demonstrated in Subsection \ref{subsubsec_unitary SUSY QFT}.
\\
\\
$\bullet$ \underline{In Section \ref{Discussions}}, we discuss possible future directions.
\\
\\
There are five Appendices. In Appendix \ref{App_Classification_UIRs D=4}, we review the classification of the $so(4,1)$ UIRs.  The rest of the Appendices focus on technical details that have been omitted from the main text.

%%%%%%%%
%%%%%%%%%%%%%%%%
\section{Background material on global dS geometry, notation, and  conventions}\label{sec_back material and conventions}

The solutions of the field equations used in this paper will be expressed in the global slicing of $dS_{4}$. In these coordinates, the line element of $dS_{4}$ is expressed as \cite{Anninos_musings}
\begin{equation} \label{dS_metric}
    ds^{2}=-dt^{2}+\cosh^{2}{t} \,d\Omega^{2}.
\end{equation}
We have denoted the line element of $S^{3}$ as $d\Omega^{2}$, which can be parameterised as
\begin{align}\label{S^3_metric}
    d\Omega^{2} = d\theta^{2}_{3} + \sin^{2}\theta_{3} \left( d\theta_{2}^{2} + \sin^{2}{\theta_{2}} \,d\theta_{1}^{2}  \right),
\end{align}
where $ 0 \leq \theta_{j} \leq \pi$ (for $j=2,3$) and $0 \leq \theta_{1} < 2 \pi$. We will also use the following notation for a point on $S^{3}$: $\bm{\theta_{3}} \equiv (\theta_{3}, \theta_{2},\theta_{1})$.  {The conformal time $\tau$ is defined by $\tan\tau = \sinh t$ ($ - \pi/2<\tau < \pi /2$), and the metric~(\ref{dS_metric}) can also be given as
\begin{equation}
    ds^2 = \sec^2\tau~(-d\tau^2 + d\Omega^2).
\end{equation}
}

The `curved space gamma matrices', $\gamma^{\mu}(x)$, are defined with the use of the vierbein fields as $\gamma^{\mu}(x)= e^{\mu}_{\hspace{3mm} b}(x) \gamma^{b}$, where $\gamma^{b}$ ($b=0,1,2,3$) are the flat-space gamma matrices. The gamma matrices $\gamma^{\mu}(x)$ satisfy the anti-commutation relations
\begin{align}
    \gamma^{\mu} \gamma^{\nu}+\gamma^{\nu} \gamma^{\mu}= 2 g^{\mu \nu }\,\bm{1},
\end{align}
where $\bm{1}$ is the 4-dimensional spinorial identity matrix. The vierbein and co-vierbein fields satisfy 
\begin{align}
    e_{\mu}{\hspace{0.2mm}}^{a} \, e_{\nu}{\hspace{0.2mm}}^{b}\eta_{ab}=g_{\mu \nu}, \hspace{4mm}e^{\mu}{\hspace{0.2mm}}_{a}\,e_{\mu}{\hspace{0.2mm}}^{b}=\delta^{b}_{a},
\end{align}
where $\eta_{ab}=diag(-1,1,1,1)$.
The fifth gamma matrix $\gamma^{5}$ is determined as~\cite{Freedman} 
\begin{align}\label{gamma_abcd-e_abcd gamma5}
  \gamma^{[a} \gamma^{b}\gamma^{c} \gamma^{d]}=- i \varepsilon^{abcd} \gamma^{5}  ,
\end{align}
where $\varepsilon_{\mu \nu \rho \sigma}$ are the components of the $dS_{4}$ volume element. In the vierbein basis, we have $\varepsilon^{0123}=-1$, while in the coordinate basis we have $\varepsilon^{t \theta_{1} \theta_{2}\theta_{3}}=-\frac{1}{\sqrt{-g}}$, where $g$ is the determinant of the dS metric. Equivalently
\begin{align}\label{def_gamma5}
   \gamma^{5}=-i\gamma^{0}\gamma^{1}\gamma^{2}\gamma^{3}.
\end{align}
 The matrix $\gamma^{5}$ anti-commutes with the other four gamma matrices, and, hence, with the Dirac operator on $dS_{4}$.

Our sign convention for the `vierbein postulate' is: 
 \begin{equation}
      \partial_{\mu} e^{\rho}\hspace{0.1mm}_{b} + {\Gamma}^{\rho}_{\mu \sigma}e^{\sigma}\hspace{0.1mm}_{b} - \omega_{\mu}\hspace{0.1mm}^{c}\hspace{0.1mm}_{b}  \,e^{\rho}\hspace{0.1mm}_{c}=0.
  \end{equation}
 The covariant derivative acts on vector-spinors as
  \begin{align}\label{covariant_deriv_vector_spinor}
      \nabla_{\nu} \Psi_{\mu} =& \left(\partial_{\nu}    + \frac{1}{4} \omega_{\nu bc} \gamma^{bc}\right)  \Psi_{\mu}- \,\Gamma^{\lambda}_{\hspace{1mm}\nu \mu} \Psi_{ \lambda} ,
  \end{align}
where $\omega_{\nu b c  }=\omega_{\nu [b c]  } =e_{\nu}{\hspace{0.2mm}}^{a}\omega_{a b c  }$ are the components of the spin connection. 
The gamma matrices are covariantly constant, $\nabla_{\nu}\gamma_{\mu} =0$. This can be easily checked by  computing their covariant derivative as
$$  \nabla_{\nu} \gamma_{\mu} = \partial_{\nu}\gamma_{\mu}    + \frac{1}{4} \omega_{\nu bc} [\gamma^{bc},  \gamma_{\mu}]- \,\Gamma^{\lambda}_{\hspace{1mm}\nu \mu} \gamma_{ \lambda}. $$
Details on the Christoffel symbols, spin connection, and vierbein on global de Sitter, as well as our representation of gamma matrices, can be found in Appendix \ref{Appendix_global dS details}.

De Sitter spacetime has ten Killing vectors, $\xi^{\mu}$,
\begin{align}   \label{Killing equation}
    \nabla_{\mu}  \xi_{\nu}   + \nabla_{\nu}   \xi_{\mu}  =  0,
\end{align}
generating the dS algebra, $so(4,1)$,
and five genuine conformal Killing vectors, $V^{\mu}$, satisfying 
\begin{align}\label{CKV equation}
  \nabla_{\mu}  V_{\nu} + \nabla_{\nu}  V_{\mu}= g_{\mu \nu}\frac{\nabla^{\alpha}    V_{\alpha}}{2}
\end{align}
with $\nabla^{\alpha}    V_{\alpha} \neq 0$. The 15-dimensional Lie algebra generated by the dS Killing vectors and the genuine conformal Killing vectors is isomorphic to the conformal algebra $so(4,2)$. The $so(4,2)$ Lie brackets are given by 
\begin{align} \label{algebra of CKVs}
   & [\xi, \xi'  ]^{\mu} = \pounds_{\xi} \xi^{'\mu},\nonumber \\
    & [\xi, V  ]^{\mu} = \pounds_{\xi} V^{\mu}, \nonumber \\
     & [V, V'  ]^{\mu} = \pounds_{V} V^{'\mu}, 
\end{align}
where $\pounds$ is the Lie derivative, $\xi^{\mu}$ and $\xi^{'\mu}$ are any two Killing vectors, $V^{\mu}$ and $V^{'\mu}$ are any two genuine conformal Killing vectors, $\pounds_{\xi}V^{\mu}$ is a genuine conformal Killing vector, while $\pounds_{V}V^{'\mu}$ is a Killing vector.
Note that each of the five genuine conformal Killing vectors of $dS_{4}$ can be expressed as:
\begin{align}\label{V=nabla phi}
    V_{\mu}= \nabla_{\mu} \phi_{V},
\end{align}
where the scalar function $\phi_{V}$ satisfies\footnote{See, e.g., Refs.~\cite{Allen}.}
{\begin{align}
   &\nabla_{\mu}V_{\nu}=~ \nabla_{\mu} \nabla_{\nu}\phi_{V} =  - g_{\mu \nu}\phi_{V}.\label{properties of phi 1} 
\end{align}}
There are five such independent functions: $\phi_{V^{(0)}}, \phi_{V^{(1)}}, \dots,\phi_{V^{(4)}}$. These functions are related to the embedding space coordinates for $dS_{4}$. Specifically, by embedding $dS_{4}$ as a hyperboloid in 5-dimensional Minkowski space, $$-(X^{0})^{2}+ \sum_{A=1}^{4} (X^{A})^{2}=1,$$ we have $\phi_{V^{(0)}}=X^{0}$, and $\phi_{V^{(A)}} = X^{A}$ (for $A=1, \dots,4$).
\\
\\
\noindent \textbf{\textit{Notation and conventions}}$-$We use the mostly plus metric sign convention for $dS_{4}$. Lowercase Greek tensor indices refer to components with respect to the ‘coordinate basis’. Coordinate basis tensor indices on $S^{3}$ are denoted as $\tilde{\mu}, \tilde{\nu},...$ . Lowercase Latin tensor indices refer to components with respect to the vielbein basis. Repeated indices are summed over. Spinor indices are suppressed, except in the case of spinorial supercharges. We denote the symmetrisation of indices with the use of round brackets, e.g., $A_{(\mu   \nu)} = (A_{\mu \nu} +A_{\nu   \mu})/2$, and the anti-symmetrisation with the use of square brackets, e.g., $A_{[\mu \nu]} = (A_{\mu \nu} -A_{\nu \mu})/2$. Complex conjugation is denoted using the symbol ${*}$ and hermitian conjugation using $\dagger$. Totally anti-symmetrised products of gamma matrices are denoted as: $ \gamma^{bc} = \gamma^{[b}   \gamma^{c]}$ , $ \gamma^{bcd} = \gamma^{[b}  \gamma^{c}  \gamma^{d]}$ and $ \gamma^{abcd} = \gamma^{[a} \gamma^{b}  \gamma^{c}  \gamma^{d]}$. For a tensor (or tensor-spinor with suppressed spinor indices) $B_{\mu_{1} \nu_{1} \dots}$ that is anti-symmetric under the exchange of the tensor indices $ \mu_{1} \leftrightarrow \nu_{1}$, the duality operation is denoted using the `wide tilde' symbol: $\widetilde{ B}_{\mu_{1} \nu_{1} \cdots} = \frac{1}{2} \varepsilon_{\mu_{1} \nu_{1}}^ {~~~~~\alpha \beta}  B_{\alpha \beta \dots}$, with $\widetilde{\widetilde{B}}_{\mu_{1} \nu_{1} \dots} = - B_{\mu_{1}  \nu_{1} \dots}$. {For} quantities that depend on two spacetime points, $x$ and $x'$, primed tensor indices are associated with point $x'$ and unprimed indices with point $x$.  The real graviton gauge potential is denoted as ${h}_{\mu \nu}$. The symbol $h$ does \textbf{not} stand for the trace of $h_{\mu \nu}$ - see, e.g., eq.~(\ref{lin Einstein operator real}).  The complex graviton gauge potential is denoted with the symbol $\frak{h}$, as $\frak{h}_{\mu \nu}$. The symbol $\frak{h}$ in this paper does \textbf{not} stand for the trace of $\frak{h}_{\mu \nu}$ - see, e.g., eq.~(\ref{EOM_ cmplx graviton general}). The superscript `$(\text{TT})$' will be used to indicate that the graviton or gravitino gauge potential is in the transverse-traceless gauge - `TT gauge' for short [see, e.g., eqs.~(\ref{Dirac_eqn_fermion_dS}) and  (\ref{EOM cmplx graviton TT})]. TT graviton mode solutions are denoted as $\varphi_{\mu \nu}$, where labels indicating  particular solutions will be also introduced - see, e.g., 
eq.~(\ref{physmodes_negative_spin_2_dS4}).
%%%%%%%%%%%%%%%%%%%%%%%%%%%%%%%%%%%%%%%%%%%%%%%%%%%%%%%%%%%%%%%%%%%%%%%%%%%%%%%%%%%%%%%%%%

%%%%%%%%%%%%%%%%%%%%%%%%%%%%%%%%%%%%%%%%%%%%%%%%%%%%%%%%%%%%%%%%%%%%%%%%%%%%%%%%%%%%%%%%%

%%%%%%%%%%%%%%%%%%%%%%%%%
\section{Free gravitino gauge potential on \texorpdfstring{$dS_{4}$}{dS4}, UIRs of \texorpdfstring{$so(4,1)$}{so(4,1)} and \texorpdfstring{$so(4,2)$}{so(4,2)}, quantisation and (anti-)self-duality}  \label{Sec_gravitino}

$$   \textbf{Background material for the gravitino on}~dS_{4} $$
Let us start with some useful observations,  some familiar and others less commonly recognised, concerning the massless Rarita-Schwinger (RS) field (gauge potential), also known as gravitino, on a fixed dS spacetime. 

The free gravitino field on $dS_{4}$ is described by a vector-spinor gauge potential satisfying the Rarita-Schwinger (RS) equation with an imaginary mass parameter\footnote{A massive RS field satisfies eq.~(\ref{RS_eqn imag mass}) with $i$ replaced by a real mass parameter $\mathcal{M}$: 
$  \gamma^{\mu \rho \sigma} \left(  \nabla_{\rho}+\frac{\mathcal{M}}{2}\gamma_{\rho} \right) \Psi_{\sigma} = 0. $}~\cite{Deser_Waldron_ArbitrarySR}
\begin{align}\label{RS_eqn imag mass}
   \gamma^{\mu \rho \sigma} \left(  \nabla_{\rho}+\frac{i}{2}\gamma_{\rho} \right) \Psi_{\sigma}=0,\hspace{10mm} 
\end{align}
where~\cite{Freedman}
\begin{align}
    \gamma^{\mu \rho \sigma}= \gamma^{[\mu}   \gamma^{\rho}   \gamma^{\sigma]} = \gamma^{\mu}  \gamma^{\rho}  \gamma^{\sigma} - g^{\mu \rho}\gamma^{\sigma} - g^{\rho \sigma} \gamma^{\mu} + g^{\mu \sigma}  \gamma^{\rho},
\end{align}
and {hence},
\begin{align*}
    \gamma^{\mu}\left( \slashed{\nabla}-i   \right)\gamma^{\beta}  \Psi_{\beta} - \gamma^{\mu} \nabla^{\beta}  \Psi_{\beta} - \nabla^{\mu} \gamma^{\beta}  \Psi_{\beta}+ \left( \slashed{\nabla}+i  \right) \Psi^{\mu} =0.
\end{align*}
Let $\overline{\Psi}_{\mu} = i\Psi^{\dagger}_{\mu}  \gamma^{0}$ be the Dirac conjugate of $\Psi_{\mu}$. The field equation for $\overline{\Psi}_{\mu}$ can be found by taking the hermitian conjugate of eq.~(\ref{RS_eqn imag mass}) as
\begin{align}\label{RS_eqn Dirac conjugate}
  \left(  \nabla_{\rho}\overline{\Psi}_{\sigma}+\frac{i}{2}\overline{\Psi}_{\sigma}\gamma_{\rho} \right)   \gamma^{\mu \rho \sigma}=0,
\end{align}
where we have used $(\gamma^{\mu})^{\dagger} = \gamma^{0}   \gamma^{\mu}   \gamma^{0}$.
The `strict masslessness' of the gravitino manifests itself by the fact that the field equation~(\ref{RS_eqn imag mass}) is invariant under infinitesimal gauge transformations of the form
\begin{align}\label{gauge_transf_spin3/2}
    \delta^{\text{gauge}}({\lambda})\, \Psi_{\mu}= \left( \nabla_{\mu}+\frac{i}{2} \,\gamma_{\mu} \right) \lambda,
\end{align}
where $\lambda$ are spinor gauge functions. Similarly, the equation for $\overline{\Psi}_{\mu}$~(\ref{RS_eqn Dirac conjugate}) is invariant under the gauge transformations 
\begin{align}\label{gauge_transf_spin3/2 conj}
    \delta^{\text{gauge}}(\lambda)\, \overline{\Psi}_{\mu }\equiv  \left(  \delta^{\text{gauge}}(\lambda)\, {\Psi}_{\mu }\right)^{\dagger} i \gamma^{0}  =  \nabla_{\mu}\overline{\lambda}+\frac{i}{2} \,\overline{\lambda}\gamma_{\mu}  ,
\end{align}
where $\overline{\lambda} = i \lambda^{\dagger}  \gamma^{0}$.
For later convenience let us define 
\begin{align}\label{RS_operator}
    \mathcal{R}^{\mu} (\Psi) \equiv \gamma^{\mu \rho \sigma} \left(  \nabla_{\rho}+\frac{i}{2}\gamma_{\rho} \right) \Psi_{\sigma} ,
\end{align}
which will be understood to be non-zero off-shell and zero on-shell. This is gauge invariant off-shell: $\mathcal{R}^{\mu}(  \delta^{\text{gauge}}({\lambda})\, \Psi) = 0  $.

\noindent \textit {\textbf{Problems of the conventional RS action functional and an alternative $-$}} The conventional RS action~\cite{Freedman}
\begin{align} \label{conventional gravitino action}
S_{RS}= -\int d^{4}x \, \sqrt{-g}\, \overline{\Psi}_{\mu}  \, \mathcal{R}^{\mu} = -\int d^{4}x \, \sqrt-{g}\, \overline{\Psi}_{\mu}\gamma^{\mu \rho \sigma} \left(  \nabla_{\rho}+\frac{i}{2}\gamma_{\rho} \right) \Psi_{\sigma}
\end{align}
is not hermitian in de Sitter spacetime because of the imaginary mass term. This non-hermiticity leads to some problematic consequences: although the Euler-Lagrange equation for $\Psi_{\mu}$ derived from the action~(\ref{conventional gravitino action}) is the desired RS equation~(\ref{RS_eqn imag mass}), the Euler-Lagrange equation for $\overline{\Psi}_{\mu}$ is
\begin{align*}
  \left(  \nabla_{\rho}\overline{\Psi}_{\sigma}\bm{-}\frac{i}{2}\overline{\Psi}_{\sigma}\gamma_{\rho} \right)   \gamma^{\mu \rho \sigma}=0,
\end{align*}
which does \textbf{not} correspond to the hermitian conjugate of~(\ref{RS_eqn imag mass}) as it has the wrong sign for the mass term [compare with eq.~(\ref{RS_eqn Dirac conjugate})].
However, the alternative action~\cite{Letsios_thesis}
\begin{align}\label{gravitino action}
    S_{\frac{3}{2}}= -\int d^{4}x \, \sqrt{-g}\, \overline{\Psi}_{\mu} \,\gamma^{5} \, \mathcal{R}^{\mu}(\Psi)
\end{align}
is hermitian and its Euler-Lagrange equations for $\Psi_{\mu}$ and $\overline{\Psi}_{\mu}$ are eqs.~(\ref{RS_eqn imag mass}) and (\ref{RS_eqn Dirac conjugate}), respectively, as consistency requires. Moreover, interestingly, the conventional RS action~(\ref{conventional gravitino action}) is \textbf{not} invariant under the gauge transformation of $\overline{\Psi}_{\mu}$~(\ref{gauge_transf_spin3/2 conj}), but the alternative action~(\ref{gravitino action}) is. 
%%%%%%%%%%%%%%%%%%%%%%%%%%%%%%%%%%%%%%%%%%%%%%%%%%%%%%%%%%%%%%%%%%%%%%%%%%%%%%%%%
\\
\\
\noindent \textit{\textbf{Incompatibility with the Majorana condition $-$ }} It is known that the gravitino equation~(\ref{RS_eqn imag mass}) is not consistent with the Majorana condition because of the imaginary mass parameter (however, a symplectic Majorana condition is possible - see e.g., Ref.~\cite{Nieuwenhuizen}). This can be easily verified by recalling the definition of the charge conjugate of $\Psi_{\mu}$ that preserves the RS equation (\ref{RS_eqn imag mass}):
\begin{align} \label{def: charge conjugation}
    \Psi_{\mu}^{C} \equiv B_{-}^{-1}\Psi^{*}_{\mu},
\end{align}
where the matrix $B_{-}$ satisfies
\begin{align}
    -\left(\gamma^{\mu}   \right)^{*}= B_{-} \gamma^{\mu}B_{-}^{-1} ,
\end{align}
and, in our conventions $B_{-} = \gamma^{0}   \gamma^{2}   \gamma^{3} = i \gamma^{5}  \gamma^{1}$ (defined up to a phase) - see also Appendix \ref{Appendix_global dS details}.
Although $\Psi_{\mu}^{C}$ satisfies the same equation as  $\Psi_{\mu}$, i.e.\  
\begin{align}\label{RS_eqn_charge conj}
   \gamma^{\mu \rho \sigma} \left(  \nabla_{\rho}+\frac{i}{2}\gamma_{\rho} \right) \Psi^{C}_{\sigma}=0, 
\end{align}
it is easy to check that one cannot use the matrix $B_{-}$ to define a consistent reality condition. In particular, by applying charge conjugation twice we find
\begin{align}
    \left(\Psi_{\mu}^{C}\right)^{C} = - \Psi_{\mu}.
\end{align}
Because of this, $\Psi_{\mu}$ cannot be Majorana. In this paper, we are considering only Dirac vector-spinors.\footnote{On the other hand, one can define charge conjugation using the matrix $B_{+} = \gamma^{5}  B_{-}$, where $ \left(\gamma^{\mu}   \right)^{*}= B_{+} \gamma^{\mu}B_{+}^{-1}$. In this case, charge conjugation is defined as: $\Psi^{C_{+}}_{\mu} \equiv B_{+}^{-1} \Psi_{\mu}^{*}$. Although the property $\left(\Psi^{C_{+}}_{\mu} \right)^{C+} = \Psi_{\mu}$ holds and a Majorana condition can be introduced, the charge conjugate, $\Psi^{C_{+}}_{\mu}$, does not preserve the field equation. To be precise, $\Psi^{C_{+}}_{\mu}$ satisfies eq.~(\ref{RS_eqn_charge conj}) with the opposite sign for the mass parameter.}

%%%%%%%%%%%%%%%%%%%%%%%%%%%%%%%%%%%%%%%%%%%%%%%%%%%%%%%%%%%%%5

  \subsection{Discrete Series UIRs of \texorpdfstring{$so(4,1)$}{so(4,1)} in the space of gravitino modes} \label{Subsec_gravitino modes so(4,1)}

Let us review how the gravitino positive frequency mode functions on global $dS_{4}$ form a direct sum of discrete series UIRs of the dS algebra, $so(4,1)$~\cite{Letsios_announce, Letsios_announce_II, Letsios_conformal-like}. As is well known, if the space of (physical) positive frequency mode solutions forms a UIR, it can be identified with the single-particle Hilbert space of the corresponding free {quantum field theory}. Thus, this Subsection sets the stage for the quantisation of the gravitino field, which is carried out in Subsection \ref{Subsec_gravitino quantisation}.

The gravitino mode solutions that form the $so(4,1)$ UIRs are solutions of the RS equation~(\ref{RS_eqn imag mass}) in the transverse-traceless (TT) gauge ($\nabla^{\alpha}\Psi^{(\text{TT})}_{\alpha}= \gamma^{\alpha}\Psi^{(\text{TT})}_{\alpha}  =0$). The field equations and the TT gauge conditions read~\cite{Deser_Waldron_ArbitrarySR, Rahman, Letsios_announce}
\begin{align}
   &\left( \slashed{\nabla}+i\right)\Psi^{(\text{TT})}_{\mu}=0, \nonumber  \\
   & \nabla^{\alpha}\Psi^{(\text{TT})}_{\alpha}=0, \hspace{4mm}  \gamma^{\alpha}\Psi^{(\text{TT})}_{\alpha}=0. \label{Dirac_eqn_fermion_dS}
\end{align}
Only a subset of the initial gauge transformations~(\ref{gauge_transf_spin3/2}) preserve eqs.~(\ref{Dirac_eqn_fermion_dS}). These are the restricted gauge transformations:
\begin{align}\label{gauge_transf_spin3/2_restricted}
    \delta^{\text{gauge}}_{\text{res}}({X})\, \Psi^{(\text{TT})}_{\mu }= \left( \nabla_{\mu}+\frac{i}{2} \,\gamma_{\mu} \right) X,
\end{align}
where the spinor gauge functions satisfy
\begin{align}\label{gauge_transf_spin3/2_restricted_gauge fun}
    \slashed{\nabla}  X= -2i\, X.
\end{align}
The generators of $so(4,1)$ (i.e.\ the Killing vectors of $dS_{4}$) act on vector-spinors $\Psi_{\mu}$ via the Lie-Lorentz derivative \cite{Ortin}
\begin{align}\label{Lie_Lorentz}
 \mathbb{L}_{{\xi}}{\Psi}_{\mu}  =~&  \xi^{\nu} \nabla_{\nu} {\Psi}_{ \mu} +(\nabla_{\mu} \xi^{\nu})\,{\Psi}_{ \nu}+ \frac{1}{4}  (\nabla_{\kappa} \xi_{\lambda})  \gamma^{\kappa \lambda}    {\Psi}_{\mu},
\end{align}
where $\xi^{\mu}$ is any Killing vector of $dS_{4}$. If $\Psi_{\mu}$ is a solution of eq.~(\ref{Dirac_eqn_fermion_dS}), then so is $\mathbb{L}_{\xi} \Psi_{\mu}$. Moreover, the Lie-Lorentz derivative preserves the Lie bracket between any two Killing vectors $\xi^{\mu}$ and $\xi^{' \mu}$~\cite{Ortin}
\begin{align}
    [\mathbb{L}_{\xi },\mathbb{L}_{\xi '}] \Psi_{\mu} = \mathbb{L}_{[\xi , \xi']} \Psi_{\mu}.
\end{align} 
This means that the space of gravitino mode solutions of eq.~(\ref{Dirac_eqn_fermion_dS}) is a representation space for the dS algebra $so(4,1)$. 
\noindent Equations~(\ref{Dirac_eqn_fermion_dS}) admit \textbf{physical} and TT \textbf{pure-gauge} mode solutions. The pure-gauge modes can be identified with zero in the solution space, while the physical modes are the ones forming the direct sum of discrete series UIRs of $so(4,1)$~\cite{Letsios_announce, Letsios_announce_II}. Some details are in order.

%%%%%%%%%%%%%%%%%%
$$ \textbf{TT pure-gauge gravitino modes}$$

 The TT pure-gauge modes are expressed in the form\footnote{We have omitted the quantum number labels from the pure-gauge modes for convenience. Details about these labels can be found in Refs.~\cite{Letsios_announce, Letsios_announce_II, Letsios_conformal-like}.}
\begin{align}\label{pure_gauge_modes_3/2}
     \psi^{(pg)}_{ \mu} =  \nabla_{\mu}X + \frac{i}{2} \gamma_{\mu}  X,
\end{align}
 where 
 \begin{align}
   &\left( \slashed{\nabla}+ 2i \right)X=0, \label{EOM_for_gauge_functions_Dirac}
\end{align}
in agreement with eqs.~(\ref{gauge_transf_spin3/2_restricted}) and (\ref{gauge_transf_spin3/2_restricted_gauge fun}). Explicit expressions for the spinors $X$ can be found in Ref.~\cite{Letsios_announce_II}.
%%%%%%%%%%%%%%%%%
%%%%%%%%%%%%%%%%%%%%%%

  $$\textbf{Physical gravitino modes}$$
 The physical modes come in two helicities: negative ($-3/2$) and positive ($+3/2$) helicity modes~\cite{Letsios_announce, Letsios_announce_II, Letsios_conformal-like}.
In global coordinates~(\ref{dS_metric}), the physical modes with negative and positive helicity are given by~\cite{Letsios_announce, Letsios_announce_II, Letsios_conformal-like}\footnote{In Ref.~\cite{Letsios_conformal-like}, the functions $\alpha_{\ell}(t)$ and $\beta_{\ell}(t)$ are denoted as $\alpha^{(1)}_{\ell}(t)$ and $\beta^{(1)}_{\ell}(t)$, respectively, while in Ref.~\cite{Letsios_announce} they are denoted as $\varPhi^{(-1)}_{M \ell}(t)$ and $\varPsi^{(-1)}_{M \ell}(t)$ (with $M=i$), respectively. In \cite{Letsios_announce, Letsios_conformal-like}, these functions are expressed in terms of the Gauss hypergeometric functions. However, in the present paper, as the hypergeometric series terminates, we have chosen to express the functions in a simpler form in our eqs.~(\ref{a(t)_spin-3/2}) and (\ref{b(t)_spin-3/2}).}
\begin{equation}\label{physmodes_negative_spin_3/2_dS4}
  {\psi}^{(phys ,\,- \ell; \,m;k)}_{t}(t,\bm{\theta}_{3})= 0,\hspace{5mm}  {\psi}^{(phys ,\,- \ell; \,m;k)}_{\tilde{\mu}}(t,\bm{\theta}_{3})=\left(\frac{\ell+2}{2(\ell+1)} \right)^{1/2} \begin{pmatrix}  \alpha_{\ell}(t) \, \tilde{\psi}_{-\tilde{\mu}}^{(\ell; m;k)} (\bm{\theta_{3}})  \\ - i  \beta_{\ell}(t) \, \tilde{\psi}^{(\ell; m;k)}_{-\tilde{\mu}} (\bm{\theta_{3}}
    )  \end{pmatrix},
\end{equation}
and
\begin{equation}\label{physmodes_positive_spin_3/2_dS4}
  {\psi}^{(phys ,\,+ \ell; \,m;k)}_{t}(t,\bm{\theta}_{3})= 0,\hspace{5mm}  {\psi}^{(phys ,\,+ \ell; \,m;k)}_{\tilde{\mu}}(t,\bm{\theta}_{3})= \left(\frac{\ell+2}{2(\ell+1)} \right)^{1/2}\begin{pmatrix} i \beta_{\ell}(t) \, \tilde{\psi}_{+\tilde{\mu}}^{(\ell; m;k)} (\bm{\theta_{3}})  \\ -   \alpha_{\ell}(t) \, \tilde{\psi}^{(\ell; m;k)}_{+\tilde{\mu}} (\bm{\theta_{3}}
    )  \end{pmatrix},
\end{equation}
respectively, where $\tilde{\mu}$ is a vector index on $S^{3}$, while $\ell, m$ and $k$ are angular momentum quantum numbers corresponding to the chain of subalgebras $so(4) \supset$ $so(3)  \supset$ $so(2)$ with $\ell \in \{1,2,...   \}$, $m \in \{1,2,...,\ell\}$ and $k \in \{ -m-1,-m,...,0,...,m   \}$. 
The functions describing the time dependence are conveniently expressed in terms of the variable
\begin{align}\label{definition of x(t)}
    x(t)= \frac{\pi}{2} - it
\end{align}
as
\begin{align}
    &\alpha_{\ell}(t)= \left( \sin{\frac{x(t)}{2}}  \right)^{\ell+1}\, \left( \cos{\frac{x(t)}{2}}  \right)^{-\ell-2}\, \left( 1 - \frac{\sin^{2}\frac{x(t)}{2}}{\ell+2}   \right), \label{a(t)_spin-3/2} \\
   &\beta_{\ell}(t)=\frac{1}{\ell+2}\, \left( \sin{\frac{x(t)}{2}}  \right)^{\ell+2}\, \left( \cos{\frac{x(t)}{2}}  \right)^{-\ell-1}  \label{b(t)_spin-3/2},
\end{align}
 where
 \begin{align}
   &\cos{\frac{x(t)}{2}}=\left(  \sin{\frac{x(t)}{2}} \right)^{*}=\frac{\sqrt{2}}{2}\left(\cosh{\frac{t}{2}} + i \sinh{\frac{t}{2}} \right)  ,  \\
   &\sin^{2}\frac{x(t)}{2}=\frac{1- i \sinh{t}}{2}.
       \end{align}
%%%%%%%%%%%%%%%%%%%%%%%%%%%%%%%%%%%%%%%%%%%%%%%%%%%%%%%%%%%%%%%%%%%%%%%%%%%%%%%%%%%%%5
We note that
\begin{align}\label{relating db(t)/dx and a(t)}
   \frac{\partial}{\partial x}  \beta_{\ell}(t) =   i \frac{\partial}{\partial t}  \beta_{\ell}(t)  = \frac{1}{2}   \alpha_{\ell}(t).
\end{align}
{It is also useful to note that with the conformal time $\tau$ defined by $\tan\tau = \sinh t$ we have
\begin{align}
    \cos\frac{x(t)}{2} & = \frac{e^{i\tau/2}}{\sqrt{2\cos\tau}},\\
    \sin\frac{x(t)}{2} & = \frac{e^{-i\tau/2}}{\sqrt{2\cos\tau}},
\end{align}
so that
\begin{align}
    \alpha_\ell(t) & = \frac{e^{-i(\ell+\frac{3}{2})\tau}}{\sqrt{2\cos\tau}}\left(2\cos\tau - \frac{e^{-i\tau}}{\ell+2}\right),\\
    \beta_\ell(t) & =  \frac{e^{-i(\ell+\frac{3}{2})\tau}}{(\ell+2)\sqrt{2\cos\tau}}.
\end{align}
}

%%%%%%%%%%%%%%5
%%%%%%%%%%%%%%%
\noindent \textbf{\textit{Transverse-traceless vector-spinor spherical harmonics on $S^{3}$}$-$} The $\bm{\theta_{3}}$-dependence of the physical modes in eqs.~(\ref{physmodes_negative_spin_3/2_dS4}) and (\ref{physmodes_positive_spin_3/2_dS4}) is given by the transverse-traceless vector-spinor spherical harmonics on $S^{3}$, $\tilde{\psi}_{\pm \tilde{\mu}}^{(\ell; m;k)} (\bm{\theta_{3}})$. These satisfy~\cite{Homma, Letsios_announce_II, Chen}
\begin{align}
   & \tilde{\slashed{\nabla}}\tilde{\psi}^{(\ell; {m};k)}_{\pm \tilde{\mu}}(\bm{\theta_{3}})= \pm i \left(\ell+\frac{3}{2}\right)  \tilde{\psi}^{(\ell; {m};k)}_{\pm \tilde{\mu}}(\bm{\theta_{3}}),\hspace{5mm}\ell \in \{1,2,...\}\nonumber \\
   &\tilde{\gamma}^{\tilde{\mu}}\tilde{\psi}^{(\ell; {m};k)}_{\pm \tilde{\mu} }(\bm{\theta_{3}})=\tilde{\nabla}^{\tilde{\mu}}\tilde{\psi}^{(\ell; {m};k)}_{\pm \tilde{\mu}}(\bm{\theta_{3}})=0 \label{vector-spinor+-eigen_S3},
\end{align}
where the tildes have been used to denote quantities on $S^{3}$. They are normalised with the standard inner product on $S^{3}$ \cite{Letsios_announce_II}:
\begin{align}\label{normlzn_S3}
    \int_{S^{3}} \sqrt{\tilde{g}}&\,d\bm{\theta_{3}} ~\tilde{g}^{\tilde{\mu}    \tilde{\nu}}~\tilde{\psi}^{(\ell'; {m}'; k')}_{\sigma '\,\tilde{\mu}}(\bm{\theta_{3}})^{\dagger}
    ~\tilde{\psi}^{(\ell; {m};k)}_{\sigma \,\tilde{\nu}}(\bm{\theta_{3}}) \nonumber\\
    &= \delta_{\sigma \sigma'}\, \delta_{\ell \ell'} \,\delta_{{m}\,   {m}'} \delta_{k k'},
\end{align}
where $\sigma , \sigma' \in \{ +, -  \}$ and $d \bm{\theta_{3}} \equiv d{\theta_{3}}  d\theta_{2}   d\theta_{1}$. For each value of $\ell \in \{1,2,...   \}$, the set $\{ \tilde{\psi}^{(\ell; {m};k)}_{+ \tilde{\mu}} \}$ forms a $so(4)$ representation with highest weight given by~\cite{Homma}:
\begin{align}\label{so(4) +weight TT spin-3/2}
   \vec{f}^{\,(+3/2)}_{\ell} = \left( \ell + \frac{1}{2}, \frac{3}{2} \right).
\end{align}
The set $\{ \tilde{\psi}^{(\ell; {m};k)}_{- \tilde{\mu}} \}$ forms a $so(4)$ representation with highest weight given by~\cite{Homma}:
\begin{align}\label{so(4) -weight TT spin-3/2}
   \vec{f}^{\,(-3/2)}_{\ell} = \left( \ell + \frac{1}{2},- \frac{3}{2} \right).
\end{align}
Let $\tilde{\varepsilon}_{\tilde{\mu} \tilde{\nu} \tilde{\alpha}}$ denote the invariant 3-form on $S^{3}$ with $\tilde{\varepsilon}_{\theta_{1}  \theta_{2}   \theta_{3}}= \sqrt{\tilde{g}}$, where $\tilde{g}$ is the determinant of the $S^{3}$ metric (\ref{S^3_metric}). Let us also introduce the duality operator (helicity operator)  acting on the vector-spinor spherical harmonics (\ref{vector-spinor+-eigen_S3}) as \cite{HiguchiLinearised}:
\begin{align}\label{duality operator vec-spinor S^3}
\frac{1}{\ell + 3/2}\,\tilde{\varepsilon}_{\tilde{\mu} \tilde{\nu} \tilde{\alpha}} \tilde{\nabla}^{\tilde{\nu}}\tilde{\psi}^{(\ell; {m};k)\tilde{\alpha}}_{\pm}.
\end{align}
This is the analogue of the flat-space helicity operator.
Using $\tilde{\varepsilon}_{\tilde{\mu} \tilde{\nu} \tilde{\alpha}}= -i \tilde{\gamma}_{[\tilde{\mu}} \tilde{\gamma}_{\tilde{\nu}} \tilde{\gamma}_{\tilde{\alpha}]}$ \cite{Freedman}, we find that the duality operator is proportional to the Dirac operator on $S^{3}$, as:
\begin{align}
\frac{1}{\ell + 3/2}\tilde{\varepsilon}_{\tilde{\mu} \tilde{\nu} \tilde{\alpha}} \tilde{\nabla}^{\tilde{\nu}}\tilde{\psi}^{(\ell; {m};k)\tilde{\alpha}}_{\pm} = -\frac{i}{\ell +3/2} \, \tilde{\slashed{\nabla}}\tilde{\psi}^{(\ell; {m};k)}_{\pm \tilde{\mu}} =  \pm \tilde{\psi}^{(\ell; {m};k)}_{\pm \tilde{\mu}}.
\end{align}
Thus, the modes $\tilde{\psi}^{(\ell; {m};k)}_{+ \tilde{\mu}}$ are self-dual, while the modes $\tilde{\psi}^{(\ell; {m};k)}_{- \tilde{\mu}}$ are anti-self-dual. This notion of (anti-)self-duality should not be confused with the notion of (anti-)self-duality defined using $\varepsilon_{\mu \nu \rho \sigma}$ on $dS_{4}$ - see 
e.g., eqs.~(\ref{anti-self constr gravitino}) and (\ref{anti-self constr graviton}).
%%%%%%%%%%%%%%%%%%%%%%%%%%%%%%%%%%%%%%%%%%%%%%%%%%%%%%%%%%%%%%%%%%%%%%%%%%%%%%
\\
\\
\noindent \textbf{\textit{Positive and negative frequency}$-$}The mode functions~(\ref{physmodes_negative_spin_3/2_dS4}) and (\ref{physmodes_positive_spin_3/2_dS4}) are the analogues of positive frequency modes, as for short wavelengths, $ \ell \gg 1$, they satisfy \cite{Letsios_conformal-like}
\begin{align}
    \frac{\partial}{\partial t} {\psi}^{(phys ,\,\pm \ell; \,m;k)}_{\mu}(t,\bm{\theta}_{3}) \sim - i \frac{\ell}{\cosh{t}} {\psi}^{(phys ,\,\pm \ell; \,m;k)}_{\mu}(t,\bm{\theta}_{3}).
\end{align}
Eq.~(\ref{Dirac_eqn_fermion_dS}) also admits physical transverse-traceless solutions that are the analogues of negative frequency modes given by \cite{Letsios_conformal-like}
\begin{equation}\label{negfreq_physmodes_negative_spin_3/2_dS4}
  {v}^{(phys ,\,- \ell; \,m;k)}_{t}(t,\bm{\theta}_{3})= 0,\hspace{5mm}  {v}^{(phys ,\,- \ell; \,m;k)}_{\tilde{\mu}}(t,\bm{\theta}_{3})=\left(\frac{\ell+2}{2(\ell+1)} \right)^{1/2} \begin{pmatrix}  i \beta^{*}_{\ell}(t) \, \tilde{\psi}_{-\tilde{\mu}}^{(\ell; m;k)} (\bm{\theta_{3}})  \\    \alpha^{*}_{\ell}(t) \, \tilde{\psi}^{(\ell; m;k)}_{-\tilde{\mu}} (\bm{\theta_{3}}
    )  \end{pmatrix},
\end{equation}
and
\begin{equation}\label{negfreq_physmodes_positive_spin_3/2_dS4}
  {v}^{(phys ,\,+ \ell; \,m;k)}_{t}(t,\bm{\theta}_{3})= 0,\hspace{5mm}  {v}^{(phys ,\,+ \ell; \,m;k)}_{\tilde{\mu}}(t,\bm{\theta}_{3})= \left(\frac{\ell+2}{2(\ell+1)} \right)^{1/2}\begin{pmatrix}  \alpha^{*}_{\ell}(t) \, \tilde{\psi}_{+\tilde{\mu}}^{(\ell; m;k)} (\bm{\theta_{3}})  \\  i \beta^{*}_{\ell}(t) \, \tilde{\psi}^{(\ell; m;k)}_{+\tilde{\mu}} (\bm{\theta_{3}})  \end{pmatrix}.
\end{equation}
The negative frequency modes can be obtained by applying charge conjugation (\ref{def: charge conjugation}) to the positive frequency modes. For short wavelengths, $\ell \gg 1$, they satisfy the generalised negative frequency condition
\begin{align}
    \frac{\partial}{\partial t} {v}^{(phys ,\,\pm \ell; \,m;k)}_{\mu}(t,\bm{\theta}_{3}) \sim + i \frac{\ell}{\cosh{t}} {v}^{(phys ,\,\pm \ell; \,m;k)}_{\mu}(t,\bm{\theta}_{3}).
\end{align}

\noindent \textbf{Note.} The field strength (\ref{def:gravitino_field-strength}) calculated for the positive frequency  modes of helicity $-3/2$, $ {\psi}^{(phys ,\,- \ell; \,m;k)}_{\mu}$, is anti-self-dual, and so is the field strength  for the negative frequency modes of helicity $+3/2$, ${v}^{(phys ,\,+ \ell; \,m;k)}_{\mu }$. Similarly, the field strength (\ref{def:gravitino_field-strength}) calculated for the positive frequency  modes of helicity $+3/2$, $ {\psi}^{(phys ,\,+ \ell; \,m;k)}_{\mu}$, is self-dual, and so is the field strength  for the negative frequency modes of helicity $-3/2$, ${v}^{(phys ,\,- \ell; \,m;k)}_{\mu}$. The mode expansion of the field strength and (anti)-self-duality are discussed further in Subsection \ref{Subsec_gravitino quantisation}.

$$ \textbf{Discrete Series UIRs of}~so(4,1)$$
The two sets of (positive frequency) physical modes $\{\psi^{(phys ,\,-\ell; \,m;k)}_{\mu} \}$ and $\{\psi^{(phys ,\,+\ell; \,m;k)}_{\mu} \}$ separately form two irreducible representations of $so(4,1)$ (and, thus, a different choice for a scalar product is allowed for each set)~\cite{Letsios_announce, Letsios_announce_II, Letsios_conformal-like}. This can be understood as follows. First, it is clear that the modes $\{\psi^{(phys ,\,+\ell; \,m;k)}_{\mu} \}$ do not mix with the modes $\{\psi^{(phys ,\,-\ell; \,m;k)}_{\mu} \}$ under any $so(4)$ transformation as they belong to different $so(4)$ representations - the former correspond to the $so(4)$ highest weights in (\ref{so(4) +weight TT spin-3/2}), while the latter to the ones in (\ref{so(4) -weight TT spin-3/2}). Moreover, under the infinitesimal isometry generated by the boost Killing vector
\begin{align}\label{dS boost}
   B = B^{\mu} \partial_{\mu} = \cos{\theta_{3}}\, \frac{\partial}{\partial t} - \tanh{t} \sin{\theta_{3}}\, \frac{\partial}{\partial {\theta_{3}}},
\end{align}
physical modes of a given helicity transform only among themselves. To be specific, they transform as~\cite{Letsios_announce_II, Letsios_conformal-like}:
\begin{align}\label{infinitesimal dS of spin-3/2 physical modes}
     \mathbb{L}_{B}{\psi}^{\left(phys ,\,\pm \ell; m; k\right)}_{{\mu}} =&-\frac{i}{2}\,\sqrt{(\ell-m+1)(\ell+m+3)}\,{\psi}^{\left(phys ,\,\pm (\ell+1)\,;m;k \right)}_{{\mu}} \nonumber\\
     &-  \frac{i}{2}\sqrt{(\ell-m)(\ell+m+2)}\,{\psi}^{\left(phys ,\,\pm (\ell-1)\,;{m;k}\right)}_{{\mu}}+(\text{pure-gauge}),
     \end{align}
     where the term `(pure-gauge)' is a TT pure-gauge mode (\ref{pure_gauge_modes_3/2}).
As the $so(4,1)$ algebra can be generated using only the $so(4)$ generators and just one dS boost, we conclude that the modes $\{\psi^{(phys ,\,-\ell; \,m;k)}_{\mu} \}$ and $\{\psi^{(phys ,\,+\ell; \,m;k)}_{\mu} \}$ separately form irreducible representations of $so(4,1)$ with the equivalence relation: if for any two physical modes, $\psi^{(1)}_{\mu}$ and $\psi^{(2)}_{\mu}$, the difference $\psi^{(1)}_{\mu} -\psi^{(2)}_{\mu}$ is a linear combination of TT pure-gauge modes, then $\psi^{(1)}_{\mu}$ and $\psi^{(2)}_{\mu}$ belong to the same equivalence class. This equivalence relation is introduced because the pure-gauge modes can be identified with zero, as will become clear shortly.  Note that eq.~(\ref{infinitesimal dS of spin-3/2 physical modes}) agrees with the expression for the infinitesimal boost matrix elements in the discrete series UIRs of $so(4,1)$ with $\Delta = 5/2$ and $s=3/2$ \cite{Schwarz, Ottoson} {in the `modern notation' for labels} -  see Appendix \ref{App_Classification_UIRs D=4} and Refs.~\cite{Yale_Thesis, Letsios_announce} for the translation between the old and modern notation for the labels of the UIRs.

%%%%%%%%%%%%%%%%%%%%%%%%

%%%%%%%%%%%%%%%%%%%%%%%%%%%%%%%%%%%%%%%%%%%%%%%%%%%%%%%%%%%%%%%%%%%%%%%%%%%%%%%
The unitarity of the {afore-mentioned} irreducible representations formed by $\{\psi^{(phys ,\,-\ell; \,m;k)}_{\mu} \}$ and $\{\psi^{(phys ,\,+\ell; \,m;k)}_{\mu} \}$ can be demonstrated as follows~\cite{Letsios_announce, Letsios_announce_II, Letsios_conformal-like}.  Let $\braket{ \psi^{(1)}| \psi^{(2)}}_{ax}$ be the following dS invariant\footnote{By `dS-invariant scalar product'  we mean that the $so(4,1)$ generators/Lie derivatives are realised as anti-hermitian operators with respect to the scalar product under consideration~\cite{STSHS}.} and time-independent scalar product~\cite{Letsios_announce, Letsios_announce_II, Letsios_conformal-like}:
\begin{align}\label{axial_scalar prod}
 \braket{ \psi^{(1)}| \psi^{(2)}}_{ax}   &= \int_{S^{3}} \sqrt{-{g}} \,d\bm{\theta_{3}}\,g^{\mu \nu}\,{\psi}^{(1)}_{\mu}(t,\bm{\theta_{3}})^{\dagger} \,\gamma^{5}\,  \psi^{(2)}_{\nu}(t,\bm{\theta_{3}}),
\end{align}
where $\psi^{(1)}_{\mu}$ and $\psi^{(2)}_{\nu}$ are any two solutions of the field equations in the TT gauge~(\ref{Dirac_eqn_fermion_dS}). The scalar product (\ref{axial_scalar prod}) is the time-independent Noether charge associated with the axial current\footnote{The axial current (\ref{axial current}) is covariantly conserved because of the imaginary mass parameter of the gravitino \cite{Letsios_announce_II, Letsios_conformal-like}. It is easy to check that in the case of a real-mass spin-3/2 field the axial current is not conserved.}  \cite{Letsios_announce_II, Letsios_conformal-like}
\begin{align}\label{axial current}
    J_{ax}^{\mu}\left(\psi^{(1)},  \psi^{(2)}   \right) = i~ \overline{\psi^{(1)}}_{\nu} \gamma^{\mu} \gamma^{5} \psi^{(2) \nu},~~~ \nabla_{\mu} J_{ax}^{\mu}\left(\psi^{(1)},  \psi^{(2)}   \right) =0,
\end{align}
specifically,
\begin{align}\label{axial scalar product in terms of current}
  \braket{ \psi^{(1)}| \psi^{(2)}}_{ax}   = \int_{S^{3}} \sqrt{-{g}} \,d\bm{\theta_{3}}~   J_{ax}^{t}\left(\psi^{(1)},  \psi^{(2)}   \right).
\end{align}
The physical modes~(\ref{physmodes_negative_spin_3/2_dS4}), (\ref{physmodes_positive_spin_3/2_dS4}), (\ref{negfreq_physmodes_negative_spin_3/2_dS4}) and (\ref{negfreq_physmodes_positive_spin_3/2_dS4}) are normalised as:
\begin{align}\label{norms of physical modes_3/2}
    \braket{\psi^{(phys ,\,\sigma \ell; \,m;k)}|\psi^{(phys ,\,\sigma' \ell'; \,m';k')}}_{ax} = (-\sigma)\times \delta_{\sigma \sigma'}\delta_{\ell \ell'}   \delta_{mm'} \delta_{kk'},
\end{align}
\begin{align}\label{norms of physical modes_3/2_neg freq}
   & \braket{v^{(phys ,\,\sigma \ell; \,m;k)}|v^{(phys ,\,\sigma' \ell'; \,m';k')}}_{ax} = (+\sigma)\times \delta_{\sigma \sigma'}\delta_{\ell \ell'}   \delta_{mm'} \delta_{kk'} \nonumber \\
 &   \braket{v^{(phys ,\,\sigma \ell; \,m;k)}|\psi^{(phys ,\,\sigma' \ell'; \,m';k')}}_{ax} = 0,
\end{align}
with $\sigma, \sigma' \in \{ +, - \}$. 
Also,
\begin{align}
    \braket{\psi^{(1)}|\psi^{(pg)}}_{ax} =0,
\end{align}
where  $\psi^{(1)}_{\mu}$ is any physical or TT pure-gauge mode, and thus, the pure-gauge modes can be identified with zero. It is interesting that, with respect to the scalar product~(\ref{axial_scalar prod}), there is indefiniteness of the norm among the positive frequency physical modes, as well as among the negative frequency physical modes~\cite{Letsios_announce_II, Letsios_conformal-like}. Moreover, we observe that the sign of the norm depends on the helicity $\sigma \in \{ +,- \}$ [see eqs.~(\ref{norms of physical modes_3/2}) and (\ref{norms of physical modes_3/2_neg freq})]. 
 Unitarity requires a positive-definite inner product that is invariant under dS transformations, i.e.\ the dS generators (Lie-Lorentz derivatives) are realised as anti-hermitian operators. Indeed the scalar product~(\ref{axial_scalar prod}) is dS invariant as, for any dS Killing vector $\xi^{\mu}$, we have~\cite{Letsios_announce_II, Letsios_conformal-like}
\begin{align} \label{anti-herm_Lie deriv_gravitino}
    \braket{\mathbb{L}_{\xi}\psi^{(1)}|\psi^{(2)}}_{ax} + \braket{\psi^{(1)}|\mathbb{L}_{\xi}\psi^{(2)}}_{ax}=0.
\end{align}
 As we mentioned earlier, according to eq.~(\ref{norms of physical modes_3/2}) the scalar product~(\ref{axial_scalar prod}) is positive definite for the physical modes $\{\psi^{(phys ,\,-\ell; \,m;k)}_{\mu}\}$ and negative definite for the physical modes $\{\psi^{(phys ,\,+\ell; \,m;k)}_{\mu}\}$.\footnote{The conventional inner product, $\int_{S^{3}} \sqrt{-{g}} \,d\bm{\theta_{3}}\,g^{\mu \nu}\,{\psi}^{(1)\dagger}_{\mu}(t,\bm{\theta_{3}}) \,  \psi^{(2)}_{\nu}(t,\bm{\theta_{3}}),$
despite its positive definiteness, is neither dS invariant nor time-independent. Therefore, it is not a `good' choice for a representation-theoretic analysis~\cite{Letsios_announce, Letsios_announce_II, Letsios_conformal-like}.} As the two sets do not mix with each other under dS transformations, we conclude:
\begin{itemize}
    \item The positive frequency physical gravitino modes with positive helicity, $\{ {\psi}^{\left(phys ,\,+ \ell; m; k\right)}_{\mu} \}$, form the discrete series UIR $D^{+}(\Delta ,s) =D^{+}(5/2 , {3}/{2})$ of $so(4,1)$ - see Appendix \ref{App_Classification_UIRs D=4}. The $so(4)$ content corresponds to the $so(4)$ highest weights~(\ref{so(4) +weight TT spin-3/2}). The $so(4,1)$-invariant inner product that is positive definite is given by the negative of eq.~(\ref{axial_scalar prod}).

    \item The positive frequency physical gravitino modes with negative helicity, $\{ {\psi}^{\left(phys ,\,- \ell; m; k\right)}_{\mu} \}$, form the discrete series UIR $D^{-}(\Delta ,s) =D^{-}(5/2 , {3}/{2})$ of $so(4,1)$ - see Appendix \ref{App_Classification_UIRs D=4}. The $so(4)$ content corresponds to the $so(4)$ highest weights~(\ref{so(4) -weight TT spin-3/2}). The $so(4,1)$-invariant inner product that is positive definite is given by eq.~(\ref{axial_scalar prod}).
\end{itemize}
Thus, the two sets of positive frequency modes, $\{ {\psi}^{\left(phys ,\,+ \ell; m; k\right)}_{\mu} \}$ and $\{ {\psi}^{\left(phys ,\,- \ell; m; k\right)}_{\mu} \}$,  with the {afore-mentioned} choice of positive-definite scalar products, form the direct sum $D^{+}(5/2 , {3}/{2})$ $ \bigoplus D^{-}(5/2 , {3}/{2})$. The negative frequency modes, $\{ {v}^{\left(phys ,\,+ \ell; m; k\right)}_{\mu} \}$ 
[eq.~(\ref{negfreq_physmodes_positive_spin_3/2_dS4})] and $\{ {v}^{\left(phys ,\,- \ell; m; k\right)}_{\mu} \}$ [eq.~(\ref{negfreq_physmodes_negative_spin_3/2_dS4})], form the same direct sum of UIRs. The transformation $\mathbb{L}_{B}{v}^{\left(phys ,\,\pm \ell; m; k\right)}_{\mu}$ is found from (\ref{infinitesimal dS of spin-3/2 physical modes}) by replacing ${\psi}^{\left(phys ,\,\pm (\ell \pm 1); m; k\right)}_{\mu}$ with ${v}^{\left(phys ,\,\pm (\ell \pm 1); m; k\right)}_{\mu}$, while the coefficients in the linear combination on the right-hand side must be replaced by the complex conjugates of the ones in (\ref{infinitesimal dS of spin-3/2 physical modes}). 

%\noindent \textbf{Note:} that the $so(4,1)$ UIRs corresponding to the gravitino, $D^{\sigma}(5/2 , {3}/{2})$ ($\sigma = \pm$), can also be realised in the space of anti-symmetric rank-2 tensor-spinor mode functions, $ \{  f^{(\sigma \ell;m;k)}_{\mu \nu}\}$, corresponding to the gravitino field strength - see (\ref{mode expansion 3/2 field strength}). Unlike the case of the gravitino gauge-potential modes, we expect that a dS-invariant inner product exists that is positive definite for both helicities for the field-strength modes $ \{  f^{(\sigma \ell;m;k)}_{\mu \nu}\}$. This expectation is justified by observing that these modes satisfy a Dirac equation with a zero mass parameter (not imaginary), $\slashed{\nabla}f^{(\sigma \ell;m;k)}_{\mu \nu} =0 $, and as a consequence, the following inner product
%$$ \int_{S^{3}} \sqrt{-g} ~d \bm{\theta_{3}}~ f^{(\sigma \ell;m;k)\dagger}_{\mu \nu}f^{(\sigma' \ell';m';k')\mu \nu}$$
%is dS-invariant and possibly positive definite for both helicities $\sigma = \pm$. This will be studied further in future work.

%%%%%%%%%%%%%%%%%%%%%%%%%%%%%%%%%%%%%%%%%%%%%%%%%%%%%%%%%%%%%%%%%%%%%%%%%%%%%%%%%

 \subsection{Conformal-like symmetry and UIRs of \texorpdfstring{$so(4,2)$}{so(4,2)}} \label{Subsec_gravitino modes so(4,2)}
It was recently found that the two sets of mode functions, $\{ {\psi}^{\left(phys ,\,+ \ell; m; k\right)}_{\mu} \}$ and $\{ {\psi}^{\left(phys ,\,- \ell; m; k\right)}_{\mu} \}$, form not only a direct sum of $so(4,1)$ UIRs but also a direct sum of $so(4,2)$ UIRs \cite{Letsios_conformal-like}. Let us review the basic findings of \cite{Letsios_conformal-like}, as these will be useful in our discussions on SUSY later on.

%%%%%%%%%%%%%%
$$ \textbf{Conformal-like symmetries of the field equations}$$
The $so(4,2)$ symmetry that preserves the solution space of 
eq.~(\ref{Dirac_eqn_fermion_dS}) is generated by the ten familiar infinitesimal dS transformations (\ref{Lie_Lorentz}) [generating the dS subalgebra of $so(4,2)$], as well as by five infinitesimal conformal-like transformations \cite{Letsios_conformal-like}:
\begin{align}\label{conf-like gravitino TT}
    \mathbb{T}_{V}\Psi_{\mu} 
    \equiv & ~\gamma^{5}\Big( V^{\rho}\nabla_{\rho}\Psi_{\mu}+ i\,  V^{\rho}\gamma_{\rho} \Psi_{\mu}  -i\, V^{\rho}\gamma_{\mu} \Psi_{\rho} - \frac{3}{2} \phi_{V} \, \Psi_{\mu}\Big ) \nonumber\\ 
   & -\frac{2}{3} \left( \nabla_{\mu} +\frac{i}{2}\gamma_{\mu}\right) \gamma^{5}\Psi_{\rho}V^{\rho},
\end{align}
 where $V^{\mu}$ is any genuine conformal Killing vector (\ref{V=nabla phi}). If $\Psi_{\mu}$ is a solution of (\ref{Dirac_eqn_fermion_dS}), i.e.\ $\Psi_{\mu} = \Psi^{(\text{TT})}_{\mu}$, then so is $\mathbb{T}_{V} \Psi^{(\text{TT})}_{\mu}$.\footnote{Note that the field-dependent gauge transformation on the second line of (\ref{conf-like gravitino TT}) is \textbf{not} a restricted gauge transformation (\ref{gauge_transf_spin3/2_restricted}), but it still is an off-shell gauge transformation (\ref{gauge_transf_spin3/2}). It is needed to preserve the TT gauge conditions \cite{Letsios_conformal-like}.}

 \noindent \textbf{Note:}   The conformal-like symmetry transformation (\ref{conf-like gravitino TT}) is also a symmetry of the non-gauge-fixed RS equation (\ref{RS_eqn imag mass}) \cite{Letsios_conformal-like}. In this case, the last term in eq.~(\ref{conf-like gravitino TT}) can be omitted as it corresponds to an off-shell gauge transformation (\ref{gauge_transf_spin3/2}) that leaves the RS equation (\ref{RS_eqn imag mass}) invariant. However, this gauge transformation cannot be omitted when working in the TT gauge, as it ensures that if $\Psi_{\mu}$ is in the TT gauge, then so is $\mathbb{T}_{V}\Psi_{\mu}$ \cite{Letsios_conformal-like}.

 The full symmetry algebra (10 dS isometries plus 5 conformal-like symmetries) closes on $so(4,2)$ up to field-dependent gauge transformations. In particular, we have \cite{Letsios_conformal-like},:
 \begin{subequations}
\begin{equation}\label{so(4,2)_algebra_gravitino_1}
   [\mathbb{L}_{\xi} , \mathbb{L}_{\xi'}] \Psi^{(\text{TT})}_{\mu} =\mathbb{L}_{[\xi,\xi']}\Psi^{(\text{TT})}_{\mu},
\end{equation}    
\begin{equation}\label{so(4,2)_algebra_gravitino_2}
    [\mathbb{L}_{\xi} , \mathbb{T}_{V}] \Psi^{(\text{TT})}_{\mu} = 
 \mathbb{T}_{[\xi,V]}\Psi^{(\text{TT})}_{\mu},
\end{equation}
\begin{equation}\label{so(4,2)_algebra_gravitino_3}
     [\mathbb{T}_{V'} , \mathbb{T}_{V}] \Psi^{(\text{TT})}_{\mu}= \mathbb{L}_{[V',V]}\Psi^{(\text{TT})}_{\mu}+\left(\nabla_{\mu}+ \frac{i}{2} \gamma_{\mu}\right)  K_{[V',V]},
\end{equation}
\end{subequations}
where
\begin{align}\label{gauge parametr in [hid,hid]}
    K_{[V',V]}=\frac{4}{9}\left((\nabla^{\lambda}-\frac{i}{2}  \,\gamma^{\lambda})\Psi^{(\text{TT})\rho}\,\nabla_{\lambda}[V',V]_{\rho}-2 \,\Psi^{(\text{TT})\rho} \,[V',V]_{\rho}   \right).
\end{align}
Here, $[\mathbb{L}_{\xi} , \mathbb{T}_{V}] \equiv \mathbb{L}_{\xi}  \mathbb{T}_{V} -  \mathbb{T}_{V}\mathbb{L}_{\xi}$, $[\mathbb{T}_{V'} , \mathbb{T}_{V}] \equiv  \mathbb{T}_{V'}  \mathbb{T}_{V} -  \mathbb{T}_{V}  \mathbb{T}_{V'}$, and so forth.
It is clear that the algebra (\ref{so(4,2)_algebra_gravitino_1})-(\ref{so(4,2)_algebra_gravitino_3}) has the structure of the conformal algebra $so(4,2)$ [eq.~(\ref{algebra of CKVs})] up to the gauge transformation in (\ref{so(4,2)_algebra_gravitino_3}). This $so(4,2)$ symmetry is the dS analogue of the $so(4,2)$ symmetry found for strictly massless gauge potentials on $AdS_{4}$ in the {unfolded} formalism by Vasiliev \cite{Vasiliev}.

%%%%%%%%%%%%%%%%%%%%%%%%%%%%%%%%%%%%%%%%%%%%%%%%%%%%%%%%%%%%%%%%%%%%%%%%%%%%%%%%%%%%%%%%%%5

$$\textbf{UIRs of $so(4,2)$ formed by gravitino modes}$$
Each of the two positive frequency single-helicity sets of modes, $\{ {\psi}^{\left(phys ,\,+ \ell; m; k\right)}_{\mu} \}$ and \\ $\{ {\psi}^{\left(phys ,\,- \ell; m; k\right)}_{\mu} \}$, forms a UIR of $so(4,2)$ \cite{Letsios_conformal-like}.  {This fact} is readily demonstrated by specialising to the following genuine conformal Killing vector
\begin{align} \label{CKV_dS dilation}
    V^{(0)}_{\mu}= \nabla_{\mu} \sinh{t},
\end{align}
i.e.\ $(V^{(0)}_{t},V^{(0)}_{\theta_{3}}, V^{(0)}_{\theta_{2}}, V^{(0)}_{\theta_{1}}) = (\cosh{t},0,0,0)$. The conformal-like transformations (\ref{conf-like gravitino TT}) generated by $V^{(0)}$ act on the physical modes as:
\begin{align}\label{conf-like transf V0 3/2-}
    \mathbb{T}_{V^{(0)}}\psi^{(phys ,\,- \ell; \,m;k)}_{\mu}&=+ i \left(\ell +\frac{3}{2}   \right)\psi^{(phys ,\,-\ell; \,m;k)}_{\mu}
\end{align}
and
\begin{align}
    \mathbb{T}_{V^{(0)}}\psi^{(phys ,\,+ \ell; \,m;k)}_{\mu}&=- i \left(\ell +\frac{3}{2}   \right)\psi^{(phys ,\,+\ell; \,m;k)}_{\mu}.
\end{align}
Thus, from eqs.~(\ref{algebra of CKVs}) and (\ref{so(4,2)_algebra_gravitino_1})-(\ref{so(4,2)_algebra_gravitino_3}) it follows that
$\{ {\psi}^{\left(phys ,\,+ \ell; m; k\right)}_{\mu} \}$ and $\{ {\psi}^{\left(phys ,\,- \ell; m; k\right)}_{\mu} \}$ separately form irreducible representations of $so(4,2)$. These representations are unitary because the conformal-like generators (\ref{conf-like gravitino TT}) are anti-hermitian with respect to the scalar product (\ref{axial_scalar prod}) \cite{Letsios_conformal-like}:
\begin{align} \label{anti-herm_conf-like deriv_gravitino}
    \braket{\mathbb{T}_{V}\psi^{(1)}|\psi^{(2)}}_{ax} + \braket{\psi^{(1)}|\mathbb{T}_{V}\psi^{(2)}}_{ax}=0,
\end{align}
for any two solutions $\psi^{(1)}_{\mu}, \psi^{(2)}_{\nu}$ of (\ref{Dirac_eqn_fermion_dS}). However, as in the $so(4,1)$ case, a different choice of a positive-definite norm is needed for each $so(4,2)$ UIR of single helicity - see the discussion below (\ref{anti-herm_Lie deriv_gravitino}).

As in the $so(4,1)$ case, the negative frequency modes (\ref{negfreq_physmodes_negative_spin_3/2_dS4}) and (\ref{negfreq_physmodes_positive_spin_3/2_dS4}) form the same $so(4,2)$ UIRs as the positive frequency ones; their conformal-like transformations under $V^{(0) \mu}$ (\ref{CKV_dS dilation}) are
\begin{align}\label{conf-like transf V0 3/2-}
    \mathbb{T}_{V^{(0)}}v^{(phys ,\,- \ell; \,m;k)}_{\mu}&=- i \left(\ell +\frac{3}{2}   \right)v^{(phys ,\,-\ell; \,m;k)}_{\mu},
\end{align}
and
\begin{align}
    \mathbb{T}_{V^{(0)}}v^{(phys ,\,+ \ell; \,m;k)}_{\mu}&=+i \left(\ell +\frac{3}{2}   \right)v^{(phys ,\,+\ell; \,m;k)}_{\mu}.
\end{align}
%%%%%%%%%%%%%%%%%%%%%%%%%%%%%%%%%%%%%%%%%%%%%%%%%%%%%%%%%%
%%%%%%%%%%%%%%%%%%%%%%%%%%%%%%%%%%%%%%%55
\subsection{Conformal-like symmetry  of the hermitian action (\ref{gravitino action})} \label{Subsec_gravitino conf-like off-shell}
Interestingly, as we will present here for the first time, the conformal-like symmetry transformation (\ref{conf-like gravitino TT}) is also an off-shell symmetry of the hermitian action (\ref{gravitino action}). To prove this, let us consider the variation of the action (\ref{gravitino action}),
\begin{align}
   \delta S_{\frac{3}{2}}= -\int d^{4}x \, \sqrt{-g}\, \left( \delta \overline{\Psi}_{\mu} \,\gamma^{5} \, \mathcal{R}^{\mu}(\Psi) +  \overline{\Psi}_{\mu} \,\gamma^{5} \, \delta\mathcal{R}^{\mu}(\Psi) \right),
\end{align}
under 
$ \delta \Psi_{\mu} = \mathbb{T}_{V}\Psi_{\mu}$, where now $\Psi_{\mu}$ is an off-shell field configuration {with no gauge conditions imposed.}  After a straightforward off-shell calculation, we {find}  the following useful quantities:
\begin{align}
    \delta \overline{\Psi}_{\mu}=& \left( \mathbb{T}_{V}\Psi_{\mu}  \right)^{\dagger}  i  \gamma^{0} \nonumber\\
    &=    -\Big( V^{\rho}\nabla_{\rho}\overline{\Psi}_{\mu}+ i\,  V^{\rho} \overline{\Psi}_{\mu}  \gamma_{\rho} -i\, V^{\rho} \overline{\Psi}_{\rho} \, \gamma_{\mu} - \frac{3}{2} \phi_{V} \, \overline{\Psi}_{\mu}\Big )  \gamma^{5}+\frac{2}{3}  \overline{\Psi}_{\rho}V^{\rho}\gamma^{5}\left(\overset{\leftarrow}{\nabla}_{\mu}  + \frac{i}{2}  \gamma_{\mu} \right),
\end{align}
and
\begin{align}
\delta \mathcal{R}^{\mu}(\Psi) =    \gamma^{\mu \rho \sigma} &(\nabla_{\rho}+\frac{i}{2}\gamma_{\rho})\,  {\delta}\Psi_{\sigma}\nonumber \\
&= \gamma^{5}\phi_{V}\frac{5}{2}\mathcal{R}^{\mu}(\Psi)+\gamma^{5}\left(-V^{\rho}\nabla_{\rho}+i V^{\rho}\gamma_{\rho}   \right) \mathcal{R}^{\mu}(\Psi)-i \gamma^{5}V^{\mu}\gamma^{\sigma}\mathcal{R}_{\sigma}(\Psi),
\end{align}
where $\mathcal{R}^{\mu}(\Psi)$ is defined in~(\ref{RS_operator})\footnote{For off-shell fields we have
$\gamma^{\sigma }\mathcal{R}_{\sigma}(\Psi)= 2 \left(  \slashed{\nabla} \gamma^{\sigma} \Psi_{\sigma} - \frac{3}{2}i\,\gamma^{\sigma} \Psi_{\sigma} - \nabla^{\sigma} \Psi_{\sigma}  \right)$.}. Then, we easily find
\begin{align}
   \delta S_{\frac{3}{2}}= -\int d^{4}x \, \sqrt{-g}\, \nabla_{\rho}\left(  -V^{\rho} \overline{\Psi}_{\mu} \mathcal{R}^{\mu}(\Psi)  \right),
\end{align}
and thus, the conformal-like transformation (\ref{conf-like gravitino TT}) is an off-shell symmetry of the hermitian action (\ref{gravitino action}). 

%%%%%%%%%%%%%%%%%%%%%%%
\subsection{Quantisation of the  gravitino field, the necessity for the anti-self-duality constraint, and UIRs in the fermionic Fock space} \label{Subsec_gravitino quantisation}

$$ \textbf{Why does the (anti-)self-duality constraint have to be imposed?}$$

So far, we have explained how the gravitino positive frequency modes furnish a direct sum of discrete series UIRs of $so(4,1)$ and a direct sum of UIRs of $so(4,2)$. Our next task is to realise these UIRs in the single-particle Hilbert space associated with the QFT of the (free) quantum gravitino field on global $dS_{4}$. 
{In this task we encounter a problem related to 
 the discussions in Subsections \ref{Subsec_gravitino modes so(4,1)} and \ref{Subsec_gravitino modes so(4,2)}.} On the one hand, the single-particle Hilbert space of the QFT must be equipped with a positive-definite and dS-invariant scalar product. On the other hand, this space is identified with the space of physical positive frequency solutions, and, in our case, there is \textbf{no} dS-invariant inner product that remains \textbf{positive-definite for mode functions of both helicities}. One could argue that since positive-helicity and negative-helicity modes separately form UIRs, two different positive-definite inner products can be used for each fixed-helicity subspace for the quantisation of the theory, as discussed in the passage below eq.~(\ref{anti-herm_Lie deriv_gravitino}). However, although this approach works at the level of the classical mode solutions, it does not seem to work in a quantum field-theoretic setting if one insists on the locality of the action functional of the theory. In particular, one can {see}  that locality requires the indefiniteness of the norm, in the sense that the negative- or positive-definiteness of the norm depends on the helicity of each state, as follows. If one decides to include both helicities in the positive frequency, as well as in the negative frequency, sectors of the quantum gravitino field (as one usually does in Minkowski spacetime, for example), then they can follow the canonical quantisation procedure using the hermitian and local Lagrangian density in (\ref{gravitino action}) to define the conjugate momentum, and impose equal-time anti-commutation relations. Then, by expanding the field in modes, one finds that the equal-time anti-commutator (\ref{eq-time anti-com from lagrangian}) between the field and its conjugate momentum requires the anti-commutators between creation and annihilation operators to have helicity-dependent signs [eq.~(\ref{NON-unitary anti-cmtrs gravitino})], leading to the indefiniteness of the norm in the Fock space of the local QFT. This is explained in more detail in the passages below eq.~(\ref{crtn and annihil ops in terms of Psi_m}). Moreover, note that insisting on keeping all propagating helicity degrees of freedom while simultaneously enforcing the positivity of the norm, by using a different positive-definite scalar product for each fixed-helicity subspace, leads to a non-local theory. This becomes evident upon observing that the two distinct scalar products - namely, the axial scalar product (\ref{axial_scalar prod}) and its negative - can essentially be re-expressed in terms of a single scalar product. This modified scalar product arises from the initial axial scalar product (\ref{axial_scalar prod}) with the insertion of the `helicity' operator (\ref{duality operator vec-spinor S^3}), which is non-local.

To avoid the appearance of negative norms, and achieve positive definiteness in the QFT Fock space, we will quantise the `chiral'  gravitino field. The positive frequency sector of this field will furnish a UIR with helicity $- 3/2$ (or $+3/2$), corresponding to the $so(4,1)$ discrete series $D^{-}(5/2, 3/2)$ (or $D^{+}(5/2, 3/2)$), while the negative frequency sector will furnish a UIR with helicity $ + 3/2$ (or $-3/2$), corresponding to $D^{+}(5/2, 3/2)$ (or $D^{-}(5/2, 3/2)$) - see Appendix \ref{App_Classification_UIRs D=4} for details on UIRs of $so(4,1)$. Note that, for the chiral field under consideration, the helicity of the positive frequency sector is opposite from the helicity of the negative frequency sector\footnote{See 
Refs.~\cite{ASHTEKAR, Penrose} for further discussions on self-dual and anti-self-dual field strengths of massless fields and their quantum theory.}. In order to restrict to a theory with particles of a single helicity $-3/2$ (or $+3/2$), we will impose a chirality constraint, i.e.\ anti-self-duality (or self-duality) constraint, on the gravitino field strength. Without loss of generality, we will work with a quantum field whose positive and negative frequency sectors contain states with helicities $-3/2$ and $+3/2$, respectively, corresponding to an anti-self-dual field strength. The anti-self-duality constraint ensures the positivity of the norm  without violating the dS invariance of the theory.

$$  \textbf{Takahashi's quantisation method}  $$
The conventional way to quantise the gravitino field is the canonical quantisation procedure, in which one makes use of a local hermitian action functional, such as (\ref{gravitino action}), in order to define the conjugate momentum, and then imposes equal-time anti-commutation relations. Here, we will follow a different method discussed in detail by Takahashi  \cite{Takahashi}\footnote{Takahashi's method refers to free quantum fields in Minkowski spacetime, but the generalisation to dS spacetime is straightforward.}, which aligns well with the emphasis we have put on the group-theoretic properties of the mode solutions, as well as with the fact that we are considering a theory with an on-shell chirality constraint - the two helicities of the gravitino cannot be split locally at the level of the action, but such a split is possible on-shell by imposing  the (anti-)self-duality condition. The starting point in Takahashi's method is the field equation. In our case, the field equation is (\ref{Dirac_eqn_fermion_dS}) accompanied by the anti-self-duality constraint on the field strength\footnote{For the reader who is interested in the quantisation of chiral theories using Takahashi's method, the treatment of the chiral (massless) Majorana spin-1/2 field can be found in Takahashi's book \cite{Takahashi}.}
\begin{align}\label{anti-self constr gravitino}
   \widetilde{F}_{\mu   \nu} = -i {F}_{\mu \nu},
\end{align}
where $\widetilde{F}_{\mu   \nu} = \frac{1}{2}\varepsilon_{\mu \nu \alpha \beta} {F}^{\alpha   \beta}$\footnote{The need to impose a (anti-)self-duality constraint was explained at the beginning of this Subsection.}. The field strength is defined as \cite{Letsios_conformal-like}
\begin{align} \label{def:gravitino_field-strength}
    F_{\mu \nu}  = \left( \nabla_{[\mu} + \frac{i}{2} \gamma_{[\mu}   \right) \, \Psi_{\nu]}.
\end{align}
 {It is} divergence-free, gamma-traceless, and satisfies \cite{Letsios_conformal-like}
\begin{align} \label{duality property of field strength 3/2}
     \widetilde{F}_{\mu   \nu} = -i \gamma^{5}F_{\mu \nu},
\end{align}
by virtue of eqs.~(\ref{gamma_abcd-e_abcd gamma5}), (\ref{Dirac_eqn_fermion_dS}) and (\ref{duality properties3/2 fieldstrngth}), i.e.\ the duality operation on $F_{\mu \nu}$ is equivalent to an infinitesimal chiral rotation - see also Appendix \ref{Append_field strengths}\footnote{Using (\ref{duality property of field strength 3/2}), it follows that the anti-self-duality constraint (\ref{anti-self constr gravitino}) is equivalent to $\gamma^{5} {F}_{\mu \nu}= +  {F}_{\mu \nu}$.}. Later we will use the superscript `$-$' to denote the field strength that satisfies the anti-self-duality constraint (\ref{anti-self constr gravitino}), as ${F}^{-}_{ \mu \nu}$ in eq.~(\ref{def:gravitino_field-strength ANTISELF}). The field strength (\ref{def:gravitino_field-strength}) is gauge invariant, and thus, the field $\Psi_{\nu}$ in (\ref{def:gravitino_field-strength}) can be replaced by the gravitino gauge potential in any gauge without affecting the form of $F_{\mu \nu}$. Here we will  {impose} the TT gauge {condition}\footnote{ {Imposing} the TT gauge {condition} allows residual gauge symmetry. To quantise the gravitino field we will fix the gauge completely - see eq.~(\ref{mode expansion gravitino}).}.  Then, the quantum gravitino field operator ${\Psi}^{\text{(TT)}}_{\mu}(t, \bm{\theta_{3}})$ that is required to satisfy eqs.~(\ref{Dirac_eqn_fermion_dS}) and (\ref{anti-self constr gravitino}) is expressed as a mode sum in terms of our previously obtained mode functions, where the expansion coefficients are promoted to creation and annihilation operators. The main objective of Takahashi's method {for the theory at hand} is to determine the operators ${\Psi}^{(\text{TT})}_{\mu}$ and ${{Q}}^{dS}_{\frac{3}{2}}{[\xi]}$ such that the Heisenberg equations of motion are satisfied:\footnote{The equality here is modulo pure-gauge TT solutions.}
\begin{align} \label{Heisenberg eom_gravitino}
    -i \, \mathbb{L}_{\xi}{\Psi}^{\text{(TT)}}_{\mu}(t, \bm{\theta_{3}}) = \left [  {\Psi}^{\text{(TT)}}_{\mu}(t, \bm{\theta_{3}})  , {{Q}}^{dS}_{\frac{3}{2}}{[\xi]}  \right],
\end{align}
where $\xi^{\mu}$ is any Killing vector of dS spacetime, and ${{Q}}^{dS}_{\frac{3}{2}}{[\xi]}$ is the hermitian quantum operator (dS charge) that generates the infinitesimal dS transformation on the QFT Fock space. The subscript `$\frac{3}{2}$' in ${{Q}}^{dS}_{\frac{3}{2}}{[\xi]}$ has been used to distinguish between the quantum generators of the chiral gravitino and of the chiral graviton - see 
eq.~(\ref{Heisenberg eom_graviton}).   In addition, quantum operators representing physical quantities must  {(anti-)commute with one another} for spacelike separations (microcausality).

\noindent \textbf{\textit{Mode expansion$-$}}Let us now quantise the chiral gravitino field following the steps outlined in the previous paragraph. From our discussions in Subsection~\ref{Subsec_gravitino modes so(4,1)} it follows that the $t$-component of the gravitino field in the TT gauge is pure-gauge, i.e.\ it can be gauged away. Thus, to isolate the propagating degrees of freedom, we consider the completely gauge-fixed field, ${\Psi}^{\text{(TT)}}_{t} =  g^{\tilde{\mu} \tilde{\nu}} \gamma_{\tilde{\mu}} {\Psi}^{\text{(TT)}}_{\tilde{\nu}}=0$ \footnote{This gauge is the analogue of the Coulomb gauge for the Maxwell gauge potential.}, and we expand it in modes as follows 
\begin{align} \label{mode expansion gravitino}
{\Psi}^{\text{(TT)}-}_{\tilde{\mu}}(t, \bm{\theta_{3}})  = \sum_{\ell =1}^{\infty}   \sum_{m=1}^{\ell} \sum_{k=-m-1}^{m} \left( {a}^{(-)}_{\ell m k}{\psi}^{\left(phys ,\,- \ell\,;m;k \right)}_{\tilde{\mu}}(t, \bm{\theta_{3}})  + {b}^{(+)\dagger}_{\ell m k}\,{v}^{\left(phys ,\,+ \ell\,;m;k \right)}_{\tilde{\mu}} (t, \bm{\theta_{3}}) \right),
\end{align}
where $\tilde{\mu}$ is a vector index on $S^{3}$. The superscript `$-$' in  ${\Psi}^{(\text{TT})-}_{\mu}$ refers to the fact that the corresponding field strength is anti-self-dual (this will be verified below). The non-zero anti-commutators between creation and annihilation operators are
\begin{align}
    \{  {a}^{(-)}_{\ell m k} , {a}^{(-)\dagger}_{\ell' m' k'}  \} = \delta_{\ell   \ell'}  \delta_{m m'}  \delta_{kk'},~~~~~~\{  {b}^{(+)}_{\ell m k} , {b}^{(+)\dagger}_{\ell' m' k'}  \} = \delta_{\ell   \ell'}  \delta_{m m'}\delta_{kk'}.
\end{align}
The vacuum is defined as the state, $\ket{0}_{\frac{3}{2}}$, in the Fock space that satisfies:
\begin{align}\label{def:chiral gravitino vacuum}
    {a}^{(-)}_{\ell m k} \ket{0}_{\frac{3}{2}} =  {b}^{(+)}_{\ell m k} \ket{0}_{\frac{3}{2}}  = 0,
\end{align}
 for all $\ell , m , k$. Using the dS-invariant inner product (\ref{axial_scalar prod}) [see eqs.~(\ref{norms of physical modes_3/2}) and (\ref{norms of physical modes_3/2_neg freq})], we find
\begin{align}\label{crtn and annihil ops in terms of Psi_m}
    {a}^{(-)}_{\ell m k}= \braket{ {\psi}^{\left(phys ,\,- \ell\,;m;k \right)} |{\Psi}^{\text{(TT)}-}}_{ax},~~~~~{b}^{(+)\dagger}_{\ell m k}= \braket{ {v}^{\left(phys ,\,+ \ell\,;m;k \right)} |{\Psi}^{\text{(TT)}-}}_{ax}.
\end{align}
The dS invariance of the vacuum follows from the dS invariance of the positive frequency solution space, and it will be further verified below by showing that the quantum dS generators annihilate the vacuum.

\noindent \textbf{\textit{What goes wrong if we include both helicities?}$-$}To demonstrate the appearance of negative norms in the Fock space  {for} a non-chiral gravitino, let us consider a completely gauge-fixed quantum gravitino field ${\Psi}_{\mu}$, as in (\ref{mode expansion gravitino}), 
{which now includes} all helicities in the mode sum as follows:
\begin{align} 
 {\Psi}^{\text{(TT)}}_{\tilde{\mu}}(t, \bm{\theta_{3}})  = \sum_{\sigma \in \{ +, -  \}} \sum_{\ell, m,  k }  \left( {a}^{(\sigma)}_{\ell m k}{\psi}^{\left(phys ,\,\sigma \ell\,;m;k \right)}_{\tilde{\mu}}(t, \bm{\theta_{3}})  + {b}^{(\sigma)\dagger}_{\ell m k}\,{v}^{\left(phys ,\,\sigma \ell\,;m;k \right)}_{\tilde{\mu}} (t, \bm{\theta_{3}}) \right),
\end{align}
where from eqs.~(\ref{norms of physical modes_3/2}) and (\ref{norms of physical modes_3/2_neg freq}) we have
\begin{align}\label{crtn and annihil ops in terms of Psi_m non-uni}
    {a}^{(\mp)}_{\ell m k}= \pm\braket{ {\psi}^{\left(phys ,\,\mp \ell\,;m;k \right)} |{\Psi}^{\text{(TT)}}}_{ax},~~~~~{b}^{(\pm)\dagger}_{\ell m k}=\pm \braket{ {v}^{\left(phys ,\,\pm \ell\,;m;k \right)} |{\Psi}^{\text{(TT)}}}_{ax}.
\end{align}
Let us also denote the vacuum annihilated by all annihilation operators as $\ket{\Omega}$.
{In this case we can proceed with the  canonical quantisation procedure using the hermitian Lagrangian in (\ref{gravitino action}).} The standard  equal-time anti-commutation relations (expressed in the form of a {$4\times 4$} spinorial matrix) are
\begin{align}\label{eq-time anti-com from lagrangian}
 & \left \{ {\Psi}^{\text{(TT)}}_{\tilde{\mu}}(t, \bm{\theta_{3}}),  {\Psi}^{\text{(TT)}}_{\tilde{\nu}'}(t, \bm{\theta_{3}}')^{\dagger}  \gamma^{5}\right \}   \nonumber \\
 & = \frac{e^{~m}_{\tilde{\mu}} e^{~n'}_{\tilde{\nu}'}}{\sqrt{-g}}\begin{pmatrix}
    \sqrt{\tilde{g}}~ \Delta^{TT}_{mn'}(\bm{\theta_{3}},\bm{\theta_{3}}') & & \bm{0} \\
    \bm{0} & &  \sqrt{\tilde{g}}~\Delta^{TT}_{m n'}(\bm{\theta_{3}},\bm{\theta_{3}}')
 \end{pmatrix} \nonumber \\
 & = \frac{1}{\cosh{t}}\begin{pmatrix}
   \Delta^{TT}_{\tilde{\mu}  \tilde{\nu}'}(\bm{\theta_{3}},\bm{\theta_{3}}') & & \bm{0} \\
    \bm{0} & &  \Delta^{TT}_{\tilde{\mu}  \tilde{\nu}'}(\bm{\theta_{3}},\bm{\theta_{3}}')
 \end{pmatrix},
\end{align}
where $$\Delta^{TT}_{\tilde{\mu}  \tilde{\nu}'}(\bm{\theta_{3}},\bm{\theta_{3}}') = \cosh^{-2}{t}~ e^{~m}_{\tilde{\mu}} e^{~n'}_{\tilde{\nu}'}\Delta^{TT}_{m  n'}(\bm{\theta_{3}},\bm{\theta_{3}}')$$
is the transverse and $\tilde{\gamma}$-traceless delta function %(\ref{def: TT delta function 3/2}) 
for vector-spinors on $S^{3}$ defined by
     \begin{align} \label{def: TT delta function 3/2}
         \Delta^{TT}_{\tilde{\mu}  \tilde{\nu}'}(\bm{\theta_{3}},\bm{\theta_{3}}') =& \sum_{\sigma \in \{ +,-  \}}  \sum_{\ell=1}^{\infty}\sum_{m , k} \tilde{\psi}_{\sigma \tilde{\mu}}^{(\ell; m;k)} (\bm{\theta_{3}})  \otimes  \tilde{\psi}_{\sigma \tilde{\nu}'}^{(\ell; m;k)} (\bm{\theta_{3}} ')^{\dagger} .
     \end{align}
In particular, if $\tilde{\psi}_{\tilde{\mu}}(\bm{\theta_{3}})$ is a vector-spinor on $S^{3}$, and $\tilde{\psi}'_{\tilde{\mu}}(\bm{\theta_{3}})$ is its divergence-free and $\tilde{\gamma}$-traceless part, then
\begin{align}
 \tilde{\psi}'_{\tilde{\mu}}(\bm{\theta_{3}}) = \int_{S^{3}}  d\bm{\theta_{3}} \sqrt{\tilde{g}}    \, \Delta^{TT}_{\tilde{\mu}  \tilde{\nu}'}(\bm{\theta_{3}},\bm{\theta_{3}}')  ~ \tilde{\psi}^{\tilde{\nu}'}(\bm{\theta_{3}}').
\end{align}
Using the expressions (\ref{crtn and annihil ops in terms of Psi_m non-uni}), as well as  the anticommutation relations (\ref{eq-time anti-com from lagrangian}), we find 
\begin{align}\label{NON-unitary anti-cmtrs gravitino}
    \{  {a}^{(\sigma)}_{\ell m k} , {a}^{(\sigma')\dagger}_{\ell' m' k'}  \} = (- \sigma) \, \delta_{\sigma \sigma'} \delta_{\ell   \ell'}  \delta_{m m'}  \delta_{kk'},~~~~~~\{  {b}^{(\sigma)}_{\ell m k} , {b}^{(\sigma')\dagger}_{\ell' m' k'}  \} =\sigma \, \delta_{\sigma \sigma'} \delta_{\ell   \ell'}  \delta_{m m'}\delta_{kk'},
\end{align}
$\sigma, \sigma' \in \{ + , -\}$, while the rest of the anti-commutators are zero. It is clear that the states ${a}^{(+)\dagger}_{\ell m k} \ket{\Omega}$ and ${b}^{(-)\dagger}_{\ell m k} \ket{\Omega}$ have negative norm, i.e.\ the theory is non-unitary. As a  verification of this fact, we can expand in modes the fields on the left-hand side of the equal-time anti-commutator (\ref{eq-time anti-com from lagrangian}). After some algebra, one can arrive at the expression in terms of the transverse-traceless delta function on the right-hand side of (\ref{eq-time anti-com from lagrangian}) only if the anti-commutators (\ref{NON-unitary anti-cmtrs gravitino}) are used. In other words, the equal-time anti-commutation relations for a gravitino field that contains both helicities require the appearance of negative-norm states in the QFT Fock space. This justifies our choice to exclude half of the helicities, i.e.\ quantise the chiral gravitino field (\ref{mode expansion gravitino}), to achieve unitarity in the QFT Fock space. We will now continue the quantisation of the chiral gravitino field.
%%%%%%%%%%%%%%%%%%%%%%%%%%%%%%%%%%%%%%%%%%%%%%%%%%%%%%%%%%%%%%%%%%%%%%%%%%%%%%%%%%%%%%%%%%%%%%%%%%%%%%%%%%%%%%%%%%%%%%%%%%%%%%%
\\
\\
\noindent  \textbf{\textit{Anti-self-duality constraint}$-$}Let us demonstrate that our choice for the mode expansion for the chiral gravitino (\ref{mode expansion gravitino}) is consistent with the anti-self-duality constraint (\ref{anti-self constr gravitino}). To be specific, we will show that the following field strength:
\begin{align} \label{def:gravitino_field-strength ANTISELF}
    F^{-}_{\mu \nu}  = \left( \nabla_{[\mu} + \frac{i}{2} \gamma_{[\mu}   \right) \, \Psi^{(\text{TT})-}_{\nu]},
\end{align}
is anti-self-dual.
Substituting the mode expansion  (\ref{mode expansion gravitino}) into (\ref{def:gravitino_field-strength ANTISELF}), we find the mode expansion for the field strength
\begin{align} \label{mode expansion 3/2 field strength}
     {F}^{-}_{{\mu}  \nu}(t, \bm{\theta_{3}})  = \sum_{\ell=1}^{\infty} \sum_{m,k} \Big( & {a}^{(-)}_{\ell m k}\, f^{(-\ell;m;k)}_{\mu \nu}(t,   \bm{\theta_{3}}) + {b}^{(+)\dagger}_{\ell m k}\, f^{(+\ell;m;k)C}_{\mu \nu}(t,   \bm{\theta_{3}})  \Big),
\end{align}
where 
\begin{align*}
    & f^{(-\ell;m;k)}_{\mu \nu}(t,   \bm{\theta_{3}}) \equiv \left( \nabla_{[\mu} + \frac{i}{2} \gamma_{[\mu}   \right){\psi}^{\left(phys,- \ell\,;m;k \right)}_{ \nu]} (t, \bm{\theta_{3}}) , \nonumber \\
     &f^{(+\ell;m;k)C}_{\mu \nu}(t,   \bm{\theta_{3}}) \equiv \left( \nabla_{[\mu} + \frac{i}{2} \gamma_{[\mu}   \right){v}^{\left(phys,+ \ell\,;m;k \right)}_{ \nu]} (t, \bm{\theta_{3}}).
\end{align*}
For convenience, let us start by showing  that the $t \tilde{\nu}$-component, ${F}^{-}_{{t}  \tilde{\nu}}$, satisfies the anti-self-duality constraint. Substituting the expressions of the mode functions (\ref{physmodes_negative_spin_3/2_dS4}) and (\ref{negfreq_physmodes_positive_spin_3/2_dS4}) into (\ref{mode expansion 3/2 field strength}), we find after a straightforward calculation
\begin{align}
 {F}^{-}_{t \tilde{\nu}}(t, \bm{\theta_{3}})  =\frac{1}{2} \sum_{\ell=1}^{\infty}\sum_{m,k} \left(\frac{\ell+2}{2(\ell+1)} \right)^{1/2}  \Bigg [ &{a}^{(-)}_{\ell m k}\,   \begin{pmatrix}  \left({\partial}_{t}\alpha_{\ell}(t) -\frac{i}{2}\beta_{\ell}(t) \right)~ \tilde{\psi}_{-\tilde{\nu}}^{(\ell; m;k)} (\bm{\theta_{3}})  \\ 0
      \end{pmatrix}  \nonumber \\
& + {b}^{(+)\dagger}_{\ell m k}\,   \begin{pmatrix}  \left({\partial}_{t}\alpha^{*}_{\ell}(t) +\frac{i}{2}\beta_{\ell}^{*}(t) \right)~ \tilde{\psi}_{+\tilde{\nu}}^{(\ell; m;k)} (\bm{\theta_{3}})  \\ 0
      \end{pmatrix} \Bigg] ,
\end{align}
where we have also  used (\ref{relating db(t)/dx and a(t)}). It is clear that ${F}^{-}_{{t}  \tilde{\nu}} $ is an eigenfunction of $\gamma^{5}$ [eq.~(\ref{gamma5 chiral rep})], $\gamma^{5}{F}^{-}_{{t}  \tilde{\nu}}  =+ {F}_{{t}  \tilde{\nu}}^{-}$. Using (\ref{duality property of field strength 3/2}), it follows that, for all the components of the field strength, we have  $\gamma^{5}{F}^{-}_{{\mu}  {\nu}}  =+ {F}^{-}_{{\mu}  {\nu}}$. This means that the anti-self-duality constraint (\ref{anti-self constr gravitino}) is satisfied.
\\
\\
%%%%%%%%%%%%%%%%%%%%%%%%%%%%%%%%%%%%%%%%%%%%%%%%%%%%%%%%%%%%%%%%%%%%%%%%%%%%%
\noindent \textbf{\textit{Quantum symmetry generators}$-$}The hermitian dS generators can be constructed in the standard way \cite{Takahashi,Yale_Thesis}, using the dS-invariant inner product (\ref{axial_scalar prod}):
\begin{align}\label{def:dS charges gravitino}
    {{Q}}^{dS}_{\frac{3}{2}}{[\xi]} =- {i} :\braket{{\Psi}^{\text{(TT)}-}| \mathbb{L}_{\xi}  {\Psi}^{\text{(TT)}-}}_{ax}:~,
\end{align}
where the symbols $: ... :$ denote normal ordering.
Here we will give the explicit expression only for the generator corresponding to the dS boost $\xi^{\mu} = B^{\mu}$ [eq.~(\ref{dS boost})]; the other dS generators can be constructed similarly. Expanding the field in modes (\ref{mode expansion gravitino}), and using eqs.~(\ref{infinitesimal dS of spin-3/2 physical modes}), (\ref{norms of physical modes_3/2}) and (\ref{norms of physical modes_3/2_neg freq}), we find:
    \begin{align} \label{boost quantum generator_3/2}
     {{Q}}^{dS}_{\frac{3}{2}}{[B]} =&-\frac{1}{2}\sum_{\ell=1}^{\infty}\sum_{m,k}\Bigg    ( \sqrt{(\ell-m+1)(\ell+m+3)}\,{a}^{(-)\dagger}_{(\ell+1) m k}{a}^{(-)}_{\ell m k} +  \sqrt{(\ell-m)(\ell+m+2)}\, {a}^{(-)\dagger}_{(\ell-1) m k}{a}^{(-)}_{\ell m k} \Bigg) \nonumber\\
     -& \frac{1}{2}\sum_{\ell=1}^{\infty} \sum_{m,k} \Bigg( \sqrt{(\ell-m+1)(\ell+m+3)}\,{b}^{(+)\dagger}_{(\ell+1) m k} {b}^{(+)}_{\ell m k} +  \sqrt{(\ell-m)(\ell+m+2)}\, {b}^{(+)\dagger}_{(\ell-1) m k}{b}^{(+)}_{\ell m k} \Bigg ).
     \end{align}
     This can be clearly expressed as a sum of two independent hermitian generators,
     $$ {{Q}}^{dS}_{\frac{3}{2}}{[B]} =  {{Q}}^{dS-}_{\frac{3}{2}}{[B]} +  {{Q}}^{dS+}_{\frac{3}{2}}{[B]},$$
     where $ {{Q}}^{dS-}_{\frac{3}{2}}{[B]}$ is given by the expression in the first line of (\ref{boost quantum generator_3/2}), while $ {{Q}}^{dS+}_{\frac{3}{2}}{[B]}$ is given by the expression in the second line. The two generators $ {{Q}}^{dS-}_{\frac{3}{2}}{[B]}$ and $ {{Q}}^{dS+}_{\frac{3}{2}}{[B]}$ act on the negative- and positive-helicity sectors, respectively (i.e.\ positive-frequency and negative-frequency sectors, respectively). In particular, the two charges generate the two discrete series UIRs of $so(4,1)$, $D^{-}(\Delta = 5/2,s= 3/2)$ and $D^{+}(\Delta= 5/2, s =3/2)$, respectively.
     The corresponding infinitesimal dS transformations of the creation operators are
     \begin{align}
      \delta_{B} {a}^{(-)\dagger}_{\ell m k} \equiv    \left[ {a}^{(-)\dagger}_{\ell m k} , \,  {{Q}}^{dS}_{\frac{3}{2}}{[B]} \right]= \left[ {a}^{(-)\dagger}_{\ell m k} , \,  {{Q}}^{dS-}_{\frac{3}{2}}{[B]} \right] 
         = & \frac{1}{2}\sqrt{(\ell-m+1)(\ell+m+3)}\,{a}^{(-)\dagger}_{(\ell+1) m k}\nonumber\\
        &+  \frac{1}{2}\sqrt{(\ell-m)(\ell+m+2)}\,{a}^{(-)\dagger}_{(\ell-1) m k}
     \end{align}
     and
     \begin{align}
       \delta_{B} {b}^{(+)\dagger}_{\ell m k}\equiv   \left[ {b}^{(+)\dagger}_{\ell m k} , \,  {{Q}}^{dS}_{\frac{3}{2}}{[B]}  \right]= \left[ {b}^{(+)\dagger}_{\ell m k} , \,  {{Q}}^{dS+}_{\frac{3}{2}}{[B]}  \right] 
         = & \frac{1}{2}\sqrt{(\ell-m+1)(\ell+m+3)}\,{b}^{(+)\dagger}_{(\ell+1) m k}\nonumber\\
        &+  \frac{1}{2}\sqrt{(\ell-m)(\ell+m+2)}\,{b}^{(+)\dagger}_{(\ell-1) m k}.
     \end{align}
     Using these expressions, it is clear that single-particle states ${a}^{(-)\dagger}_{\ell m k} \ket{0}_{\frac{3}{2}}$ transform as the corresponding positive frequency modes (\ref{infinitesimal dS of spin-3/2 physical modes}), i.e.\ they furnish the $so(4,1)$ discrete series UIR $D^{-}(\Delta = 5/2, s = 3/2)$. Similarly, single-particle states ${b}^{(+)\dagger}_{\ell m k} \ket{0}_{\frac{3}{2}}$ furnish the $so(4,1)$ discrete series UIR  $D^{+}(\Delta= 5/2, s =3/2)$ - see Appendix \ref{App_Classification_UIRs D=4}.
     Finally, it is straightforward to find that the quantum field operator transforms as 
     \begin{align}
         \left[ {\Psi}^{\text{(TT)}-}_{\mu},  {{Q}}^{dS}_{\frac{3}{2}}{[B]}  \right] = -i \, \mathbb{L}_{B}  {\Psi}^{\text{(TT)}-}_{\mu},
     \end{align}
     modulo pure-gauge TT solutions, where $\mathbb{L}_{B}$ is the Lie-Lorentz derivative (\ref{Lie_Lorentz}) with respect to the dS boost Killing vector $B$ [eq.~(\ref{dS boost})], in agreement with the Heisenberg equations of motion (\ref{Heisenberg eom_gravitino}).

   %%%%%%%%%%%%%%%%%%%%%%%%%%%%%%%%%%%%%%%%%%%%%%%%%%%%%%%%%%%%%%%%%%%%%%%%%

     It is also interesting to note that we can construct the five hermitian generators of the conformal-like symmetry (\ref{conf-like gravitino TT}) in the same way,
     \begin{align}\label{def:conf charges gravitino}
    {{Q}}^{conf}_{\frac{3}{2}}{[V]} =- {i} :\braket{{\Psi}^{\text{(TT)}-}| \mathbb{T}_{V}  {\Psi}^{\text{(TT)}-}}_{ax}: ~,
\end{align}
  such that the Heisenberg equations of motion are again satisfied
     \begin{align} 
    -i \, \mathbb{T}_{V}{\Psi}^{\text{(TT)}-}_{\mu}(t, \bm{\theta_{3}}) = [ ~ {\Psi}^{\text{(TT)}-}_{\mu}(t, \bm{\theta_{3}})  ,  {{Q}}^{conf}_{\frac{3}{2}}{[V]}~],
\end{align}
{modulo pure-gauge TT solutions}, where $\mathbb{T}_{V}$ is the conformal-like transformation (\ref{conf-like gravitino TT}).
This can be easily checked for the conformal-like symmetry generated by the genuine conformal Killing vector $V^{(0)\mu}$ [eq.~(\ref{CKV_dS dilation})], for which the quantum generator is  found to be 
      \begin{align} 
     {{Q}}^{conf}_{\frac{3}{2}}{[V^{(0)}]} =\, \sum_{\ell =1}^{\infty}\sum_{m,k} \left(\ell +\frac{3}{2}\right)\left({a}^{(-)\dagger}_{\ell m k}{a}^{(-)}_{\ell m k}   - {b}^{(+)\dagger}_{\ell m k}{b}^{(+)}_{\ell m k} \right).
     \end{align}
     This conformal-like quantum charge is expressed as a sum of two independent conformal-like charges,
 \begin{align} \label{conf-like charge_3/2 = pos freq + neg freq}
    &{{Q}}^{conf}_{\frac{3}{2}}{[V^{(0)}]} = {{Q}}^{conf-}_{\frac{3}{2}}{[V^{(0)}]} + {{Q}}^{conf+}_{\frac{3}{2}}{[V^{(0)}]}, \nonumber \\
    &{{Q}}^{conf-}_{\frac{3}{2}}{[V^{(0)}]} =  \sum_{\ell =1}^{\infty}\sum_{m,k} \left(\ell +\frac{3}{2}\right){a}^{(-)\dagger}_{\ell m k}{a}^{(-)}_{ \ell m k} ,\nonumber\\
    &{{Q}}^{conf+}_{\frac{3}{2}}{[V^{(0)}]} = - \sum_{\ell =1}^{\infty}\sum_{m,k} \left(\ell +\frac{3}{2}\right){b}^{(+)\dagger}_{\ell m k}{b}^{(+)}_{ \ell m k}.
     \end{align}
     The charges ${{Q}}^{conf-}_{\frac{3}{2}}{[V^{(0)}]}$ and ${{Q}}^{conf+}_{\frac{3}{2}}{[V^{(0)}]}$  generate $so(4,2)$ UIRs on the  spaces of negative-helicity and positive-helicity states, respectively.
     Note that the vacuum of the chiral gravitino, $\ket{0}_{\frac{3}{2}}$, is  invariant under the whole conformal-like symmetry, $so(4,2)$.
%%%%%%%%%%%%%%%%%%%%%%%%%%%%%%%%%%%%%%%%%%%%%%%%%%%%%%%%%%%%%%%%%%%%
\\
\\
  \noindent \textbf{\textit{Microcausality}$-$} {Finally,} we will demonstrate the microcausality of the theory by computing the anti-commutator between two gauge invariant quantities, { the anti-self-dual field strength and its hermitian conjugate,}  at two spacelike separated points:
     $$   \left\{ {F}^{-}_{\mu \nu}(t, \bm{\theta_{3}}), {F}^{-}_{\alpha' \beta'}(t', \bm{\theta_{3}}')^{\dagger}       \right\}. $$
     For convenience, let us start by choosing two equal-time points $(t, \bm{\theta_{3}})$ and $(t, \bm{\theta_{3}'})$, and compute the equal-time anti-commutator for the following components of the field strengths:
     \begin{align}
         \left\{ {F}^{-}_{t \tilde{\mu}}(t, \bm{\theta_{3}}), {F}^{-}_{t \tilde{\nu}'}(t, \bm{\theta_{3}}')^{\dagger}       \right\}.
     \end{align}
        Expanding the field strengths in modes [as in (\ref{mode expansion 3/2 field strength})], we find\footnote{The anti-commutator (\ref{local eq-time anti-com 3/2 special}) is a $4$-dimensional bi-spinorial matrix and each of its components are bi-vectors on $S^{3}$. The vector index $\tilde{\mu}$ refers to the tangent space at the point $(t , \bm{\theta_{3}})$, while $\tilde{\nu}'$ to the tangent space at $(t , \bm{\theta_{3}}')$. }
    \begin{align} \label{local eq-time anti-com 3/2 special}
    \left\{ {F}^{-}_{t \tilde{\mu}}(t, \bm{\theta_{3}}), {F}^{-}_{t \tilde{\nu}'}(t, \bm{\theta_{3}}')^{\dagger}       \right\} = -\frac{1}{ 4 \,\cosh^{3}{t} }   
     \begin{pmatrix}
        \left( \tilde{\slashed{\nabla}}^{2}  + \frac{1}{4}\right)  {\Delta^{TT}_{\tilde{\mu}  \tilde{\nu}'}(\bm{\theta_{3}},\bm{\theta_{3}}')} &&~~~~~ \bm{0} \\
      ~    \bm{0} && ~~~~~\bm{0}
        \end{pmatrix},
    \end{align}
     where ${\Delta^{TT}_{\tilde{\mu}  \tilde{\nu}'}(\bm{\theta_{3}},\bm{\theta_{3}}')}$ is
     %we have introduced 
     the transverse and $\tilde{\gamma}$-traceless delta function in the space of vector-spinors on $S^{3}$
     defined by eq.~(\ref{def: TT delta function 3/2}).
 %
 %   \begin{align} %\label{def: TT delta function 3/2}
 %        \Delta^{TT}_{\tilde{\mu}  \tilde{\nu}'}(\bm{\theta_{3}},\bm{\theta_{3}}') =& \sum_{\sigma \in \{ +,-  \}}  \sum_{\ell=1}^{\infty}\sum_{m , k} \tilde{\psi}_{\sigma \tilde{\mu}}^{(\ell; m;k)} (\bm{\theta_{3}})  \otimes  \tilde{\psi}_{\sigma \tilde{\nu}'}^{(\ell; m;k)} (\bm{\theta_{3}} ')^{\dagger} .
 %    \end{align}
     %
 In Appendix \ref{Appenix_Locality and TT-delta function}  we show that, while $ \Delta^{TT}_{\tilde{\mu}  \tilde{\nu}'}(\bm{\theta_{3}},\bm{\theta_{3}}') $ is non-local, the quantity $ \left( \tilde{\slashed{\nabla}}^{2}  + \frac{1}{4}\right)  {\Delta^{TT}_{\tilde{\mu}  \tilde{\nu}'}(\bm{\theta_{3}},\bm{\theta_{3}}')}$ in (\ref{local eq-time anti-com 3/2 special}) is local, i.e.\ it vanishes for $\bm{\theta_{3}} \neq \bm{\theta_{3}}'$ [see eq.~(\ref{(diff op) on TT delta function 3/2 final})]. This means that the anti-commutator (\ref{local eq-time anti-com 3/2 special}) is also local. Then, taking the dual of (\ref{local eq-time anti-com 3/2 special}) and using (\ref{duality property of field strength 3/2}), it is easy to conclude that all equal-time anti-commutators of the form
     \begin{align}\label{local eq-time anti-com 3/2 general}
         \left\{ {F}^{-}_{\mu {\nu}}(t, \bm{\theta_{3}}), {F}^-_{\alpha' {\beta'}}(t, \bm{\theta_{3}}')^{\dagger}       \right\}
     \end{align}
    are also local, for any value of the tensor indices $\mu, \nu$, $\alpha' , \beta'$. {Finally}, the locality of the anti-commutator for any two causally disconnected (not necessarily equal-time) points can be easily demonstrated by exploiting the following observation:
      any two points $(t, \bm{\theta_{3}})$ and $(t', \bm{\theta_{3}}')$ that are causally disconnected can be moved to the same equal-time Cauchy surface by a suitable dS transformation. Thus, the locality of (\ref{local eq-time anti-com 3/2 general}) implies that all anti-commutators of the form
      \begin{align*}
     \left\{ {F}^-_{\mu {\nu}}(t, \bm{\theta_{3}}), {F}^-_{\alpha' {\beta'}}(t', \bm{\theta_{3}}')^{\dagger}  \right\} 
     \end{align*}
      {vanish} for any two points $(t, \bm{\theta_{3}})$ and $(t', \bm{\theta_{3}}')$ that are spacelike separated. Then, it follows that Grassmann-even observables, such as  the currents $J^{\rho}(t,\bm{\theta_{3}})=\overline{F}^{\,-}_{\mu {\nu}}(t, \bm{\theta_{3}}) \gamma^{\rho} {F}^{-\mu {\nu}}(t, \bm{\theta_{3}})$, commute for spacelike separations. This shows the microcausality of the theory.   {This concludes the discussion of} 
     the quantisation of the chiral gravitino field.

%%%%%%%%%%%%%%%%%%%%%%%%%%%%%%%%%%%%%%%%%%%%%%%%%%%%%%%%%%%%%%%%%%%%%%%%%%%%%%%%%%%%%%%%%%%5

\section{Free graviton gauge potential on \texorpdfstring{$dS_{4}$}{dS4}, UIRs of \texorpdfstring{$so(4,1)$}{so(4,1)} and \texorpdfstring{$so(4,2)$}{so(4,2)}, quantisation and (anti-)self-duality}  \label{Sec_graviton}
The graviton and its unitarity in de Sitter spacetime have been studied more extensively than the gravitino \cite{Yale_Thesis, HiguchiLinearised, Higuchi_forbidden, Higuchi_Instab}. 
The linearised Einstein-Hilbert action around a $dS_{4}$ background describes a real massless spin-2 field, $h_{\mu \nu} = h_{(\mu \nu)}$, propagating on a fixed dS spacetime. The linearised Einstein-Hilbert action (after some integrations by parts) can be expressed as \cite{Higuchi_forbidden}
\begin{align}\label{real graviton action}
    S_{EH}= -\frac{1}{4}\int d^{4}x \, \sqrt{-g}\,  h^{\mu \nu}\, H_{\mu \nu}(h),
\end{align}
with
\begin{align}\label{lin Einstein operator real}
     H_{\mu \nu}(h)\equiv &\, \nabla_{\mu} \nabla_{\alpha}h^{\alpha}_{\nu}+ \nabla_{\nu} \nabla_{\alpha}h^{\alpha}_{\mu}- \Box h_{\mu \nu}+g_{\mu \nu}\,\Box h^{\alpha}_{~\alpha} -\nabla_{\mu} \nabla_{\nu}h^{\alpha}_{~\alpha} \nonumber \\
    &- g_{\mu \nu}\, \nabla^{\alpha}\nabla^{\beta}h_{\alpha \beta} +2\, h_{\mu \nu}+g_{\mu \nu}h^{\alpha}_{~\alpha},
\end{align}
where $\Box = g^{\mu  \nu}\nabla_{\mu} \nabla_{\nu}$ is the Laplace-Beltrami operator on $dS_{4}$. {The symbol $h$ in $H^{\mu \nu}(h)$ does \textbf{not} stand for the trace of $h_{\alpha \beta}$.}
The equations of motion for linearised gravity on $dS_{4}$ are 
\begin{align}\label{EOM_graviton general}
     H_{\mu \nu}(h)=0.
\end{align}
The action (\ref{real graviton action}) is invariant under gauge transformations of the form
%
%\textcolor{red}{[There was a misprint here! Observed it after submission to JHEP]}
%
\begin{align}\label{gauge_transf_spin2_real}
    \delta^{\text{gauge}}(Z)\, h_{\mu \nu}= \nabla_{(\mu} Z_{\nu)},
\end{align}
where $Z_{\nu}$ is an arbitrary real vector gauge function. Note that, for any $Z_{\nu}$, we have $$ \,H^{\mu \nu}(\delta^{\text{gauge}}(Z)h) = 0,$$
corresponding to the well-known gauge invariance of the linearised Einstein equations.
As {is well known (see, e.g., \cite{HiguchiLinearised})}, one can  {impose} the transverse-traceless (TT) gauge {condition}, in which the field equations are
\begin{align}\label{EOM graviton TT}
  &  \Box h^{(\text{TT})}_{\mu \nu } = 2 h^{(\text{TT})}_{\mu \nu}, \nonumber \\
  & \nabla^{\mu}h^{(\text{TT})}_{\mu \nu} =0,~~ ~h^{(\text{TT})\alpha}_{~ \alpha} = 0.
\end{align}
These equations still enjoy invariance under restricted gauge transformations 
\begin{align}\label{gauge_transf_spin2_real_res}
    \delta^{\text{gauge}}_{\text{res}}(A)\, h^{(\text{TT})  }_{\mu \nu}= \nabla_{(\mu} A_{\nu)},
\end{align}
with
\begin{align}\label{gauge_function_spin2_real_res}
   & \Box A_{\nu} = - 3 A_{\nu}, \nonumber\\
   & \nabla^{\nu}A_{\nu} = 0.
\end{align}
The dS Killing vectors $\xi^{\mu}$ act on tensors 
$h_{\mu \nu}$ via the Lie derivative 
\begin{align}\label{Lie_derivative}
 \pounds_{{\xi}}{h}_{\mu \nu}  =~&  \xi^{\rho} \nabla_{\rho} {h}_{\mu \nu} +(\nabla_{\mu} \xi^{\rho})\,{h}_{\rho \nu}+(\nabla_{\nu} \xi^{\rho})\,{h}_{\mu \rho}.
\end{align}
 If $h_{\mu \nu}$ is a solution of eq.~(\ref{EOM graviton TT}) (or (\ref{EOM_graviton general})), then so is $\pounds_{\xi} h_{\mu \nu}$. Since Lie derivatives preserve the Lie brackets between Killing vectors, $ [\pounds_{\xi},\pounds_{\xi '}] h_{\mu \nu} = \pounds_{[\xi , \xi']} h_{\mu \nu}$, the space of solutions of eq.~(\ref{EOM graviton TT}) is a representation space for the dS algebra $so(4,1)$. 
As in the case of the gravitino discussed earlier, a key aspect of our analysis will be how the mode solutions of (\ref{EOM graviton TT}) on global $dS_{4}$  form discrete series UIRs of $so(4,1)$. 
This has been discussed in detail in Refs.~\cite{Yale_Thesis, HiguchiLinearised}. In the following Subsection, we briefly review the main results from these references.
%%%%%%%%%%%%%%%%%%%%%%%%%%%%%%%%%%%%%%%%%%%%%%%%%%%
%%%%%%%%%%%%%%%%%%%%%%%%%%%%%%%%%%%%%%%%%%%%%%%%%%%%%%%%%%%%%%%%%%%%%%%%%%%%%%%
\subsection{Discrete Series UIRs of \texorpdfstring{$so(4,1)$}{so(4,1)} in the  space of graviton modes} \label{Subsec_graviton modes so(4,1)}
 A general classical TT solution $h^{(\text{TT})}_{\mu \nu}$ of (\ref{EOM graviton TT}) can be expressed as a linear combination of physical modes, ${\varphi}^{(phys ,\,\pm L; \,M;K)}_{\mu \nu}$, and pure-gauge modes, $\varphi^{(pg)}_{ \mu \nu}$ \cite{HiguchiLinearised}. Let us present the form of these modes.

$$ \textbf{TT pure-gauge graviton modes}$$
The pure-gauge solutions of (\ref{EOM graviton TT}) are expressed in the form
\begin{align}\label{pure_gauge_modes_2}
     \varphi^{(pg)}_{ \mu \nu} =  \nabla_{(
     \mu}A_{\nu)},
\end{align}
 where  $A_{\nu}$ satisfies (\ref{gauge_function_spin2_real_res}).
 %%%%%%%%%%%%

  $$\textbf{Physical graviton modes}$$
In global coordinates~(\ref{dS_metric}), the physical modes of eq.~(\ref{EOM graviton TT}), with negative ($-2$) and positive ($+2$) helicity, are given by \cite{HiguchiLinearised}
\begin{align}\label{physmodes_negative_spin_2_dS4}
&  {\varphi}^{(phys ,\,- L; \,M;K)}_{t \mu}(t,\bm{\theta}_{3})= 0,\nonumber  ~~~~~~~~ \mu \in \{t, \theta_{3} , \theta_{2}, \theta_{1} \}\\
&{\varphi}^{(phys ,\,- L; \,M;K)}_{\tilde{\mu} \tilde{\nu}}(t,\bm{\theta}_{3})=\left(\frac{{2}(L+1)}{L(L+2)} \right)^{1/2} \, \kappa_{L}(t) \,   \tilde{T}^{(-;L M;K)}_{\tilde{\mu} \tilde{\nu}} (\bm{\theta_{3}}) ,
\end{align}
and
\begin{align}\label{physmodes_positive_spin_2_dS4}
&  {\varphi}^{(phys ,\,+ L; \,M;K)}_{t \mu}(t,\bm{\theta}_{3})= 0,\nonumber  ~~~~~~~~ \mu \in \{t, \theta_{3} , \theta_{2}, \theta_{1} \}\\
&{\varphi}^{(phys ,\,+ L; \,M;K)}_{\tilde{\mu} \tilde{\nu}}(t,\bm{\theta}_{3})=\left(\frac{{2}(L+1)}{L(L+2)} \right)^{1/2} \, \kappa_{L}(t) \,   \tilde{T}^{(+;L M;K)}_{\tilde{\mu} \tilde{\nu}} (\bm{\theta_{3}}) ,
\end{align}
respectively\footnote{There is an extra factor of $2^{1/2}$ in the normalisation factor of the mode functions (\ref{physmodes_negative_spin_2_dS4}) and (\ref{physmodes_positive_spin_2_dS4}) relative to the mode functions in \cite{HiguchiLinearised} because of our different convention for the Klein-Gordon scalar product (\ref{def: KG inner product spin-2}).}, where $\tilde{\mu}, \tilde{\nu}$ are tensor indices on $S^{3}$. The labels $L, M$ and $K$ are angular momentum quantum numbers corresponding to the chain of subalgebras $so(4) \supset$ $so(3)  \supset$ $so(2)$ with $L \in \{2, 3,...   \}$, $M \in \{2,3,...,L\}$ and $K \in \{ -M,...,0,...,M \}$. 
The function $\kappa_{L}(t)$ is given by 
\begin{align}
    \kappa_{L}(t)=& ~2 \left( \sin{\frac{x(t)}{2}}  \right)^{L+2}\, \left( \cos{\frac{x(t)}{2}}  \right)^{-L}\, \left( 1 + \frac{\cos{\left(x(t) \right)}}{L+1}   \right) \nonumber\\
    =& \cosh{t} \left( 1 + \frac{i \sinh{t}}{L+1}   \right)  \left(  \frac{1 - i \sinh{t}}{1 + i \sinh{t}}\right)^{(L+1)/2} , \label{k_L(t)_spin-2} 
\end{align}
where the variable $x(t)$ is defined in (\ref{definition of x(t)}).  {With the conformal time $\tau$ this can be given as
\begin{align}
    \kappa_L(t) & = \frac{1}{\cos\tau}\left( 1+ \frac{i\tan\tau}{L+1}\right)e^{-i(L+1)\tau}.
\end{align}
}
%%%%%%%%%%%%%%%%%%%%%%%%%%%%%%%%%%%%%%%%%%%%%%%%%%%%%%%%%%%%%%%%%%%%%%%%%%%%%%%%%%%%%5
\\
\\
\noindent \textbf{\textit{Symmetric rank-2 tensor spherical harmonics on $S^{3}$}$-$}The $\bm{\theta_{3}}$-dependence of the physical modes in eqs.~(\ref{physmodes_negative_spin_2_dS4}) and (\ref{physmodes_positive_spin_2_dS4}) is given by the rank-2 tensor spherical harmonics on $S^{3}$, $\tilde{T}_{ \tilde{\mu} \tilde{\nu}}^{(\pm; L; M;K)} (\bm{\theta_{3}}) = \tilde{T}_{( \tilde{\mu} \tilde{\nu})}^{(\pm; L; M;K)} (\bm{\theta_{3}})$. These satisfy
\begin{align}\label{spin-2 spherical harmonics S3}
   &\tilde{\Box}\tilde{T}_{ \tilde{\mu} \tilde{\nu}}^{(\pm; L; M;K)} = \left( -L(L+2)+2  \right)\, \tilde{T}_{ \tilde{\mu} \tilde{\nu}}^{(\pm; L; M;K)}, \nonumber \\
   &\tilde{\nabla}^{\tilde{\mu}}\tilde{T}_{ \tilde{\mu} \tilde{\nu}}^{(\pm; L; M;K)} = 0, ~~ \tilde{g}^{\tilde{\mu}  \tilde{\nu}} \tilde{T}_{ \tilde{\mu} \tilde{\nu}}^{(\pm; L; M;K)} =0,
\end{align}
where $\tilde{\Box}= \tilde{g}^{\tilde{\alpha}  \tilde{\beta}}  \tilde{\nabla}_{\tilde{\alpha}} \tilde{\nabla}_{\tilde{\beta}} $ is the Laplace-Beltrami operator on $S^{3}$. The spherical harmonics $\tilde{T}_{ \tilde{\mu} \tilde{\nu}}^{(+; L; M;K)}$ and $\tilde{T}_{ \tilde{\mu} \tilde{\nu}}^{(-; L; M;K)}$ are self-dual and anti-self-dual\footnote{This notion of (anti-)self-duality should not be confused with the notion of (anti-)self-duality defined using $\varepsilon_{\mu \nu \rho \sigma}$ on $dS_{4}$ - see e.g., eqs.~(\ref{anti-self constr gravitino}) and (\ref{anti-self constr graviton}).}, respectively, in the sense that they are eigenfunctions of the duality operator (helicity operator), as \cite{HiguchiLinearised}
\begin{align}\label{duality properties spin-2 S^3}
\frac{1}{L + 1}\tensor{\tilde{\varepsilon}}{_{\tilde{\mu}}^{\tilde{\alpha} \tilde{\beta}}} \tilde{\nabla}_{\tilde{\alpha}}\tilde{T}^{(\pm; L; M;K)}_{\tilde{\beta} \tilde{\nu}} =  \pm \tilde{T}^{(\pm; L; M;K)}_{\tilde{\mu} \tilde{\nu} }.
\end{align}
{(The anti-symmetric part of the left-hand side vanishes
because $\tilde{T}^{(\pm;L;M;K)}_{\tilde{\mu}\tilde{\nu}}$ are transverse
and traceless.)  We note that
\begin{align} 
\tensor{\tilde{\varepsilon}}{_{\tilde{\mu}}^{ \tilde{\alpha}\tilde{\beta}}}\tilde{\nabla}_{\tilde{\alpha}}
\left(\tensor{\tilde{\varepsilon}}{_{\tilde{\beta}}^{\tilde{\lambda}
\tilde{\kappa}}} \tilde{\nabla}_{\tilde{\lambda}}
\tilde{T}_{\tilde{\kappa} \tilde{\nu}}^{(\sigma; L; M;K)}\right)
= (-\tilde{\nabla}_{\tilde{\alpha}}\tilde{\nabla}^{\tilde{\alpha}}
+ 3)T_{\tilde{\mu}\tilde{\nu}}^{(\sigma; L; M;K)},~~\sigma = \pm.
\end{align}
Thus, the duality operator is a square-root of the operator
$-\tilde{\nabla}_{\tilde{\alpha}}\tilde{\nabla}^{\tilde{\alpha}}
+ 3$ on the TT spin-2 tensors on $S^3$.}
{The TT spin-2 tensor spherical harmonics} are normalised with respect to the standard inner product on $S^{3}$ \cite{STSHS}:
\begin{align}\label{normlzn_S3 spin-2}
    \int_{S^{3}} \sqrt{\tilde{g}}&\,d\bm{\theta_{3}} ~\tilde{g}^{\tilde{\mu}    \tilde{\nu}} \tilde{g}^{\tilde{\alpha}    \tilde{\beta}}  ~\tilde{T}_{ \tilde{\mu} \tilde{\alpha}}^{(\sigma; L; M;K)*}(\bm{\theta_{3}})
    ~\tilde{T}_{ \tilde{\nu} \tilde{\beta}}^{(\sigma '; L'; M';K')}(\bm{\theta_{3}}) \nonumber\\
    &= \delta_{\sigma \sigma'}\, \delta_{L L'} \,\delta_{{M}\,   {M}'} \delta_{K K'},
\end{align}
where $\sigma , \sigma' \in \{ +, -  \}$. For each value of $L \in \{2,3,...   \}$, the set $\{ \tilde{T}_{ \tilde{\mu} \tilde{\nu}}^{(+; L; M;K)} \}$ forms a $so(4)$ representation with highest weight given by~\cite{Homma, Yale_Thesis}:
\begin{align}\label{so(4) +weight TT spin-2}
   \vec{f}^{\,(+2)}_{L} = \left( L , +{2} \right),
\end{align}
while the set $\{ \tilde{T}_{ \tilde{\mu} \tilde{\nu}}^{(-; L; M;K)} \}$ forms a $so(4)$ representation with highest weight given by~\cite{Homma, Yale_Thesis}:
\begin{align}\label{so(4) -weight TT spin-2}
   \vec{f}^{\,(-2)}_{L} = \left( L,-{2} \right).
\end{align}
%%%%%%%%%%%%%%%%%%%%%%%%%%%%%%%%%%%%%%%%%%%%%%%%%%%%%%%%%%%%%%%%%%%%%%%%%%%%%%%%%%%%%%%%%%%%%%%%%%%%%%%%%%%%%%%%%%%%%%%%%%%%%%%%%%%%%%%%%%%%%%%%%%%%%%%%%%%%%%%%%%

\noindent \textbf{\textit{Positive and negative frequency}$-$}The physical graviton mode functions~(\ref{physmodes_negative_spin_2_dS4}) and (\ref{physmodes_positive_spin_2_dS4}) are the analogues of positive frequency modes, as for short wavelengths, $ L \gg 1$, they satisfy~\cite{HiguchiLinearised}
\begin{align}
    \frac{\partial}{\partial t}  {\varphi}^{(phys ,\,\pm L; \,M;K)}_{\mu \nu}(t,\bm{\theta}_{3}) \sim - i \frac{L}{\cosh{t}} {\varphi}^{(phys ,\,\pm L; \,M;K)}_{\mu \nu}(t,\bm{\theta}_{3}).
\end{align}
Eqs.~(\ref{EOM graviton TT}) also admits physical TT solutions that are the analogues of negative frequency modes. 
{The negative frequency graviton modes $\varphi^{(phys ,\pm  L; \,M;K)\star}_{\mu \nu}$ are obtained from     
the positive frequency graviton modes  $\varphi^{(phys ,\,\pm L; \,M;K)}_{\mu \nu}$ given by  eqs.~(\ref{physmodes_negative_spin_2_dS4}) and (\ref{physmodes_positive_spin_2_dS4}) by replacing $\kappa_L(t)$ with its complex conjugate.  That is,
\begin{align}
    {\varphi}^{(phys ,\,\pm L; \,M;K)\star}_{\tilde{\mu} \tilde{\nu}}(t,\bm{\theta}_{3})=\left(\frac{{2}(L+1)}{L(L+2)} \right)^{1/2} \, \kappa_{L}^*(t) \,   \tilde{T}^{(\pm;L M;K)}_{\tilde{\mu} \tilde{\nu}} (\bm{\theta_{3}}).
\end{align}
}
\\
\\
\noindent \textbf{Note.} The field strength (\ref{def:graviton_field-strength}) calculated for the positive frequency  modes of helicity $-2$, $ {\varphi}^{(phys ,\,- L; \,M;K)}_{\mu \nu}$, is anti-self-dual, and so is the field strength  for the negative frequency modes of helicity $+2$, ${\varphi}^{(phys ,\,+ L; \,M;K)\star}_{\mu \nu}$. Similarly, the field strength (\ref{def:graviton_field-strength}) calculated for the positive frequency  modes of helicity $+2$, $ {\varphi}^{(phys ,\,+ L; \,M;K)}_{\mu \nu}$, is self-dual, and so is the field strength  for the negative frequency modes of helicity $-2$, ${\varphi}^{(phys ,\,- L; \,M;K)\star}_{\mu \nu}$. See Subsection \ref{Subsec_graviton quantisation} for more details on the mode expansion of the field strength and (anti-)self-duality.
%%%%%%%%%%%%%%%%%

$$ \textbf{Graviton Discrete Series UIRs of}~so(4,1)$$
The two sets of (positive frequency) physical modes $\{ {\varphi}^{(phys ,\,- L; \,M;K)}_{ \mu \nu} \}$ and $\{ {\varphi}^{(phys ,\,+ L; \,M;K)}_{ \mu \nu} \}$ separately form two discrete series UIRs of $so(4,1)$ ~\cite{HiguchiLinearised, Yale_Thesis}. In particular, they form the direct sum: $D^{+}(3 , 2)$ $ \bigoplus D^{-}(3 , {2})$ - see Appendix \ref{App_Classification_UIRs D=4} for details on our notation of the UIRs. 
{It can be seen that each of these two sets of modes forms an UIR as follows.} 
The two sets $\{ {\varphi}^{(phys ,\,- L; \,M;K)}_{ \mu \nu} \}$ and  $\{ {\varphi}^{(phys ,\,+ L; \,M;K)}_{ \mu \nu} \}$ do not mix with each other under any $so(4)$ transformation as they belong to different $so(4)$ representations 
[eqs.~(\ref{so(4) +weight TT spin-2}) and (\ref{so(4) -weight TT spin-2}), respectively]. Also, they do not mix with each other under any dS boost, as under (\ref{dS boost}) they transform as \cite{HiguchiLinearised}: 
\begin{align}\label{infinitesimal dS of spin-2 physical modes}
     \pounds_{B}{\varphi}^{(phys ,\,\pm L; \,M;K)}_{\mu \nu}  =&-\frac{i}{2}\,\sqrt{(L-M+1)(L+M+2)}\,{\varphi}^{(phys ,\,\pm \left(L+1 \right); \,M;K)}_{ \mu \nu}  \nonumber\\
     &-  \frac{i}{2}\sqrt{(L-M)(L+M+1)}\,{\varphi}^{(phys ,\,\pm \left(L-1\right); \,M;K)}_{ \mu \nu} +(\text{pure-gauge}),
     \end{align}
     where the term `(pure-gauge)' is a TT pure-gauge mode (\ref{pure_gauge_modes_2}).
One can thus conclude that $\{ {\varphi}^{(phys ,\,- L; \,M;K)}_{ \mu \nu} \}$ and  $\{ {\varphi}^{(phys ,\,+ L; \,M;K)}_{ \mu \nu} \}$ separately form irreducible representations. As the pure-gauge modes are orthogonal to themselves and to all physical modes with respect to the Klein-Gordon inner product (this inner product will be introduced shortly) \cite{HiguchiLinearised}, the physical modes  form representations with the following equivalence relation: if for any two physical modes, $\varphi^{(1)}_{\mu \nu}$ and $\varphi^{(2)}_{\mu \nu}$, the difference $\varphi^{(1)}_{\mu \nu} -\varphi^{(2)}_{\mu \nu}$ is a linear combination of pure-gauge modes, then $\varphi^{(1)}_{\mu \nu}$ and $\varphi^{(2)}_{\mu \nu}$ belong to the same equivalence class.\footnote{Eq.~(\ref{infinitesimal dS of spin-2 physical modes}) agrees with the expression for the infinitesimal boost matrix elements in the discrete series UIRs of $so(4,1)$ with $\Delta = 3$ and $s=2$ \cite{Schwarz, Ottoson}. See Appendix \ref{App_Classification_UIRs D=4} and Refs.~\cite{Yale_Thesis, Letsios_announce} for the translation between the old and modern notation for the labels of the $so(4,1)$ UIRs.}
%%%%%%%%%%%%%%%%%%%%%%%%
%%%%%%%%%%%%%%%%%%%%%%%%%%%%%%%%%%%%%%%%%%%%%%%%%%%%%%%%%%%%%%%%%%%%%%%%%%%%%%%
These irreducible representations are unitary because the Klein-Gordon inner product:
\begin{align}\label{def: KG inner product spin-2}
     \braket{ \varphi^{(1)}| \varphi^{(2)}}_{KG} = \frac{i}{{4}} \int_{S^{3}} d\bm{\theta_{3}} \sqrt{-g} \left(\varphi^{(1) \mu \nu *} ~\frac{\partial}{\partial t}\varphi^{(2)}_{\mu \nu} - \varphi^{(2)}_{\mu \nu}~\frac{\partial}{\partial t}\varphi^{(1) \mu \nu *}  \right),
\end{align}
is both positive definite (for physical positive frequency modes) and dS invariant, where $\varphi^{(1)}$ and $\varphi^{(2)}$ are any two classical solutions of eqs.~(\ref{EOM graviton TT}). 
The Klein-Gordon inner product is related to the Klein-Gordon current,
\begin{align} \label{KG current}
    J^{\mu}_{KG}\left(\varphi^{(1)}, \varphi^{(2)}\right) = -\frac{i}{{4}}\left(\varphi^{(1) \alpha \beta *}~{\nabla^{\mu}}\varphi^{(2)}_{\alpha \beta} - \varphi^{(2)}_{\alpha \beta}~{\nabla^{\mu}}\varphi^{(1) \alpha \beta *}  \right),~~~\nabla_{\mu}J^{\mu}_{KG}\left(\varphi^{(1)}, \varphi^{(2)}\right) =0,
\end{align}
as 
\begin{align}\label{KG inner product in terms of current}
    \braket{ \varphi^{(1)}| \varphi^{(2)}}_{KG} =  \int_{S^{3}} d\bm{\theta_{3}} \sqrt{-g}~J^{t}_{KG}\left(\varphi^{(1)}, \varphi^{(2)}\right) .
\end{align}
The positive definiteness of the Klein-Gordon inner product in the positive frequency sector - and negative definiteness in the negative frequency sector - has been explicitly verified in Refs.~\cite{HiguchiLinearised, STSHS}, as:
\begin{align}\label{norms of physical modes_2}
    \braket{{\varphi}^{(phys ,\,\sigma L; \,M;K)}|{\varphi}^{(phys ,\,\sigma' L'; \,M';K')}}_{KG} =  \delta_{\sigma \sigma'}\delta_{L L'}   \delta_{MM'} \delta_{KK'},
\end{align}
\begin{align}\label{norms of physical modes_2_neg freq}
  &  \braket{{\varphi}^{(phys ,\,\sigma L; \,M;K)\star}|{\varphi}^{(phys ,\,\sigma' L'; \,M';K')\star}}_{KG} = - \delta_{\sigma \sigma'}\delta_{L L'}   \delta_{MM'} \delta_{KK'}, \nonumber \\
   &  \braket{{\varphi}^{(phys ,\,\sigma L; \,M;K)\star}|{\varphi}^{(phys ,\,\sigma' L'; \,M';K')}}_{KG} =0
\end{align}
with $\sigma, \sigma' \in \{ +, - \}$. 
Also,
\begin{align}
    \braket{\varphi^{(1)}|\varphi^{(pg)}}_{KG} =0,
\end{align}
where  $\varphi^{(1)}_{\mu \nu}$ is any physical or pure-gauge mode, and thus, the pure-gauge modes can be identified with zero as they are orthogonal to all modes, including themselves. Moreover, the anti-hermiticity of the generators (Lie derivatives) is known \cite{HiguchiLinearised, STSHS}, as:
\begin{align} \label{anti-herm_Lie deriv_graviton}
    \braket{\pounds_{\xi}\varphi^{(1)}|\varphi^{(2)}}_{KG} + \braket{\varphi^{(1)}|\pounds_{\xi}\varphi^{(2)}}_{KG}=0,
\end{align}
for any two solutions $\varphi^{(1)}, \varphi^{(2)}$ of eqs.~(\ref{EOM graviton TT}) and any Killing vector $\xi^{\mu}$. To conclude:
\begin{itemize}
    \item The positive frequency physical graviton modes with positive helicity, $\{{\varphi}^{(phys ,\,+ L; \,M;K)}_{\mu \nu} \}$, form the discrete series UIR $D^{+}(\Delta ,s) =D^{+}(3 , {2})$ of $so(4,1)$ - see Appendix \ref{App_Classification_UIRs D=4}. The $so(4)$ content corresponds to the $so(4)$ highest weights~(\ref{so(4) +weight TT spin-2}). The $so(4,1)$-invariant inner product that is positive definite is given by ~(\ref{def: KG inner product spin-2}).

    \item The positive frequency physical graviton modes with negative helicity, $\{{\varphi}^{(phys ,\,- L; \,M;K)}_{\mu \nu} \}$, form the discrete series UIR $D^{-}(\Delta ,s) =D^{-}(3 , {2})$ of $so(4,1)$ - see Appendix \ref{App_Classification_UIRs D=4}. The $so(4)$ content corresponds to the $so(4)$ highest weights~(\ref{so(4) -weight TT spin-2}). The $so(4,1)$-invariant inner product that is positive definite is again given by eq.~(\ref{def: KG inner product spin-2}).
\end{itemize}
The negative frequency modes, $\{{\varphi}^{(phys ,\,- L; \,M;K)\star}_{\mu \nu} \}$ and $\{{\varphi}^{(phys ,\,+ L; \,M;K)\star}_{\mu \nu} \}$, also form the direct sum $D^{+}(3 , 2)$ $ \bigoplus D^{-}(3 , {2})$ of discrete series UIRs of $so(4,1)$, where the positive-definite inner product is given by the negative of the Klein-Gordon inner product. The transformation of the negative frequency modes under the dS boost (\ref{dS boost}) is found by replacing the coefficients on the right-hand side of (\ref{infinitesimal dS of spin-2 physical modes}) with their complex conjugates.

%%%%%%%%%%%%%%%%%%%%%%%%%%%%%%%%%%%%%%%%
\subsection{Conformal-like symmetry for the (real and complex) graviton and UIRs of \texorpdfstring{$so(4,2)$}{so(4,2)}} \label{Subsec_graviton modes so(4,2)}

In this Subsection, we discuss a conformal-like symmetry of the graviton gauge potential generated by the genuine conformal Killing vectors (\ref{V=nabla phi}) of $dS_{4}$, akin to the conformal-like symmetry of the gravitino discussed in Subsections \ref{Subsec_gravitino modes so(4,2)} and \ref{Subsec_gravitino conf-like off-shell}. This symmetry is the dS analogue of the symmetry found for strictly massless gauge potentials on $AdS_{4}$ in the unfolded formalism by Vasiliev \cite{Vasiliev}. We will present new details on how the conformal-like symmetry acts on graviton mode functions on $dS_{4}$ and how these form UIRs of $so(4,2)$. We will also investigate the invariance of the action functional (\ref{complex graviton action}) under conformal-like transformations. Before proceeding to the technical details and mathematical expressions, let us give some details on the outline of this Subsection since there are certain subtleties concerning the reality properties of the graviton - see also \cite{Vasiliev}.

\begin{align*}
   &\textbf{Outline and subtleties concerning the conformal-like symmetries} \\
    &\textbf{of mode solutions, and of graviton field theory: } 
\end{align*}
We will start by discussing the conformal-like transformation for the graviton, $T_{V}h_{\mu \nu}$ [eq.~(\ref{conf-like graviton diff operator compact})], which is a symmetry (a map from solutions to other solutions) for both the full linearised Einstein equations (\ref{EOM_graviton general}) and the graviton equations in the TT gauge (\ref{EOM graviton TT}).  We will show that the conformal-like transformations $T_{V}$ enlarge the symmetry of the field equations from $so(4,1)$ to $so(5,1)$, but the $so(5,1)$ algebra closes up to gauge transformations of the graviton. However, when the transformation $T_{V}$ acts on TT mode solutions it fails to preserve the Klein-Gordon inner product (\ref{def: KG inner product spin-2}), and thus, the graviton mode solutions cannot form UIRs of the enlarged algebra $so(5,1)$. Moreover, the conformal-like transformation fails to be a symmetry of the linearised Einstein-Hilbert action (\ref{real graviton action}).  
%%%%%%%%%%%%%%%%%%%%%%%%%%%%%%%%%%%%%%%%%%%%%%%%%%%%%%%%%%%%%%%%%%%%%%%%%%%%%%%%%%%%
Interestingly,  introducing a modified version of the conformal-like transformation by inserting a factor of $i=\sqrt{-1}$ as $\mathcal{T}_{V} \equiv i T_{V}$ 
[eq.~(\ref{conf-like graviton complex})], we will show the modified transformation is a symmetry of not only the field equations (\ref{EOM_graviton general}) and (\ref{EOM graviton TT}) but also of the Klein-Gordon inner product. The Lie brackets will also  be modified so that the full algebra closes on $so(4,2)$ (up to gauge transformations), instead of $so(5,1)$. Once this modification of the conformal-like transformation has been made, we will show that the positive frequency mode functions, $\{{\varphi}^{(phys ,\,- L; \,M;K)}_{\mu \nu} \}$ and $\{{\varphi}^{(phys ,\,+ L; \,M;K)}_{\mu \nu} \}$, separately form UIRs of the conformal-like algebra $so(4,2)$, as in the case of the gravitino modes discussed in Subsection \ref{Subsec_gravitino modes so(4,2)}. We will also show that there is a hermitian action (\ref{complex graviton action}) functional for the complex graviton $\mathfrak{h}_{\mu \nu}$ which enjoys invariance under the conformal-like transformations $\mathcal{T}_{V} \mathfrak{h}_{\mu \nu}$. 
%%%%%%%%%%%%%%%%%%%%%%%%%%%%%%%%%%%%%%%%%%%%%%%%%%%%%%%%%%%%%%%%%%%%%%%%%%%%%%%%%%%%%%%%%%%%%%%%%%%

\noindent \textbf{\textit{Subtleties concerning the conformal-like symmetries and the reality of the graviton}$-$}At this point, certain subtleties  {need to} be discussed concerning the reality properties of the graviton {related to} the {afore-mentioned} introduction of a factor of $i$. 
{The transformation $\mathcal{T}_V$ cannot act on the real field $h_{\mu\nu}$ 
because of the factor of $i$.  Thus, the field $h_{\mu\nu}$ needs to be 
replaced by a complex field, which we denote by $\mathfrak{h}_{\mu\nu}$.  The negative
frequency part of $\mathfrak{h}_{\mu\nu}$ describes the `anti-particle' whereas
the positive frequency part describes the `particle'. 
 Both parts acquire the 
same phase factor under the transformation $\mathcal{T}_V$.  This means that
the phase factor for the positive frequency modes for the `anti-particle' is the
complex conjugate of the phase factor of the positive frequency modes for
the `particle'.} 
{Some basic properties of the transformations $T_V$ and $\mathcal{T}_V$ are summarised in \textbf{Tables 1 and 2}.}

\begin{table}\label{table 1}
 \begin{center}
 \scalebox{0.85}{
\begin{tabular}{ |c|c|c| } 
 \hline
 \textbf{Conformal-like transformation on modes}& \textbf{Is ($\checkmark$)/ Is not ($\times$) a symmetry of} & \textbf{Algebra}  \\ 
 $T_{V}{\varphi}^{(phys ,\,\pm L; \,M;K)}_{ \mu \nu}$ (\ref{conf-like graviton diff operator compact}) & Field equation (\ref{EOM graviton TT}) $\checkmark$. Inner product (\ref{def: KG inner product spin-2}) $\times$ & $so(5,1)$ \\  
  $\mathcal{T}_{V}{\varphi}^{(phys ,\,\pm L; \,M;K)}_{ \mu \nu}$ (\ref{conf-like graviton complex}) & Field equation (\ref{EOM graviton TT}) $\checkmark$. Inner product (\ref{def: KG inner product spin-2}) $\checkmark$ & $so(4,2)$ \\  
 \hline
\end{tabular}}
\caption{Conformal-like symmetry and graviton mode solutions.}
\end{center}
\end{table}

%%%%%%%%%%%%%%%%%%%%%%%%%%%%%%%%%%%%%%%%%%%%%%%%%%%%%%%%%%%%%%%%%%%%%%%%%%%%%%%%%%%%%%%%%%%%%%%%%%
\begin{table}\label{table 2}
 \begin{center}
 \scalebox{0.85}{
\begin{tabular}{ |c|c| } 
 \hline
 \textbf{Conformal-like transformation of the:}& \textbf{Is ($\checkmark$)/ Is not ($\times$) a symmetry of}   \\ 
 Real graviton, $T_{V}h_{\mu \nu}$ (\ref{conf-like graviton diff operator compact}) & Field equations (\ref{EOM_graviton general}), (\ref{EOM graviton TT}) $\checkmark$. ~Action (\ref{real graviton action}) $\times$ \\  
  Complex graviton, $\mathcal{T}_{V} \mathfrak{h}_{\mu \nu}$ (\ref{conf-like graviton complex}) & Field equations (\ref{EOM_ cmplx graviton general}), (\ref{EOM cmplx graviton TT}) $\checkmark$. ~Action (\ref{complex graviton action}) $\checkmark$  \\  
 \hline
\end{tabular}}
\caption{Conformal-like symmetry: real vs. complex graviton field theory.}
\end{center}
\end{table}
%%%%%%%%%%%%%%%%%%%%%%%%%%%%%%%%%%%%%%%%%%%%%%%%%%%%%%%%%%%%%%%%%%%%%%%%%%%%%%%%%%%%%%%
%%%%%%%%%%%%%%%%%%%%%%%%%%%%%%%%%%%%%%%%%%%%%%%%%%%%%%%%%%%%%%%%%%%%%%%%%%%%%%%%%%%%%%

\noindent \textbf{Note:}  The complex graviton will be relevant in our discussion on SUSY  in Section \ref{Sec_SUSY}. In particular, in our unitary supersymmetric model in Subsection  \ref{Subsec_CHIRAL theory SUSY}, the super-partner of a chiral gravitino is a chiral graviton. Both of these fields have anti-self-dual field strengths, and thus, must be complex \cite{Penrose}.

 \subsubsection{Real graviton field theory, conformal-like transformation and \texorpdfstring{$so(5,1)$}{so(5,1)}}\label{subsubsec_real graviton and so(5,1)}

 The differential operator underlying the conformal-like symmetry transformation is 
\begin{align}\label{conf-like graviton diff operator}
   _{V}D_{ \mu \nu}^{~~~\alpha \beta}  = \frac{1}{2}V^{\rho}  \varepsilon_{\rho \sigma \lambda (\mu}  \left( \delta_{\nu)}^{\beta}~g^{\lambda \alpha}~\nabla^{\sigma}  +\delta_{\nu)}^{\alpha}~g^{\lambda \beta}~\nabla^{\sigma} \right),
\end{align}
where $V^{\rho}$ is any genuine conformal Killing vector (\ref{V=nabla phi}). The conformal-like transformation acts on generic symmetric spin-2 fields, $B_{\mu \nu}$, as\footnote{Note the similarity of the expression (\ref{conf-like graviton diff operator compact}) with the hidden symmetry transformation of the Maxwell gauge potential in Minkowski spacetime, given by equation (41) in \cite{Letsios_Zilches}.}
\begin{align}\label{conf-like graviton diff operator compact}
   T_{V}B_{\mu \nu} \equiv ~ _{V}D_{ \mu \nu}^{~~~\alpha \beta} \, B_{\alpha \beta}  = V^{\rho}  \varepsilon_{\rho \sigma \lambda (\mu}  \nabla^{\sigma}B^{\lambda}_{~ \nu)}.
\end{align}
\noindent \textbf{\textit{Conformal-like invariance of real graviton field equations}$-$}We will first show that the conformal-like transformation $T_{V}h_{\mu \nu}$ for the real graviton is a symmetry of the standard linearised Einstein equations {$H_{\mu\nu}(h) = 0$ [see eq.~ (\ref{EOM_graviton general})].} For the sake of generality, let us work with  {a} symmetric spin-2 field $B_{\mu \nu}$ {which may not obey the linearised Einstein equations.}
Using the expressions (\ref{useful quantities conf-like graviton}), it is easy to show that $ H_{\mu \nu}(T_{V}B)$ 
    can be expressed as
    \begin{align}\label{conf-like covariance of lin Einstein eqs}
  H_{\mu \nu}(T_{V}B)= V^{\rho}\varepsilon_{\rho \sigma \lambda (\mu}  \nabla^{\sigma}H^{\lambda}_{~ \nu)}(B) = T_{V} H_{\mu \nu}(B),
\end{align}
for any symmetric spin-2 field $B_{\mu \nu}$.
This means that if $B_{\mu \nu} = h_{\mu \nu}$ satisfies the linearised Einstein equations, $H_{\mu \nu}(h)=0$ [eq.~(\ref{EOM_graviton general})], then $T_{V}h_{\mu \nu}$ also satisfies the same equations, i.e.\ $H_{\mu \nu}(T_{V}h) = 0$. In other words, the conformal-like transformation (\ref{conf-like graviton diff operator compact}) of the real graviton is a symmetry of the linearised Einstein equations (\ref{EOM_graviton general}). {Furthermore}, as  eqs.~(\ref{useful quantities conf-like graviton}) hold for any symmetric spin-2 field, we can also apply them to the case of the real graviton in the TT gauge $B_{\mu \nu}= h^{(\text{TT})}_{\mu \nu}$. We thus  find that if $h^{(\text{TT})}_{\mu \nu}$ satisfies eqs.~(\ref{EOM graviton TT}), then so does $T_{V}h^{(\text{TT})}_{\mu \nu}$, {i.e.\ $T_{V}h^{(\text{TT})}_{\mu \nu}$ is a TT solution to the linearised Einstein equations.}  Thus, the conformal-like transformation (\ref{conf-like graviton diff operator compact}) is a symmetry of the field equations of the real graviton in the TT gauge [eqs.~(\ref{EOM graviton TT})] - this is also true, of course, for the case of the TT graviton mode solutions which are complex.
\\
\\
%%%%%%%%%%%%%%%%%%%%%%%%%%%%%%%%%%%%%%%%%%%%%%%%%%%%%%%%%%%%%%%%%%%%%%%%%%%%%%%%%%%%
\noindent \textbf{\textit{$\bm{so(5,1)}$ symmetry for real graviton}$-$}Consider a TT solution of (\ref{EOM graviton TT}), $h^{(\text{TT})}_{\mu \nu}$. The structure of the full symmetry algebra, generated by the ten dS isometries (\ref{Lie_derivative}) and the five conformal-like symmetries (\ref{conf-like graviton diff operator compact}), is described by the following commutation relations: 
\begin{subequations}
\begin{equation}\label{so(5,1)_algebra_graviton_1}
   [\pounds_{\xi} , \pounds_{\xi'}] h^{(\text{TT})}_{\mu \nu} =\pounds_{[\xi,\xi']}h^{(\text{TT})}_{\mu \nu},
\end{equation}    
\begin{equation}\label{so(5,1)_algebra_graviton_2}
    [\pounds_{\xi} , T_{V}] h^{(\text{TT})}_{\mu \nu} =T_{[\xi,V]} h^{(\text{TT})}_{\mu \nu},
\end{equation}
\begin{equation}\label{so(5,1)_algebra_graviton_3}
     [T_{V'} , T_{V}] h^{(\text{TT})}_{\mu \nu}= - \pounds_{[V',V]} h^{(\text{TT})}_{\mu \nu}  + \nabla_{(\mu}\left[ - \frac{1}{2} \left(\nabla^\kappa h^{(\text{TT})}_{~\nu)  \sigma} \right)\nabla_\kappa [V',V]^\sigma + [V',V]^\sigma h^{(\text{TT})}_{ \nu)\sigma}\right],
\end{equation}
\end{subequations}
where $\xi^{\mu}, \xi^{'\mu}$ are any two dS Killing vectors, while $V^{\mu}, V^{'\mu}$ are any two genuine conformal Killing vectors (\ref{V=nabla phi}). The commutators (\ref{so(5,1)_algebra_graviton_1})-(\ref{so(5,1)_algebra_graviton_3}) coincide with  the commutation relations of $so(5,1)$ up the field-dependent gauge transformation in (\ref{so(5,1)_algebra_graviton_3}). If the minus sign in front of $ \pounds_{[V',V]}$ in (\ref{so(5,1)_algebra_graviton_3}) gets flipped, then the commutation relations will be the ones of $so(4,2)$. This will be the case when {we consider} the modified conformal-like transformation acting on complex gravitons {later}.
\\
\\
\noindent \textbf{\textit{Non-invariance of Klein-Gordon inner product}$-$}Let $\varphi^{(1)}_{\mu \nu}$ and $\varphi^{(2)}_{\mu \nu}$ be any two TT graviton mode solutions of (\ref{EOM graviton TT}). A straightforward calculation shows that the infinitesimal change of the Klein-Gordon inner product under $T_{V}$ [(\ref{conf-like graviton diff operator compact})] is \textbf{not} zero
\begin{align*} 
    \braket{T_{V}\varphi^{(1)}|\varphi^{(2)}}_{KG} + \braket{\varphi^{(1)}| T_{V}\varphi^{(2)}}_{KG} \neq 0.
\end{align*}
In other words, the conformal-like transformations $T_{V}$ are not anti-hermitian, and thus, the corresponding $so(5,1)$ representation cannot be unitary.  {In fact,} the conformal-like transformations $T_{V}h_{\mu \nu}$ are hermitian.
\\
\\
\noindent \textbf{\textit{Non-invariance of the linearised Einstein-Hilbert action (\ref{real graviton action})}}$-$Using (\ref{conf-like covariance of lin Einstein eqs}), it is easy to show that the variation of the action (\ref{real graviton action}) under $\delta{h}_{\mu \nu} = T_{V}h_{\mu \nu}$, does \textbf{not} vanish,
\begin{align}\label{non-invariance conf-like EH action}
    \delta S_{EH}= -\frac{1}{4}\int d^{4}x \, \sqrt{-g}\, \left( T_{V}h^{\mu \nu}\, H_{\mu \nu}(h) +  h^{\mu \nu}\, H_{\mu \nu}(T_{V}h)\right)  \neq 0.
\end{align}
Also, $\delta S_{EH}$ is \textbf{not} equal to the integral of a total divergence. We conclude that $T_{V}h_{\mu \nu}$ is \textbf{not} a symmetry of the Einstein-Hilbert action.
%%%%%%%%%%%%%%%
%%%%%%%%%%%%%%%%%%%%%%%%%%%%%%%%%%%%%%%%%%%%%%%%%%%%%%%%%%%%%%%%%%%%%%%%%%%%%%%%%%%%%%%%%%%%%%%%%%%%%%%%%%%%%
 \subsubsection{Complex graviton field theory, conformal-like transformation and \texorpdfstring{$so(4,2)$}{so(4,2)}} \label{subsubsec_cmplx graviton and so(4,2)}
Consider a modified version of the conformal-like transformation of the real graviton by introducing a factor of $i$, as $\mathcal{T}_{V} \equiv i T_{V}$. We will also refer to $\mathcal{T}_{V}$ as a conformal-like transformation. As we explained earlier, although $T_{V}$ [eq.~(\ref{conf-like graviton diff operator compact})] can act on both real and complex graviton fields, which we denote as $h_{\mu \nu}$ and $\mathfrak{h}_{\mu \nu}$, respectively, the  transformation $\mathcal{T}_{V}$ acts only on the complex graviton field, as
\begin{align}\label{conf-like graviton complex}
   \mathcal{T}_{V} \mathfrak{h}_{\mu \nu}  \equiv  i T_{V} \mathfrak{h}_{\mu \nu} =i V^{\rho}  \varepsilon_{\rho \sigma \lambda (\mu}  \nabla^{\sigma} \mathfrak{h}^{\lambda}_{~ \nu)},
\end{align}
where $V^{\mu}$ is any genuine conformal Killing vector (\ref{V=nabla phi}). Let us {first} give some details for the complex graviton theory.

\noindent \textbf{\textit{The complex graviton field}$-$}The on-shell complex graviton field satisfies the linearised Einstein equations (\ref{EOM_graviton general}) - with $h_{\mu \nu}$ replaced by $\mathfrak{h}_{\mu \nu}$ - as
\begin{align}\label{EOM_ cmplx graviton general}
     H_{\mu \nu}(\mathfrak{h})=0,
\end{align}
where
\begin{align}\label{lin Einstein operator complex}
     H_{\mu \nu}(\mathfrak{h})\equiv &\, \nabla_{\mu} \nabla_{\alpha}\mathfrak{h}^{\alpha}_{\nu}+ \nabla^{\nu} \nabla_{\alpha}\mathfrak{h}^{\alpha}_{\mu}- \Box \mathfrak{h}_{\mu \nu}+g_{\mu \nu}\,\Box \mathfrak{h}^{\alpha}_{~\alpha} -\nabla_{\mu} \nabla_{\nu}\mathfrak{h}^{\alpha}_{~\alpha} \nonumber \\
    &- g_{\mu \nu}\, \nabla^{\alpha}\nabla^{\beta}\mathfrak{h}_{\alpha \beta} +2\, \mathfrak{h}_{\mu \nu}+g_{\mu \nu}\mathfrak{h}^{\alpha}_{~\alpha}.
\end{align}
Eq.~(\ref{EOM_ cmplx graviton general})  is invariant under complex gauge transformations of the form
\begin{align}\label{gauge_transf_spin2_cmplx}
    \delta^{\text{gauge}}(\mathcal{Z})\, \mathfrak{h}_{\mu \nu}= \nabla_{(\mu} \mathcal{Z}_{\nu)},
\end{align} 
where $\mathcal{Z}_{\nu}$ is an arbitrary complex vector gauge function.  In fact the gauge invariance of the field equation follows from the off-shell property: $ H_{\mu \nu}(  \delta^{\text{gauge}}(\mathcal{Z}) \frak{h} ) = 0$.
In the TT gauge, the field equations for the complex graviton are
\begin{align}\label{EOM cmplx graviton TT}
  &  \Box\mathfrak{h}^{(\text{TT})}_{\mu \nu } = 2 ~\mathfrak{h}^{(\text{TT})}_{\mu \nu}, \nonumber \\
  & \nabla^{\mu} \mathfrak{h}^{(\text{TT})}_{\mu \nu} =0,~~ ~ \mathfrak{h}^{(\text{TT})\alpha}_{~ \alpha} = 0,
\end{align}
and they  enjoy invariance under restricted gauge transformations 
\begin{align}\label{gauge_transf_spin2_cmplx_res}
    \delta^{\text{gauge}}_{\text{res}}(\mathfrak{A})\, \mathfrak{h}^{(\text{TT})  }_{\mu \nu}= \nabla_{(\mu} \mathfrak{A}_{\nu)},
\end{align}
where the complex gauge function satisfies
\begin{align}\label{gauge_function_spin2_cmplx_res}
   & \Box \, \mathfrak{A}_{\nu} = - 3 ~\mathfrak{A}_{\nu}, \nonumber\\
   & \nabla^{\nu}\mathfrak{A}_{\nu} = 0.
\end{align}
\\
\\
\noindent \textbf{\textit{Conformal-like invariance of complex graviton field equations}$-$}As eq.~(\ref{conf-like covariance of lin Einstein eqs}) holds for any (complex or real) symmetric spin-2 field, it follows that if $\mathfrak{h}_{\mu \nu}$ is a solution of the field equation (\ref{EOM_ cmplx graviton general}), then so is $\mathcal{T}_{V}\mathfrak{h}_{\mu \nu} =i \, {T}_{V}\mathfrak{h}_{\mu \nu}$.  {In} the TT gauge it is easy to show {by using (\ref{useful quantities conf-like graviton})} that if $\mathfrak{h}^{(\text{TT})}_{\mu \nu}$ is a solution of 
eq.~(\ref{EOM cmplx graviton TT}), then so is $\mathcal{T}_{V}\mathfrak{h}^{(\text{TT})}_{\mu \nu}$. Thus, $\mathcal{T}_{V}$ is a symmetry of the complex graviton field equations both in their non-gauge-fixed form (\ref{EOM_ cmplx graviton general}) and in the TT gauge (\ref{EOM cmplx graviton TT}).  
%%%%%%%%%%%%%%%%%%%%%%%%%%%%%%%%%%%%%%%%%%%%%%%%%%%%%
\\
\\
\noindent \textit{\textbf{$\bm{so(4,2)}$ symmetry  for complex graviton}}$-$Consider a TT solution of (\ref{EOM cmplx graviton TT}), $\mathfrak{h}^{(\text{TT})}_{\mu \nu}$. The commutators for the full symmetry algebra, generated by the ten dS isometries (\ref{Lie_derivative}) and the five conformal-like symmetries (\ref{conf-like graviton complex}),   can be found by {multiplying $V$ and $V'$ by $i$ and 
replacing $h_{\mu\nu}$ by $\mathfrak{h}_{\mu\nu}$}  in 
eqs.~(\ref{so(5,1)_algebra_graviton_1})-(\ref{so(5,1)_algebra_graviton_3}). 
We find in this manner,
\begin{subequations}
\begin{equation}\label{so(4,2)_algebra_cmplx_graviton_1}
   [\pounds_{\xi} , \pounds_{\xi'}] \mathfrak{h}^{(\text{TT})}_{\mu \nu} =\pounds_{[\xi,\xi']} \mathfrak{h}^{(\text{TT})}_{\mu \nu},
\end{equation}    
\begin{equation}\label{so(4,2)_algebra_cmplx_graviton_2}
    [\pounds_{\xi} , \mathcal{T}_{V}] \mathfrak{h}^{(\text{TT})}_{\mu \nu} =\mathcal{T}_{[\xi,V]} \mathfrak{h}^{(\text{TT})}_{\mu \nu},
\end{equation}
\begin{equation}\label{so(4,2)_algebra_cmplx_graviton_3}
     [\mathcal{T}_{V'} , \mathcal{T}_{V}] \mathfrak{h}^{(\text{TT})}_{\mu \nu}= + \pounds_{[V',V]} \mathfrak{h}^{(\text{TT})}_{\mu \nu}  - \nabla_{(\mu}\left[ - \frac{1}{2} \left(\nabla^\kappa \mathfrak{h}^{(\text{TT})}_{~\nu)  \sigma} \right)\nabla_\kappa [V',V]^\sigma + [V',V]^\sigma \mathfrak{h}^{(\text{TT})}_{ \nu)\sigma}\right],
\end{equation}
\end{subequations}
where $\xi^{\mu}, \xi^{'\mu}$ are any two dS Killing vectors, and $V^{\mu}, V^{'\mu}$ are any two genuine conformal Killing vectors (\ref{V=nabla phi}). The commutators (\ref{so(4,2)_algebra_cmplx_graviton_1})-(\ref{so(4,2)_algebra_cmplx_graviton_3}) coincide with  the commutation relations of $so(4,2)$ up the field-dependent gauge transformation in (\ref{so(4,2)_algebra_cmplx_graviton_3}), as in the gravitino case in (\ref{so(4,2)_algebra_gravitino_1})-(\ref{so(4,2)_algebra_gravitino_3}). 
  {The sign difference between} eqs.~(\ref{so(5,1)_algebra_graviton_3}) and (\ref{so(4,2)_algebra_cmplx_graviton_3}),  {corresponds to} the difference between $so(5,1)$ and $so(4,2)$. 
%%%%%%%%%%%%%%%%%%%%%%%%%%%%%%%%%%%%%%%%%%%%%%
\\
\\
\noindent \textbf{\textit{Conformal-like invariance of Klein-Gordon inner product}}$-$Let $\varphi^{(1)}_{\mu \nu}$ and $\varphi^{(2)}_{\mu \nu}$ be any two TT graviton mode solutions - these are solutions of both (\ref{EOM graviton TT}) and (\ref{EOM cmplx graviton TT}).
 We will show that the Klein-Gordon inner product (\ref{def: KG inner product spin-2}) is invariant under the (complex) conformal-like transformations [eq.~(\ref{conf-like graviton complex})]
\begin{align*}
   \mathcal{T}_{V} \varphi^{(1,2)}_{\mu \nu}   =\,i\, V^{\rho}  \varepsilon_{\rho \sigma \lambda (\mu}  \nabla^{\sigma} \varphi^{(1,2)\lambda}_{~ \nu)}.
\end{align*}
 Let us start by considering  the Klein-Gordon current $  J^{\mu}_{KG}\left(\varphi^{(1)}, \varphi^{(2)}\right)$ [eq.~(\ref{KG current})].
After a straightforward calculation, the infinitesimal change $ \delta_{V}J_{KG}^{\mu}\left(\varphi^{(1)}, \varphi^{(2)}\right)$ under $\mathcal{T}_{V}$ is found to be equal to the divergence of an rank-2 antisymmetric tensor {as follows:} 
\begin{align}
    \delta_{V}J_{KG}^{\mu}\left(\varphi^{(1)}, \varphi^{(2)}\right) &= J_{KG}^{\mu}\left( \mathcal{T}_{V}\varphi^{(1)}, \varphi^{(2)}\right) + J_{KG}^{\mu}\left(\varphi^{(1)}, \mathcal{T}_{V}\varphi^{(2)}\right) \nonumber \\
    &=- \frac{1}{{2}}\nabla_{\sigma} \left( V^{\rho} ~\varphi^{(1)\lambda *}_{~\beta}~\varepsilon_{\rho \lambda \alpha}^{~~~~[\sigma} ~ \nabla^{\mu]}\varphi^{(2) \alpha \beta}  + V^{\rho} ~\varphi^{(2)\lambda }_{~\beta}~\varepsilon_{\rho \lambda \alpha}^{~~~~[\sigma} ~ \nabla^{\mu]}\varphi^{(1) \alpha \beta*}  \right).
\end{align}
It immediately follows that the Klein-Gordon inner product (\ref{def: KG inner product spin-2}) remains invariant under infinitesimal conformal-like transformations, as
\begin{align}
   \delta_{V} \braket{ \varphi^{(1)}| \varphi^{(2)}}_{KG} = \int_{S^{3}} d\bm{\theta_{3}} \sqrt{-g}~ ~  \delta_{V}J_{KG}^{t}\left(\varphi^{(1)}, \varphi^{(2)}\right) =0,
\end{align}
for any genuine conformal Killing vector $V^{\mu}$ (\ref{V=nabla phi}).
This implies  the anti-hermiticity of all five conformal-like generators 
\begin{align}\label{anti-herm_conf-like KG}
    \braket{\mathcal{T}_{V} \varphi^{(1)}| \varphi^{(2)}}_{KG} +\braket{ \varphi^{(1)}| \mathcal{T}_{V} \varphi^{(2)}}_{KG} =0.
\end{align}
Since the requirements of positive-definiteness of the Klein-Gordon inner product and anti-hermiticity of all 15 $so(4,2)$ generators (10 isometries+5 conformal-like symmetries) are satisfied, the physical graviton  modes must form UIRs of not only $so(4,1)$ but also $so(4,2)$, as in the case of gravitino modes discussed in Subsection \ref{Subsec_gravitino modes so(4,2)}. Let us elaborate on this further.

%%%%%%%%%%%%%%%%%%%%%%%%%%%%%%%%%%%%%%%%%%%%%%%%%%%%%%%%%%%%%%%
\subsubsection{UIRs of $so(4,2)$ formed by graviton modes}\label{subsubsec_cmplx graviton and so(4,2) ON MODES}
In the previous Subsections we showed that the space of graviton mode solutions is a representation space for $so(4,2)$ - see the commutation relations (\ref{so(4,2)_algebra_cmplx_graviton_1})-(\ref{so(4,2)_algebra_cmplx_graviton_3}). Here we will show, for the first time, that each of the two positive frequency single-helicity sets of physical graviton modes,  $\{{\varphi}^{(phys ,\,- L; \,M;K)}_{\mu \nu} \}$ and $\{{\varphi}^{(phys ,\,+ L; \,M;K)}_{\mu \nu} \}$, forms a UIR of $so(4,2)$.

 As in the gravitino case discussed in Subsection \ref{Subsec_gravitino modes so(4,2)}, it is sufficient to study the conformal-like transformation generated by one (out of five) genuine conformal Killing vectors, specifically the genuine conformal Killing vector $V^{(0) \mu}$ (\ref{CKV_dS dilation}). From eq.~(\ref{conf-like graviton complex}), we have:
 \begin{align}
     \mathcal{T}_{V^{(0)}}{\varphi}^{(phys ,\,\pm L; \,M;K)}_{\mu \nu} = -i\, \cosh{t}~~\,  \varepsilon_{t \sigma \lambda (\mu}  \nabla^{\sigma} {\varphi}^{(phys ,\,\pm L; \,M;K)\lambda}_{~ \nu)}.
 \end{align}
Using the explicit expressions of the physical modes (\ref{physmodes_negative_spin_2_dS4}) and (\ref{physmodes_positive_spin_2_dS4}), as well as
 $\varepsilon_{t \tilde{\alpha}   \tilde{\beta}  \tilde{\gamma}}= \cosh^{3}{t}~ \tilde{\varepsilon}_{ \tilde{\alpha}   \tilde{\beta}  \tilde{\gamma}}$ and (\ref{duality properties spin-2 S^3}), we find 
 \begin{align}\label{conf-like transf V0 2-}
     \mathcal{T}_{V^{(0)}}{\varphi}^{(phys ,\,- L; \,M;K)}_{\mu \nu} = +i(L+1)\, {\varphi}^{(phys ,\,- L; \,M;K)}_{\mu \nu}
 \end{align}
 and
 \begin{align}\label{conf-like transf V0 2+}
     \mathcal{T}_{V^{(0)}}{\varphi}^{(phys ,\,+ L; \,M;K)}_{\mu \nu} = -i(L+1)\, {\varphi}^{(phys ,+ L; \,M;K)}_{\mu \nu}.
 \end{align}
Thus, from eqs.~(\ref{conf-like transf V0 2-}) and (\ref{conf-like transf V0 2+}), as well as (\ref{so(4,2)_algebra_cmplx_graviton_1})-(\ref{so(4,2)_algebra_cmplx_graviton_3}), it follows that
$\{{\varphi}^{(phys ,\,+ L; \,M;K)}_{\mu \nu} \}$ and $\{{\varphi}^{(phys ,\,- L; \,M;K)}_{\mu \nu} \}$ separately form irreducible representations of $so(4,2)$.
These representations are unitary because the Klein-Gordon inner product (\ref{def: KG inner product spin-2}) is positive definite and all $so(4,2)$ generators are anti-hermitian (\ref{anti-herm_conf-like KG}). Similarly, one can show that the negative frequency modes $\{{\varphi}^{(phys ,\,+ L; \,M;K)\star}_{\mu \nu} \}$ and $\{{\varphi}^{(phys ,\,- L; \,M;K)\star}_{\mu \nu} \}$ separately form UIRs of $so(4,2)$ with positive-definite inner product given by the negative of the Klein-Gordon inner product.

%%%%%%%%%%%%%%%%%%%%%%%%%%%%%%%%%%%%%%%%%%%%%%%%%%%%%%%%%%%%%

\subsubsection{A hermitian action for the complex graviton and its conformal-like invariance} \label{Subsubsec_cmplx graviton conf-like action}

A hermitian action for the complex graviton, which gives rise to the desired Euler-Lagrange equations (\ref{EOM_ cmplx graviton general}), is  
\begin{align}\label{complex graviton action}
    S_{2}= -\frac{1}{4}\int d^{4}x \, \sqrt{-g}\,  \mathfrak{h}^{\dagger}_{\mu \nu}\, H^{\mu \nu}(\mathfrak{h}).
\end{align}
This action is invariant under the gauge transformations in (\ref{gauge_transf_spin2_cmplx}) and their complex conjugates
\begin{align}
    \delta^{\text{gauge}}(\mathcal{Z})\, \mathfrak{h}^{\dagger}_{\mu \nu}= \nabla_{(\mu} \mathcal{Z}^{\dagger}_{\nu)}.
\end{align} 
For later convenience, let us introduce the conserved symplectic current~\cite{Friedman,WaldZoupas}  of the theory as follows. The covariant conjugate momentum {current} density for $\frak{h}_{\lambda \nu}^{\dagger}$ is defined as
\begin{align}
    p^{\mu\nu\lambda} & = \frac{1}{\sqrt{-g}}\frac{\delta S_{2}}{\delta \nabla_\mu \frak{h}_{\nu\lambda}^{\dagger}} \notag \\
    & = - \frac{1}{4}\nabla^\mu \frak{h}^{\nu\lambda} + \frac{1}{4}( 2 g^{\mu(\nu}\nabla_\alpha \frak{h}^{\lambda)  \alpha}  - g^{\nu\lambda}\nabla_\alpha \frak{h}^{\alpha\mu}) \notag \\
    & \quad - \frac{1}{4}g^{\mu(\nu}\nabla^{\lambda)} \frak{h}^{\alpha}_{\alpha} + \frac{1}{4}g^{\nu\lambda} \nabla^\mu \frak{h}^{\alpha}_{\alpha}\,.
\end{align}
Thus, the conserved symplectic current between two complex classical solutions $\frak{h}^{(1)}_{\nu\lambda}$ and $\frak{h}^{(2)}_{\nu\lambda}$ of eq.~(\ref{EOM_ cmplx graviton general}) is 
\begin{align}\label{sym-current}
   J^\mu_{symp}(\frak{h}^{(1)}, \frak{h^{(2)}}) & = i \left( \frak{h}^{(1)*}_{\nu\lambda}~p^{(2)\mu\nu\lambda} - p^{(1)\mu\nu\lambda *} ~\frak{h}^{(2)}_{\nu\lambda} \right) \notag \\
& = -\frac{i}{4} \Bigg(   \frak{h}^{(1)*}_{\nu\lambda}\nabla^\mu \frak{h}^{(2)\nu\lambda}
   -  2{\frak{h}^{(1)\mu*}}_\lambda\nabla_\alpha \frak{h}^{(2)\alpha\lambda}+ \frak{h}^{(1)\beta*}_{\beta} \nabla_\alpha \frak{h}^{(2)\alpha\mu} \notag \\
   & \quad + {\frak{h}^{(1)\mu*}}_\lambda \nabla^\lambda \frak{h}^{(2) \alpha}_{\alpha} - \frak{h}^{(1)\beta *}_{\beta} \nabla^\mu \frak{h}^{(2) \alpha}_{\alpha} \notag \\
   & \quad - \frak{h}^{(2)}_{\nu\lambda}\nabla^\mu \frak{h}^{(1)\nu\lambda*}
   + 2{\frak{h}^{(2)\mu}}_\lambda\nabla_\alpha \frak{h}^{(1)\alpha\lambda *}- \frak{h}^{(2) \beta}_{\beta} \nabla_\alpha \frak{h}^{(1)\alpha\mu *} \notag \\
   & \quad - {\frak{h}^{(2)\mu}}_\lambda \nabla^\lambda \frak{h}^{(1)\alpha *}_{\alpha} + \frak{h}^{(2)\beta}_{\beta} \nabla^\mu \frak{h}^{(1)\alpha *}_{\alpha}  \Bigg)\,,
\end{align}
see, e.g., Refs.~\cite{Faizal,Fewster}.  The  time-independent (pre-)symplectic scalar product between $\frak{h}^{(1)}_{\nu \lambda }$ and $\frak{h}^{(2)}_{\nu \lambda }$ is 
\begin{align} \label{pre-sympl scalar prod}
 \braket{ \frak{h}^{(1)}| \frak{h}^{(2)}}_{symp} =  \int_{S^{3}} d\bm{\theta_{3}} \sqrt{-g}~J^{t}_{symp}\left(\frak{h}^{(1)}, \frak{h}^{(2)}\right).
\end{align}
Importantly, the scalar product (\ref{pre-sympl scalar prod}) is gauge-independent \cite{Fischer,Arms,Faizal,Fewster}, and thus invariant under gauge transformations (\ref{gauge_transf_spin2_cmplx}), as
\begin{align} \label{gauge-invariance symp prod}
 &\braket{ \delta^{\text{gauge}}(\mathcal{Z})\frak{h}^{(1)}| \frak{h}^{(2)}}_{symp}  = \int_{S^{3}} d\bm{\theta_{3}} \sqrt{-g}~J^{t}_{symp}\left(\delta^{\text{gauge}}(\mathcal{Z})\frak{h}^{(1)}, \frak{h}^{(2)}\right) = 0 \nonumber \\
 &   \braket{ \frak{h}^{(1)}| \delta^{\text{gauge}}(\mathcal{Z})\frak{h}^{(2)}}_{symp} = \int_{S^{3}} d\bm{\theta_{3}} \sqrt{-g}~J^{t}_{symp}\left(\frak{h}^{(1)}, \delta^{\text{gauge}}(\mathcal{Z}) \frak{h}^{(2)}\right) = 0.
\end{align}
Indeed it is straightforward to show that
\begin{align}\label{gauge-invariance of sym curent}
J^{\mu}_{symp}\left(\frak{h}^{(1)},  \delta^{\text{gauge}}(\mathcal{Z})\frak{h}^{(2)}\right) = \nabla_{\rho} A^{\rho \mu},
\end{align}
where  $A^{\rho \mu} =  A^{[\rho  \mu]}$ is an anti-symmetric rank-2 tensor depending on the gauge parameter $\mathcal{Z}$ and on  $\frak{h}^{(1)*}_{\mu \nu}$ (and their derivatives) - clearly, a similar statement also holds for $J^{\mu}_{symp}\left( \delta^{\text{gauge}}(\mathcal{Z}) \frak{h}^{(1)},\frak{h}^{(2)}\right)$.
Thus, pure-gauge solutions, i.e.\ complex solutions of the form $\frak{h}_{\mu \nu} = \nabla_{(\mu}  \mathcal{Z}_{\nu)}$ for any $\mathcal{Z}_{\nu}$, are orthogonal to themselves and to all other solutions, with respect to the (pre-)symplectic scalar product (\ref{pre-sympl scalar prod}).
 Note that, if one {imposes} the TT gauge {condition} (\ref{EOM cmplx graviton TT}) for \textbf{both} arguments of the symplectic current (\ref{sym-current}) , then the symplectic current coincides with the Klein-Gordon current (\ref{KG current}). Thus, the (pre-)symplectic scalar product (\ref{pre-sympl scalar prod}) coincides with  the Klein-Gordon inner product (\ref{KG inner product in terms of current}) on the space of TT solutions.
\\
\\
\noindent \textbf{\textit{Conformal-like invariance of the action (\ref{complex graviton action})}}$-$The hermitian action in (\ref{complex graviton action}) is not only invariant under dS transformations but also under conformal-like transformations (\ref{conf-like graviton complex}), unlike the linearised Einstein-Hilbert action (\ref{real graviton action}) for the real graviton which is \textbf{not} invariant under the corresponding conformal-like transformations (\ref{non-invariance conf-like EH action}). The conformal-like invariance of the action (\ref{complex graviton action}) can be readily checked as follows. Computing the variation 
\begin{align}
     \delta_{V}S_{2}= -\frac{1}{4}\int d^{4}x \, \sqrt{-g}\, \left(  \delta_{V}\mathfrak{h}^{\dagger}_{\mu \nu}\, H^{\mu \nu}(\mathfrak{h}) + \mathfrak{h}^{\dagger}_{\mu \nu}\, H^{\mu \nu}(\delta_{V}\mathfrak{h}) \right),
\end{align}
under $\delta_{V} \mathfrak{h}_{\mu \nu} = \mathcal{T}_{V} \mathfrak{h}_{\mu \nu}= i T_{V}\mathfrak{h}_{\mu \nu}$
and $\delta_{V} \mathfrak{h}^{\dagger}_{\mu \nu} = \left(\mathcal{T}_{V} \mathfrak{h}_{\mu \nu}\right)^{\dagger}$, and using (\ref{conf-like covariance of lin Einstein eqs}) with $B_{\mu \nu} = \mathfrak{h}_{\mu \nu}$, we find
\begin{align}
   \delta_{V}S_{2}= \int d^{4}x \, \sqrt{-g} ~\nabla^{\sigma}  \left(\frac{i}{4}~  V^{\rho}   \varepsilon_{\rho  \sigma   \lambda   \gamma} \, h^{\dagger  \lambda}_{\nu}~H^{\gamma 
  \nu}(\mathfrak{h})\right),
\end{align}
which demonstrates the conformal-like invariance of the action $S_{2}$. Notice that  $\left(\mathcal{T}_{V} \mathfrak{h}_{\mu \nu}\right)^{\dagger}$ has an extra minus sign relative to $\mathcal{T}_{V} \mathfrak{h}_{\mu \nu}$ (see (\ref{conf-like graviton complex})), which plays a crucial role in the calculation.
%%%%%%%%%%%%%%%%%%%%%%%%%%%%%%%%%%%%%%%%%%%%%%%%%%%%%%%%%%%%%%%%%%%%%%%%%%%%%%%
\subsection{Quantisation of the chiral graviton field, anti-self-duality constraint, and UIRs in the bosonic Fock space} \label{Subsec_graviton quantisation}
Here we will discuss a particular case of a chiral graviton field, the graviton with anti-self-dual field strength as this will be the superpartner of the chiral gravitino, as discussed in Section \ref{Sec_SUSY}.\footnote{Chiral gravitational tensor perturbations around de Sitter spacetime in terms of Ashtekar variables have been discussed in Refs.~\cite{Magueijo1, Magueijo2}.} The chiral graviton is described by a complex symmetric spin-2 field $\mathfrak{h}_{\mu \nu}$, as the one discussed in \ref{subsubsec_cmplx graviton and so(4,2)}, but with the extra restriction of anti-self-duality on the field strength. In the present Subsection, we will quantise the chiral graviton following Takahashi's method, as we also did for the chiral gravitino in Subsection \ref{Subsec_gravitino quantisation}. 

%%%%%%%%%%%%%%%%%%%%%%%%%%%%%%%%%%%%%%%%%%%%%%%%%%%%%%%%%%%%%%%%%%%%%%55

Following Takahashi's method \cite{Takahashi}, we take as our starting point the field equation  (\ref{EOM cmplx graviton TT}) accompanied by the anti-self-duality constraint on the (complex) graviton field strength
\begin{align}\label{anti-self constr graviton}
   \widetilde{U}_{\alpha \beta \mu   \nu} = -i \, {U}_{\alpha \beta \mu \nu},
\end{align}
where $\widetilde{U}_{\alpha \beta \mu   \nu} = \frac{1}{2}\varepsilon_{\alpha \beta \gamma \delta} {U}^{\gamma   \delta}_{~~~~\mu \nu}$. The complex graviton field strength, which we also call `complex linearised Weyl tensor', is defined as in the case of the real graviton \cite{Kouris}:
\begin{align} \label{def:graviton_field-strength}
    U_{\alpha \beta \mu \nu} =\Big( -\nabla_{\mu}\nabla_{[ \alpha}\mathfrak{h}_{\beta]\nu}-g_{\mu [\alpha} \mathfrak{h}_{\beta] \nu}  \Big)-(\mu \leftrightarrow \nu),
\end{align}
and is invariant under the gauge transformations (\ref{gauge_transf_spin2_cmplx}). The field strength (\ref{def:graviton_field-strength}) has the symmetries of the Riemann tensor. {The anti-self-dual linearised Weyl tensor, i.e.\ the linearised Weyl tensor field that satisfies the anti-self-duality constraint (\ref{anti-self constr graviton}), will be later denoted as $U^{-}_{\alpha   \beta  \mu \nu}$ - see 
eq.~(\ref{def:anti-self-dual Weyl}).} The complex graviton field $\mathfrak{h}_{\mu \nu}$ in (\ref{def:graviton_field-strength}) can, of course, be chosen to be in any gauge without affecting $U_{\alpha \beta \mu \nu}$. {To}  proceed with the quantisation we will choose to work in the TT gauge. If $\mathfrak{h}_{\mu \nu}$ satisfies the field equations (\ref{EOM_ cmplx graviton general}) or (\ref{EOM cmplx graviton TT}), then it is easy to show that
\begin{align} \label{properties of Weyl}
 g^{\alpha \mu}  U_{\alpha \beta \mu \nu} =  g^{\beta \nu}  U_{\alpha \beta \mu \nu} = 0,~~~~~~~\nabla^{\alpha} U_{\alpha \beta \mu \nu}  = \nabla^{\mu} U_{\alpha \beta \mu \nu}  = 0.
\end{align}
As we mentioned earlier, the main objective of Takahashi's method {applied here} is to determine the quantum field operator ${\mathfrak{h}}^{(\text{TT})}_{\mu \nu}$ and the hermitian quantum dS generators $ {{Q}}^{dS}_{{2}}{[\xi]}$ such that the Heisenberg equations of motion are satisfied:
\begin{align} \label{Heisenberg eom_graviton}
    -i \, {\pounds}_{\xi}{\mathfrak{h}}^{\text{(TT)}}_{\mu \nu}(t, \bm{\theta_{3}}) = \left [  {\mathfrak{h}}^{\text{(TT)}}_{\mu \nu}(t, \bm{\theta_{3}})  ,{{Q}}^{dS}_{{2}}{[\xi]}  \right],
\end{align}
{up to pure-gauge TT solutions}, for any dS Killing vector $\xi^{\mu}$. The subscript `${2}$' in ${{Q}}^{dS}_{{2}}{[\xi]}$ has been used to distinguish between the quantum dS generators of the chiral graviton and of the chiral gravitino - see 
eq.~(\ref{Heisenberg eom_gravitino}).
 As an additional requirement, quantum operators representing physical quantities  must  {commute with one another} for spacelike separations (microcausality).
\\
\\
%%%%%%%%%%%%%%%%%%%%%%%%%%%%%%%%%%%%%%%%%%%%%%%%%%%%%%%%%%%
\noindent \textbf{\textit{Mode expansion}}$-$Let us now quantise the chiral graviton field\footnote{The real graviton field on global $dS_{4}$ has been quantised in, e.g., \cite{Higuchi_Instab}.}. From Subsection~\ref{Subsec_graviton modes so(4,1)}, it follows that the components $ {\mathfrak{h}}^{\text{(TT)}}_{t \nu}$ are {nonzero only for pure-gauge modes} (see also Ref.~\cite{HiguchiLinearised}). To isolate the  {physical} degrees of freedom, we fix the gauge completely by imposing the gauge conditions: ${\mathfrak{h}}^{\text{(TT)}}_{t \mu} =0$, and $g^{\tilde{\alpha} \tilde{\mu}} \nabla_{\tilde{\alpha}} {\mathfrak{h}}^{\text{(TT)}}_{\tilde{\mu}  \tilde{\nu}}=0$. We then expand the chiral graviton gauge potential in modes as
\begin{align} \label{mode expansion graviton}
    & {\mathfrak{h}}^{(\text{TT})-}_{t \mu}(t, \bm{\theta_{3}})  = 0, \nonumber\\
    &  {\mathfrak{h}}^{\text{(TT)}-}_{\tilde{\mu}   \tilde{\nu}}(t, \bm{\theta_{3}})  = \sum_{L =2}^{\infty}   \sum_{M=2}^{L} \sum_{K=-M}^{M} \left( {c}^{(-)}_{LMK}{\varphi}^{(phys ,\,- L; \,M;K)}_{\tilde{\mu} \tilde{\nu}}(t, \bm{\theta_{3}})  + {d}^{(+)\dagger}_{LMK}\,{\varphi}^{(phys ,\,+ L; \,M;K)\star}_{\tilde{\mu} \tilde{\nu}} (t, \bm{\theta_{3}}) \right),
\end{align}
where $\mu$ is a tensor index on $dS_{4}$, while $\tilde{\mu}$ and $\tilde{\nu}$ are  tensor indices on $S^{3}$. The superscript `$-$' in  $\mathfrak{h}^{(\text{TT})-}_{\mu \nu}$ refers to the fact that the field strength is anti-self-dual (anti-self-duality is demonstrated below). The non-zero commutators are
\begin{align}
[  {c}^{(-)}_{LMK} , {c}^{(-)\dagger}_{L' M' K'}  ] = \delta_{L   L'}  \delta_{MM'}  \delta_{KK'},~~~~~~[  {d}^{(+)}_{LMK} , {d}^{(+)\dagger}_{L' M' K'}  ] = \delta_{L   L'}  \delta_{MM'}  \delta_{KK'}.
\end{align}
The dS-invariant vacuum is the state $\ket{0}_{2}$ in the Fock space that satisfies:
\begin{align}\label{def:chiral graviton vacuum}
    {c}^{(-)}_{LMK} \ket{0}_{2} =  {d}^{(+)}_{ L M K } \ket{0}_{2}  = 0,
\end{align}
 for all $L,M,K$. Using the Klein-Gordon scalar product (\ref{def: KG inner product spin-2}), we find
\begin{align}\label{crtn and annihil ops in terms of h_mn}
    {c}^{(-)}_{LMK}= \braket{ {\varphi}^{(phys ,\,- L; \,M;K)}|{\mathfrak{h}}^{\text{(TT)}-}}_{KG},~~~~~{d}^{(+)\dagger}_{LMK}= -\braket{ {\varphi}^{(phys ,\,+ L; \,M;K)\star}|{\mathfrak{h}}^{\text{(TT)}-}}_{KG}
\end{align}
[see eqs.~(\ref{norms of physical modes_2}) and (\ref{norms of physical modes_2_neg freq})].
\\
\\
\noindent  \textbf{\textit{Anti-self-duality constraint}}$-$Let us verify that the mode expansion for the chiral graviton (\ref{mode expansion graviton}) is consistent with the anti-self-duality constraint (\ref{anti-self constr graviton}). In other words, we will show that the following field strength
\begin{align} \label{def:anti-self-dual Weyl}
    {U}^{-}_{\alpha \beta \mu \nu}=\Big( -\nabla_{\mu}\nabla_{[ \alpha}{\mathfrak{h}}^{(\text{TT})-}_{\beta]\nu}-g_{\mu [\alpha} {\mathfrak{h}}^{\text{(TT)}-}_{\beta] \nu}  \Big)-(\mu \leftrightarrow \nu),
\end{align}
is anti-self-dual.
For convenience, let us demonstrate this for the component ${U}^{-}_{\tilde{\rho}   \tilde{\gamma} 
 t   \tilde{\nu}}$, where $\tilde{\rho},   \tilde{\gamma}$ and $\tilde{\nu}$ are spatial indices - the calculation for the rest of the components is similar. In the global coordinates (\ref{dS_metric}), we have
 \begin{align}\label{anti-self constr graviton part I}
  {U}^{-}_{\tilde{\rho}   \tilde{\gamma} 
 t   \tilde{\nu}}=\Big( 
-{\frac{\partial\ }{\partial t}}+ 2 \tanh{t} \Big) \tilde{\nabla}_{[\tilde{\rho}}{\mathfrak{h}}^{\text{(TT)}-}_{\tilde{\gamma}] \tilde{\nu}} ,
 \end{align}
where $\tilde{\nabla}_{\tilde{\rho}}$ is the covariant derivative on $S^{3}$. On the other hand, for the $\tilde{\rho}   \tilde{\gamma} 
 t   \tilde{\nu}$-component of the dual field strength, we have
\begin{align}\label{anti-self constr graviton part II}
   \frac{1}{2} \varepsilon_{\tilde{\rho}   \tilde{\gamma} 
 \alpha \beta }~{U}^{-\alpha \beta}_{~~~~~
 t   \tilde{\nu}}=  \varepsilon_{\tilde{\rho}   \tilde{\gamma} 
 t \tilde{\delta} }~{U}^{-t \tilde{\delta}}_{~~~~~
 t   \tilde{\nu}} &=-\cosh{t}~\tilde{\varepsilon}_{\tilde{\rho}\tilde{\gamma} \tilde{\delta}}~ \tilde{g}^{\tilde{\delta} \tilde{\mu}}~{U}^{-}_{t \tilde{\mu}   
 t   \tilde{\nu}} \nonumber \\
 & = -\cosh{t}~ \left( -\frac{\partial^{2}}{\partial t^{2}}  + 2 \tanh{t} + \frac{2}{\cosh^{2}{t}}  \right) \frac{\tilde{\varepsilon}_{\tilde{\rho}\tilde{\gamma}  \tilde{\delta}} }{2} ~\tilde{g}^{\tilde{\delta} \tilde{\mu}}~ {\mathfrak{h}}^{\text{(TT)}-}_{\tilde{\mu} \tilde{\nu}}.
\end{align}
We want to show that eq.~(\ref{anti-self constr graviton part II}) is equal to $-i \, \times $ (\ref{anti-self constr graviton part I}). Indeed, substituting the mode expansion (\ref{mode expansion graviton})  {into} eqs.~(\ref{anti-self constr graviton part I}) and (\ref{anti-self constr graviton part II}), and making use of (\ref{duality properties spin-2 S^3}), one finds that the anti-self-duality constraint (\ref{anti-self constr graviton}) is satisfied, as
\begin{align}
    \frac{1}{2} \varepsilon_{\tilde{\rho}   \tilde{\gamma} 
 \alpha \beta }~{U}^{-\alpha \beta}_{~~~~~
 t   \tilde{\nu}} = -i \,  {U}^{-}_{\tilde{\rho}   \tilde{\gamma} 
 t   \tilde{\nu}}.
\end{align}
This can be similarly verified for the rest of the components of the field strength (\ref{def:anti-self-dual Weyl}).
%%%%%%%%%%%%%%%%%%%%%%%%%%%%%%%%%%%%%%%%%%%%%%%%%%%%%%%%%%%%%%%%%%%%%%%%%%%%%%%%%%%%%%%%%%%%%%%%%%%%%%%%

\noindent \textbf{\textit{Quantum symmetry generators}}$-$The hermitian dS generators for the chiral graviton can be constructed in the standard way \cite{Takahashi,Yale_Thesis}:
\begin{align}\label{def:dS charges graviton}
   {{Q}}^{dS}_{{2}}{[\xi]} =- {i} :\braket{{\mathfrak{h}}^{\text{(TT)}-}| {\pounds}_{\xi}  {\mathfrak{h}}^{\text{(TT)}-}}_{KG}:~,
\end{align}
where $\xi^{\mu}$ is any dS Killing vector.
Here we will give the explicit expression only for the dS boost $\xi^{\mu} = B^{\mu}$ [eq.~(\ref{dS boost})]. Expanding the field in modes (\ref{mode expansion graviton}), and using eqs.~(\ref{infinitesimal dS of spin-2 physical modes}), (\ref{norms of physical modes_2}) and (\ref{norms of physical modes_2_neg freq}), we find:
\begin{align}
    {{Q}}^{dS}_{{2}}{[B]} =  {{Q}}^{dS-}_{{2}}{[B]} +  {{Q}}^{dS+}_{{2}}{[B]},
\end{align}
where
    \begin{align} \label{boost quantum generator_2 pos freq}
     {{Q}}^{dS-}_{{2}}{[B]} =-\frac{1}{2}\sum_{L=2}^{\infty}\sum_{M.K}& \Bigg    ( \sqrt{(L-M+1)(L+M+2)}\,{c}^{(-)\dagger}_{(L+1) M K}{c}^{(-)}_{L M K}\nonumber \\
     &+  \sqrt{(L-M)(L+M+1)}\, {c}^{(-)\dagger}_{(L-1) M K}{c}^{(-)}_{LMK} \Bigg ) \end{align}
and
     \begin{align}\label{boost quantum generator_2 pos freq}
    {{Q}}^{dS+}_{{2}}{[B]} =  - \frac{1}{2}\sum_{L=2}^{\infty} \sum_{M,K}&\Bigg    ( \sqrt{(L-M+1)(L+M+2)}\,{d}^{(+)\dagger}_{(L+1) M K} {d}^{(+)}_{L   M   K} \nonumber \\
    &+  \sqrt{(L-M)(L+M+2)}\, {d}^{(+)\dagger}_{(L-1) M K}{d}^{(+)}_{L M K } \Bigg ).
     \end{align}
     As in the gravitino case, the dS charge has two independent parts: $ {{Q}}^{dS}_{{2}}{[B]} =  {{Q}}^{dS-}_{{2}}{[B]} +  {{Q}}^{dS+}_{{2}}{[B]}$, where $ {{Q}}^{dS-}_{{2}}{[B]}$ and $ {{Q}}^{dS+}_{{2}}{[B]}$ act on the negative- and positive-helicity sectors, respectively. In other words, they generate the two discrete series UIRs of $so(4,1)$, $D^{-}(\Delta= 3,s= 2)$ and $D^{+}(\Delta= 3, s =2)$, respectively.
     The charge ${{Q}}^{dS}_{{2}}{[B]}$  generates the following dS transformations of  creation operators:
     \begin{align}
      \delta_{B} {c}^{(-)\dagger}_{LMK} \equiv    \left[ {c}^{(-)\dagger}_{LMK} , \,  {{Q}}^{dS}_{{2}}{[B]} \right]= \left[ {c}^{(-)\dagger}_{LMK} , \,  {{Q}}^{dS-}_{{2}}{[B]}  \right] 
         = & \frac{1}{2}\sqrt{(L-M+1)(L+M+2)}\,{c}^{(-)\dagger}_{(L+1) M K}\nonumber\\
        &+  \frac{1}{2}\sqrt{(L-M)(L+M+1)}\,{c}^{(-)\dagger}_{(L-1) M K }
     \end{align}
     and
     \begin{align}
       \delta_{B} {d}^{(+)\dagger}_{L M K}\equiv   \left[ {d}^{(+)\dagger}_{LMK} , \,  {{Q}}^{dS}_{{2}}{[B]}  \right]= \left[ {d}^{(+)\dagger}_{LMK} , \,  {{Q}}^{dS+}_{{2}}{[B]}  \right] 
         = & \frac{1}{2}\sqrt{(L-M+1)(L+M+2)}\,{d}^{(+)\dagger}_{(L+1) M K}\nonumber\\
        &+  \frac{1}{2}\sqrt{(L-M)(L+M+1)}\,{d}^{(+)\dagger}_{(L-1) M K}.
     \end{align}
     Using these expressions, it is clear that negative-helicity single-particle states ${c}^{(-)\dagger}_{LMK} \ket{0}_{2}$ transform as the corresponding positive frequency modes (\ref{infinitesimal dS of spin-2 physical modes}), thus furnishing the $so(4,1)$ discrete series UIR $D^{-}(\Delta=3,s= 2)$. Similarly, positive-helicity single-particle states ${d}^{(+)\dagger}_{LMK} \ket{0}_{2}$ furnish the $so(4,1)$ discrete series UIR  $D^{+}(\Delta=3, s=2)$ - see Appendix \ref{App_Classification_UIRs D=4}. It is now straightforward to show that the Heisenberg equation of motion (\ref{Heisenberg eom_graviton}) is satisfied
     \begin{align}
         \left[ {\mathfrak{h}}^{\text{(TT)}-}_{\mu \nu},  {{Q}}^{dS}_{{2}}{[B]}  \right] = -i \, {\pounds}_{B}  {\mathfrak{h}}^{\text{(TT)}-}_{\mu \nu},
     \end{align}
     {module a pure-gauge TT solution}, where $\pounds_{B}$ is the Lie derivative with respect to the dS boost Killing vector $B^{\mu}$ (\ref{dS boost}).

   %%%%%%%%%%%%%%%%%%%%%%%%%%%%%%%%%%%%%%%%%%%%%%%%%%%%%%%%%%%%%%%%%%%%%%%%%

     As in the gravitino case, apart from the ten dS charges we can also construct the five hermitian charges corresponding to the conformal-like symmetry (\ref{conf-like graviton complex}),
     \begin{align}\label{def:conf charges graviton}
     {{Q}}^{conf}_{{2}}{[V]} =- {i} :\braket{{\frak{h}}^{\text{(TT)}-}| \mathcal{T}_{V}  {\frak{h}}^{\text{(TT)}-}}_{KG}: ~,
\end{align}
      such that the Heisenberg equations of motion are again satisfied 
     \begin{align} 
    -i \, \mathcal{T}_{V}{\frak{h}}^{\text{(TT)}-}_{\mu \nu}(t, \bm{\theta_{3}}) = [ ~ {\frak{h}}^{\text{(TT)}-}_{\mu \nu}(t, \bm{\theta_{3}})  ,  {{Q}}^{conf}_{{2}}{[V]}~],
\end{align}
 {module a pure-gauge TT solution}, where $\mathcal{T}_{V}$ is the conformal-like transformation (\ref{conf-like graviton complex}) with respect to any genuine conformal Killing vector $V^{\mu}$ (\ref{V=nabla phi}). An easy way to verify the Heisenberg equations of motion is to focus on the conformal-like symmetry generated by the genuine conformal Killing vector $V^{(0)\mu}$ [eq.~(\ref{CKV_dS dilation})], for which the quantum generator is readily found to be 
      \begin{align} 
    {{Q}}^{conf}_{{2}}{[V^{(0)}]} =\, \sum_{L =2}^{\infty}\sum_{M,K} \left(L +{1}\right)\left({c}^{(-)\dagger}_{L M K}{c}^{(-)}_{ LMK}   - {d}^{(+)\dagger}_{LMK}{d}^{(+)}_{LMK} \right).
     \end{align}
     This quantum generator consists of two independent conformal-like charges, as
 \begin{align} \label{conf-like charge_2 = pos freq + neg freq}
    &{{Q}}^{conf}_{{2}}{[V^{(0)}]} = {{Q}}^{conf-}_{{2}}{[V^{(0)}]} + {{Q}}^{conf+}_{{2}}{[V^{(0)}]}, \nonumber \\
    &{{Q}}^{conf-}_{{2}}{[V^{(0)}]} =  \sum_{L =2}^{\infty}\sum_{M,K} \left(L +{1}\right){c}^{(-)\dagger}_{L M K}{c}^{(-)}_{ LMK} ,\nonumber\\
    &{{Q}}^{conf+}_{{2}}{[V^{(0)}]} = - \sum_{L =2}^{\infty}\sum_{M,K} \left(L +{1}\right){d}^{(+)\dagger}_{L M K}{d}^{(+)}_{ LMK},
     \end{align}
     acting on negative-helicity and positive-helicity states, respectively. 
     Using this expression for ${{Q}}^{conf}_{{2}}{[V^{(0)}]}$, as well as the mode expansion for ${\frak{h}}^{\text{(TT)}-}_{\mu \nu}$ (\ref{mode expansion graviton}), one can readily verify the Heisenberg equations of motion.
     {Finally}, it is also easy to verify that, as in the gravitino case, the chiral graviton vacuum $\ket{0}_{2}$ is also invariant under the whole  $so(4,2)$ symmetry.
%%%%%%%%%%%%%%%%%%%%%%%%%%%%%%%
\\
\\
\noindent \textbf{\textit{Microcausality}}$-$We will show the microcausality of the chiral graviton theory by demonstrating that, for any two spacelike separated points, $(t, \bm{\theta_{3}})$ and $(t', \bm{\theta_{3}}')$, the commutator
 \begin{align}\label{def: chiral Weyl-Weyl commutator}
     \left [ {U}^-_{\mu {\nu} \rho \sigma}(t, \bm{\theta_{3}}), {U}^-_{\alpha' {\beta'} \gamma'   \delta'}(t', \bm{\theta_{3}}')^{\dagger}  \right] 
     \end{align}
     vanishes.
This can be inferred from the standard theory of the real graviton \cite{Higuchi_Instab, HiguchiLinearised}, as follows. Let ${h}^{(\text{TT})}_{\mu \nu}$ be the real graviton  gauge potential that has been completely gauge-fixed as ${{h}}^{\text{(TT)}}_{t \mu} =0$, and $g^{\tilde{\alpha} \tilde{\mu}} \nabla_{\tilde{\alpha}} {{h}}^{\text{(TT)}}_{\tilde{\mu}  \tilde{\nu}}=0$. {This} field can be expanded in terms of the Bunch-Davies mode functions of both helicities (\ref{physmodes_negative_spin_2_dS4}) and (\ref{physmodes_positive_spin_2_dS4}) \cite{Higuchi_Instab, HiguchiLinearised}. The real graviton is related to our chiral graviton potential as 
$${h}^{(\text{TT})}_{\mu \nu}(t, \bm{\theta_{3}}) = {\frak{h}}^{(\text{TT})-}_{\mu \nu}(t, \bm{\theta_{3}}) + \left( {\frak{h}}^{(\text{TT})-}_{\mu \nu} (t, \bm{\theta_{3}}) \right)^{\dagger}.$$
Let also ${U}^{(real)}_{\mu \nu \rho \sigma}$ be the field strength of ${h}^{(\text{TT})}_{\mu \nu}$, i.e.\ real linearised Weyl tensor, which has the symmetries of the Riemann tensor and satisfies eq.~(\ref{properties of Weyl}). The real linearised Weyl tensor can be expressed in terms of the field strength of our chiral graviton as
\begin{align}
   {U}^{(real)}_{\mu \nu \rho \sigma}(t, \bm{\theta_{3}}) = {U}^{-}_{\mu \nu \rho \sigma}(t, \bm{\theta_{3}}) +    {U}^{-}_{\mu \nu \rho \sigma}(t, \bm{\theta_{3}})^{\dagger}, 
\end{align}
where the anti-self-dual part of ${U}^{(real)}_{\mu \nu \rho \sigma}$ is
\begin{align}\label{anti-self-dual part of Weyl}
    {U}^{-}_{\mu \nu \rho \sigma}(t, \bm{\theta_{3}}) = \frac{1}{2}  \left(   {U}^{(real)}_{\mu \nu \rho \sigma}(t, \bm{\theta_{3}}) 
  + i {\widetilde{U}}^{(real)}_{\mu \nu \rho \sigma}(t, \bm{\theta_{3}})\right),
\end{align}
 while its self-dual part is
 \begin{align}\label{self-dual part of Weyl}
    {U}^{-}_{\mu \nu \rho \sigma}(t, \bm{\theta_{3}})^{\dagger} = \frac{1}{2}  \left(   {U}^{(real)}_{\mu \nu \rho \sigma}(t, \bm{\theta_{3}}) 
  - i {\widetilde{U}}^{(real)}_{\mu \nu \rho \sigma}(t, \bm{\theta_{3}})\right) \equiv  {U}^{+}_{\mu \nu \rho \sigma}(t, \bm{\theta_{3}}) .
\end{align}
 Because of the microcausality of the real graviton field  on $dS_{4}$, the Weyl-Weyl commutator between any two causally disconnected points vanishes\footnote{See 
 Refs.~\cite{Kouris, Faizal} for related discussions.}:
\begin{align}\label{real Weyl-Weyl commutator local}
     \left [ {U}^{(real)}_{\mu {\nu} \rho \sigma}(t, \bm{\theta_{3}}), {U}^{(real)}_{\alpha' {\beta'} \gamma'   \delta'}(t', \bm{\theta_{3}}')  \right] =0,~\text{for spacelike separated points}~~(t,\bm{\theta_{3}}), ~(t',\bm{\theta_{3}'}).
     \end{align}
It is now easy to explain how this implies the locality of the commutator (\ref{def: chiral Weyl-Weyl commutator}) that is relevant to the chiral graviton theory. By taking the dual of the Weyl tensor on the left slot of the commutator in (\ref{real Weyl-Weyl commutator local}), we have
\begin{align}\label{real dualWeyl-Weyl commutator local}
     \left [ {\widetilde{U}}^{(real)}_{\mu {\nu} \rho \sigma}(t, \bm{\theta_{3}}), {U}^{(real)}_{\alpha' {\beta'} \gamma'   \delta'}(t', \bm{\theta_{3}}')  \right] =0,~\text{for spacelike separated points}~~(t,\bm{\theta_{3}}), ~(t',\bm{\theta_{3}'}).
     \end{align}
     Then, by adding (\ref{real Weyl-Weyl commutator local}) $+ i \times$(\ref{real dualWeyl-Weyl commutator local}), and with the use of eq.~(\ref{anti-self-dual part of Weyl}), we find
\begin{align}\label{antiself-dualWeyl-Weyl commutator local}
     \left [ {{U}}^{-}_{\mu {\nu} \rho \sigma}(t, \bm{\theta_{3}}), {U}^{(real)}_{\alpha' {\beta'} \gamma'   \delta'}(t', \bm{\theta_{3}}')  \right] =0,~\text{for spacelike separated points}~~(t,\bm{\theta_{3}}), ~(t',\bm{\theta_{3}'}).
     \end{align}
Then, by taking the dual of the Weyl tensor on the right slot of the commutator in (\ref{antiself-dualWeyl-Weyl commutator local}), and working similarly, we find
\begin{align}\label{antiself-dualWeyl-antiself-dualWeyl commutator local}
     \left [ {{U}}^{-}_{\mu {\nu} \rho \sigma}(t, \bm{\theta_{3}}), {U}^{-}_{\alpha' {\beta'} \gamma'   \delta'}(t', \bm{\theta_{3}}')^{\dagger}  \right] =0,~\text{for spacelike separated points}~~(t,\bm{\theta_{3}}), ~(t',\bm{\theta_{3}'}).
     \end{align}
This demonstrates the microcausality of the chiral graviton theory.

{In the quantisation presented above, we started from a complex graviton field and restricted it to its chiral (anti-self-dual) part, which might seem puzzling at first.  However, we could have started from the linearised Einstein-Hilbert action with a {\it real} graviton field 
and defined its anti-self-dual part, which is a complex field, to construct the chiral graviton field.  Then,
the completely gauge-fixed real field and the corresponding linearised Weyl tensor would be identical to, respectively,  
${h}^{(\text{TT})}_{\mu \nu}(t, \bm{\theta_{3}})$ and 
${{U}}^{(real)}_{\mu {\nu} \rho \sigma}(t, \bm{\theta_{3}})$ discussed above.  That is, there would be
no need to start from a complex graviton field if our only purpose was to define the chiral graviton field.  
However, as we shall see,
the SUSY transformation on the real graviton field would be highly non-local: it is a complex (i.e., non-real)
transformation and the anti-self-dual part and its complex conjugate, the self-dual part, transform differently.
In particular, the $U(1)$ transformation, which is part of the superalgebra, assigns the opposite charges to the
self-dual and anti-self-dual gravitons.  In contrast, as we shall see, it is possible to define a simple 
SUSY transformation
on a non-chiral complex graviton field, which can be restricted to a chiral graviton field.  For this reason we
started from a complex graviton field to construct the chiral graviton field.
}

\section{Complex Killing spinors on \texorpdfstring{$dS_{4}$}{dS_4} and their conformal-like symmetry}\label{Subsec_Killing spinors dS4}

 Let us  review the basics about Killing spinors on $dS_{4}$ - see also, e.g., 
 Ref.~\cite{Anous}. Killing spinors, $\epsilon_{+}$ and $\epsilon_{-}$, on $dS_{4}$ satisfy
\begin{align} \label{Killing spinor eqn}
  \nabla_{\mu}\epsilon_{\pm}  = \pm \frac{i}{2} \gamma_{\mu}     \epsilon_{\pm}.
\end{align}
The Killing spinors with the two different signs in eq.~(\ref{Killing spinor eqn}) are related to each other as $\epsilon_{-} = \gamma^{5} \epsilon_{+}$.\footnote{The Dirac adjoint $\bar{\epsilon}_{\pm}$ of a Killing spinor $\epsilon_{\pm}$ satisfies $\nabla_{\mu} \bar{\epsilon}_{\pm}~ {\mp} \tfrac{i}{2}~ \bar{\epsilon}_{\pm}~ \gamma_{\mu}=0$. }
There are no Majorana Killing spinors\footnote{However, eq.~(\ref{Killing spinor eqn}) admits symplectic Majorana Killing spinor solutions.} satisfying 
eq.~(\ref{Killing spinor eqn}) - the explanation is similar to the one for the absence of a Majorana condition in the case of the gravitino, see the passage below eq.~(\ref{def: charge conjugation}). There are four independent complex (Dirac) Killing spinors $\epsilon_{+}$, and four independent $\epsilon_{-}$. The Killing spinors $\epsilon_{+}$ and  $\epsilon_{-}$ form equivalent finite-dimensional (non-unitary) representations of the dS algebra. In what follows, we will \textbf{only use the Killing spinors $\epsilon_{-}$}, and therefore, we will omit the subscript `$-$', denoting them as $\epsilon$. 

The Killing spinors $\epsilon$  form a 4-dimensional  non-unitary representation of $so(4,1)$. The dS generators act on Killing spinors in terms of the Lie-Lorentz derivative
\begin{align}\label{Lie_Lorentz_kil spinor}
 \mathbb{L}_{{\xi}}{\epsilon}  =~&  \xi^{\nu} \nabla_{\nu} {\epsilon} + \frac{1}{4}  (\nabla_{\kappa} \xi_{\lambda})  \gamma^{\kappa \lambda}    {\epsilon},
\end{align}
where $\xi^{\mu}$ is a dS Killing vector. For later convenience, note that using any two Killing spinors, $\epsilon_{1}$ and $\epsilon_{2}$, satisfying eq.~(\ref{Killing spinor eqn}) with the `$-$' sign, one can construct the following  bilinears:
\begin{itemize}
    \item the real Killing vectors 
    \begin{align} \label{real KV from K spinors}
    \xi^{\mu}_{(\epsilon)}=\frac{1}{4}\overline{\epsilon}_{2} \gamma^{5}\gamma^{\mu} \epsilon_{1} - \frac{1}{4}\overline{\epsilon}_{1} \gamma^{5}\gamma^{\mu} \epsilon_{2}=& \frac{1}{4}\overline{\epsilon}_{2} \gamma^{5}\gamma^{\mu} \epsilon_{1}+ \frac{1}{4} \left( \overline{\epsilon}_{2} \gamma^{5}\gamma^{\mu} \epsilon_{1}  \right)^{\dagger} , 
\end{align}
where $\nabla_{\mu}\xi_{(\epsilon)}^{\mu} = 0 $ and $\nabla_{\mu}  \xi_{(\epsilon)\nu}+ {\left(\mu \leftrightarrow \nu\right)} =0$. The factors of $\frac{1}{4}$ in (\ref{real KV from K spinors}) have been inserted for later convenience.

\item the real genuine conformal Killing vectors [see eq.~(\ref{V=nabla phi})]
   \begin{align} \label{real CKV from K spinors}
    V^{\mu}_{(\epsilon)}&=\frac{1}{4} \overline{\epsilon}_{2} \gamma^{\mu} \epsilon_{1}- \frac{1}{4} \overline{\epsilon}_{1} \gamma^{\mu} \epsilon_{2}  = \frac{1}{4}\overline{\epsilon}_{2} \gamma^{\mu} \epsilon_{1}+\frac{1}{4} \left( \overline{\epsilon}_{2} \gamma^{\mu} \epsilon_{1}  \right)^{\dagger} \nonumber \\
   &= \frac{1}{4}\nabla^{\mu}  \left({i}\,\overline{\epsilon}_{2} \epsilon_{1} -{i}\,\overline{\epsilon}_{1} \epsilon_{2}  \right)= \nabla^{\mu} \phi_{V_{(\epsilon)}},~~~~~~\phi_{V_{(\epsilon)}} \equiv \frac{i}{4}\,\overline{\epsilon}_{2} \epsilon_{1} -\frac{i}{4}\,\overline{\epsilon}_{1} \epsilon_{2},
\end{align}
where $\nabla_{\mu} V_{(\epsilon)\nu}  = -g_{\mu \nu} \phi_{V_{(\epsilon)}} = \frac{1}{4}g_{\mu \nu} \nabla^{\alpha}  V_{(\epsilon)\alpha}$. The factors of $\frac{1}{4}$ in (\ref{real CKV from K spinors}) have been inserted for later convenience.
\end{itemize}
%%%%%%%%%%%%%%%%%%%%%%%%%%%%%%%%%%%%%%
The {afore-mentioned} real Killing spinor bilinears will appear in the commutators of  SUSY transformations [eqs.~(\ref{commutator of SUSY on spin-2}) and (\ref{commutator of SUSY on spin-3/2})] in the following Subsections. Complex Killing vectors and complex genuine conformal Killing vectors are given by 
\begin{align}\label{cmplx_Kil_vec_from_KS}
   \xi^{(2,1)\mu}_{\mathbb{C}} = \frac{1}{4} \overline{\epsilon}_{2} \gamma^{5}\gamma^{\mu} \epsilon_{1},~~~~ \xi^{(1,2)\mu}_{\mathbb{C}} = \frac{1}{4} \overline{\epsilon}_{1} \gamma^{5}\gamma^{\mu} \epsilon_{2}= - \left(   \xi^{(2,1)\mu}_{\mathbb{C}}\right)^{*}
\end{align}
and
\begin{align}\label{cmplx_conf_Kil_vec_from_KS}
   V^{(2,1)\mu}_{\mathbb{C}} = \frac{1}{4} \overline{\epsilon}_{2} \gamma^{\mu} \epsilon_{1} \equiv \nabla^{\mu} \phi_{V^{(2,1)}_{\mathbb{C}}}, ~~~~V^{(1,2)\mu}_{\mathbb{C}} = \frac{1}{4} \overline{\epsilon}_{1} \gamma^{\mu} \epsilon_{2} \equiv \nabla^{\mu} \phi_{V^{(1,2)}_{\mathbb{C}}}= - \left(  V^{(2,1)\mu}_{\mathbb{C}} \right)^{*}, 
\end{align}
 respectively. However,  as we will show below,  only their real parts will appear in the commutators (\ref{commutator of SUSY on spin-2}) and (\ref{commutator of SUSY on spin-3/2}) of two SUSY transformations. 

\noindent   \textbf{\textit{Killing spinors and their conformal-like symmetry}}$-$Something that is not widely known, and to the best of our knowledge will be presented here for the first time, is that dS Killing spinors enjoy a conformal-like $so(4,2)$ symmetry akin to the conformal-like symmetry for the graviton and gravitino  discussed earlier. In particular, the Killing spinor equation (\ref{Killing spinor eqn}) is invariant under the following conformal-like transformations:
\begin{align}\label{extra symmetry Killing spinor}
   \mathbb{T}_{V}\,\epsilon=& \gamma^{5} \left( V^{\rho}\nabla_{\rho}\epsilon +\frac{1}{2} \phi_{V}\epsilon \right) , 
\end{align}
where $V_{\mu} = \nabla_{\mu} \phi_{V}$ is any genuine conformal Killing vector (\ref{V=nabla phi}). 
It can be readily verified that if $\epsilon$ satisfies the Killing spinor equation~(\ref{Killing spinor eqn}), then ${\mathbb{T}}_{V}\,\epsilon$ satisfies the same equation. The $so(4,2)$ commutation relations are given by
\begin{subequations}
\begin{equation}
   [\mathbb{L}_{\xi} , \mathbb{L}_{\xi'}] \epsilon =\mathbb{L}_{[\xi,\xi']}\epsilon,
\end{equation}    
\begin{equation}
    [\mathbb{L}_{\xi} , \mathbb{T}_{V}] \epsilon = 
 \mathbb{T}_{[\xi,V]}\epsilon,
\end{equation}
\begin{equation}
     [\mathbb{T}_{V'} , \mathbb{T}_{V}] \epsilon= \mathbb{L}_{[V',V]}\epsilon ,
\end{equation}
\end{subequations}
where $\xi^{\mu}$ and $\xi^{'\mu}$ are any two dS Killing vectors, while $V^{\mu}$ and $V^{'\mu}$ are any two genuine conformal Killing vectors (\ref{V=nabla phi}). Note that the Dirac adjoint of $\mathbb{T}_{V} \epsilon$ (\ref{extra symmetry Killing spinor}) is
\begin{align}\label{extra symmetry Killing spinor adjoi nt}
   \overline{\mathbb{T}_{V}\,\epsilon}=&-  \left( V^{\rho}\nabla_{\rho}\overline{\epsilon} +\frac{1}{2} \phi_{V}   \overline{\epsilon} \right)\gamma^{5} .
\end{align}
%%%%%%%%%%%%%%%%%%%%

\noindent  \textbf{\textit{Explicit expressions for Killing spinors on $dS_{4}$}}$-$Explicit expressions for the Killing spinors $\epsilon(t, \bm{\theta_{3}})$  on global $dS_{4}$ can be found by analytically continuing the Killing spinors on $S^{4}$ - see Ref.~\cite{Letsios} for details on the analytic continuation of spinor eigenfunctions of the Dirac operator from $S^{4}$ to $dS_{4}$.
The line element on the unit $S^{4}$ is 
\begin{align}
    d \Omega_{(4)}^{2} = d \theta_{4}^{2}+\sin^{2}{\theta_{4}} ~ d\Omega^{2},
\end{align}
where $\pi \geq \theta_{4} \geq 0$, and $d \Omega^{2}$ is the line element (\ref{S^3_metric}) of $S^{3}$. It is well-known that one can analytically continue the line element of $S^{4}$ to obtain the line element of global $dS_{4}$ (\ref{dS_metric}) by making the replacement \cite{STSHS}
\begin{align}  \label{replacement S4->dS4}
    \theta_{4}  \rightarrow  \frac{\pi}{2}  - i t.
\end{align}
It is  also known that Killing spinors on $S^{4}$ are eigenfunctions of the Dirac operator with the lowest allowed eigenvalue
%%%%%%%%%%%%%%%%
\footnote{The spinor eigenfunctions of the Dirac operator on $S^{4}$ have two different signs for their eigenvalues: $\slashed{\nabla} \psi_{n} = - i (n +2) \psi_{n}$ and  $\slashed{\nabla} \psi'_{n} = + i (n +2) \psi'_{n}$, where $n=0,1,2,...$ \cite{Camporesi}. The two families of eigenfunctions, $\psi_{n}$ and $\psi'_{n}$, form equivalent representations of $so(5)$ for each fixed $n$. The two families are related to each other as $\psi'_{n} = \gamma^{5} \psi_{n}$. For $n=0$, the spinors $\psi_{0}\equiv \psi$ are Killing spinors on $S^{4}$ satisfying $\nabla_{\mu} \psi = -\frac{i}{2} \gamma_{\mu}\psi$, as we show in the main text. The spinors $\psi'_{0} = \gamma^{5} \psi_{0}$ are also Killing spinors that satisfy $\nabla_{\mu} \psi'_{0} = +\frac{i}{2} \gamma_{\mu}\psi'_{0}$.}
%%%%%%%%%%%
\begin{align}
    \slashed{\nabla} \,{\psi}(\theta_{4}, \bm{\theta_{3}}) = - \,  2 i \, {\psi}(\theta_{4},\bm{\theta_{3}}).
\end{align}
There are four such independent spinor eigenfunctions \cite{Camporesi, K_spinors_spheres} forming a 4-dimensional representation of $so(5)$ with highest weight given by $\tau = (\frac{1}{2}, \frac{1}{2})$ \cite{Camporesi}. It can be easily verified that the spinor eigenfunctions $\psi$  of the Dirac operator satisfy the Killing spinor equation on $S^{4}$, as follows. Define the following vector-spinors on $S^{4}$:
\begin{align}
    {\zeta}_{\mu} \equiv \left( \nabla_{\mu} + \frac{i}{2} \gamma_{\mu}  \right) {\psi}.
\end{align}
It can be shown that these vector-spinors are identically zero by  computing their norm using the standard inner product on $S^{4}$ \cite{Letsios_announce_II}
\begin{align}
\int_{S^{4}} \sin^{3}{\theta_{4}}~\sqrt{ \tilde{g}}~ d\theta_{4} ~ d\bm{\theta_{3}}~~\zeta_{ \mu}(\theta_{4},\bm{\theta_{3}})^{\dagger} ~  ~\zeta^{\mu}(\theta_{4},\bm{\theta_{3}})   ,
\end{align}
where $\sin^{3}{\theta_{4}}~\sqrt{ \tilde{g}}$ is the square root of the determinant of the $S^{4}$ metric, $\tilde{g}$ is the determinant of the $S^{3}$ metric (\ref{S^3_metric}), and $d \bm{\theta_{3}} \equiv d\theta_{3} \, d \theta_{2} \, d \theta_{1}$. The computation of the norm is straightforward and it involves some integration by parts, and one also has to use \cite{Camporesi}
\begin{align}
    \Box  \, \psi = \left( \slashed{\nabla}^{2} + \frac{R}{4} \right) \psi,
\end{align}
where $R= 12$ is the scalar curvature of the unit $S^{4}$. As the inner product on $S^{4}$ is positive definite, the vanishing of the norm implies $\zeta_{\mu} = 0$, and thus
\begin{align}\label{Killing spinor eqn on S^4}
 \nabla_{\mu}   \psi =  - \frac{i}{2} \gamma_{\mu}  {\psi},
\end{align}
which is the Killing spinor equation on $S^{4}$.
In other words, the eigenfunctions  $\psi$   of the Dirac operator with the lowest eigenvalue on $S^{4}$ are Killing spinors - their explicit expressions can be found in \cite{Camporesi, K_spinors_spheres}.

Now, one can use the replacement (\ref{replacement S4->dS4}) to analytically continue the Killing spinor equation (\ref{Killing spinor eqn on S^4}) on $S^{4}$ to  the Killing spinor equation (\ref{Killing spinor eqn}) (with the `$-$' sign) on $dS_{4}$. In particular, making the replacement (\ref{replacement S4->dS4}), the $S^{4}$ Killing spinors $\psi(\theta_{4} , \bm{\theta_{3}})$ are analytically continued to $dS_{4}$ Killing spinors $\epsilon(t, \bm{\theta_{3}})$. In this manner, we find that there are four Killing spinors on $dS_{4}$. Two of them have `positive helicity' and the other two have `negative helicity'. The four Killing spinors on global $dS_{4}$ are given by
\begin{align}\label{dS Killing spinors explicit}
  \epsilon^{(-;q)}(t, \bm{\theta_{3}}) = \begin{pmatrix}
      \cos{\left(  \frac{\pi/2 - i t}{2} \right)} ~\tilde{\epsilon}_{-,q}(\bm{\theta_{3}}) \\
-i \sin{\left(  \frac{\pi/2 - i t}{2} \right) ~\tilde{\epsilon}_{-,q}(\bm{\theta_{3}})}
  \end{pmatrix},  ~~~~\epsilon^{(+;q)}(t, \bm{\theta_{3}}) = \begin{pmatrix}
      i \, \sin{\left(  \frac{\pi/2 - i t}{2} \right)} ~\tilde{\epsilon}_{+,q}(\bm{\theta_{3}}) \\
- \cos{\left(  \frac{\pi/2 - i t}{2} \right) ~\tilde{\epsilon}_{+,q}(\bm{\theta_{3}})}
  \end{pmatrix},
\end{align}
where $\tilde{\epsilon}_{\pm,q}(\bm{\theta_{3}})$ are Killing spinors on the unit $S^{3}$ satisfying
\begin{align} \label{Killing spinor eqn S^3 MAIN TXT}
\tilde{\nabla}_{\tilde{\mu}}\tilde{\epsilon}_{\pm,q}(\bm{\theta_{3}})  = \pm \frac{i}{2} \tilde{\gamma}_{\tilde{\mu}}~\tilde{\epsilon}_{\pm,q}(\bm{\theta_{3}}),
\end{align}
and the meaning of the label $q$ will be explained shortly. The Killing spinors $\epsilon^{(\sigma;q)}(t , \bm{\theta_{3}})$  and $\tilde{\epsilon}_{\sigma,q}(\bm{\theta_{3}})$ ($\sigma = \pm$) will be treated as commuting; Grassmann-odd Killing spinors will be discussed below.
The two `helicity' labels $\pm$ in the dS Killing spinors (\ref{dS Killing spinors explicit}) stem from the Killing spinors on $S^{3}$ and their behaviour under $so(4)$ rotations. In particular, 
on $S^{3}$, there are two independent `positive-helicity' Killing spinors $\tilde{\epsilon}_{+,q}(\bm{\theta_{3}})$ and two independent  `negative-helicity' Killing spinors  $\tilde{\epsilon}_{-,q}(\bm{\theta_{3}})$. The Killing spinors on $S^{3}$  coincide with the spinor eigenfunctions of the Dirac operator on $S^{3}$ with the lowest eigenvalue
\begin{align*} 
\tilde{\slashed{\nabla}} \tilde{\epsilon}_{\pm,q}(\bm{\theta_{3}})  = \pm ~\frac{3}{2}{i}\, \tilde{\epsilon}_{\pm,q}(\bm{\theta_{3}}),
\end{align*}
and their explicit expressions can be found from \cite{Camporesi}.
The label $q = 0,-1$ is a $S^{1}$ angular momentum quantum number, related to $so(2)$ rotations generated by $\partial_{\theta_{1}}$  in the coordinates (\ref{S^3_metric}) - see also eq.~(\ref{Killing spinor eqn S^3}). 
Specifically, $\partial_{\theta_{1}} \tilde{\epsilon}_{\pm,q } = i (q +1/2) \tilde{\epsilon}_{\pm,q }$ since, according to the construction of \cite{Camporesi}, the label $q \in \{ 0,-1\}$ determines the $\theta_{1}$-dependence  for the Killing spinors in the coordinates (\ref{S^3_metric}), as $ \tilde{\epsilon}_{\pm,q}(\bm{\theta_{3}} )\equiv \tilde{\epsilon}_{\pm,q}({\theta_{3}, \theta_{2}, \theta_{1}})$   $\propto e^{i (q+1/2) \theta_{1}}$.   The two Killing spinors    $\{\tilde{\epsilon}_{+,q} \}_{q=0,-1}$ on $S^{3}$ form the 2-dimensional representation of $so(4)$ with highest weight $\tilde{\tau}^{+} = ( \frac{1}{2}, \frac{1}{2})$ \cite{Camporesi}. Similarly, the two Killing spinors  $\{\tilde{\epsilon}_{-,q} \}_{q=0,-1}$ on $S^{3}$ form the 2-dimensional representation of $so(4)$ with highest weight $\tilde{\tau}^{-} = ( \frac{1}{2}, -\frac{1}{2})$. For later convenience,  note that  the scalar quantities  $\tilde{\epsilon}^{\dagger}_{\pm,q}(\bm{\theta_{3}}) \tilde{\epsilon}_{\pm,q'}(\bm{\theta_{3}})$ are constant on $S^{3}$ for any $q,q' \in \{0,-1 \}$ - this is easy to check. Here, we normalise the Killing spinors on $S^{3}$ such that
\begin{align}
 \tilde{\epsilon}^{\dagger}_{\pm,q}(\bm{\theta_{3}}) \tilde{\epsilon}_{\pm,q'}(\bm{\theta_{3}}) = \delta_{qq'} \frac{1}{2 \, \pi^{2}} ,
\end{align}
and thus,
\begin{align}
 \int_{S^{3}} 
  \sqrt{\tilde{g}} d \bm{\theta_{3}}~\tilde{\epsilon}^{\dagger}_{\pm,q}(\bm{\theta_{3}}) \tilde{\epsilon}_{\pm,q'}(\bm{\theta_{3}}) = \delta_{qq'}\frac{1}{2 \pi^{2}}  \times \left( \int_{S^{3}} 
  \sqrt{\tilde{g}} d \bm{\theta_{3}}\right)= \delta_{qq'}.
\end{align}

To sum up, in total, there are four independent  Killing spinors (\ref{dS Killing spinors explicit}) on $dS_{4}$:  
$\epsilon^{(+;-1)}(t, \bm{\theta_{3}})$.   $ \epsilon^{(+;0)}(t, \bm{\theta_{3}})$, $\epsilon^{(-;-1)}(t, \bm{\theta_{3}})$ and $\epsilon^{(-;0)}(t, \bm{\theta_{3}})$. Each of these Killing spinors can be re-expressed in the form of a spacetime-dependent spinorial matrix acting on a constant  spinor $\eta^{(\sigma;q)}$, as in Ref.~\cite{K_spinors_spheres}. To be specific,
\begin{align}\label{dS Killing spinors explicit_ e=S(t,x)n}
 \epsilon^{(\sigma;q)}(t, \bm{\theta_{3}}) =    S(t,\bm{\theta_{3}}) \, \eta^{(\sigma;q)},
\end{align}
where the spinorial matrix $S(t, \bm{\theta_{3}})$ is given by
\begin{align}
   S(t,\bm{\theta_{3}}) =  e^{-\frac{\pi/2 - it}{2} \gamma^{0} } ~  e^{-\frac{   i \, \theta_{3}}{2} \gamma^{03}} ~e^{\frac{    \, \theta_{2}}{2} \gamma^{32}} ~e^{\frac{    \, \theta_{1}}{2} \gamma^{21}},
\end{align}
and the constant spinors are:
\begin{align}
   & \eta^{(-;-1)} = \frac{1+i}{2 \pi}\begin{pmatrix}
       1 \\0\\0\\0
   \end{pmatrix},~~ \eta^{(-;0)} = \frac{-1+i}{2 \pi}\begin{pmatrix}
       0 \\1\\0\\0
   \end{pmatrix}, \\
       & \eta^{(+;-1)}= \frac{-1-i}{2 \pi}\begin{pmatrix}
       0 \\0\\1\\0
   \end{pmatrix} ,~~ \eta^{(+;0)}= \frac{1-i}{2 \pi}\begin{pmatrix}
       0 \\0\\0\\1
   \end{pmatrix}.
\end{align}
As we mentioned earlier, the Killing spinors denoted as $\epsilon^{(\sigma;q)}$  will be treated as commuting, and thus, the constant spinors $\eta^{(\sigma;q)}$ in  
eq.~(\ref{dS Killing spinors explicit_ e=S(t,x)n}) are  also commuting. Grassmann-odd Killing spinors are also expressed as
\begin{align}\label{dS Killing spinors e=S(t,x)n_anti-com}
 \epsilon(t, \bm{\theta_{3}}) =    S(t,\bm{\theta_{3}}) \, \eta,
\end{align}
but now $\eta$ is a Grassmann-odd constant spinor parameter.
%%%%%%%%%%%%%%%%%%%%%%%%%%%%%%%%%%%%%%%%%%%%%%%%%%%%%%%%%%%%%%%%%%%%%%%
 \section{Unitary rigid SUSY for the supermultiplet of the chiral graviton and  chiral gravitino} \label{Sec_SUSY}

\subsection{Non-unitary SUSY representation for complex (non-chiral)  graviton and  gravitino} \label{Subsec_non-chiral theory SUSY}

In this Subsection, we will start by demonstrating that the multiplet consisting of the complex graviton and the complex gravitino   on $dS_{4}$, each with 2 complex propagating degrees of freedom, carries a  non-unitary representation of global SUSY. Then, in Subsection \ref{Subsec_CHIRAL theory SUSY}, we will specialise to the case where both the graviton and the gravitino are chiral - i.e.\ their corresponding field strengths are anti-self-dual - and we will show that the supermultiplet consisting of these two fields  carries a representation of global SUSY which is unitary. 

As we show below, the SUSY transformations for the supermultiplet of the complex graviton $\frak{h}_{\mu \nu}$ and the Dirac gravitino $\Psi_{\mu}$ on $dS_{4}$ are:
\begin{align}
   & \delta^{susy}({\epsilon})\Psi_{\mu}=\frac{1}{4}  \left( i\, \frak{h}_{\mu \sigma}\gamma^{\sigma}+\nabla_{\lambda}  \frak{h}_{\mu \sigma} \gamma^{\sigma  \lambda}      \right)\epsilon, \label{SUSY_transf_spin3/2->spin2}\\
   & \delta^{susy}({\epsilon}) \frak{h}_{\mu \nu}=\frac{\overline{\epsilon}}{2}\gamma^{5}  \left(  \gamma_{\mu}  \Psi_{\nu}+\gamma_{\nu}\Psi_{\mu}    \right),\label{SUSY_transf_spin2->spin3/2}
\end{align}
where  $\epsilon$ is an anti-commuting complex Killing spinor satisfying eq.~(\ref{Killing spinor eqn}) with the `$-$' sign. SUSY transformations with commuting Killing spinors will be also used when {we consider}  their action on mode solutions in Subsections \ref{subsubsec_unitary SUSY modes} and \ref{subsubsec_unitary SUSY QFT}.  
These transformations are gauge invariant.  That is, if we consider pure-gauge solutions,
\begin{align}
\Psi_\mu^{(pg)} & = \left( \nabla_\mu + \frac{i}{2}\gamma_\mu\right)X,\\
\mathfrak{h}_{\mu\nu}^{(pg)} & = \nabla_{(\mu}\mathcal{Z}_{\nu)}, 
\end{align}
then
\begin{align}
    (\delta^{susy}(\epsilon)\Psi_{\mu})^{(pg)}
    & = \frac{1}{8}\left(\nabla_\mu + \frac{i}{2}\gamma_\mu\right)
    \{ [2i\mathcal{Z}_\sigma\gamma^\sigma
    + (\nabla_\lambda\mathcal{Z}_\sigma)\gamma^{\sigma\lambda}]\epsilon \},\\
    (\delta^{susy}(\epsilon)\mathfrak{h}_{\mu\nu})^{(pg)}
    & = \nabla_{(\mu}(\overline{\epsilon}\gamma^5\gamma_{\nu)}X).
\end{align}

It is easy to find the Dirac conjugate of $\delta^{susy}({\epsilon})\Psi_{\mu}$ and the hermitian conjugate of $\delta^{susy}({\epsilon}) \frak{h}_{\mu \nu}$, as:
\begin{align*}
   & \overline{\delta^{susy}({\epsilon})\Psi}_{\mu}=-\frac{\overline{\epsilon}}{4}  \left( -i\, \frak{h}^{\dagger}_{\mu \sigma}\gamma^{\sigma}+\nabla_{\lambda}  \frak{h}^{\dagger}_{\mu \sigma} \gamma^{\sigma  \lambda}      \right), \\
   & \left(\delta^{susy}({\epsilon}) \frak{h}_{\mu \nu}\right)^{\dagger}=\frac{1}{2}  \left(    \overline{\Psi}_{\nu}\gamma_{\mu}+ \overline{\Psi}_{\mu} \gamma_{\nu}   \right) \gamma^{5}\epsilon.
\end{align*}
Let us emphasise that the SUSY transformations (\ref{SUSY_transf_spin3/2->spin2}) and (\ref{SUSY_transf_spin2->spin3/2}) are relevant to two different supersymmetric theories:
\begin{itemize}
    \item The first corresponds to the theory of a complex graviton and a complex gravitino, each with two complex propagating {helicity} degrees of freedom. \ This theory is non-unitary as it involves a gravitino field containing all of its propagating helicity degrees of freedom which leads to the appearance of negative-norm states, as explained in Subsection \ref{Subsec_gravitino quantisation}. 

    \item 
The second theory is {obtained from} the first
{by a simple projection}, and is our theory of interest. It consists of a chiral graviton and a chiral gravitino, each with one complex propagating {helicity} degree of freedom (the field strength of each gauge potential is anti-self-dual). {For the unitarity of this theory it is crucial} to demonstrate that the SUSY transformations are consistent with the anti-self-duality constraint [eqs.~(\ref{anti-self constr gravitino}) and (\ref{anti-self constr graviton})]. In other words, we have to show that the anti-self-dual gravitino field strength (\ref{mode expansion 3/2 field strength}) transforms only into the anti-self-dual linearised Weyl tensor (\ref{def:anti-self-dual Weyl}) and vice versa - see Subsection \ref{subsubsec_duality+SUSY}. This means that gravitons with helicity $-2$ ($+2$) transform into gravitini with helicity $- \frac{3}{2}$ ($+\frac{3}{2}$) and vice versa. Once the compatibility  of the SUSY transformations with the anti-self-duality constraint has been verified (this can be done only on-shell), one can rewrite the SUSY transformations (\ref{SUSY_transf_spin3/2->spin2}) and (\ref{SUSY_transf_spin2->spin3/2}) in a form that refers explicitly to the supermultiplet of a chiral graviton and a chiral gravitino, as
\begin{align}
   & \delta^{susy}({\epsilon})\Psi^{-}_{\mu}=\frac{1}{4}  \left( i\, \frak{h}^{-}_{\mu \sigma}\gamma^{\sigma}+\nabla_{\lambda}  \frak{h}^{-}_{\mu \sigma} \gamma^{\sigma  \lambda}      \right)\epsilon, \label{SUSY_transf_spin3/2->spin2 chiral}\\
   & \delta^{susy}({\epsilon}) \frak{h}^{-}_{\mu \nu}=\frac{\overline{\epsilon}}{2}\gamma^{5}  \left(  \gamma_{\mu}  \Psi^{-}_{\nu}+\gamma_{\nu}\Psi^{-}_{\mu}    \right),\label{SUSY_transf_spin2->spin3/2 chiral}
\end{align}
where $\frak{h}_{\mu \nu}^{-}$ and $\Psi_{\mu}^{-}$ are the chiral graviton and gravitino gauge potentials\footnote{See eqs.~(\ref{mode expansion graviton}) and (\ref{mode expansion gravitino}), respectively, for the mode expansion of the completely gauge-fixed version of the chiral gauge potentials.}.
The \textbf{main result} of this paper is that the supermultiplet consisting of the chiral graviton and the chiral gravitino $\left(\frak{h}^{-}_{\mu \nu}, \Psi^{-}_{\mu} \right)$, with their corresponding field strengths satisfying the anti-self-duality constraints (\ref{anti-self constr graviton}) and (\ref{anti-self constr gravitino}), respectively, forms a unitary representation of global SUSY that is also unitarily realised on the QFT Fock space. 

\noindent \textbf{Note.} One can instead consider the chiral supermutliplet $\left(\frak{h}^{+}_{\mu \nu}, \Psi^{+}_{\mu} \right)$, with  corresponding field strengths being self-dual instead of anti-self-dual. This theory also realises a unitary representation of global SUSY where the SUSY transformations are given by (\ref{SUSY_transf_spin3/2->spin2 chiral}) and (\ref{SUSY_transf_spin2->spin3/2 chiral}) with $\frak{h}^{-}_{\mu \nu}$ and $\Psi^{-}_{\mu}$ replaced by $\frak{h}^{+}_{\mu \nu}$ and $\Psi^{+}_{\mu}$, respectively. The two supermutliplets $\left(\frak{h}^{-}_{\mu \nu}, \Psi^{-}_{\mu} \right)$ and $\left(\frak{h}^{+}_{\mu \nu}, \Psi^{+}_{\mu} \right)$ separately form unitary representations of global SUSY in $dS_{4}$. Although in this paper we  show the unitarity of the supermultiplet  $\left(\frak{h}^{-}_{\mu \nu}, \Psi^{-}_{\mu} \right)$,  the unitarity of the supermultiplet $\left(\frak{h}^{+}_{\mu \nu}, \Psi^{+}_{\mu} \right)$ can be shown in the same way. However, if the two theories are combined together to form the supermultiplet $\left(\frak{h}_{\mu \nu}, \Psi_{\mu} \right)$ with $\frak{h}_{\mu \nu}=\frak{h}^{-}_{\mu \nu}+ \frak{h}^{+}_{\mu \nu}$ and $\Psi_{\mu} = \Psi^{-}_{\mu}  +   \Psi^{+}_{\mu}$, then the resulting theory would be non-unitary because, in this case, the gravitino field contains all of its helicities, giving rise to negative norms in the QFT Fock space - see Subsection \ref{Subsec_graviton quantisation}.
\end{itemize}
The `chiral' SUSY transformations (\ref{SUSY_transf_spin3/2->spin2 chiral}) and (\ref{SUSY_transf_spin2->spin3/2 chiral}), which are the transformations relevant to our theory of interest, are a special case of the initial {non-chiral} SUSY transformations (\ref{SUSY_transf_spin3/2->spin2}) and (\ref{SUSY_transf_spin2->spin3/2}). We will show that the latter {non-chiral} transformations are symmetries at the level of both the hermitian action (\ref{lin_SUGRA_action}) of the theory and the field equations. On the other hand, the theory that contains {only} a chiral graviton and a chiral gravitino has no local action principle, as it is not possible to split the helicities in a local way at the level  of the action. However, the `chiral' SUSY transformations (\ref{SUSY_transf_spin3/2->spin2 chiral}) and (\ref{SUSY_transf_spin2->spin3/2 chiral}) are symmetries at the level of the equations of motion - we will show that this follows from the invariance of the equations of motion under the (non-chiral) SUSY transformations (\ref{SUSY_transf_spin3/2->spin2}) and (\ref{SUSY_transf_spin2->spin3/2}). In other words, we will show that the  `chiral graviton-chiral gravitino' supermultiplet $(\frak{h}^{-}_{\mu \nu},  \Psi^{-}_{\mu})$ carries a representation of SUSY. The unitarity of this representation will be demonstrated in the next Subsection. 

Let  us start discussing the general theory of a complex graviton and gravitino, each with two complex propagating {helicity} degrees of freedom, and specialise to our chiral theory later.
%%%
%%%%%%%%%%%%%%%%%%%%55

\subsubsection{SUSY invariance of non-gauge-fixed field equations}

Let $\frak{h}_{\mu \nu}$ and $\Psi_{\mu}$ be complex off-shell field configurations, and let us consider the differential operators appearing in the field equations, $H_{\mu \nu}(\frak{h})$ [eq.~(\ref{lin Einstein operator complex})] and $\mathcal{R}(\Psi)$ [eq.~(\ref{RS_operator})], respectively, acting on the off-shell fields.
After a straightforward, but lengthy, off-shell calculation, one can show that $H_{\mu \nu}(\frak{h})$ and $\mathcal{R}(\Psi)$ transform into each other under the SUSY transformations~(\ref{SUSY_transf_spin3/2->spin2}) and (\ref{SUSY_transf_spin2->spin3/2}), as
\begin{align}\label{SUSY_transf_EOM_spin2->spin3/2}
  \delta^{susy}({\epsilon}) H^{\mu \nu}(\frak{h})  \equiv  H^{\mu \nu}(\delta^{susy}({\epsilon})\frak{h}) &=~\overline{\epsilon}\gamma^{5} \Big( {\frac{5}{2}i\,}\gamma^{(\mu}+\nabla^{(\mu}-\gamma^{(\mu}\slashed{\nabla} \Big)\mathcal{R}^{\nu)}(\Psi) \nonumber\\
    &=~\overline{\epsilon}\gamma^{5} \Big({\frac{5}{2}i\,} \gamma^{(\mu}+\gamma^{\lambda (\mu}\nabla_{\lambda}\Big)\mathcal{R}^{\nu)}(\Psi),
\end{align}
and
\begin{align}\label{SUSY_transf_EOM_spin3/2->spin2}
   \delta^{susy}({\epsilon}) \mathcal{R}^{\mu}(\Psi) \equiv \mathcal{R}^{\mu}(\delta^{susy}(\epsilon)\Psi)  =~\frac{1}{4}\gamma_{\alpha}\epsilon \,H^{\mu \alpha}(\frak{h}).
\end{align}
(These equations hold for both commuting and anti-commuting Killing spinors.)
This shows that the solution spaces of equations $\mathcal{R}^{\mu}(\Psi)=0$ 
[eq.~(\ref{RS_eqn imag mass})] and  $H^{\mu \nu}(\frak{h}) =0$ [eq.~(\ref{EOM_ cmplx graviton general})] transform into each other under the SUSY transformations (\ref{SUSY_transf_spin3/2->spin2}) and (\ref{SUSY_transf_spin2->spin3/2}). Thus, the supermultiplet of the complex graviton and the complex gravitino carries a representation of global SUSY.  As we mentioned earlier, this SUSY representation is bound to be non-unitary, but  
{unitarity will be achieved} by restricting to the `chiral graviton-chiral gravitino' supermultiplet in Subsection \ref{Subsec_CHIRAL theory SUSY}.
%%%%%%%%%%%%%%%%%%%%%%%%%%%%%%%%%%%%%%%%%%%%%%%%%%%%%%%%%%%%%%%%%%%%%%%%%%%%%%%%%%%%%%%5

\subsubsection{SUSY invariance of the hermitian action and supercurrents}

The hermitian action for the theory consisting of the complex graviton and Dirac gravitino is given by the sum of the free  actions (\ref{gravitino action}) and (\ref{complex graviton action}):
\begin{align}\label{lin_SUGRA_action}
    S&=~S_{2}+S_{\frac{3}{2}} =~ \int d^{4}x \, \sqrt{-g}\, \left( -\frac{1}{4} \frak{h}^{\dagger}_{\mu \nu}   \, H^{\mu \nu}(\frak{h})- \overline{\Psi}_{\mu} {\gamma}^{5} \mathcal{R}^{\mu}(\Psi)\right).
\end{align}
It is  useful to show that the action (\ref{lin_SUGRA_action}) is SUSY-invariant, as this will allow us to find the conserved Noether currents associated with SUSY.  Varying the action (\ref{lin_SUGRA_action}) under $\delta^{susy}(\epsilon) \Psi_{\mu}$ and $(\delta^{susy}(\epsilon) \frak{h}_{\mu \nu})^{\dagger}$, we find
\begin{align}
    \delta S=&~ \int d^{4}x \, \sqrt{-g}\, \left( -\frac{1}{4} \left(\delta^{susy}(\epsilon)\frak{h}_{\mu \nu} \right)^\dagger   \, H^{\mu \nu}(\frak{h})- \overline{\Psi}_{\mu} {\gamma}^{5} ~\mathcal{R}^{\mu}(\delta^{susy}(\epsilon)\Psi) \right)=0,
\end{align}
where $\mathcal{R}^{\mu}(\delta^{susy}(\epsilon) \Psi)$ is given by 
eq.~(\ref{SUSY_transf_EOM_spin3/2->spin2}). Also, varying the action under
$\overline{\delta^{susy}(\epsilon) \Psi}_{\mu}$ and $\delta^{susy}(\epsilon) \frak{h}_{\mu \nu}$, we find that $\delta S$ is equal to the integral of a total divergence, as
\begin{align}
    \delta S=&~ \int d^{4}x \, \sqrt{-g}\,  \left( -\frac{1}{4} \frak{h}^{\dagger}_{\mu \nu}   \, H^{\mu \nu}(\delta^{susy}(\epsilon)\frak{h})- \overline{\delta^{susy}(\epsilon) \Psi}_{\mu} \,{\gamma}^{5} \mathcal{R}^{\mu}(\Psi)\right) \nonumber \\
    =&~ \int d^{4}x \, \sqrt{-g}\,~  \nabla_{\lambda} \left(\frac{\overline{\epsilon}}{4} \gamma^{5}   \gamma^{\sigma  \lambda}  \mathcal{R}^{\nu}(\Psi)   ~\frak{h}^{\dagger}_{\nu \sigma}\right),
\end{align}
where $H^{\mu \nu}(\delta^{susy}(\epsilon)\frak{h})$ is given by 
eq.~(\ref{SUSY_transf_EOM_spin2->spin3/2}).

The covariantly conserved  Noether vector currents arising from  the SUSY invariance of the action are easily found as
\begin{align}\label{def: SUSY Noether currents}
  & \left(\mathcal{J}^{\mu }_{(\epsilon)}(\frak{h}, \Psi)   \right)^{\dagger}= \frac{i}{4} \overline{\Psi}_{\kappa}   \gamma^{\kappa   \mu \sigma} \gamma^{5}  \left( i \frak{h}_{\sigma  \nu}\gamma^{\nu} + \gamma^{\nu \rho}  \nabla_{\rho}\frak{h}_{\sigma  \nu}    \right)\epsilon = i ~\overline{\Psi}_{\kappa} \gamma^{\kappa \mu \sigma} \gamma^{5}~ \delta^{susy}(\epsilon)\Psi_{\sigma} , \nonumber \\
  &  \mathcal{J}^{\mu}_{(\epsilon)}(\frak{h},\Psi)= \frac{i}{4} \overline{\epsilon}  \left(i \frak{h}^{\dagger}_{\sigma \nu} \gamma^{\nu}- \gamma^{ \nu  \rho} \nabla_{\rho}  \frak{h}^{\dagger}_{\sigma \nu} \right)\, \gamma^{\sigma \mu \kappa} \gamma^{5}\Psi_{\kappa} = i~ ~ \overline{\delta^{susy}(\epsilon)\Psi}_{\sigma}~ \gamma^{\sigma \mu \kappa}   \gamma^{5}  \Psi_{\kappa}.
\end{align}
The Grassmann-odd fermionic supercurrents $\frak{J}^{\mu}_{A}$ and $\overline{\frak{J}}^{\mu \, A}$ are related to the SUSY Noether currents as 
$$\mathcal{J}^{\mu}_{(\epsilon)} = \overline{\epsilon}^{A}~\frak{J}^{\mu}_{A},~~~~~~~~~\mathcal{J}^{\mu \dagger}_{(\epsilon)} = \overline{\frak{J}}^{\mu \, A} \epsilon_{A},$$
where $A=1,...,4$ is a spinor index.
The time-independent (complex)  Noether charges associated to the vector currents (\ref{def: SUSY Noether currents}) are defined as \cite{Freedman}
\begin{align} \label{def: SUSY noether charge}
 & Q^{susy}{[\epsilon]} =  \int_{S^{3}} d\bm{\theta_{3}} \sqrt{-g}~\mathcal{J}^{t }_{(\epsilon)}(\frak{h},\Psi), ~~~~~~~~~\left( Q^{susy}{[\epsilon]}\right)^{\dagger} =  \int_{S^{3}} d\bm{\theta_{3}} \sqrt{-g}~\left(\mathcal{J}^{t }_{(\epsilon)}(\frak{h},\Psi) \right)^{\dagger}.
\end{align}
We refer to  $\mathcal{J}^{\mu}_{(\epsilon)}$ and $\mathcal{J}^{\mu \dagger}_{(\epsilon)}$ as SUSY Noether currents, and to $Q^{susy}{[\epsilon]}$, $Q^{susy}{[\epsilon]}^{\dagger}$ as SUSY Noether charges. Since the Killing spinors $\epsilon$ are Grassmann-odd, then $\mathcal{J}^{\mu}_{(\epsilon)}$ and $Q^{susy}{[\epsilon]}$ are Grassmann-even. If we   use commuting Killing spinors $\epsilon^{(\sigma;q)}$ [eq.~(\ref{dS Killing spinors explicit_ e=S(t,x)n})], then $\mathcal{J}^{\mu}_{(\epsilon^{(\sigma;q)})}$ and $Q^{susy}{[\epsilon^{(\sigma;q)}]}$ are also conserved, and they are Grassmann-odd. In Subsection \ref{subsubsec_unitary SUSY QFT}, it will be convenient to work with these Grassmann-odd SUSY Noether currents and charges.

%%%%%%%%%%%%%%%%%%%%%%%%%%%%%%%%%%%%%%%%%%%%%%%%%%%%%%%%%%%%%%%%%%%%
%%%%%%%%%%%%%%%%%%%%%%%%%%%%
\subsubsection{SUSY algebra  {with} complex Killing spinors}

After a straightforward calculation, the commutator of two SUSY transformations 
[eqs.~(\ref{SUSY_transf_spin3/2->spin2}) and (\ref{SUSY_transf_spin2->spin3/2})] on the complex graviton  is found to be
\begin{align}\label{commutator of SUSY on spin-2}
    \,\left[\,\delta^{susy}(\epsilon_{2}),\delta^{susy}(\epsilon_{1})\,\right]\,\mathfrak{h}_{\mu \nu} =&- \,\pounds_{\xi_{(\epsilon)}  }\mathfrak{h}_{\mu \nu}+\,\mathcal{T}_{V_{(\epsilon)}}\,\mathfrak{h}_{\mu \nu }-i\,\left(  \frac{1}{4}\overline{\epsilon}_{2} \gamma^{5} \epsilon_{1} -\frac{1}{4}\overline{\epsilon}_{1} \gamma^{5} \epsilon_{2} \right) \mathfrak{h}_{\mu \nu } \nonumber\\
    &+ \nabla_{(\mu} \left[ \mathfrak{h}_{\nu) \sigma} \,\xi^{\sigma}_{(\epsilon)}  \right],
\end{align}
where no use of the equations of motion was made. The first term, $-\pounds_{\xi_{(\epsilon)}} \frak{h}_{\mu \nu}$, on the right-hand {side} of
eq.~(\ref{commutator of SUSY on spin-2}) is an infinitesimal dS transformation (Lie derivative) generated by the  Killing vector $\xi^{\mu}_{(\epsilon)}$ defined in 
eq.~(\ref{real KV from K spinors}). The second term, $\mathcal{T}_{V_{(\epsilon) }}  \frak{h}_{\mu \nu}$, is a conformal-like transformation (\ref{conf-like graviton complex}) generated by the  genuine conformal Killing vector $V_{ (\epsilon)}^{\mu}$ defined in eq.~(\ref{real CKV from K spinors}). The third term is an infinitesimal $u(1)$ transformation [the phase factor $ \frac{1}{4}\overline{\epsilon}_{2} \gamma^{5} \epsilon_{1} -\frac{1}{4}\overline{\epsilon}_{1} \gamma^{5} \epsilon_{2}$ is real as $  -\overline{\epsilon}_{1} \gamma^{5} \epsilon_{2}  =  (\overline{\epsilon}_{2} \gamma^{5} \epsilon_{1})^{\dagger}$, and constant, as $\nabla_{\mu}\left( \overline{\epsilon}_{2} \gamma^{5} \epsilon_{1}  \right)  = \nabla_{\mu}\left( \overline{\epsilon}_{1} \gamma^{5} \epsilon_{2}  \right) =0$, as expected for $u(1)$ transformations]. The last term is a field-dependent gauge transformation akin to the gauge transformation appearing in linearised Supergravity in Minkowski spacetime - see e.g., Ref.~\cite{West}. We conclude that the even subalgebra of the SUSY algebra closes on $so(4,2) \bigoplus u(1)$ up to gauge transformations\footnote{The $so(4,2)$ algebra generated by infinitesimal dS transformations and conformal-like transformations was studied in Subsections \ref{Subsec_gravitino modes so(4,2)} and \ref{subsubsec_cmplx graviton and so(4,2)}, for the gravitino and the graviton, respectively.}.

Calculating the commutator of two SUSY transformations on the Dirac gravitino $\Psi_{\mu}$ is much more tedious than the complex graviton case above. Moreover, one has to make use of the equations of motion, as well as of the Fierz rearrangement identities \cite{Freedman}. The result is\footnote{We made use of the Mathematica tensor computer algebra package FieldsX \cite{Frob} to simplify certain parts of the calculation that involved products of (generalised) gamma matrices.}
\begin{align}\label{commutator of SUSY on spin-3/2}
    \left[\,\delta^{susy}(\epsilon_{2}),\delta^{susy}(\epsilon_{1})\,\right]\,\Psi_{\mu } =&- \,\mathbb{L}_{\xi_{(\epsilon) } }\Psi_{\mu }+ \mathbb{T}'_{V_{(\epsilon)}}\,\Psi_{\mu}-\frac{5i}{2}\,\left( \frac{1}{4} \overline{\epsilon}_{2} \gamma^{5} \epsilon_{1} -\frac{1}{4}\overline{\epsilon}_{1} \gamma^{5} \epsilon_{2} \right)\Psi_{\mu  } \nonumber\\
    &+ \left(\nabla_{\mu}+ \frac{i}{2} \gamma_{\mu}  \right)\frac{A_{(\epsilon) }}{2},
\end{align}
where $A_{(\epsilon)}$ is a field-dependent spinor gauge function given by
\begin{align}
   A_{(\epsilon) }=& \frac{3}{4}\,  \left( \frac{1}{4}\overline{\epsilon_{2}} \gamma^{5}  \epsilon_{1}- \frac{1}{4}   \overline{\epsilon_{1}} \gamma^{5}  \epsilon_{2} \right) \,  \gamma^{\alpha}  \Psi_{\alpha}  +\xi_{ (\epsilon)}^{\alpha} \Psi_{\alpha} +\left(  \frac{1}{4} \xi_{ (\epsilon)}^{\rho} \gamma_{\rho}-\frac{i}{8} \nabla^{\lambda} \xi^{\rho}_{(\epsilon)} \gamma_{\lambda \rho} \right) \gamma^{\alpha} \Psi_{\alpha} \nonumber \\
  & - \gamma^{5} V^{\alpha}_{ (\epsilon)}  \Psi_{\alpha} -\frac{1}{4} \gamma^{5} V^{\rho}_{ (\epsilon)}  \gamma_{\rho}  \gamma^{\alpha}  \Psi_{\alpha} 
   -\frac{3i}{4}  \phi_{V_{(\epsilon)  }}\, \gamma^{\alpha} \Psi_{\alpha}  .
\end{align}
%%%%%%%%%%%%%%%%%%%%%%%%%%%%%%%%%%%%%%%%%%%%%%%%%%%%%%%%%%%%%%%%%%%
The terms that appear on the right-hand side of eq.~(\ref{commutator of SUSY on spin-3/2}) are similar to the terms appearing in the complex graviton case (\ref{commutator of SUSY on spin-2}). In particular, the first term, $-\mathbb{L}_{\xi_{ (\epsilon)}}\Psi_{\mu}$, is an infinitesimal dS transformation (\ref{Lie_Lorentz}) generated by the Killing vector $\xi^{\mu}_{(\epsilon) }$ defined in eq.~(\ref{real KV from K spinors}). The second term, $\mathbb{T}'_{V_{(\epsilon)}} \Psi_{\mu}$, is given by a conformal-like transformation $\mathbb{T}_{V_{(\epsilon) }} \Psi_{\mu}$ (\ref{conf-like gravitino TT}) plus a gauge transformation that cancels the last term of the conformal-like transformation (\ref{conf-like gravitino TT}), where $V^{\mu}_{(\epsilon) }$ is the genuine conformal Killing vector defined in eq.~(\ref{real CKV from K spinors}). To be specific, $\mathbb{T}'_{V_{ (\epsilon)  }} \Psi_{\mu} = \mathbb{T}_{V_{ (\epsilon)  }} \Psi_{\mu} + \frac{2}{3} \left( \nabla_{\mu} +\frac{i}{2}\gamma_{\mu}\right) \gamma^{5}\Psi_{\rho}V_{ (\epsilon)
 }^{\rho}$. The third term on the right-hand side of (\ref{commutator of SUSY on spin-3/2}) is a $u(1)$ transformation, while the last term is a field-dependent gauge transformation. It is thus clear that the even subalgebra of the SUSY algebra closes on $so(4,2) \bigoplus u(1)$ up to gauge transformations.
%%%%%%%%%%%%%%%%%%%%%%%%%%%%%%%%%%%%%%%%%%%%%%%%%%%%%%%%%%%%%%%%%%%%%%%%%%%%%%%%%%%%%%%%%%55
\subsubsection{SUSY transformations of the field strengths and of their duals}\label{subsubsec_duality+SUSY}
Let us give here again the expressions of the field strengths for the complex graviton [eq.~(\ref{def:graviton_field-strength})] and complex gravitino 
[eq.~(\ref{def:gravitino_field-strength})]:
\begin{align*}
    U_{\alpha \beta \mu \nu} =\Big( -\nabla_{\mu}\nabla_{[ \alpha}\mathfrak{h}_{\beta]\nu}-g_{\mu [\alpha} \mathfrak{h}_{\beta] \nu}  \Big)-(\mu \leftrightarrow \nu),~~ ~~~F_{\mu \nu}  = \left( \nabla_{[\mu} + \frac{i}{2} \gamma_{[\mu}   \right) \, \Psi_{\nu]}.
\end{align*}
Their properties are summarised in Appendix \ref{Append_field strengths}.
\\
\\ 
\noindent \textbf{\textit{SUSY transformations of field strengths}}$-$The SUSY transformations of the field strengths can be obtained   by direct calculation using the SUSY transformations of the gauge potentials [eqs.~(\ref{SUSY_transf_spin3/2->spin2}) and (\ref{SUSY_transf_spin2->spin3/2})], as
\begin{align*} 
   \delta^{susy}({\epsilon}) F_{\mu \nu}  = \left( \nabla_{[\mu} + \frac{i}{2} \gamma_{[\mu}   \right) \, \delta^{susy}({\epsilon})\Psi_{\nu]},
\end{align*}
and
\begin{align*}
   \delta^{susy}({\epsilon}) U_{\alpha \beta \mu \nu} =\Big( -\nabla_{\mu}\nabla_{[ \alpha}\delta^{susy}({\epsilon})\mathfrak{h}_{\beta]\nu}-g_{\mu [\alpha} \delta^{susy}({\epsilon})\mathfrak{h}_{\beta] \nu}  \Big)-(\mu \leftrightarrow \nu).
\end{align*}
The result is
\begin{align} \label{SUSY_transf of spin3/2 fieldstrength}
  \delta^{susy}({\epsilon})F_{\mu \nu}=&~\frac{1}{8} \gamma^{{\kappa \lambda}} \epsilon \, U_{\kappa \lambda \mu \nu} ,  \\
  \delta^{susy}({\epsilon})U^{\alpha  \beta}_{\hspace{3mm}\mu \nu}=&~\bar{\epsilon}\gamma^{5} \Big(  (\gamma^{[\alpha}\nabla^{\beta]}-\frac{i}{2}\gamma^{\alpha \beta})F_{\mu \nu}+(\gamma_{[\mu}\nabla_{\nu]}-\frac{i}{2}\gamma_{\mu \nu})F^{\alpha \beta}   \nonumber\\
 &+2i \gamma_{[\mu}^{\hspace{2mm}[\alpha}F^{\beta]}_{\hspace{2mm}\nu]} \Big),  \label{SUSY_transf of spin2 fieldstrength}
\end{align}
where no use of the equations of motion was made. 
\\
\\
\noindent \textbf{\textit{Duality commutes with SUSY transformations}}$-$It is convenient to re-write the SUSY transformation (\ref{SUSY_transf of spin2 fieldstrength}) as
\begin{align}\label{SUSY_transf of spin2 fieldstrength FINAL}
    \delta^{susy}({\epsilon})U_{\alpha  \beta \mu \nu}=&~\bar{\epsilon}\gamma^{5} \Big(  (\gamma_{[\alpha}\nabla_{\beta]}-{i}\gamma_{\alpha \beta})F_{\mu \nu}+(\gamma_{[\mu}\nabla_{\nu]}-{i}\gamma_{\mu \nu})F_{\alpha \beta}\Big).
\end{align}
To derive eq.~(\ref{SUSY_transf of spin2 fieldstrength FINAL}) from 
eq.~(\ref{SUSY_transf of spin2 fieldstrength}), we have used
\begin{align} \label{simplifier of Weyl SUSY transf}
 2 \gamma_{[\mu}^{\hspace{2mm}[\alpha}F^{\beta]}_{\hspace{2mm}\nu]} =  -\frac{1}{2}\gamma^{\alpha \beta} F_{\mu \nu}  -\frac{1}{2}\gamma_{\mu \nu} F^{\alpha \beta} ,
\end{align}
which can be proved by using the 
on-shell properties of the spin-3/2 field strength, as well as properties of products of (generalised) gamma matrices - see Appendix \ref{Append_field strengths}. Given the SUSY transformations (\ref{SUSY_transf of spin3/2 fieldstrength}) and (\ref{SUSY_transf of spin2 fieldstrength FINAL}), a straightforward calculation {using some formulae in Appendix \ref{Append_field strengths}} shows that the duality operation commutes with them, as
\begin{align} \label{SUSY_transf of spin3/2 fieldstrength DUAL}
  \delta^{susy}({\epsilon})\widetilde{F}_{\mu \nu}=&~\frac{1}{8} \gamma^{{\kappa\lambda}} \epsilon \, \widetilde{U}_{\kappa \lambda \mu \nu} ,  \\
  \delta^{susy}({\epsilon}) \widetilde{U}_{\alpha  \beta \mu \nu}=&~\bar{\epsilon}\gamma^{5} \Big(  (\gamma_{[\alpha}\nabla_{\beta]}-{i}\gamma_{\alpha \beta}) \widetilde{F}_{\mu \nu}+(\gamma_{[\mu}\nabla_{\nu]}-{i}\gamma_{\mu \nu})\widetilde{F}_{\alpha \beta}    \Big).\label{SUSY_transf of spin2 fieldstrength DUAL}
\end{align}
\\
\\
\noindent   \textbf{\textit{SUSY algebra for the field strengths}}$-$The commutator of two SUSY transformations acting on $U_{\alpha \beta \mu \nu}$ is
\begin{align}\label{commutator of SUSY on Weyl}
    \left[\,\delta^{susy}(\epsilon_{2}),\delta^{susy}(\epsilon_{1})\,\right]\,U_{\alpha \beta \mu \nu} =&- \,\pounds_{\xi_{(\epsilon)  } }U_{\alpha \beta \mu \nu}+ \mathcal{T}_{V_{ (\epsilon) }}\,U_{ \alpha \beta \mu \nu }-i\,\left(\frac{1}{4}  \overline{\epsilon}_{2} \gamma^{5} \epsilon_{1} -\frac{1}{4}\overline{\epsilon}_{1} \gamma^{5} \epsilon_{2} \right)U_{\alpha \beta \mu \nu } ,
\end{align}
where the Killing vector $\xi^{\mu}_{ (\epsilon) }$ and the genuine conformal Killing vector $V^{\mu}_{(\epsilon)  }$ are defined in eqs.~(\ref{real KV from K spinors}) and (\ref{real CKV from K spinors}), respectively. The interpretation of the terms on the right-hand side of eq.~(\ref{commutator of SUSY on Weyl}) is the same as in the case of the complex graviton gauge potential (\ref{commutator of SUSY on spin-2}), except, of course, for the gauge transformation term in (\ref{commutator of SUSY on spin-2}), which drops out. The conformal-like transformation $\mathcal{T}_{V}\,U_{ \alpha \beta \mu \nu }$, generated by genuine conformal Killing vectors, is given by the product of a conventional infinitesimal conformal transformation times a duality transformation (times $i$), as 
\begin{align}\label{extra symmetry Weyl}
  \mathcal{T}_{V}U_{\alpha   \beta \mu \nu} =i \left( \pounds_{V}  - \frac{1}{4}\nabla^{\rho}V_{\rho}\right) \,   \widetilde{U}_{ \alpha \beta \mu \nu }  =i \left( V^{\rho} \nabla_{\rho}-3 \phi_{V}   \right)\,  \widetilde{U}_{ \alpha \beta \mu \nu }.
\end{align}
For the sake of completeness, let us also compute the commutator of a SUSY variation and a conformal-like variation:
$\left[ \delta^{susy}(\epsilon),\delta_{V}   \right]U_{\alpha \beta \mu   \nu} $,
where $\delta_{V} U_{\alpha \beta \mu   \nu} = \mathcal{T}_{V} U_{\alpha \beta \mu   \nu} $ and $V^{\mu}$ is any genuine conformal Killing vector (\ref{V=nabla phi}). 
We find
\begin{align}\label{[SUSY,conf-like]spin-2 fieldstrength}
    \left[ \delta^{susy}(\epsilon),\delta_{V}   \right]U_{\alpha \beta \mu \nu}=& \overline{\mathbb{T}_{V}\epsilon}~\gamma^{5} \Big(  (\gamma_{[\alpha}\nabla_{\beta]}-{i}\gamma_{\alpha \beta})F_{\mu \nu}+(\gamma_{[\mu}\nabla_{\nu]}-{i}\gamma_{\mu \nu})F_{\alpha \beta}\Big), \nonumber \\
    =& \delta^{susy}\Big( \, \mathbb{T}_{V}\epsilon \Big)~U_{\alpha \beta \mu \nu},
\end{align}
which is a SUSY variation of $U_{\alpha \beta \mu \nu}$ (\ref{SUSY_transf of spin2 fieldstrength FINAL}), but with the Killing spinor ${\epsilon}$ replaced by its  conformal-like-transformed version, $\mathbb{T}_{V} \epsilon$ [eq.~(\ref{extra symmetry Killing spinor})]. One similarly finds the commutator between a SUSY variation and a dS variation,
\begin{align}\label{[SUSY,isometry]spin-2 fieldstrength}
  \left[ \delta^{susy}(\epsilon),\delta_{\xi}   \right]U_{\alpha \beta \mu   \nu} = \delta^{susy}(\mathbb{L}_{\xi}\epsilon) U_{\alpha \beta \mu   \nu} ,   
\end{align}
where $\delta_{\xi} U_{\alpha \beta \mu   \nu} = \pounds_{\xi} U_{\alpha \beta \mu   \nu} $ and $\xi^{\mu}$ is any  Killing vector.

The commutator of two SUSY transformations for $F_{ \mu \nu}$ is
\begin{align}\label{commutator of SUSY on spin3/2 field-strength}
    \left[\,\delta^{susy}(\epsilon_{2}),\delta^{susy}(\epsilon_{1})\,\right]\,F_{ \mu \nu} =&- \,\mathbb{L}_{\xi_{( \epsilon  )} }F_{ \mu \nu}+\mathbb{T}_{V_{(\epsilon)}}\,F_{ \mu \nu }-\frac{5i}{2}\,\left( \frac{1}{4} \overline{\epsilon}_{2} \gamma^{5} \epsilon_{1} -\frac{1}{4}\overline{\epsilon}_{1} \gamma^{5} \epsilon_{2} \right)F_{\mu \nu } ,
\end{align}
where the interpretation of the terms on the right-hand side  is  as in the case of the complex gravitino gauge potential (\ref{commutator of SUSY on spin-3/2}). The conformal-like transformation of the gravitino field strength, generated by any genuine conformal Killing vector $V^{\mu}$ (\ref{V=nabla phi}), is given by \cite{Letsios_conformal-like}
\begin{align}\label{extra symmetry spin3/2 fieldstrength}
 \mathbb{T}_{V}F_{\mu \nu} =i \left( \mathbb{L}_{V}+\frac{1}{8} \nabla_{\rho}V^{\rho}   \right)\,\widetilde{F}_{\mu \nu }= \gamma^{5} \left( V^{\rho} \nabla_{\rho}-\frac{5}{2} \phi_{V}   \right)\,F_{\mu \nu }.
\end{align}
The commutator between a SUSY variation and a conformal-like variation, as well as the commutator between a SUSY variation and a dS variation, are given by expressions similar to (\ref{[SUSY,conf-like]spin-2 fieldstrength}) and (\ref{[SUSY,isometry]spin-2 fieldstrength}), respectively.

%%%%%%%%%%%%%%%%%%%
\subsubsection{SUSY representation on the TT solution spaces and non-unitarity of the non-chiral theory}\label{Subsubsec_TT gauge SUSY}

The TT gauge is a particularly convenient gauge as the field equations have a simple form. In addition, the TT mode solutions and their representation-theoretic properties are known - see Subsections \ref{Subsec_gravitino modes so(4,1)} and \ref{Subsec_graviton modes so(4,1)}.
For convenience, let us write here again the field equations for the complex graviton and the Dirac gravitino in the TT gauge [eqs.~(\ref{EOM cmplx graviton TT}) and (\ref{Dirac_eqn_fermion_dS})]
\begin{align*}
  &  \Box\mathfrak{h}^{(\text{TT})}_{\mu \nu } = 2 ~\mathfrak{h}^{(\text{TT})}_{\mu \nu}, \nonumber \\
  & \nabla^{\mu} \mathfrak{h}^{(\text{TT})}_{\mu \nu} =0,~~ ~ \mathfrak{h}^{(\text{TT})\alpha}_{~ \alpha} = 0,
\end{align*}
\begin{align*}
   &\left( \slashed{\nabla}+i\right)\Psi^{(\text{TT})}_{\mu}=0, \nonumber  \\
   & \nabla^{\alpha}\Psi^{(\text{TT})}_{\alpha}=0, \hspace{4mm}  \gamma^{\alpha}\Psi^{(\text{TT})}_{\alpha}=0. 
\end{align*}
To achieve the compatibility of the SUSY transformations (\ref{SUSY_transf_spin3/2->spin2}) and (\ref{SUSY_transf_spin2->spin3/2}) with the TT conditions - $\nabla^{\mu} \frak{h}^{(\text{TT})}_{\mu \nu} = g^{\alpha \beta} \frak{h}^{(\text{TT})}_{\alpha \beta} = 0$ and $\nabla^{\mu}\Psi^{(\text{TT})}_{\mu}  = \gamma^{\mu}\Psi^{(\text{TT})}_{\mu}=0$ - we have to modify the SUSY transformation of the graviton (\ref{SUSY_transf_spin2->spin3/2}) by introducing a gauge transformation. 
{To be specific, we modify the SUSY transformation $\delta^{susy}({\epsilon}) \frak{h}_{\mu \nu}$ given by eq.~(\ref{SUSY_transf_spin2->spin3/2}) by adding a gauge transformation term with the field-dependent gauge parameter $-\frac{i}{3} \overline{\epsilon}\gamma^{5} \Psi_{\nu}$
to ensure that if the fields  $\frak{h}_{\mu \nu}$ and $\Psi_{\mu}$ are in the TT gauge, then the SUSY-transformed graviton remains in the TT gauge (no such gauge correction is required for the gravitino). Thus,}
the SUSY transformations that preserve the solution space of the TT field equations (\ref{EOM cmplx graviton TT}) and (\ref{Dirac_eqn_fermion_dS}) are:
\begin{align}
    \delta^{susy}({\epsilon})\Psi^{(\text{TT})}_{\mu} & =\frac{1}{4}  \left( i\, \frak{h}^{(\text{TT})}_{\mu \sigma}\gamma^{\sigma}+\nabla_{\lambda}  \frak{h}^{(\text{TT})}_{\mu \sigma} \gamma^{\sigma  \lambda}      \right)\epsilon, \label{SUSY_transf_spin3/2->spin2 TT}\\
    \delta^{susy}({\epsilon})' \frak{h}^{(\text{TT})}_{\mu \nu}&=
    \delta^{susy}({\epsilon}) \frak{h}^{(\text{TT})}_{\mu \nu} + \delta^{gauge}\left( -\frac{i}{3} \overline{\epsilon}\gamma^{5} \Psi^{(\text{TT})}\right)  \frak{h}^{(\text{TT})}_{\mu \nu} \label{SUSY_transf_spin2->spin3/2 TT}   \\
    &=\frac{\overline{\epsilon}}{2}\gamma^{5}  \left(  \gamma_{\mu}  \Psi^{(\text{TT})}_{\nu}+\gamma_{\nu} \Psi^{(\text{TT})}_{\mu}    \right)
   + \nabla_{(\mu}\left( -\frac{i}{3} \overline{\epsilon}\gamma^{5} \Psi_{\nu)}^{(\text{TT})}  \right) \nonumber   \\
   &= \frac{7}{6} {\overline{\epsilon}} \gamma^{5}    \gamma_{(\mu}  \Psi^{(\text{TT})}_{\nu)} -\frac{i}{3}{\overline{\epsilon}} \gamma^{5}    \nabla_{(\mu}  \Psi^{(\text{TT})}_{\nu)}, \nonumber
\end{align}
where $\epsilon$ is a Grassmann-odd Killing spinor satisfying eq.~(\ref{Killing spinor eqn}) with the `$-$' sign. 
Note that the complex graviton gauge transformation in (\ref{SUSY_transf_spin2->spin3/2 TT}) is \textbf{not} a restricted gauge transformation (\ref{gauge_transf_spin2_cmplx_res}) as it is not divergence-free. In particular, 
$$ \nabla^{\mu}\nabla_{(\mu}\left( -\frac{i}{3} \overline{\epsilon}\gamma^{5} \Psi_{\nu)}^{(\text{TT})}\right) = -\nabla^{\mu} \delta^{susy}({\epsilon}) \frak{h}^{(\text{TT})}_{\mu \nu} = -i~ \overline{\epsilon}  \gamma^{5} \Psi^{(\text{TT})}_{\nu},$$
 leading to $\nabla^{\mu} \delta^{susy}({\epsilon})' \frak{h}^{(\text{TT})}_{\mu \nu} =0$.
 {It follows from the gauge invariance of the SUSY transformations that}
 the commutators of the SUSY transformations in the TT gauge (\ref{SUSY_transf_spin3/2->spin2 TT}) and (\ref{SUSY_transf_spin2->spin3/2 TT}) takes the following form:
\begin{align}\label{commutator of SUSY on spin-2 TT}
   & \,\left(\,\delta^{susy}(\epsilon_{2})\delta^{susy}(\epsilon_{1})'- \delta^{susy}(\epsilon_{1})\delta^{susy}(\epsilon_{2})'\right)\,\mathfrak{h}^{(\text{TT})}_{\mu \nu} = \nonumber\\
    &- \,\pounds_{\xi_{ (\epsilon) }  }\mathfrak{h}^{(\text{TT})}_{\mu \nu}+\,\mathcal{T}_{V_{ (\epsilon)
  }}\,\mathfrak{h}^{(\text{TT})}_{\mu \nu }-i\,\left(  \frac{1}{4}\overline{\epsilon}_{2} \gamma^{5} \epsilon_{1} -\frac{1}{4}\overline{\epsilon}_{1} \gamma^{5} \epsilon_{2} \right) \mathfrak{h}^{(\text{TT})}_{\mu \nu } + \left(  \text{{pure-gauge term}} \right),
\end{align}
\begin{align}\label{commutator of SUSY on spin-3/2 TT}
   & \left(\,\delta^{susy}(\epsilon_{2})'\delta^{susy}(\epsilon_{1}) - \delta^{susy}(\epsilon_{1})'\delta^{susy}(\epsilon_{2})\,\right)\,\Psi^{(\text{TT})}_{\mu} \nonumber\\
   =&- \,\mathbb{L}_{\xi_{ (\epsilon) } }\Psi^{(\text{TT})}_{\mu }+ \mathbb{T}_{V_{ (\epsilon)
 }}\,\Psi^{(\text{TT})}_{\mu}-\frac{5i}{2}\,\left( \frac{1}{4} \overline{\epsilon}_{2} \gamma^{5} \epsilon_{1} -\frac{1}{4}\overline{\epsilon}_{1} \gamma^{5} \epsilon_{2} \right)\Psi^{(\text{TT})}_{\mu  }  + \left(  \text{{pure-gauge term}} \right).
\end{align}
%%%%%%%%%%%%%%%
This is the same algebra structure as in the non-gauged-fixed case 
[eqs.~(\ref{commutator of SUSY on spin-2}) and (\ref{commutator of SUSY on spin-3/2})], with the only difference being that each of the field transformations in 
eqs.~(\ref{commutator of SUSY on spin-2 TT}) and (\ref{commutator of SUSY on spin-3/2 TT}) preserves the TT gauge conditions.
\\
\\
\noindent  \textbf{Note.}  The TT SUSY transformations (\ref{SUSY_transf_spin3/2->spin2 TT}) and (\ref{SUSY_transf_spin2->spin3/2 TT}) also describe symmetries of the TT field equations if one uses commuting Killing spinors $\epsilon^{(\sigma ;q)}$ [eq.~(\ref{dS Killing spinors explicit_ e=S(t,x)n})] instead of Grassmann-odd Killing spinors. Moreover, if one uses commuting Killing spinors, the commutator of two SUSY transformations on the TT graviton is the same as in (\ref{commutator of SUSY on spin-2 TT}). However, the commutator of two SUSY transformations on the TT gravitino will be given by eq.~(\ref{commutator of SUSY on spin-3/2 TT}) with opposite signs on the right-hand side. These comments also apply to the case of the SUSY transformations of the non-gauge-fixed fields [eqs.~(\ref{SUSY_transf_spin3/2->spin2}) and (\ref{SUSY_transf_spin2->spin3/2})].
\\
\\
\noindent   \textbf{\textit{The question of unitarity of SUSY and its failure in the QFT Fock space of the non-chiral theory}}$-$Until now, we have demonstrated the existence of a  representation of our SUSY algebra on the solution space of the {classical} field equations of the complex graviton and Dirac gravitino, for both their non-gauge-fixed version and in the TT gauge. However,  we have not addressed the question of unitarity. Let us specialise to the TT gauge. Our representation space is the direct sum of a bosonic and a fermionic solution space, i.e.\ the direct sum of the Hilbert spaces of the graviton and gravitino mode solutions $\mathcal{H}_{2} \bigoplus \mathcal{H}_{\frac{3}{2}}$.  The solution spaces of positive frequency {mode solutions are}: 
$$ \mathcal{H}_{2} =\mathcal{H}^{+}_{2}  \bigoplus \mathcal{H}^{-}_{2}= \{ {\varphi}^{(phys ,\,+ L; \,M;K)}_{{\mu}{\nu}}\} \bigoplus  \{{\varphi}^{(phys ,\,- L; \,M;K)}_{{\mu}{\nu}} \}, $$
$$\mathcal{H}_{\frac{3}{2}} =\mathcal{H}^{+}_{\frac{3}{2}} \bigoplus \mathcal{H}^{-}_{\frac{3}{2}}  = \{ {\psi}^{(phys ,\,+ \ell; \,m;k)}_{\mu}\} \bigoplus \{ {\psi}^{(phys ,\,- \ell; \,m;k)}_{\mu}  \},$$
where the TT pure-gauge modes are identified with zero, as discussed in  Sections \ref{Sec_graviton} and \ref{Sec_gravitino}, respectively. As an equivalent representation space one can choose the space of negative frequency mode solutions, denoted as $\mathcal{H}_{{2}}^{*}$   and $\mathcal{H}_{\frac{3}{2}}^{*}$.
The bosonic and fermionic solution spaces, $ \mathcal{H}_{2}$ and $ \mathcal{H}_{\frac{3}{2}}$, are equipped with $so(4,2)$-invariant scalar products, $\braket{\cdot| \cdot}_{KG}$ (\ref{def: KG inner product spin-2}) and $\braket{\cdot| \cdot}_{ax}$ (\ref{axial_scalar prod}), respectively, and we have already explained how the mode solutions form UIRs of  $so(4,2)$ in Subsections \ref{Subsec_gravitino modes so(4,2)} and \ref{Subsec_graviton modes so(4,2)}. A interesting feature, discussed in Section \ref{Sec_gravitino}, is that the positive-definite scalar product in $\mathcal{H}^{-}_{\frac{3}{2}}$ is the axial scalar product (\ref{axial_scalar prod}), while the positive-definite scalar product in $\mathcal{H}^{+}_{\frac{3}{2}}$ is the \textbf{negative} of the axial scalar product. One has the freedom to choose a different positive-definite scalar product for each solution space,  $\mathcal{H}^{-}_{\frac{3}{2}}$ and $\mathcal{H}^{+}_{\frac{3}{2}}$, as each of these two spaces  separately forms a UIR. However, when {we quantised} the gravitino theory in Subsection \ref{Subsec_gravitino quantisation}, it became clear that the requirement of the positivity of the norm in the QFT Fock space forces the gravitino to be chiral, and thus, not both $\mathcal{H}^{-}_{\frac{3}{2}}$ and $\mathcal{H}^{+}_{\frac{3}{2}}$ can be part of the positive frequency sector of the QFT - one has to use only one of them. Thus, as the notion of unitarity is tied to the positivity of the norm, it is clear that the SUSY representation realised on the QFT Fock space of the complex non-chiral graviton and gravitino is non-unitary because the  positive frequency sector of the gravitino contains both $\mathcal{H}^{-}_{\frac{3}{2}}$ and $\mathcal{H}^{+}_{\frac{3}{2}}$.
 \\
 \\
 \noindent \textbf{Note.}  One can construct two different SUSY UIRs at the level of classical mode solutions: one UIR formed by  $\mathcal{H}^{-}_{{2}} \bigoplus \mathcal{H}^{-}_{\frac{3}{2}}$ and another  one formed by 
 $\mathcal{H}^{+}_{{2}} \bigoplus \mathcal{H}^{+}_{\frac{3}{2}}$. However, for the same reasons as in the gravitino case in Subsection \ref{Subsec_gravitino quantisation}, the unitary supersymmetric QFT of a chiral graviton and a chiral gravitino in Subsection \ref{Subsec_CHIRAL theory SUSY} will have a positive frequency sector consisting only of  $\mathcal{H}^{-}_{{2}} \bigoplus \mathcal{H}^{-}_{\frac{3}{2}}$. The space $\mathcal{H}^{+}_{{2}} \bigoplus \mathcal{H}^{+}_{\frac{3}{2}}$ will be excluded from the Hilbert space with the help of the anti-self-duality constraints.

    %%%%%%%%%%%%%%%%%%%%%%
   
\subsection{Unitary  SUSY for the chiral graviton and chiral gravitino}\label{Subsec_CHIRAL theory SUSY}

%%%%%%%%%%%%%%%%

\subsubsection{Unitarity of SUSY in the space of chiral 
 mode solutions}\label{subsubsec_unitary SUSY modes}

In this Subsection, we demonstrate how the mode solutions that are relevant to the supersymmetric theory of the chiral graviton and chiral gravitino form UIRs of SUSY. 
In particular, 
we work in the TT gauge and  show that the SUSY transformations (\ref{SUSY_transf_spin3/2->spin2 TT}) and (\ref{SUSY_transf_spin2->spin3/2 TT}) generate a UIR of SUSY that is realised on the space of classical TT mode solutions of positive frequency with helicities $-2$ and $-3/2$, $\mathcal{H}^{-}_{{2}} \bigoplus \mathcal{H}^{-}_{\frac{3}{2}}$.  We also show that another UIR of SUSY is formed in the space of negative frequency modes $\mathcal{H}^{*+}_{{2}} \bigoplus \mathcal{H}^{*+}_{\frac{3}{2}}$. We focus on these representation spaces, i.e.\ spaces on which the field strengths are anti-self-dual, because the axial scalar product (\ref{axial_scalar prod}) is positive definite in both of them [see eqs.~(\ref{norms of physical modes_3/2}) and (\ref{norms of physical modes_3/2_neg freq})].\footnote{One can similarly show that two SUSY UIRs are separately formed by the TT solution spaces $\mathcal{H}^{+}_{{2}} \bigoplus \mathcal{H}^{+}_{\frac{3}{2}}$ and  $\mathcal{H}^{*-}_{{2}} \bigoplus \mathcal{H}^{*-}_{\frac{3}{2}}$, in which  the axial scalar product (\ref{axial_scalar prod})  is negative definite, 
i.e.\ the \textbf{negative} of the axial scalar product is positive definite. We do not give the corresponding representation-theoretic details because they are similar to the ones presented in the main text, as well as because the  solution spaces $\mathcal{H}^{+}_{{2}} \bigoplus \mathcal{H}^{+}_{\frac{3}{2}}$ and  $\mathcal{H}^{*-}_{{2}} \bigoplus \mathcal{H}^{*-}_{\frac{3}{2}}$  are omitted in the unitary quantum theory of the chiral graviton and chiral gravitino.}  To proceed with the representation-theoretic discussion, we recall that the bosonic and fermionic solution spaces that form the SUSY representation are equipped with the Klein-Gordon, $\braket{\cdot| \cdot}_{KG}$ (\ref{def: KG inner product spin-2}), and axial, $\braket{\cdot| \cdot}_{ax}$ (\ref{axial_scalar prod}), scalar products, respectively.

Let us give the definition of unitarity for representations of our superalgebra - the structure of the superalgebra is determined by eqs.~(\ref{commutator of SUSY on spin-2 TT}) and (\ref{commutator of SUSY on spin-3/2 TT}). A unitary representation of SUSY must satisfy the following three conditions simultaneously~\cite{Furutsu}: 
\begin{enumerate}
    \item Positivity of the norm in both the bosonic and  fermionic solution spaces. \label{def: susy unit 1}

\item  Invariance of the inner products under the generators of the even subalgebra [$so(4,2)$ $\bigoplus u(1)$ in our case], i.e.\ anti-hermiticity of even generators.
\label{def: susy unit 2}

    \item  SUSY-invariance of the inner products, in the sense that, for any  TT solution  $\psi_{\mu}$  of eq.~(\ref{Dirac_eqn_fermion_dS}), and any  TT  solution $\varphi_{\mu \nu}$  of eq.~(\ref{EOM cmplx graviton TT}), the following {equation}  {holds}: \label{def: susy unit 3}
    \begin{align}\label{SUSY invariance of scalar product}
       \braket{\delta^{susy}(\epsilon)\psi| \psi}_{ax} =  \braket{\varphi| \delta^{susy}(\epsilon)'\varphi}_{KG}.
    \end{align}
     According to eqs.~(\ref{SUSY_transf_spin3/2->spin2 TT}) and (\ref{SUSY_transf_spin2->spin3/2 TT}), the SUSY transformations of the TT  solutions are 
\begin{align}
    \delta^{susy}({\epsilon})\psi_{\mu} & =\frac{1}{4}  \left( i\, \varphi_{\mu \sigma}\gamma^{\sigma}+\nabla_{\lambda}  \varphi_{\mu \sigma} \gamma^{\sigma  \lambda}      \right)\epsilon,  \label{SUSY_transf_spin3/2->spin2 modes} \\
    \delta^{susy}({\epsilon})' \varphi_{\mu \nu}&=
    \delta^{susy}({\epsilon}) {\varphi}_{\mu \nu} + \delta^{gauge} \left( -\frac{i}{3} \overline{\epsilon}\gamma^{5} \psi\right)  \varphi_{\mu \nu} \nonumber  \\
    &=\frac{\overline{\epsilon}}{2}\gamma^{5}  \left(  \gamma_{\mu}  \psi_{\nu}+\gamma_{\nu} \psi_{\mu}    \right)
   + \nabla_{(\mu}\left( -\frac{i}{3} \overline{\epsilon}\gamma^{5} \psi_{\nu)}  \right).\label{SUSY_transf_spin2->spin3/2 modes}
\end{align}
Condition \ref{def: susy unit 3} can be proved for both commuting and Grassmann-odd Killing spinors. However, in this Subsection, we will focus on the commuting Killing spinors $\epsilon^{(\sigma,q)}$ (\ref{dS Killing spinors explicit_ e=S(t,x)n}).

\end{enumerate}
  Let us now demonstrate that each of the conditions \ref{def: susy unit 1}, \ref{def: susy unit 2} and \ref{def: susy unit 3} is satisfied for the SUSY representation furnished by the positive frequency solution space $\mathcal{H}^{-}_{{2}} \bigoplus \mathcal{H}^{-}_{\frac{3}{2}}$. The following analysis can be straightforwardly generalised to the case of the SUSY UIR furnished by the negative frequency solution space $\mathcal{H}^{*+}_{{2}} \bigoplus \mathcal{H}^{*+}_{\frac{3}{2}}$,  {as} will be discussed briefly {later}.
\\
\\
\noindent \textbf{\textit{1. Positivity of the norm}}$-$The positivity of the norm for the physical gravitino modes of helicity $-3/2$, $\mathcal{H}^{-}_{\frac{3}{2}}  = \{{\psi}^{(phys ,\,- \ell; \,m;k)}_{\mu} \}$, 
has been demonstrated in eq.~(\ref{norms of physical modes_3/2}). Also, the positivity of the norm for the physical graviton modes of helicity $-2$, $\mathcal{H}^{-}_{{2}}  = \{{\varphi}^{(phys ,\,- L; \,M;K)}_{\mu \nu}\}$, has been demonstrated in eq.~(\ref{norms of physical modes_2}). Thus, it is clear that the norm is positive in the direct sum of spaces $\mathcal{H}^{-}_{{2}}  \bigoplus \mathcal{H}^{-}_{\frac{3}{2}}$.

\noindent Let us also verify that the SUSY transformations (\ref{SUSY_transf_spin3/2->spin2 TT}) and (\ref{SUSY_transf_spin2->spin3/2 TT}) preserve the space $\mathcal{H}^{-}_{{2}}  \bigoplus \mathcal{H}^{-}_{\frac{3}{2}}$.
 This is important because if there are SUSY transformations acting on positive frequency graviton modes of helicity $-2$  by transforming them into positive frequency gravitino modes of helicity $+3/2$, then negative norms will appear, as the axial scalar product is negative definite in $\mathcal{H}^{+}_{\frac{3}{2}}$ [eq.~(\ref{norms of physical modes_3/2})]. Thus, we want to ensure that graviton modes in $\mathcal{H}^{-}_{{2}}$ transform under SUSY only into gravitino modes in  $\mathcal{H}^{-}_{\frac{3}{2}}$, and vice versa.    {It can be seen that SUSY transformations do not mix the spaces $\mathcal{H}_2^- \bigoplus \mathcal{H}_{\frac{3}{2}}^-$ and
 $\mathcal{H}_2^+ \bigoplus \mathcal{H}_{\frac{3}{2}}^+$ because}
 the SUSY transformations of the field strengths commute with duality transformations, as shown in Subsection \ref{subsubsec_duality+SUSY}. To be specific, eqs.~(\ref{SUSY_transf of spin3/2 fieldstrength DUAL}) and (\ref{SUSY_transf of spin2 fieldstrength DUAL}) imply that the anti-self-dual spin-3/2 field strength transforms into the anti-self-dual spin-2 field strength and vice versa. {  However, this observation does not rule out the
possibility that some SUSY transformations mix the space  $\mathcal{H}^{-}_{{2}}  \bigoplus \mathcal{H}^{-}_{\frac{3}{2}}$ with the space
 $\mathcal{H}^{*+}_{{2}}  \bigoplus \mathcal{H}^{*+}_{\frac{3}{2}}$ since both spaces
 consist of anti-self-dual mode solutions.  We prove that the space $\mathcal{H}_2^- \bigoplus \mathcal{H}_{\frac{3}{2}}^-$ is indeed fixed under the SUSY transformations by investigating how the individual TT graviton and gravitino modes transform.}
 
\noindent {First,} we determine the gravitino SUSY transformation (\ref{SUSY_transf_spin3/2->spin2 TT}) generated by the commuting Killing spinors $\epsilon^{(\sigma;q)}$ [eq.~(\ref{dS Killing spinors explicit_ e=S(t,x)n})], by working  at the level of mode solutions in our representation space, $\mathcal{H}^{-}_{{2}}  \bigoplus \mathcal{H}^{-}_{\frac{3}{2}}$.  {More specifically,} we will  substitute the modes ${\varphi}^{(phys ,\,- L; \,M;K)}_{\mu \nu}(t, \bm{\theta_{3}})$ [eq.~(\ref{physmodes_negative_spin_2_dS4})] into the right-hand side of the SUSY transformation (\ref{SUSY_transf_spin3/2->spin2 modes}), and then we will re-express the  {transformed modes} in terms of gravitino modes.

As both gravitino and graviton modes are expressed in terms of TT spherical harmonics {on $S^3$} of spin 3/2  (\ref{vector-spinor+-eigen_S3})  and spin 2 (\ref{spin-2 spherical harmonics S3}), respectively, it is 
{useful first} to clarify how SUSY acts on them. In particular, given a TT spin-2  spherical harmonic $\tilde{T}^{(\sigma; L; M;K)}_{\tilde{\mu} \tilde{\nu}}(\bm{\theta_{3}})$ ($\sigma = \pm$) and Killing spinors $\tilde{\epsilon}_{\pm,q} (\bm{\theta_{3}})$ (\ref{Killing spinor eqn S^3 MAIN TXT}) on $S^{3}$, we can construct TT spin-3/2 spherical harmonics on $S^{3}$ as
\begin{align}
    \tilde{T}^{(\sigma; L; M;K)}_{\tilde{\mu} \tilde{\nu}}  \tilde{\gamma}^{\tilde{\nu}} \tilde{\epsilon}_{+,q},~~~\text{and}~~~\tilde{T}^{(\sigma; L; M;K)}_{\tilde{\mu} \tilde{\nu}}  \tilde{\gamma}^{\tilde{\nu}} \tilde{\epsilon}_{-,q}.
\end{align}
These can be viewed as SUSY  {transforms}
   of TT spin-3/2 spherical harmonics on $S^{3}$. Indeed it can be readily verified that these are eigenfunctions of the Dirac operator on $S^{3}$ as 
\begin{align}\label{Dirac eigenvalues of S3 SUSY}
    \tilde{\slashed{\nabla}}  \left( \tilde{T}^{(\sigma; L; M;K)}_{\tilde{\mu} \tilde{\nu}}  \tilde{\gamma}^{\tilde{\nu}} \tilde{\epsilon}_{\mp,q}\right) = ~i \, \sigma \,  \Bigg( L- \delta_{\sigma, \mp} + \frac{3}{2}   \Bigg) ~\tilde{T}^{(\sigma; L; M;K)}_{\tilde{\mu} \tilde{\nu}}  \tilde{\gamma}^{\tilde{\nu}} \tilde{\epsilon}_{\mp,q},
\end{align}
where $\delta_{+ , +} = \delta_{-,-} =1$ and $\delta_{+, -} = \delta_{-,+} = 0$. We can thus identify $\tilde{T}^{(\sigma; L; M;K)}_{\tilde{\mu} \tilde{\nu}}  \tilde{\gamma}^{\tilde{\nu}} \tilde{\epsilon}_{+,q}$ and $\tilde{T}^{(\sigma; L; M;K)}_{\tilde{\mu} \tilde{\nu}}  \tilde{\gamma}^{\tilde{\nu}} \tilde{\epsilon}_{-,q}$ with linear combinations {for} TT spin-3/2 spherical harmonics  on $S^{3}$ (\ref{vector-spinor+-eigen_S3}). 
{Thus,} 
%
%\ah{[AH comment: some (possibly all) primes on the label $M'$ etc. can be omitted, I think.  If this is OK with you,
%acI prefer omitting the primes where possible.]}
%
{\begin{align}\label{SUSY on S^3 1}
& \tilde{T}^{({\pm}; L; M;K)}_{\tilde{\mu} \tilde{\nu}}(\bm{\theta_{3}})  \tilde{\gamma}^{\tilde{\nu}} \tilde{\epsilon}_{{\pm},q}(\bm{\theta_{3}}) = {\sum_{m'=M-1}^M}\tilde{\beta}^{({\pm},\ell',m',k';M)}_{{\pm},q}~ \tilde{\psi}^{(\ell';~ {m}'; k')}_{{\pm} \,\tilde{\mu}}(\bm{\theta_{3}}),~~\ell'=L-1,~~{k'=K+q} \nonumber  ,\\
& \tilde{T}^{({\pm}; L; M;K)}_{\tilde{\mu} \tilde{\nu}} (\bm{\theta_{3}}) \tilde{\gamma}^{\tilde{\nu}} \tilde{\epsilon}_{{\mp},q}(\bm{\theta_{3}}) = {\sum_{m'=M-1}^M}\tilde{\beta}^{({\pm},\ell',m',k';M)}_{{\mp},q}~ \tilde{\psi}^{(\ell';~ {m}'; k')}_{{\pm} \,\tilde{\mu}}(\bm{\theta_{3}}),~~\ell'=L,~~{k'=K+q} ,
\end{align}}
{where
\begin{align}\label{beta tilde by integration}
    \tilde{\beta}^{(\sigma,\ell',m',k';M)}_{\pm,q}
    = \int_{S^3}\sqrt{\tilde{g}}~d\bm{\theta_3}~
    \left(\tilde{\psi}^{(\ell';~ {m}'; k')\tilde{\mu}}_{{\sigma}}(\bm{\theta_{3}})\right)^{\dagger}~\tilde{T}^{(\sigma; L; M;K)}_{\tilde{\mu} \tilde{\nu}}(\bm{\theta_{3}})  \tilde{\gamma}^{\tilde{\nu}} \tilde{\epsilon}_{\pm,q}(\bm{\theta_{3}}).
\end{align}
}
{We find $k'=K+q$ by considering how both sides depend on $\theta_1$.  The range of $m'$ is found by noting
that the tensors $\tilde{T}^{(\pm; L; M;K)}_{\tilde{\mu} \tilde{\nu}}(\bm{\theta_{3}})$ and each of 
the sets of Killing spinors,
$\{\tilde{\epsilon}_{+,q}\}_{q=-1,0}$ and $\{\tilde{\epsilon}_{-,q}\}_{q=-1,0}$ form $so(3)$ representations 
with highest weights $M$ and $1/2$, respectively, and that the vector-spinors $\tilde{\psi}^{(\ell';~ {m}'; k')}_{\pm \,\tilde{\mu}}(\bm{\theta_{3}})$ form a representation with highest weight $m'+1/2$.  If the vector-spinor with the label
$(\ell',m',k')$ does not exist, then the corresponding coefficient is set to $0$. 
We do not need the explicit form of the nonzero coefficients $\tilde{\beta}^{(\sigma,\ell',m',k';M)}_{\pm,q}$.}}

Similarly, we can construct TT spin-2 spherical harmonics from TT spin-3/2 spherical harmonics and Killing spinors on $S^3$, as 
\begin{align}\label{definition of mathcal S}
\mathcal{S}_{(\pm,q;\sigma)\tilde{\mu}\tilde{\nu}}^{(\ell;~m;k)} = \tilde{\epsilon}_{\pm,q}^\dagger
{\left(\sigma\,\tilde{\gamma}_{\tilde{\mu}\tilde{\lambda}}\tilde{\nabla}^{\tilde{\lambda}} \tilde{\psi}^{(\ell;~ {m}; k)}_{\sigma \,\tilde{\nu}} 
\mp 2i\,\tilde{\gamma}_{\tilde{\mu}}\tilde{\psi}_{\sigma\,\tilde{\nu}}^{(\ell;~m;k)}\right)+ (\tilde{\mu}\leftrightarrow \tilde{\nu}).
}
\end{align}
These can be viewed as SUSY transforms for TT spin-2 spherical
harmonics on $S^3$. They are eigenfunctions of the duality
operator defined by eq.~(\ref{duality properties spin-2 S^3}):
\begin{align}
    \tensor{\tilde{\varepsilon}}{_{\tilde{\mu}}^{\tilde{\alpha}\tilde{\beta}}}
\tilde{\nabla}_{\tilde{\alpha}}~\mathcal{S}_{(\pm,q;\sigma)\tilde{\beta}\tilde{\nu}}^{(\ell;~m;k)}
& = \sigma (\ell + \delta_{\sigma,\pm} + 1)
\mathcal{S}_{(\pm,q;\sigma)\tilde{\mu}\tilde{\nu}}^{(\ell;~m;k)}.
\end{align}
As a result, we have that $\mathcal{S}_{(\pm,q;\sigma)\tilde{\beta}\tilde{\nu}}^{(\ell;~m;k)}$ are eigenfunctions of the Laplace-Beltrami operator on $S^3$, as
\begin{align} 
(-\tilde{\nabla}_{\tilde{\alpha}}\tilde{\nabla}^{\tilde{\alpha}}+3)
\mathcal{S}_{(\pm,q;\sigma)\tilde{\mu}\tilde{\nu}}^{(\ell;~m;k)}
= (\ell + \delta_{\sigma,\pm} + 1)^2
\mathcal{S}_{(\pm,q;\sigma)\tilde{\mu}\tilde{\nu}}^{(\ell;~m;k)},
\end{align}
where $\delta_{+,+} = \delta_{-,-} = 1$ and $\delta_{+,-} = \delta_{-,+} = 0$ as defined after eq.~(\ref{Dirac eigenvalues of S3 SUSY}).
(For $\ell=1$ one has $\pm(\ell+\delta_{\pm,\mp}+1) = \pm 2$.  There are no TT spin-2 spherical harmonics with these
eigenvalues for the duality operator. This implies that
$\mathcal{S}_{(\pm,q;\mp)\tilde{\mu}\tilde{\nu}}^{(\ell=1;~m;k)} = 0$.) 
Hence, one can express $\mathcal{S}_{(\pm,q;\sigma)\tilde{\mu}\tilde{\nu}}^{(\ell;~m;k)}$ as  linear combinations of 
TT spin-2 spherical harmonics, as
\begin{align}\label{SUSY on S^3 1 Atsushi}
\mathcal{S}_{({\pm},q;\,{\mp})\tilde{\mu}\tilde{\nu}}^{(\ell;~m;k)}(\bm{\theta_3})
    =  \sum_{M'=m}^{m+1}\check{\beta}^{({\mp},L',M',K';m)}_{{\pm},q}~T^{({\mp};L';M';K')}_{\tilde{\mu}\tilde{\nu}}(\bm{\theta_{3}}),~~L'=\ell,~~K'=k-q, \nonumber \\
\mathcal{S}_{({\pm},q;\,{\pm})\tilde{\mu}\tilde{\nu}}^{(\ell;~m;k)}(\bm{\theta_3})
    =  \sum_{M'=m}^{m+1}\check{\beta}^{({\pm},L',M',K';m)}_{{\pm},q}~T^{({\pm};L';M';K')}_{\tilde{\mu}\tilde{\nu}}(\bm{\theta_{3}}),~~L'=\ell+1,~~K'=k-q,
\end{align}
where 
\begin{align}\label{beta check by integration}
\check{\beta}^{({\sigma},L',M',K';m)}_{\pm,q}
= \int_{S^3}~\sqrt{\tilde{g}}~ d\bm{\theta_3} ~
T^{({\sigma};L';M';K')*}_{\tilde{\mu}\tilde{\nu}}
(\bm{\theta_3})~\mathcal{S}_{(\pm, q;{\sigma})}^{(\ell;~m;k)\tilde{\mu}\tilde{\nu}}(\bm{\theta_3})\,.
\end{align}
The relation $K'=k-q$ and the range of $M'$ have been determined as before.  Again, if there is no TT spin-2 spherical
harmonic with the label $(L',M',K')$, we let $\check{\beta}^{(\sigma,L',M',K';m)}_{\pm,q}=0$.  {From
eqs.~(\ref{beta tilde by integration}) and (\ref{beta check by integration}) we find the
following relations:
\begin{align}
\check{\beta}^{(\pm,L,M,K;m)}_{\mp,q}
& =  2iL\tilde{\beta}^{(\pm,\ell,m,k;M)*}_{\mp,q},~~L=\ell,~~K=k-q,\label{beta beta relation +}\\
\check{\beta}^{(\pm,L,M,K;m)}_{\pm,q}
& =  2i(L+2)\tilde{\beta}^{(\pm,\ell,m,k;M)*}_{\pm,q},~~L=\ell+1,~~K=k-q. \label{beta beta relation -}
\end{align}
}

Having sketched how SUSY acts on spherical harmonics on $S^{3}$, we can readily compute the  SUSY transformation {[eqs.~(\ref{SUSY_transf_spin3/2->spin2 modes}) and (\ref{SUSY_transf_spin2->spin3/2 modes})]} using the explicit expressions of the physical {spin-3/2 and} spin-2 modes, {$\psi^{(phys,-\ell; m ; k)}_{\nu}$ and} 
${\varphi}^{(phys ,\,- L; \,M;K)}_{\mu \nu}$ [eqs.~(\ref{physmodes_negative_spin_3/2_dS4}) and (\ref{physmodes_negative_spin_2_dS4})] and the expressions of the  Killing spinors (\ref{dS Killing spinors explicit}) on $dS_{4}$. The result is
\begin{align}\label{SUSY_transf RULE_3/2->2, epsilon^-}
 \left(\delta^{susy} (\epsilon^{(-;q)} )\psi\right)^{(-L; \,M;K)}_{\mu} & \equiv \frac{1}{4}  \left( i\, \varphi^{(phys ,\,- L; \,M;K)}_{\mu \sigma}~\gamma^{\sigma}+\nabla_{\lambda}  \varphi^{(phys ,\,- L; \,M;K)}_{\mu \sigma}~ \gamma^{\sigma  \lambda}      \right)\epsilon^{(-;q)} \nonumber  \\
 & = {{\frac{i}{2}\sqrt{L+2}~\sum_{m'=M-1}^M~\tilde{\beta}^{(-,\ell',m',k';M)}_{-,q}}~ \psi^{(phys ,\,- \ell'; \,m';k')}_{\mu}},\nonumber \\
 &~~\ell'=L-1,~~k'=K+q,
\end{align}
and
\begin{align}\label{SUSY_transf RULE_3/2->2, epsilon^+}
 \left( \delta^{susy}(\epsilon^{(+;q)}) \psi\right)^{( -L; \,M;K)}_{\mu}  & \equiv \frac{1}{4}  \left( i\, \varphi^{(phys ,\,- L; \,M;K)}_{\mu \sigma}~\gamma^{\sigma}+\nabla_{\lambda}  \varphi^{(phys ,\,- L; \,M;K)}_{\mu \sigma}~ \gamma^{\sigma  \lambda}      \right)\epsilon^{(+;q)} \nonumber \\
  & =- {\frac{1}{2}\sqrt{L}~{\sum_{m'=M-1}^M~\tilde{\beta}^{(-,\ell',m',k';M)}_{+,q}} ~ \psi^{(phys ,\,- \ell'; \,m';k')}_{\mu}},\nonumber \\
  &~~\ell'=L,~~k'=K+q,
\end{align}
where $q \in \{ -1, 0 \}$. The coefficients $\tilde{\beta}^{(-,\ell',m',k';M)}_{\pm,q} $, and the  angular momentum quantum numbers $m'$ and $k'$, have been introduced in eq.~(\ref{SUSY on S^3 1}).
Equations (\ref{SUSY_transf RULE_3/2->2, epsilon^-}) and (\ref{SUSY_transf RULE_3/2->2, epsilon^+}) describe the transformation rules for the gravitino modes under SUSY  generated by the four Killing spinors (\ref{dS Killing spinors explicit}) of $dS_{4}$.

%%%%%%%%%%%%%%%%%%%%%%%%%%%%%%%%%%%%%%%%%%%%%%%%%%
One can similarly obtain the SUSY transformation rules for the graviton modes using eq.~(\ref{SUSY_transf_spin2->spin3/2 modes}), as
\begin{align}\label{SUSY_transf RULE_2->3/2, epsilon^-}
 & \left( \delta^{susy}(\epsilon^{(-;q)})' \varphi\right)^{(-\ell; m ; k)}_{\mu \nu} \nonumber \\
 &\equiv \frac{ \overline{\epsilon^{(-;q)}} }{2}\gamma^{5}  \left(  \gamma_{\mu}  \psi^{(phys,-\ell; m ; k)}_{\nu}+\gamma_{\nu} \psi^{(phys,-\ell; m ; k)}_{\mu}    \right)
   + \nabla_{(\mu}\left( -\frac{i}{3}   ~ \overline{\epsilon^{(-;q)}}\gamma^{5} \psi^{(phys,-\ell; m ; k)}_{\nu)}  \right) \nonumber \\
   & {=-\frac{i}{2}\sqrt{L'+2}\sum_{M'=m}^{m+1}~\tilde{\beta}^{(-,\ell,m,k;M')*}_{-,q}~\varphi^{(phys,-L';M';K')}_{\mu \nu}} \nonumber\\
   &~~~+ \text{(TT pure-gauge graviton mode)},~~ ~~~ L' = \ell +1,~~~K'=k-q,
\end{align}
and
\begin{align}\label{SUSY_transf RULE_2->3/2, epsilon^+}
 & \left( \delta^{susy}(\epsilon^{(+;q)})' \varphi\right)^{(-\ell; m ; k)}_{\mu \nu} \nonumber \\
 &\equiv \frac{ \overline{\epsilon^{(+;q)}} }{2}\gamma^{5}  \left(  \gamma_{\mu}  \psi^{(phys,-\ell; m ; k)}_{\nu}+\gamma_{\nu} \psi^{(phys,-\ell; m ; k)}_{\mu}    \right)
   + \nabla_{(\mu}\left( -\frac{i}{3}   ~ \overline{\epsilon^{(+;q)}}\gamma^{5} \psi^{(phys,-\ell; m ; k)}_{\nu)}  \right) \nonumber \\
  & { = -\frac{1}{2}\sqrt{L'}\sum_{M'=m}^{m+1}~\tilde{\beta}^{(-,\ell,m,k;M')*}_{+,q}~\varphi^{(phys,-L';M';K')}_{\mu \nu}} \nonumber\\
   &~~~+ \text{(TT pure-gauge graviton mode)},~~ ~~~ L' = \ell,~~~K'=k-q,
\end{align}
{with}  $q \in \{ -1,0\}$, {where eqs.~(\ref{beta beta relation +}) and (\ref{beta beta relation -}) have been used.} 
{Equations (\ref{SUSY_transf RULE_2->3/2, epsilon^-}) and (\ref{SUSY_transf RULE_2->3/2, epsilon^+}) describe the transformation rules for the graviton modes under SUSY  generated by the four Killing spinors (\ref{dS Killing spinors explicit}) of $dS_{4}$. 
}

The transformation rules (\ref{SUSY_transf RULE_3/2->2, epsilon^-})-(\ref{SUSY_transf RULE_2->3/2, epsilon^+})   
{prove} that positive frequency gravitino modes of helicity $-3/2$ and positive frequency graviton modes of helicity $-2$ transform among themselves.     Thus, condition \ref{def: susy unit 1} is satisfied for the representation space $\mathcal{H}^{-}_{{2}}  \bigoplus \mathcal{H}^{-}_{\frac{3}{2}}$.
%%%%%%%%%%%%%%%%%%%
\\
\\
\noindent \textbf{Note on negative frequency modes.} Our  results concerning the positivity of the norm and the irreducibility of the SUSY representation formed by the positive frequency solution space   $\mathcal{H}^{-}_{{2}}  \bigoplus \mathcal{H}^{-}_{\frac{3}{2}}$ can be readily {adapted} to the case of the negative frequency space  $\mathcal{H}^{*+}_{{2}}  \bigoplus \mathcal{H}^{*+}_{\frac{3}{2}} = \{\varphi^{(phys ,\,+ L; \,M;K)\star}_{\mu \nu} \} \bigoplus \{ v^{(phys ,\,+\ell; \,m;k)}_{\mu}  \} $. In other words,  the SUSY representation formed by $\mathcal{H}^{*+}_{{2}}  \bigoplus \mathcal{H}^{*+}_{\frac{3}{2}} $ satisfies condition \ref{def: susy unit 1}. Let us also write the  SUSY gravitino transformation rules at the level of  the corresponding negative frequency mode solutions: 
\begin{align}\label{SUSY_transf RULE_3/2->2, NEGFREQ epsilon^+}
 \left(\delta^{susy} (\epsilon^{(+;q)} )v\right)^{(+ L; M;K)}_{\mu}  & \equiv \frac{1}{4}  \left( i\, \varphi^{(phys ,\,+ L; \,M;K){\star}}_{\mu \sigma}~\gamma^{\sigma}+\nabla_{\lambda}  \varphi^{(phys ,\,+ L; \,M;K){\star}}_{\mu \sigma}~ \gamma^{\sigma  \lambda}      \right)\epsilon^{(+;q)} \nonumber  \\
  & = {\frac{1}{2}\sqrt{L+2}\sum_{m'=M-1}^M\tilde{\beta}^{(+,\ell',m',k';M)}_{+,q} ~ v^{(phys ,\,+ \ell'; \,m';k')}_{\mu}},\nonumber\\
  &\ell'=L-1,~k'=K+q
\end{align}
\begin{align}\label{SUSY_transf RULE_3/2->2, NEGFREQ epsilon^-}
 \left( \delta^{susy}(\epsilon^{(-;q)}) v  \right)^{( + L; M;K)}_{\mu}  & \equiv \frac{1}{4}  \left( i\, \varphi^{(phys ,\,+ L; \,M;K){\star}}_{\mu \sigma}~\gamma^{\sigma}+\nabla_{\lambda}  \varphi^{(phys ,\,+ L; \,M;K){\star}}_{\mu \sigma}~ \gamma^{\sigma  \lambda}      \right)\epsilon^{(-;q)} \nonumber \\
   & =- {\frac{i}{2}\sqrt{L} \sum_{m'=M-1}^M\,\tilde{\beta}^{(+,\ell',m',k';M)}_{-,q} ~ v^{(phys ,\,+ \ell'; \,m';k')}_{\mu}},\nonumber\\
   &\ell'=L,~k'=K+q,
\end{align}
where the coefficients $\tilde{\beta}^{(+,\ell',m',k';M)}_{\pm,q}$ have been introduced in (\ref{SUSY on S^3 1}). 
%%%
Similarly, we find the graviton SUSY transformation for the negative frequency modes 
\begin{align}
 & \left( \delta^{susy}(\epsilon^{(+;q)})' \varphi\right)^{(+\ell; m ; k)\star}_{\mu \nu} \nonumber \\
 &\equiv \frac{ \overline{\epsilon^{(+;q)}} }{2}\gamma^{5}  \left(  \gamma_{\mu}  v^{(phys,+\ell; m ; k)}_{\nu}+\gamma_{\nu} v^{(phys,+\ell; m ; k)}_{\mu}    \right)
   + \nabla_{(\mu}\left( -\frac{i}{3}   ~ \overline{\epsilon^{(+;q)}}\gamma^{5} v^{(phys,+\ell; m ; k)}_{\nu)}  \right) \nonumber \\
&{=~{-\frac{1}{2}\sqrt{L'+2}}\sum_{M'=m}^{m+1}~\tilde{\beta}^{(+,\ell,m,k;M')*}_{+,q}}~~\varphi^{(phys,+L';M';K')\star}_{\mu \nu}\nonumber\\
   &~~+ \text{(TT pure-gauge graviton mode)},~~ ~~~ L' = \ell +1,~~K'=k-q
\end{align}
and
\begin{align}
 & \left( \delta^{susy}(\epsilon^{(-;q)})' \varphi\right)^{(+\ell; m ; k)\star}_{\mu \nu} \nonumber \\
 &\equiv \frac{ \overline{\epsilon^{(-;q)}} }{2}\gamma^{5}  \left(  \gamma_{\mu}  v^{(phys,+\ell; m ; k)}_{\nu}+\gamma_{\nu} v^{(phys,+\ell; m ; k)}_{\mu}    \right)
   + \nabla_{(\mu}\left( -\frac{i}{3}   ~ \overline{\epsilon^{(-;q)}}\gamma^{5} v^{(phys,+\ell; m ; k)}_{\nu)}  \right) \nonumber \\
&{=~{-\frac{i}{2}\sqrt{L'}}\sum_{M'=m}^{m+1}~\tilde{\beta}^{(+,\ell,m,k;M')*}_{-,q}}~\varphi^{(phys,+L';M';K')\star}_{\mu \nu}\nonumber\\
   &~~+ \text{(TT pure-gauge graviton mode)},~~~~~ L' = \ell ,~~K'=k-q
\end{align}
where $q \in \{ -1,0\}$. Here, {we have used eqs.~(\ref{beta beta relation +}) and (\ref{beta beta relation -}) again.} 
\\
\\
\noindent \textbf{\textit{2. Anti-hermiticity of even generators}}$-$The even generators of our SUSY algebra are those of $so(4,2) \bigoplus u(1)$. In the case of gravitino modes, the anti-hermiticity of all  $so(4,2)$ generators with respect to the axial scalar product has been demonstrated
 in eqs.~(\ref{anti-herm_Lie deriv_gravitino}) and (\ref{anti-herm_conf-like deriv_gravitino}). In the case of graviton modes, the anti-hermiticity of all $so(4,2)$ generators with respect to the Klein-Gordon scalar product has been demonstrated in eqs.~(\ref{anti-herm_Lie deriv_graviton}) and (\ref{anti-herm_conf-like KG}). In particular, we have already established that each of the solution spaces $\mathcal{H}^{-}_{\frac{3}{2}}$
and $\mathcal{H}^{-}_{{2}}$ furnishes a UIR of $so(4,2)$ in Subsections \ref{Subsec_gravitino modes so(4,2)} and \ref{Subsec_graviton modes so(4,2)}, respectively. In these Subsections, it was also shown that each of the negative frequency  spaces $\mathcal{H}^{*+}_{\frac{3}{2}}$
and $\mathcal{H}^{*+}_{{2}}$ furnishes a UIR of $so(4,2)$. {Finally}, it is  easy to check that both the axial scalar product and the Klein-Gordon scalar product are $u(1)$-invariant. Thus, condition \ref{def: susy unit 2} is satisfied for the SUSY representation formed by $\mathcal{H}^{-}_{{2}}  \bigoplus \mathcal{H}^{-}_{\frac{3}{2}} $, as well as for the SUSY representation formed by $\mathcal{H}^{*+}_{{2}}  \bigoplus \mathcal{H}^{*+}_{\frac{3}{2}} $.
%%%%%%%%%%%%%%%%%%%%%%%%%%%%%%%%%%%%%%%%%%%%%%%%%%%%%%5
\\
\\
\noindent \textbf{\textit{3. SUSY-invariance of inner products}}$-$Our aim is to show that condition \ref{def: susy unit 3} is satisfied, which means that we have to  prove eq.~(\ref{SUSY invariance of scalar product}). Equivalently we can show that the axial  (\ref{axial current}) and Klein-Gordon (\ref{KG current}) currents satisfy
\begin{align}\label{SUSY invariance of currents}
    J^{\mu}_{ax} \left( \delta^{susy}(\epsilon) \psi, \psi  \right) = &~J^{\mu}_{KG}\left(  \varphi,\delta^{susy}(\epsilon)'   \varphi  \right)  \nonumber \\
    &+ \text{(total divergence of rank-2 anti-symmetric tensor)},
\end{align}
as by integrating the $t$-component of eq.~(\ref{SUSY invariance of currents}) over $S^{3}$ we find the desired eq.~(\ref{SUSY invariance of scalar product}). Here, $\varphi_{\mu \nu}$ is any TT graviton  solution and $\psi_{\mu}$ is any TT gravitino solution, where  $\delta^{susy}(\epsilon)\psi_{\mu}$ and  $\delta^{susy}(\epsilon)'\varphi_{\mu \nu} = \delta^{susy}(\epsilon)\varphi_{\mu \nu} +  \delta^{gauge}(-\frac{i}{3}  \overline{\epsilon} 
 \gamma^{5}\psi)\varphi_{\mu \nu}$ are given by eqs.~(\ref{SUSY_transf_spin3/2->spin2 modes}) and (\ref{SUSY_transf_spin2->spin3/2 modes}), respectively. Let us first observe that, since both  $\delta^{susy}(\epsilon) \psi_{\mu}$ and $\psi_{\mu}$ are TT gravitino solutions, the SUSY Noether current (\ref{def: SUSY Noether currents}) is directly expressed as
\begin{align}\label{SUSY current=axial current}
    \mathcal{J}^{\mu}_{(\epsilon)}(\varphi,\psi) = J^{\mu}_{ax}\left( \delta^{susy}(\epsilon) \psi , \psi  \right).
\end{align}
This expression relates the SUSY Noether current with the axial current. The next step is to re-write the SUSY Noether current $\mathcal{J}^{\mu}_{(\epsilon)}(\varphi,\psi)$ in terms of the Klein-Gordon current. A straightforward calculation gives
\begin{align} \label{SUSY current = sym current}
   \mathcal{J}^{\mu}_{(\epsilon)}(\varphi,\psi) =  J^{\mu}_{symp} \left(\varphi, \delta^{susy}(\epsilon)\varphi \right) - \nabla_{\rho} \left( \frac{i}{4} \overline{\epsilon}\gamma^{5} \gamma^{\nu \mu \rho} \psi^{\sigma} ~\varphi^{*}_{\sigma \nu}  + \frac{i}{2} \overline{\epsilon} \gamma^{5} \gamma^{[\rho} \psi_{\sigma} ~\varphi^{\mu] \, \sigma *} \right),
\end{align}
where the second term is the divergence of an anti-symmetric tensor, and $J^{\mu}_{symp}$ is the symplectic current (\ref{sym-current}) with
\begin{align} \label{sym current(phi,Delta phi)}
    J^{\mu}_{symp}& \left(\varphi, \delta^{susy}(\epsilon)\varphi \right) \nonumber \\
    &= - \frac{i}{4}  \left(  \varphi_{\nu \lambda}^{*} ~ \nabla^{\mu} \delta^{susy}(\epsilon)\varphi^{\nu \lambda}  - \delta^{susy}(\epsilon)\varphi^{\nu \lambda}~ \nabla^{\mu}\varphi_{\nu \lambda}^{*}    -2 \varphi^{\mu \, *}_{~~\lambda} ~\nabla_{\alpha} \delta^{susy}(\epsilon)\varphi^{\alpha \lambda}  \right).
\end{align}
Note that the graviton SUSY transformation $\delta^{susy}(\epsilon)\varphi_{\mu \nu}$ appearing in eqs.~(\ref{SUSY current = sym current}) and (\ref{sym current(phi,Delta phi)}) is \textbf{not} a TT graviton solution, but it is related to the TT SUSY transformation 
$\delta^{susy}(\epsilon)'\varphi_{\mu \nu}$ through eq.~(\ref{SUSY_transf_spin2->spin3/2 modes}). Then, using  eq.~(\ref{SUSY_transf_spin2->spin3/2 modes}) to express $\delta^{susy}(\epsilon) \varphi_{\mu \nu}$ in terms of $\delta^{susy}(\epsilon)' \varphi_{\mu \nu}$, and recalling that the Klein-Gordon current coincides with the symplectic current when both arguments are TT solutions\footnote{See the passage below eq.~(\ref{gauge-invariance of sym curent}).}, we can re-express $  J^{\mu}_{symp} \left(\varphi, \delta^{susy}(\epsilon)\varphi \right) $ as
\begin{align}\label{sym current=KG current+gauge}
      J^{\mu}_{symp} \left(\varphi, \delta^{susy}(\epsilon)\varphi \right) =   J^{\mu}_{KG} \left(\varphi, \delta^{susy}(\epsilon)'\varphi \right) -  J^{\mu}_{symp}& \Bigg(\varphi, ~\delta^{gauge}\left(-\frac{i}{3} \overline{\epsilon}  \gamma^{5} \psi\right)\varphi \Bigg).
\end{align}
Then, comparing eqs.~(\ref{SUSY current=axial current}) and (\ref{SUSY current = sym current}), and making use of eq.~(\ref{sym current=KG current+gauge}), we find
\begin{align}
 J^{\mu}_{ax}\left( \delta^{susy}(\epsilon) \psi , \psi  \right) =&   ~J^{\mu}_{KG} \left(\varphi, \delta^{susy}(\epsilon)'\varphi \right)-  J^{\mu}_{symp} \Bigg(\varphi, ~\delta^{gauge}\left(-\frac{i}{3} \overline{\epsilon}  \gamma^{5} \psi\right)\varphi \Bigg) \nonumber \\
 &- \nabla_{\rho} \left( \frac{i}{4} \overline{\epsilon}\gamma^{5} \gamma^{\nu \mu \rho} \psi^{\sigma} ~\varphi^{*}_{\sigma \nu}  + \frac{i}{2} \overline{\epsilon} \gamma^{5} \gamma^{[\rho} \psi_{\sigma} ~\varphi^{\mu] \, \sigma *} \right).
\end{align}
Finally, we find that this equation takes the desired form (\ref{SUSY invariance of currents}) by using eq.~(\ref{gauge-invariance of sym curent}). We have thus shown that condition \ref{def: susy unit 3} is satisfied.  {This condition
can also be verified directly by noting that the coefficients in eqs.~(\ref{SUSY_transf RULE_2->3/2, epsilon^-})
and (\ref{SUSY_transf RULE_2->3/2, epsilon^+}) are the complex conjugates of those in   
eqs.~(\ref{SUSY_transf RULE_3/2->2, epsilon^-}) and
(\ref{SUSY_transf RULE_3/2->2, epsilon^+}), respectively.}

%%%%%%%%%%%%%%%%%%%%%%%%%%%%%%%%%%%%%%%%%%%%%%%%%%%%%

\subsubsection{Unitary  SUSY in the QFT Fock space of the chiral graviton and  chiral gravitino} \label{subsubsec_unitary SUSY QFT}
In the previous Subsection we showed that the space of TT positive frequency modes  $ \mathcal{H}^{-}_{2}  \bigoplus \mathcal{H}^{-}_{\frac{3}{2}}= \{ {\varphi}^{(phys ,\,- L; \,M;K)}_{{\mu}{\nu}}\} \bigoplus  \{{\psi}^{(phys ,\,- \ell; \,m;k)}_{{\mu}} \} $ forms a UIR of SUSY with SUSY transformations given by 
eqs.~(\ref{SUSY_transf_spin3/2->spin2 modes}) and (\ref{SUSY_transf_spin2->spin3/2 modes}) - or equivalently by eqs.~(\ref{SUSY_transf_spin3/2->spin2 TT}) and (\ref{SUSY_transf_spin2->spin3/2 TT}). The commutator of two SUSY transformations is given in eqs.~(\ref{commutator of SUSY on spin-2 TT}) and (\ref{commutator of SUSY on spin-3/2 TT}) and the even part of the superalgebra is isomorphic to $so(4,2) \bigoplus u(1)$. 
We also showed that the space of TT negative frequency modes  $ \mathcal{H}^{*+}_{2}  \bigoplus \mathcal{H}^{*+}_{\frac{3}{2}}= \{ {\varphi}^{(phys ,\,+ L; \,M;K)\star}_{{\mu}{\nu}}\} \bigoplus  \{{v}^{(phys ,\,+ \ell; \,m;k)}_{{\mu}} \} $  forms a UIR  of the same SUSY algebra but with opposite helicities relative to the positive frequency modes.  Now we will study the realisation of unitary SUSY in the QFT Fock space of the chiral graviton and chiral gravitino. Recall that by `chiral'  we mean that the corresponding field strengths are anti-self-dual. 
Let us start by reviewing the main features of the chiral graviton and chiral gravitino from the previous Sections:

\begin{itemize}
    
\item \textbf{Chiral gravitino.} The completely gauge-fixed chiral gravitino field $\Psi^{(\text{TT})-}_{\mu}$ was quantised in Subsection \ref{Subsec_gravitino quantisation}.  For convenience let us give here again the mode expansion (\ref{mode expansion gravitino}):
\begin{align*}
    & {\Psi}^{(\text{TT})-}_{t}(t, \bm{\theta_{3}})  = 0, \nonumber\\
    &  {\Psi}^{\text{(TT)}-}_{\tilde{\mu}}(t, \bm{\theta_{3}})  = \sum_{\ell =1}^{\infty}   \sum_{m, k} \left( {a}^{(-)}_{\ell m k}{\psi}^{\left(phys ,\,- \ell\,;m;k \right)}_{\tilde{\mu}}(t, \bm{\theta_{3}})  + {b}^{(+)\dagger}_{\ell m k}\,{v}^{\left(phys ,\,+ \ell\,;m;k \right)}_{\tilde{\mu}} (t, \bm{\theta_{3}}) \right),
\end{align*}
where 
\begin{align*}
    \{  {a}^{(-)}_{\ell m k} , {a}^{(-)\dagger}_{\ell' m' k'}  \} = \delta_{\ell   \ell'}  \delta_{m m'}  \delta_{kk'},~~~~~~\{  {b}^{(+)}_{\ell m k} , {b}^{(+)\dagger}_{\ell' m' k'}  \} = \delta_{\ell   \ell'}  \delta_{m m'}\delta_{kk'}.
\end{align*}
The field strength of the chiral gravitino [eq.~(\ref{def:gravitino_field-strength ANTISELF})] satisfies the anti-self-duality constraint (\ref{anti-self constr gravitino}).
The chiral gravitino vacuum is denoted as $\ket{0}_{\frac{3}{2}}$ and satisfies ${a}^{(-)}_{\ell m k} \ket{0}_{\frac{3}{2}} =  {b}^{(+)}_{\ell m k} \ket{0}_{\frac{3}{2}}  = 0$, 
 for all allowed values of  $\ell , m , k$. The vacuum is invariant under $so(4,1)$ and the single-particle Hilbert spaces of the QFT furnish a direct sum of two $\Delta= 5/2$ discrete series UIRs of $so(4,1)$ with opposite helicities. The vacuum is also invariant under $so(4,2)$ and the single-particle UIRs of $so(4,1)$ extend to a direct sum of $so(4,2)$ UIRs with opposite helicities - see Subsection \ref{Subsec_gravitino quantisation}. It is easy to show that these statements also extend from $so(4,2)$ to $so(4,2)\bigoplus u(1)$.

\item  \textbf{Chiral graviton.} The completely gauge-fixed chiral graviton field $\frak{h}^{(\text{TT})-}_{\mu \nu}$ has been quantised in Subsection \ref{Subsec_graviton quantisation}.  Let us present here again the mode expansion  for the chiral graviton (\ref{mode expansion graviton}):
\begin{align*}
    & {\mathfrak{h}}^{(\text{TT})-}_{t \mu}(t, \bm{\theta_{3}})  = 0, \nonumber\\
    &  {\mathfrak{h}}^{\text{(TT)}-}_{\tilde{\mu}   \tilde{\nu}}(t, \bm{\theta_{3}})  = \sum_{L =2}^{\infty}   \sum_{M,K} \left( {c}^{(-)}_{LMK}{\varphi}^{(phys ,\,- L; \,M;K)}_{\tilde{\mu} \tilde{\nu}}(t, \bm{\theta_{3}})  + {d}^{(+)\dagger}_{LMK}\,{\varphi}^{(phys ,\,+ L; \,M;K)\star}_{\tilde{\mu} \tilde{\nu}} (t, \bm{\theta_{3}}) \right),
\end{align*}
with
\begin{align*}
[  {c}^{(-)}_{LMK} , {c}^{(-)\dagger}_{L' M' K'}  ] = \delta_{L   L'}  \delta_{MM'}  \delta_{KK'},~~~~~~[  {d}^{(+)}_{LMK} , {d}^{(+)\dagger}_{L' M' K'}  ] = \delta_{L   L'}  \delta_{MM'}  \delta_{KK'}.
\end{align*}
The field strength of the chiral graviton [eq.~(\ref{def:anti-self-dual Weyl})] satisfies the anti-self-duality constraint (\ref{anti-self constr graviton}).
The chiral graviton vacuum  $\ket{0}_{2}$ satisfies ${c}^{(-)}_{LMK} \ket{0}_{2} =  {d}^{(+)}_{ L M K } \ket{0}_{2}  = 0$,
 for all allowed values of $L,M,K$. The vacuum is invariant under $so(4,1)$ and the single-particle Hilbert spaces of the QFT furnish a direct sum of two $\Delta= 3$ discrete series UIRs of $so(4,1)$ with opposite helicities. The vacuum is also invariant under $so(4,2)$ and the single-particle UIRs of $so(4,1)$ extend to a direct sum of $so(4,2)$ UIRs with opposite helicities - see Subsection \ref{Subsec_graviton quantisation}. Again, it is easy to show that these statements also extend from $so(4,2)$ to $so(4,2)\bigoplus u(1)$.
 \end{itemize}
%%%%%%%%%%%%%%%%%%%%%%%%%%%%%%%%%%%%%%%%%%%%%%%%%%%%%%%%%%%%

Let us also recall what we know so far about the SUSY representation carried by the chiral graviton and chiral gravitino. First, as the SUSY transformations of the field strengths commute with duality transformations (see Subsections \ref{subsubsec_duality+SUSY} and \ref{subsubsec_unitary SUSY modes}), it is clear that the chiral graviton and chiral gravitino gauge potentials $({\frak{h}}^{(\text{TT})-}_{\mu \nu}, \Psi^{(\text{TT})-}_{\mu} )$ form a supermultiplet. The SUSY transformations of the chiral gauge potentials are given by 
eqs.~(\ref{SUSY_transf_spin3/2->spin2 TT}) and (\ref{SUSY_transf_spin2->spin3/2 TT}) with $\frak{h}^{(\text{TT})}_{\mu \nu}$ and ${\Psi}^{(\text{TT})}_{\mu }$ replaced by $\frak{h}^{(\text{TT})-}_{\mu \nu}$ and ${\Psi}^{(\text{TT})-}_{\mu }$,  respectively. The commutators of two SUSY variations are given again by 
eqs.~(\ref{commutator of SUSY on spin-2 TT}) and (\ref{commutator of SUSY on spin-3/2 TT}), but with $\frak{h}^{(\text{TT})}_{\mu \nu}$ and ${\Psi}^{(\text{TT})}_{\mu }$ replaced by $\frak{h}^{(\text{TT})-}_{\mu \nu}$ and ${\Psi}^{(\text{TT})-}_{\mu }$,  respectively.  The TT gauge transformations in eqs.~(\ref{commutator of SUSY on spin-2 TT}) and (\ref{commutator of SUSY on spin-3/2 TT}) are identified with zero (recall that the UIRs formed by mode solutions were defined in terms of equivalence classes of mode solutions). We are allowed to do this because the transformations of quantum fields are attributed to transformations of the creation and annihilation operators and these have gauge-invariant definitions\footnote{In particular,  eq.~(\ref{crtn and annihil ops in terms of Psi_m}) implies the gauge independence of the gravitino creation and annihilation operators as the axial scalar product is invariant under TT gauge transformations (\ref{gauge_transf_spin3/2_restricted}).  Similarly, 
eq.~(\ref{crtn and annihil ops in terms of h_mn}) implies the invariance of the graviton creation and annihilation operators under TT gauge transformation (\ref{gauge_transf_spin2_cmplx_res}).}. We also know from the results of Subsection \ref{subsubsec_unitary SUSY modes} that the space of positive frequency modes  $\mathcal{H}^{-}_{2}  \bigoplus \mathcal{H}^{-}_{\frac{3}{2}}$ and the space of negative frequency modes $\mathcal{H}^{*+}_{2}  \bigoplus \mathcal{H}^{*+}_{\frac{3}{2}}$ separately form UIRs of our superalgebra.
%%%%%%%%%%%%%%%%%%%%%%%%%%%%%%%%%%%%%%%%%%%%%%%%%%%%%%%%%%%%%%%%%%%%%%%%%%%%%%%%%%%%%%%%%%%%%%%%%%%%%%%%%%%%%%%%%%%%%%%%%%%%%%%%%%%%%%%%%%%%%%%%%%%%%%%%%%%%%%%%%%%%%%%%%%%%%%%%%%%%%%%%%%%%%%%%%%%%%%%%%%%%%%%%%%%%%%%%%%%%%%%%%%%%%%%%%%%%%%%%%%%%%%%%%%%%%%

{Let us now} show that SUSY is realised unitarily in the QFT Fock space of the chiral graviton and chiral gravitino.  In fact, unitarity follows from the analysis we have already presented, as we have explicitly constructed the QFT Fock space and we have shown that the norm is positive. In addition, we have shown that the single-particle Hilbert space ($\cong$ Hilbert space of TT mode solutions) carries a direct sum of UIRs of our superalgebra - see Subsection \ref{subsubsec_unitary SUSY modes}.  {Nevertheless,} for the sake of completeness, we will
construct the quantum operators corresponding to the SUSY Noether charges (\ref{def: SUSY noether charge}) and we will show that they generate unitary representations of our superalgebra in the QFT Fock space.
%%%%%%%%%%%%%%%%%%%5%%%%%%%%%%%%%%%%%%%%%%%%%%%%%%%55%%%%%%%%%%%%%%%%%%%%%%%%%%%%%%%%%%%%%%%%%%%%%%%%%%%%%%%%%%%%%%%%%%%%%%%%%%%%%%%%%%%%%%%%%%%%%
\\
\\
\noindent \textbf{\textit{Quantum SUSY generators}}$-$    The quantum SUSY Noether charges that are relevant to our theory, $Q^{susy}{[\epsilon]}$, are found by replacing $\frak{h}^{\dagger}_{\mu \nu}$ and $\Psi_{\mu}$ in  eq.~(\ref{def: SUSY noether charge}) with the chiral quantum fields $\frak{h}^{(\text{TT})-\dagger}_{\mu \nu}$ and $\Psi^{(\text{TT})-}_{\mu}$, respectively. 
The standard approach is to use Grassmann-odd Killing spinors, rendering the SUSY  Noether charges   $Q^{susy}{[\epsilon]} = \overline{\eta}^{A} \, Q_{A}$ Grassmann-even, where $\eta$ is the constant Grassmann-odd spinor parameter in eq.~(\ref{dS Killing spinors e=S(t,x)n_anti-com}). The anti-commutators of the spinorial supercharges, $\left\{ Q_{A}, Q^{B\,\dagger} \right\}$, are  then encoded in the commutators   $\left[Q^{susy}{[\epsilon]}, Q^{susy}{[\epsilon']}^{\dagger} \right]$. Here, we will adopt an alternative approach where we will use the commuting Killing spinors $\epsilon^{(\sigma ; q)} (t , \bm{\theta_{3}})= S(t , \bm{\theta_{3}}) \eta^{(\sigma ; q)}$ [eq.~(\ref{dS Killing spinors explicit_ e=S(t,x)n})]. We will thus work with the \textit{Grassmann-odd SUSY Noether charges} $Q^{susy}{[\epsilon^{(\sigma;q)}]}$ - see the discussion below eq.~(\ref{def: SUSY noether charge}). Now, the SUSY algebra is determined by anti-commutators 
$$\left\{Q^{susy}{[\epsilon^{(\sigma ; q)}]}, Q^{susy}{[\epsilon^{(\sigma' \, ; q')}]}^{\dagger} \right \}.$$

Let us now re-express the Grassmann-odd SUSY Noether charges $Q^{susy}{[\epsilon^{(\sigma \, ; q)}]}$ in a more convenient form. Using eqs.~(\ref{SUSY current=axial current}) and (\ref{SUSY invariance of currents}), we re-express 
eq.~(\ref{def: SUSY noether charge}) as:
\begin{align}\label{quantum SUSY Noether charge}
    Q^{susy}{[\epsilon^{(\sigma \, ; q)}]}   =\braket{\delta^{susy}(\epsilon^{(\sigma \, ; q)})\Psi^{(\text{TT})-}| \Psi^{(\text{TT})-}}_{ax} =  \braket{\frak{h}^{(\text{TT})-}| ~\delta^{susy}(\epsilon^{(\sigma  ; q)})' ~\frak{h}^{(\text{TT})-}}_{KG}.
    \end{align}
    The quantum charge $Q^{susy}{[\epsilon^{(\sigma;q)}]}^{\dagger}$ is given by the hermitian conjugate of this expression.
     There are four independent SUSY Noether charges, one charge for each Killing spinor $\epsilon^{(+;-1)}(t, \bm{\theta_{3}}), \epsilon^{(+;0)}(t, \bm{\theta_{3}})$,  $\epsilon^{(-;-1)}(t, \bm{\theta_{3}})$, and $\epsilon^{(-;0)}(t, \bm{\theta_{3}})$ - see eq.~(\ref{dS Killing spinors explicit}). 
   Below we express $Q^{susy}[\epsilon^{(\sigma ; q)}]$ 
    in terms of creation and annihilation operators.
    
    Let us first recall that the transformations of the field operators are attributed to transformations of their creation and annihilation operators.
    By expanding the fields in modes, eq.~(\ref{quantum SUSY Noether charge}) gives
    \begin{align}\label{quantum SUSY Noether charge def2}
    Q^{susy}{[\epsilon^{(\sigma;q)}]}   =&\sum_{L=2}^{\infty}\sum_{M,K} \left( c^{(-)\dagger}_{LMK} ~\delta^{susy}(\epsilon^{(\sigma;q)})' c^{(-)}_{LMK} - d^{(+)}_{LMK} ~\delta^{susy}(\epsilon^{(\sigma;q)})' d^{(+)\dagger}_{LMK} \right) \\
    = & \sum_{\ell=1}^{\infty}\sum_{m,k} \left( \delta^{susy}(\epsilon^{(\sigma;q)}) a^{(-)\dagger}_{\ell m k} ~a^{(-)}_{\ell m k} + \delta^{susy}(\epsilon^{(\sigma;q)}) b^{(+)}_{\ell m k } ~b^{(+)\dagger}_{\ell m k} \right).
    \end{align}
By construction, the quantum SUSY Noether charges generate the desired SUSY transformations [eqs.~(\ref{SUSY_transf_spin3/2->spin2 TT}) and (\ref{SUSY_transf_spin2->spin3/2 TT})] on our chiral quantum fields , as
\begin{align}
 [  ~ \frak{h}^{(\text{TT})-}_{\mu \nu}, Q^{susy}{[\epsilon^{(\sigma;q)}]}  ~  ]     = \delta^{susy}(\epsilon^{(\sigma;q)})'  \frak{h}^{(\text{TT})-}_{\mu \nu},
\end{align}
\begin{align}
 \left \{   {\Psi}^{(\text{TT})-}_{\mu }, Q^{susy}{[\epsilon^{(\sigma;q)}]}^{\dagger}   \right \}     = \delta^{susy}(\epsilon^{(\sigma;q)})  {\Psi}^{(\text{TT})-}_{\mu }.
\end{align}
    Let us now obtain explicit expressions for the SUSY transformations of the creation and annihilation operators of the chiral graviton\footnote{The SUSY transformations of gravitino creation and annihilation operators can be obtained similarly.}. From eq.~(\ref{crtn and annihil ops in terms of h_mn}) we find
\begin{align}
   \delta^{susy}(\epsilon^{(\sigma;q)})' {c}^{(-)}_{LMK}&= \braket{ {\varphi}^{(phys ,\,- L; \,M;K)}|\delta^{susy}(\epsilon^{(\sigma;q)})'{\mathfrak{h}}^{\text{(TT)}-}}_{KG}  \nonumber\\
  &=\braket{ \left(\delta^{susy}(\epsilon^{(\sigma;q)}) {\psi}\right)^{(- L; \,M;K)}|{{\Psi}}^{\text{(TT)}-}}_{ax},  
  \end{align}
and
  \begin{align}
  \delta^{susy}(\epsilon^{(\sigma;q)})'{d}^{(+)\dagger}_{LMK}&= -\braket{ {\varphi}^{(phys ,\,+ L; \,M;K)\star}|\delta^{susy}(\epsilon^{(\sigma;q)})'{\mathfrak{h}}^{\text{(TT)}-}}_{KG} \nonumber \\
  &=  -\braket{ \left(\delta^{susy}(\epsilon^{(\sigma;q)}) {v}\right)^{(+ L; \,M;K)}|{{\Psi}}^{\text{(TT)}-}}_{ax},
\end{align}
where we have made use of the SUSY-invariance of the axial and Klein-Gordon inner products [eq.~(\ref{SUSY invariance of scalar product})]. The SUSY transformation of the  positive frequency gravitino mode $\left(\delta^{susy}(\epsilon^{(\sigma;q)}) {\psi}\right)^{(- L; \,M;K)}_{\mu}$ is given in eqs.~(\ref{SUSY_transf RULE_3/2->2, epsilon^-}) and (\ref{SUSY_transf RULE_3/2->2, epsilon^+}). The SUSY transformation of the negative frequency gravitino mode $\left(\delta^{susy}(\epsilon^{(\sigma;q)}) {v}\right)^{(+ L; \,M;K)}_{\mu}$ is given in eqs.~(\ref{SUSY_transf RULE_3/2->2, NEGFREQ epsilon^+}) and (\ref{SUSY_transf RULE_3/2->2, NEGFREQ epsilon^-}). We expect to find {that $\delta^{susy}(\epsilon^{(\sigma;q)})' {c}^{(-)}_{LMK}$ and $\delta^{susy}(\epsilon^{(\sigma;q)})' {d}^{(+)\dagger}_{LMK}$ are proportional to
a negative-helicity gravitino annihilation operator and a positive-helicity gravitino creation operator, respectively.}
Indeed, using eqs.~(\ref{SUSY_transf RULE_3/2->2, epsilon^-}), (\ref{SUSY_transf RULE_3/2->2, epsilon^+}), (\ref{SUSY_transf RULE_3/2->2, NEGFREQ epsilon^+}) and (\ref{SUSY_transf RULE_3/2->2, NEGFREQ epsilon^-}), as well as eq.~(\ref{crtn and annihil ops in terms of Psi_m}), we find the SUSY transformation formulae for the creation and annihilation operators of the chiral graviton, as  
\begin{align}
  & \delta^{susy}(\epsilon^{(-;q)})' {c}^{(-)}_{LMK} = - \frac{i}{2}\sqrt{L+2}~\sum_{m'=M-1}^M\tilde{\beta}^{(-,\ell',m',k';M)*}_{-,q} ~a^{(-)}_{\ell'\,m'\,k'}, ~~~\ell'=L-1,~k'=K+q ,  \nonumber 
 \\
 & \delta^{susy}(\epsilon^{(-;q)})'{d}^{(+)\dagger}_{LMK} = - {\frac{i}{2}\sqrt{L}~\sum_{m'=M-1}^M\tilde{\beta}^{(+,\ell',m',k';M)*}_{-,q}}~b^{(+)\dagger}_{\ell'\,m'\,k'},~~\ell'=L,~~k'=K+q,
\end{align}
and
\begin{align}
  & \delta^{susy}(\epsilon^{(+;q)})' {c}^{(-)}_{LMK} = - {\frac{1}{2}\sqrt{L} ~\sum_{m'=M-1}^M\tilde{\beta}^{(-,\ell',m',k';M)*}_{+,q}}~a^{(-)}_{\ell'm'k'}, ~~~\ell'=L,~~k'=K+q   ,\nonumber 
 \\
 & \delta^{susy}(\epsilon^{(+;q)})'{d}^{(+)\dagger}_{LMK} = - {\frac{1}{2}\sqrt{L+2}\, \sum_{m'=M-1}^M\tilde{\beta}^{(+,\ell',m',k';M)*}_{+,q}} ~b^{(+)\dagger}_{\ell'm'k'},~~~\ell'=L-1,~~k'=K+q.
\end{align}
The  coefficients $\tilde{\beta}^{(\sigma,\ell',m',k';M)}_{\pm,q} $ ($\sigma = \pm$) and the angular momentum quantum numbers $m'$ and $ k'$ have been introduced in 
eq.~(\ref{SUSY on S^3 1}). The label $q \in \{ 0,-1\}$ is a $so(2)$ quantum number labelling the Killing spinors of $dS_{4}$ - see eq.~(\ref{dS Killing spinors explicit}).  It is clear  that annihilation/creation operators of the chiral graviton transform into annihilation/creation operators of the chiral gravitino, demonstrating the SUSY invariance of the vacuum $\ket{0}_{2}  \otimes \ket{0}_{\frac{3}{2}}$.

Now that we have determined the SUSY transformation formulae for $ \delta^{susy}(\epsilon^{(\pm;q)})'{c}^{(-)}_{LMK} $ and  $ \delta^{susy}(\epsilon^{(\pm;q)})'{d}^{(+)\dagger}_{LMK} $, let us substitute them into 
eq.~(\ref{quantum SUSY Noether charge def2}). The quantum SUSY Noether charges are then found to be:
\begin{align}\label{quantum SUSY Noether charge explct epsilon^-}
    Q^{susy}{[\epsilon^{(-;q)}]}   =-\frac{i}{2}\sum_{L=2}^{\infty}\sum_{M,K} \Bigg ( & \,  {\sqrt{L+2}~~c^{(-)\dagger}_{LMK}~~\sum_{m'=M-1}^M\tilde{\beta}^{(-,L-1,m',K+q;M)*}_{-,q}} a^{(-)}_{L-1, \, m', \, K+q} \nonumber \\
    &- {\sqrt{L}} ~~ d^{(+)}_{LMK}   ~\sum_{m'=M-1}^M\tilde{\beta}^{(+,L,m',K+q;M)*}_{-,q}~b^{(+)\dagger}_{L, \, m' ,\, K+q} \Bigg),
    \end{align}
and
\begin{align}\label{quantum SUSY Noether charge explct epsilon^+}
    Q^{susy}{[\epsilon^{(+;q)}]}   =-\frac{1}{2}\sum_{L=2}^{\infty}\sum_{M,K} \Bigg(&\,{\sqrt{L} }  ~~c^{(-)\dagger}_{LMK}    ~\sum_{m'=M-1}^M~\tilde{\beta}^{(-,L,m',K+q;M)*}_{+,q}~a^{(-)}_{L, \, m', \,K+q} \nonumber\\
    &-  {\sqrt{L+2}}~~ d^{(+)}_{LMK}~\sum_{m'=M-1}^M  \tilde{\beta}^{(+,L-1,m',K+q;M)*}_{+,q}~b^{(+)\dagger}_{L-1,\,m',\,K+q} \Bigg),
    \end{align}
$q \in \{ -1, 0 \}$, where it is clear that they annihilate the vacuum $\ket{0}_{2} 
 \otimes  \ket{0}_{\frac{3}{2}}$. Note that the quantum SUSY Noether charges (\ref{quantum SUSY Noether charge explct epsilon^-}) and (\ref{quantum SUSY Noether charge explct epsilon^+}) can be expressed as a sum of two independent  charges that anti-commute with each other; one  charge generates a SUSY UIR in the positive-frequency sector, and the other generates a SUSY UIR (of opposite helicity) in the negative-frequency sector.

 The unitary realisation of SUSY in our QFT Fock space is now manifest as it is easy to check that single-particle states furnish the UIRs of our superalgebra presented in Subsection \ref{subsubsec_unitary SUSY modes}. For example,  graviton single-particle states, such as  $c^{(-)\dagger}_{LMK}\ket{0}_{2} 
 \otimes    \ket{0}_{\frac{3}{2}}$,  transform under SUSY as
 \begin{align}
  Q^{susy}{[\epsilon^{(\sigma;q)}]^{\dagger}}~\left(
 c^{(-)\dagger}_{LMK}\ket{0}_{2} 
 \otimes    \ket{0}_{\frac{3}{2}}\right) =& \left[~Q^{susy}{[\epsilon^{(\sigma;q)}]^{\dagger}}~,c^{(-)\dagger}_{LMK}   \right]
 \left(\ket{0}_{2} 
 \otimes \ket{0}_{\frac{3}{2}}\right) \nonumber \\
 =&~\ket{0}_{2} 
 \otimes \delta^{susy}(\epsilon^{(\sigma;q)})' c^{(-)\dagger}_{LMK} \ket{0}_{\frac{3}{2}}.
 \end{align}
According to our analysis in the previous paragraphs, this gives
\begin{align}
Q^{susy}{[\epsilon^{(-;q)}]}^{\dagger}~\left(
 c^{(-)\dagger}_{LMK}\ket{0}_{2} 
 \otimes    \ket{0}_{\frac{3}{2}}\right) =   {\frac{i}{2}\sqrt{L+2}~ \sum_{m'=M-1}^M~\tilde{\beta}^{(-,L-1,m',K+q;M)}_{-,q}}~
%  \frac{i \, \tilde{\beta}^{(-,L,M,K)}_{-,q}}{2} \sqrt{L+2}~~~
 \ket{0}_{2} 
 \otimes a^{(-)\dagger}_{L-1,m',K+q}\ket{0}_{\frac{3}{2}},
 \end{align}
and
\begin{align}
Q^{susy}{[\epsilon^{(+;q)}]}^{\dagger}~\left(
 c^{(-)\dagger}_{LMK}\ket{0}_{2} 
 \otimes    \ket{0}_{\frac{3}{2}}\right) =   - {\frac{1}{2}\sqrt{L}~  \sum_{m'=M-1}^M~\tilde{\beta}^{(-,L,m',K+q;M)}_{+,q}}~
 \ket{0}_{2}  
 \otimes a^{(-)\dagger}_{L,m',K+q}\ket{0}_{\frac{3}{2}},
 \end{align}
$q \in \{-1,0 \}$, in agreement with the transformation rules (\ref{SUSY_transf RULE_3/2->2, epsilon^-}) and (\ref{SUSY_transf RULE_3/2->2, epsilon^+}), respectively, of the mode functions forming the SUSY UIRs. 
One can similarly find the SUSY transformations of gravitino single-particle states, such as $\ket{0}_{2} 
 \otimes a^{(-)\dagger}_{\ell m k}   \ket{0}_{\frac{3}{2}}$, by using the following: 
 \begin{align}
      Q^{susy}{[\epsilon^{(\sigma;q)}]}~\left(
 \ket{0}_{2} 
 \otimes a^{(-)\dagger}_{\ell m k}   \ket{0}_{\frac{3}{2}}\right) &=\{Q^{susy}{[\epsilon^{(\sigma;q)}]},a^{(-)\dagger}_{\ell m k}   \}
 \left(\ket{0}_{2} 
 \otimes \ket{0}_{\frac{3}{2}}\right)\nonumber\\
 &= \delta^{susy}(\epsilon^{(\sigma;q)}) a^{(-)\dagger}_{\ell m k}   
 \ket{0}_{2} 
 \otimes \ket{0}_{\frac{3}{2}}.
 \end{align}
%%%%%%%%%%%%%%%%%%%%%%%%%%%%%%%%%%%%%%%%%%%%%%%%%%%%%%%%%%%%%%%%%%%%%%%%%%%%%%%%%%%%%%%%%%%%%%%%%%%%%%55%%%%%%%%%%%%%%%%55%%%%%%%%%%%%%%%%%%%%%%%%%%%%%%%%%%%%%%%%%%%%%%%%%%%%%%%%%%%55%%%%%%%%%5%%%%%%%%%%%%%%%%%%%%%%%%%%
\\
\\
\noindent \textbf{\textit{Extra check for unitarity}$-$}As mentioned in the Introduction, in the cases where the unitarity of global SUSY on a fixed $dS_{4}$ background fails \cite{Nieuwenhuizen, Lukierski}, the main obstacle is that the {sum of anti-commutators} of spinorial supercharges $Q_{A}$, 
{
\begin{align}\label{operator-to-be-shown-positive}
\sum_{A}\{Q_{A} ,Q^{A\dagger}\},
\end{align}
}
{which must be positive-definite, is shown to vanish identically using the de~Sitter superalgebra.}
  If this anti-commutator vanishes in a theory that carries a non-trivial representation of global SUSY, then negative-norm states must exist, rendering the theory non-unitary. We will show that, in our supersymmetric theory of the chiral graviton and gravitino, this anti-commutator is positive, as required by unitarity.

Let us start by introducing the Grassmann-odd spinorial supercharges, $Q_{A}$, of our theory. Using the expression (\ref{dS Killing spinors explicit_ e=S(t,x)n}) for  the commuting  Killing spinors, and the definition (\ref{def: SUSY noether charge}) of the Grassmann-odd SUSY Noether charges, we have
\begin{align}
  Q^{susy}{[\epsilon^{(\sigma;q)}]} &=  \int_{S^{3}} d\bm{\theta_{3}} \sqrt{-g}~~\overline{\epsilon}^{(\sigma;q)\,B}~~\frak{J}^{t}_{B}=\overline{\eta}^{(\sigma;q)\,A}~ \int_{S^{3}} d\bm{\theta_{3}} \sqrt{-g}~~\left(-\gamma^{0} {S(t,\bm{\theta_{3}})^{\dagger}} \gamma^{0} \right)_{A}^{~~B}~~\frak{J}^{t}_{B} \nonumber\\
  &\equiv \overline{\eta}^{(\sigma;q)\,A} ~Q_{A}.
\end{align}
Now, let us show that 
{the operator given in eq.~(\ref{operator-to-be-shown-positive}) is proportional to} the following {sum of anti-commutators} between SUSY Noether charges:
\begin{align}
    \sum_{\sigma \in \{ +, -\}} \sum_{q \in \{0,-1\} }  
  \left \{Q^{susy}[\epsilon^{(\sigma;q)}] ,~  Q^{susy}[\epsilon^{(\sigma;q)}]^{\dagger}\right \}.  
\end{align}
We straightforwardly have
\begin{align}
  \sum_{\sigma \in \{ +, -\}} \sum_{q \in \{0,-1\} }  
  \left \{Q^{susy}[\epsilon^{(\sigma;q)}] ,~  Q^{susy}[\epsilon^{(\sigma;q)}]^{\dagger}\right \} &=   \sum_{\sigma \in \{ +, -\}} \sum_{q \in \{0,-1\} } 
  ~\left \{\overline{\eta}^{(\sigma;q)A}Q_{A} ,~  \overline{Q}^{\,B}{\eta}^{(\sigma;q)}_{B} \right \}  \nonumber \\
  &=   \sum_{\sigma \in \{ +, -\}} \sum_{q \in \{0,-1\} } 
  ~\left \{ \overline{\eta}^{(\sigma;q)A} Q_{A} ,~  {Q}^{B \dagger }\left(\overline{\eta}^{(\sigma;q)}\right)_{B}^{\dagger} \right \} .
\end{align}
Using the explicit expressions for the Killing spinors (\ref{dS Killing spinors explicit_ e=S(t,x)n}), we find
\begin{align}\label{trace-NoetherQ=SUSYcharge}
  \sum_{\sigma \in \{ +, -\}} \sum_{q \in \{0,-1\} }  
  \left \{Q^{susy}[\epsilon^{(\sigma;q)}] ,~  Q^{susy}[\epsilon^{(\sigma;q)}]^{\dagger}\right \} &=  \frac{1}{2 \pi^{2}} \sum_{A=1}^{4} 
  ~\left \{  Q_{A} ,~  {Q}^{A \dagger } \right \}.
\end{align}
To determine $ \sum_{\sigma,q }   
  \left \{Q^{susy}[\epsilon^{(\sigma;q)}] ,~  Q^{susy}[\epsilon^{(\sigma;q)}]^{\dagger}\right \} $, it is convenient to study the action of two consecutive SUSY variations on our chiral quantum fields. These are expressed as 
\begin{align}\label{consecutive SUSY graviton quantum}
 \delta^{susy}(\epsilon^{(\sigma';q')})\delta^{susy}(\epsilon^{(\sigma;q)})'  \frak{h}^{(\text{TT})-}_{\mu \nu}&= \left\{  ~ \left[\frak{h}^{(\text{TT})-}_{\mu \nu},  Q^{susy}{[\epsilon^{(\sigma;q)}]} \right], Q^{susy}{[\epsilon^{(\sigma';q')}]}^{\dagger}  ~  \right\}  \nonumber\\
 &= \left[  ~ \frak{h}^{(\text{TT})-}_{\mu \nu}, \left\{ Q^{susy}{[\epsilon^{(\sigma;q)}]}, Q^{susy}{[\epsilon^{(\sigma';q')}]}^{\dagger} \right\} ~  \right]     ,
\end{align}
\begin{align} \label{consecutive SUSY gravitino quantum}
\delta^{susy}(\epsilon^{(\sigma';q')})' \, \delta^{susy}(\epsilon^{(\sigma;q)})  {\Psi}^{(\text{TT})-}_{\mu } &= \left[  ~ \left\{{\Psi}^{(\text{TT})-}_{\mu},  Q^{susy}{[\epsilon^{(\sigma;q)}]^{\dagger}} \right\}, Q^{susy}{[\epsilon^{(\sigma';q')}]}  ~  \right] \nonumber\\
&= \left[  ~ {\Psi}^{(\text{TT})-}_{\mu }, \left\{ Q^{susy}{[\epsilon^{(\sigma;q)}]}^{\dagger}, Q^{susy}{[\epsilon^{(\sigma';q')}]} \right\} ~  \right]     ,
\end{align}
where we have used $\left\{{\Psi}^{(\text{TT})-}_{\mu},  Q^{susy}{[\epsilon^{(\sigma;q)}]} \right\}$ $= \left [ \frak{h}^{(\text{TT})-}_{\mu \nu},  Q^{susy}{[\epsilon^{(\sigma;q)}]^{\dagger}} \right ] = 0$ for any $\sigma$ and $q$.
Then, using the explicit expressions for the SUSY transformations in 
eqs.~(\ref{SUSY_transf_spin3/2->spin2 TT}) and (\ref{SUSY_transf_spin2->spin3/2 TT}), we can also  express the consecutive SUSY transformations in the following form:
\begin{align}\label{consecutive SUSY graviton}
    \,\delta^{susy}(\epsilon^{(\sigma';q')})\delta^{susy}(\epsilon^{(\sigma;q)})'\,\mathfrak{h}^{(\text{TT})-}_{\mu \nu} =&   \, \pounds_{\xi^{((\sigma;q),(\sigma' ; q'))}_{\mathbb{C}}}\mathfrak{h}^{(\text{TT})-}_{\mu \nu}-\, \mathcal{T}_{V^{((\sigma;q),(\sigma' ; q'))}_{\mathbb{C}}}\,\mathfrak{h}^{(\text{TT})-}_{\mu \nu }\nonumber\\
    &+i\, \frac{\overline{\epsilon}^{(\sigma;q)} \gamma^{5} \epsilon^{(\sigma';q')}  }{4}  \mathfrak{h}^{(\text{TT})-}_{\mu \nu },
\end{align}
\begin{align}\label{consecutive SUSY gravitino}
   \delta^{susy}(\epsilon^{(\sigma';q')})'\delta^{susy}(\epsilon^{(\sigma;q)})\,\Psi^{(\text{TT})-}_{\mu} 
   =& \, \mathbb{L}_{\xi^{((\sigma';q'),(\sigma;q))}_{\mathbb{C}}}\Psi^{(\text{TT})-}_{\mu }-  \mathbb{T}_{V^{((\sigma';q'),(\sigma;q))}_{\mathbb{C}}}\,\Psi^{(\text{TT})-}_{\mu}\nonumber\\
   &+\frac{5i}{2}\, \frac{\overline{\epsilon}^{(\sigma';q')} \gamma^{5} \epsilon^{(\sigma;q)}}{4}\Psi^{(\text{TT})-}_{\mu  } .
\end{align}
{Here, the complex Killing vector $\xi^{((\sigma';q'),(\sigma;q))}_{\mathbb{C}}$ 
and genuine conformal Killing vector $V^{((\sigma';q'),(\sigma;q))}_{\mathbb{C}}$
are given by complex Killing spinor bilinears as follows:}
\begin{align}\label{KS bilinears complex collecvly}
     &\xi_{\mathbb{C}}^{((\sigma;q),(\sigma';q'))\mu}=\frac{1}{4} \overline{\epsilon}^{(\sigma;q)} \gamma^{5}\gamma^{\mu} \epsilon^{(\sigma';q')}= - (\xi_{\mathbb{C}}^{((\sigma' ; q'),(\sigma ; q))\mu})^{*},\nonumber\\
     &V_{\mathbb{C}}^{((\sigma;q),(\sigma';q'))\mu}=\frac{1}{4} \overline{\epsilon}^{(\sigma;q)} \gamma^{\mu} \epsilon^{(\sigma';q')}=- (V_{\mathbb{C}}^{((\sigma' ; q'),(\sigma ; q))\mu})^{*}.
\end{align}
{Note also}
\begin{align}
\overline{\epsilon}^{(\sigma;q)} \gamma^{5} \epsilon^{(\sigma';q')}= - (\overline{\epsilon}^{(\sigma' ; q')} \gamma^{5}  \epsilon^{(\sigma ; q)} )^{\dagger}. 
\end{align}
The complex Killing vectors $\xi_{\mathbb{C}}^{((\sigma;q),(\sigma' ; q'))\mu}$ were first introduced in eq.~(\ref{cmplx_Kil_vec_from_KS}), in a slightly different notation. Similarly, the complex genuine conformal Killing vectors $V_{\mathbb{C}}^{((\sigma ; q),(\sigma' ; q'))\mu}$ were first introduced in 
eq.~(\ref{cmplx_conf_Kil_vec_from_KS}). The scalars  $ \overline{\epsilon}^{(\sigma' ; q')} \gamma^{5}  \epsilon^{(\sigma ; q)}$ are constant and, in general, they are complex.  

From eqs.~(\ref{consecutive SUSY graviton}), (\ref{consecutive SUSY gravitino}) and eqs.~(\ref{consecutive SUSY graviton quantum}), (\ref{consecutive SUSY gravitino quantum}), we {see} that we can determine the 
`traced' anti-commutator in eq.~(\ref{trace-NoetherQ=SUSYcharge}) by summing over all dS Killing spinors (\ref{dS Killing spinors explicit_ e=S(t,x)n}), as 
\begin{align}\label{consecutive cl=quant SUSY graviton}
 \sum_{\sigma \in \{+, -\}}\sum_{q \in \{ 0,-1\} }\delta^{susy}(\epsilon^{(\sigma;q)})\delta^{susy}(\epsilon^{(\sigma;q)})'  \frak{h}^{(\text{TT})-}_{\mu \nu} = \frac{1}{2 \pi^{2}}\left[  ~ \frak{h}^{(\text{TT})-}_{\mu \nu}, \sum_{A=1}^{4}\left\{ Q_{A}, Q^{A\dagger} \right\} ~  \right]     ,
\end{align}
\begin{align} \label{consecutive cl=quant SUSY gravitino}
\sum_{\sigma \in \{+, -\}}\sum_{q \in \{ 0,-1\} }\delta^{susy}(\epsilon^{(\sigma;q)})' \, \delta^{susy}(\epsilon^{(\sigma;q)})  {\Psi}^{(\text{TT})-}_{\mu } 
&= \frac{1}{2 \pi^{2}}\left[  ~ {\Psi}^{(\text{TT})-}_{\mu },\sum_{A=1}^{4} \left\{ Q_{A}, Q^{A\dagger} ~  \right\} \right]     .
\end{align}
%%%%%%%%%%%%%%%%%%%%%%%%%%%%%%%%%%%%%%%%%%%%%%%%%%%%%%%%%%%%%%%%%%%%%%%%%%%%%%%%%%%%%%%%%%%%%%%%%%%%%%%%%%%%%%%%%%%%%%%%%%%%%%%%%%%%%%%%%%%%%%%%%%%%%%%%%%%%%%%%%%%%%%%%
 {Thus,} we are interested in the case where the two Killing spinors are equal to each other, $\epsilon^{(\sigma ; q)} = \epsilon^{(\sigma' ; q')}$ (i.e.\ $\sigma = \sigma'$ and $q=q'$), in eqs.~(\ref{consecutive SUSY graviton}) and (\ref{consecutive SUSY gravitino}). The Killing spinor bilinears in 
 eqs.~(\ref{consecutive SUSY graviton}) and (\ref{consecutive SUSY gravitino}) are imaginary for $\epsilon^{(\sigma ; q)} = \epsilon^{(\sigma' ; q')}$. This means that the complex Killing vector $\xi_{\mathbb{C}}^{((\sigma ; q),(\sigma ; q))\mu}$ can be expressed as $i=\sqrt{-1}$ times a   real Killing vector. Similarly, the complex genuine conformal Killing vector $V_{\mathbb{C}}^{((\sigma ; q),(\sigma ; q))\mu}$ can be expressed as $i$ times a  real genuine conformal Killing vector. 
   Explicit expressions for these complex Killing spinor bilinears {can be found}
by using the explicit expressions for the Killing spinors (\ref{dS Killing spinors explicit_ e=S(t,x)n}):

\begin{itemize}
    \item For the complex Killing vectors $\xi_{\mathbb{C}}^{((\sigma ; q),(\sigma ; q))\mu}$ , we find
\begin{align}
 &  \xi_{\mathbb{C}}^{((\sigma;-1),(\sigma;-1))\,t}=  \xi_{\mathbb{C}}^{((\sigma;0),(\sigma;0))\,t}=0, \nonumber \\
& \xi_{\mathbb{C}}^{((\sigma;-1),(\sigma;-1)) \, \tilde{\mu}}  \partial_{\tilde{\mu}}=-\xi_{\mathbb{C}}^{((\sigma;0),(\sigma;0)) \, \tilde{\mu}}  \partial_{\tilde{\mu}} =-\frac{i}{4}\frac{1}{2 \pi^{2}} \left( \cos{\theta_{2}} \frac{\partial}{\partial \theta_{3}} - \cot{\theta_{3}} \sin{\theta_{2}} \frac{\partial}{\partial \theta_{2}}  -\sigma \frac{\partial}{\partial \theta_{1}}\right),
\end{align}
for $\sigma = \pm $. We conclude that  $\xi_{\mathbb{C}}^{((\sigma;-1),(\sigma;-1))\,\mu} = -\xi_{\mathbb{C}}^{((\sigma;0),(\sigma;0))\,\mu} $ is equal to $i$ times a linear combination of Killing vectors of $S^{3}$, and we observe that
$$\sum_{\sigma \in \{ +,- \}}  \sum_{q  \in   \{ 0,-1 \}} \xi_{\mathbb{C}}^{((\sigma;q),(\sigma;q))\,\mu}=0.
$$

\item For the complex genuine conformal Killing vectors $V_{\mathbb{C}}^{((\sigma ; q),(\sigma ; q))\mu}$ , we find
\begin{align}
    V_{\mathbb{C}}^{((\sigma;-1),(\sigma;-1))\,\mu} =  V_{\mathbb{C}}^{((\sigma;0),(\sigma;0))\,\mu} =\frac{i}{4} \frac{1}{2 \pi^{2}} ~\partial^{\mu}\sinh{t} = \frac{i}{4} \frac{1}{2 \pi^{2}} V^{(0) \mu},
\end{align}
where $V^{(0)\mu}$ is the real genuine conformal Killing vector in eq.~(\ref{CKV_dS dilation}). We observe that
$$\sum_{\sigma \in \{ +,- \}}  \sum_{q  \in   \{ 0,-1 \}} V_{\mathbb{C}}^{((\sigma;q),(\sigma;q))\,\mu}= \frac{i}{2 \pi^{2}} V^{(0) \mu}.
$$
%
%%%%%%%%%%%%%%%%%%%%%%%%%%%%%%%%%%%%%%%%%%%%%%%%%%%%%%%%%%%%%%%%%%%%%%
\item  For the constant scalars  $\overline{\epsilon}^{(\sigma;q)} \gamma^{5} \epsilon^{(\sigma;q)}$ we have

\begin{align}
    \overline{\epsilon}^{(\sigma;-1)} \gamma^{5} \epsilon^{(\sigma;-1)} = \overline{\epsilon}^{(\sigma;0)} \gamma^{5} \epsilon^{(\sigma;0)} = \sigma  \frac{i}{2 \pi^{2}} ,~~~\text{for}~~~~\sigma = \pm ,
\end{align}
and thus,
$$\sum_{\sigma \in \{ +,- \}}  \sum_{q  \in   \{ 0,-1 \}} \overline{\epsilon}^{(\sigma;q)} \gamma^{5} \epsilon^{(\sigma;q)} = 0.
$$

\end{itemize}
%%%%%%%%%%%%%%%%%%%%%%%%%%%%%%%%%%%%%%%%%%%%%%%%%%%%%%%%%%%%%%%%%%%%%%%%%%%%%%%%%%%%%%%%%%%%%%%%%%%%%%%%%%%%%%%%%%%%%%%%%%%%%%%%%%%%%%%
{We use these properties of the complex Killing spinor bilinears and 
eqs.~(\ref{consecutive SUSY graviton}) and (\ref{consecutive SUSY gravitino}) to evaluate the left-hand side of eqs.~(\ref{consecutive cl=quant SUSY graviton}) and (\ref{consecutive cl=quant SUSY gravitino}).  Thus, we find}
\begin{align}
\left[  ~ \frak{h}^{(\text{TT})-}_{\mu \nu}, \sum_{A=1}^{4}\left\{ Q_{A}, Q^{A\dagger} \right\} ~  \right]  = - ~ \mathcal{T}_{i\,V^{(0)}} \frak{h}^{(\text{TT})-}_{\mu \nu}   ,
\end{align}
\begin{align} 
 \left[  ~ {\Psi}^{(\text{TT})-}_{\mu },\sum_{A=1}^{4} \left\{ Q_{A}, Q^{A\dagger}  \right\} \right]    =- ~     \,\mathbb{T}_{i\,V^{(0)}} {\Psi}^{(\text{TT})-}_{\mu }  .
\end{align}
  It is straightforward to find explicit expressions for  $\mathbb{T}_{i \, V^{(0)}} {\Psi}^{(\text{TT})-}_{\mu }$ and $\mathcal{T}_{i \, V^{(0)}} \frak{h}^{(\text{TT})-}_{\mu \nu}$ by expanding the field in modes and calculating the action of the transformations on the mode functions. In practice, the factor of $i$ in $i V^{(0)\mu}$ results in an imaginary phase rotation of the mode functions.  Working as in Subsections \ref{Subsec_gravitino modes so(4,2)} and \ref{subsubsec_cmplx graviton and so(4,2) ON MODES}, we find
\begin{align}
    & -\mathbb{T}_{i \, V^{(0)}} {\Psi}^{\text{(TT)}-}_{{\mu}}(t, \bm{\theta_{3}}) \nonumber \\
    &= ~\sum_{\ell =1}^{\infty}   \sum_{m, k} \left(\ell +\frac{3}{2} \right)\left( {a}^{(-)}_{\ell m k}{\psi}^{\left(phys ,\,- \ell\,;m;k \right)}_{{\mu}}(t, \bm{\theta_{3}})  -{b}^{(+)\dagger}_{\ell m k}\,{v}^{\left(phys ,\,+ \ell\,;m;k \right)}_{{\mu}} (t, \bm{\theta_{3}}) \right) \\
&= \left[ {\Psi}^{\text{(TT)}-}_{{\mu}}(t, \bm{\theta_{3}}),~ Q^{conf-}_{\frac{3}{2}}[V^{(0)}]  - Q^{conf+}_{\frac{3}{2}}[V^{(0)}]~ \right],
\end{align}
and
\begin{align}
    &  -\mathcal{T}_{i \, V^{(0)}}{\mathfrak{h}}^{\text{(TT)}-}_{{\mu}   {\nu}}(t, \bm{\theta_{3}}) \nonumber\\
    &= \sum_{L =2}^{\infty}   \sum_{M,K} \left( L+1 \right)\left( {c}^{(-)}_{LMK}{\varphi}^{(phys ,\,- L; \,M;K)}_{{\mu} {\nu}}(t, \bm{\theta_{3}})  - {d}^{(+)\dagger}_{LMK}\,{\varphi}^{(phys ,\,+ L; \,M;K)\star}_{{\mu} {\nu}} (t, \bm{\theta_{3}}) \right)\\
&= \left[ \frak{h}^{\text{(TT)}-}_{{\mu \nu}}(t, \bm{\theta_{3}}),~ Q^{conf-}_{{2}}[V^{(0)}]  - Q^{conf+}_{{2}}[V^{(0)}]~ \right],
\end{align}
where $Q^{conf\pm}_{\frac{3}{2}}[V^{(0)}]$ are the quantum conformal-like charges (\ref{conf-like charge_3/2 = pos freq + neg freq}) of the chiral gravitino acting on states of helicity $\pm 3/2$, while $Q^{conf\pm}_{{2}}[V^{(0)}]$ are the quantum conformal-like charges (\ref{conf-like charge_2 = pos freq + neg freq}) of the chiral graviton acting on states of helicity $\pm 2$. Thus, we identify the trace of the supercharge anti-commutator as
\begin{align}\label{traced anti-comtr final result}
\sum_{A=1}^{4} \left\{ Q_{A}, Q^{A\dagger}  \right\} 
= ~Q^{conf-}_{\frac{3}{2}}[V^{(0)}]  - Q^{conf+}_{\frac{3}{2}}[V^{(0)}] +   Q^{conf-}_{{2}}[V^{(0)}]  - Q^{conf+}_{{2}}[V^{(0)}],
\end{align}
which is clearly positive, i.e.\ its expectation values are always greater than or equal to zero - see eqs.~(\ref{conf-like charge_3/2 = pos freq + neg freq}) and (\ref{conf-like charge_2 = pos freq + neg freq}).
\\
\\
\noindent \textbf{Note.} From eqs.~(\ref{quantum SUSY Noether charge explct epsilon^-}) and (\ref{quantum SUSY Noether charge explct epsilon^+}) it follows that the spinorial supercharges consist of two independent parts, $Q_{A} = Q_{A}^{-} + Q^{+}_{A}$, with $ \{ Q_{A}^{-}, Q^{{+B}}\}=$   $ \{ Q_{A}^{-}, Q^{+B \dagger}\}=$ $ \{ Q_{A}^{+}, Q^{-B \dagger}\}=0$, which separately generate the two SUSY UIRs with helicities $(-2,-{3}/{2})$ and $(+2, +3/2)$, respectively. We thus have
\begin{align}
  \sum_{A=1}^{4} \left\{ Q_{A}, Q^{A\dagger}  \right\}  = \sum_{A=1}^{4} \left\{ Q^{-}_{A}, Q^{-A\dagger}  \right\}  + \sum_{A=1}^{4} \left\{ Q^{+}_{A}, Q^{+A\dagger}  \right\} ,
\end{align}
where each of the two traced anti-commutators on the right-hand side is separately positive.
%%%%%%%%%%%%%%%%%%%%%%%%%%%%%%%%%%%%%%%%%%%%%%%%%%%%%%%%%%%%%%%%%%%%%%%%%%%%%%%%%%%%%%%%%%%%%%%%%%%%%%%%%%%%%%%%%%%%%%%%%%%%%%%%%%%%%%%%
 \\
\\
\noindent   \textbf{\textit{SUSY algebra in terms of quantum charges}$-$}Using the expressions for the consecutive SUSY transformations (\ref{consecutive SUSY graviton}) and (\ref{consecutive SUSY gravitino}), as well as eqs.~(\ref{consecutive SUSY graviton quantum}) and (\ref{consecutive SUSY gravitino quantum}), one can re-express the commutators (\ref{commutator of SUSY on spin-2 TT}) and (\ref{commutator of SUSY on spin-3/2 TT}) of two SUSY variations in terms of anti-commutators of quantum SUSY Noether charges (\ref{quantum SUSY Noether charge explct epsilon^-}) and (\ref{quantum SUSY Noether charge explct epsilon^+}), as 
\begin{align}\label{SUSY algebra in terms of quant charges (e2,e1)-(e1,e2)}
  & \left\{ Q^{susy}{[\epsilon^{(\sigma;q)}]}, Q^{susy}{[\epsilon^{(\sigma';q')}]}^{\dagger} \right\} - ((\sigma;q) \leftrightarrow (\sigma';q'))\nonumber\\
  =& -2\,i Q^{dS}\left[Re(\xi^{((\sigma';q'),(\sigma ;q))}_{\mathbb{C}})  \right] + 2\,i  Q^{conf}\left[Re(V^{((\sigma';q'),(\sigma ;q))}_{\mathbb{C}}) \right] \nonumber\\
  &- i Q^{u(1)} \left[\frac{1}{4}\overline{\epsilon}^{(\sigma';q')}  \gamma^{5}  \epsilon^{(\sigma;q)} - \frac{1}{4}\overline{\epsilon}^{(\sigma;q)}  \gamma^{5}  \epsilon^{(\sigma';q')}\right].
\end{align}
Here, the  even  hermitian generators 
\begin{align}
   & Q^{dS}\left[Re(\xi^{((\sigma';q'),(\sigma ;q))}_{\mathbb{C}})  \right]\equiv Q^{dS-}\left[Re(\xi^{((\sigma';q'),(\sigma ;q))}_{\mathbb{C}})  \right] +Q^{dS+}\left[Re(\xi^{((\sigma';q'),(\sigma ;q))}_{\mathbb{C}})  \right] , \nonumber\\
&~~\text{with}~~Q^{dS\pm}\left[Re(\xi^{((\sigma';q'),(\sigma ;q))}_{\mathbb{C}})  \right]\equiv \sum_{s \in \{2, \frac{3}{2} \}} Q^{dS\pm}_{s}\left[Re(\xi^{((\sigma';q'),(\sigma ;q))}_{\mathbb{C}})  \right], 
\end{align}
are the quantum dS charges,  (\ref{def:dS charges gravitino}) and (\ref{def:dS charges graviton}), associated with the real Killing vector $Re(\xi^{((\sigma';q'),(\sigma ;q))}_{\mathbb{C}})^{\mu} = {[\overline{\epsilon}^{(\sigma';q')}  \gamma^{5}  \gamma^{\mu}\epsilon^{(\sigma ;q)} - \overline{\epsilon}^{(\sigma;q)}  \gamma^{5}  \gamma^{\mu}\epsilon^{(\sigma' ;q')}]/8}$ [see eq.~(\ref{KS bilinears complex collecvly})]. The `$-$' dS charges and the `$+$' dS charges commute with each other, and they generate discrete series UIRs of $so(4,1)$ with negative and positive helicity, respectively. The  even hermitian generators
\begin{align}
   & Q^{conf}\left[Re(V^{((\sigma';q'),(\sigma ;q))}_{\mathbb{C}})  \right] \equiv Q^{conf-}\left[Re(V^{((\sigma';q'),(\sigma ;q))}_{\mathbb{C}})  \right] +Q^{conf+}\left[Re(V^{((\sigma';q'),(\sigma ;q))}_{\mathbb{C}})  \right] , \nonumber\\
&~\text{with}~~Q^{conf\pm}\left[Re(V^{((\sigma';q'),(\sigma ;q))}_{\mathbb{C}})  \right]= \sum_{s \in \{ 2 ,\frac{3}{2}\}} Q^{conf\pm}_{s}\left[Re(V^{((\sigma';q'),(\sigma ;q))}_{\mathbb{C}})  \right]
\end{align}
 are the quantum conformal-like charges, (\ref{def:conf charges gravitino}) and (\ref{def:conf charges graviton}), associated with the real genuine conformal Killing vector  $Re(V^{((\sigma';q'),(\sigma ;q))}_{\mathbb{C}})^{\mu} = {[\overline{\epsilon}^{(\sigma';q')}    \gamma^{\mu}\epsilon^{(\sigma;q)} - \overline{\epsilon}^{(\sigma;q)}    \gamma^{\mu}\epsilon^{(\sigma';q')}]/8}$ [see eq.~(\ref{KS bilinears complex collecvly})]. Again, the `$-$' conformal-like charges {commute with} the `$+$' conformal-like charges.  The charges $(Q^{dS\mp}$, $Q^{conf\mp})$ generate UIRs of $so(4,2)$ with $\mp$ helicities.  As mentioned in the previous Sections, given a real Killing vector $\xi^{\mu}$, and a real genuine conformal Killing vector $V^{\mu}$,  {these} hermitian charges generate the following transformations:
\begin{align}\label{even quant dS generators final}
[~ \frak{h}^{(\text{TT})-}_{\mu \nu} ,Q^{dS}[\xi]  ~] = -i \, \pounds_{\xi} \frak{h}^{(\text{TT})-}_{\mu \nu},~~[~ {\Psi}^{(\text{TT})-}_{\mu } ,Q^{dS}[\xi]  ~] = -i \, \mathbb{L}_{\xi} {\Psi}^{(\text{TT})-}_{\mu},
\end{align}
\begin{align}\label{even quant conf-like generators final}
[~ \frak{h}^{(\text{TT})-}_{\mu \nu} ,Q^{conf}[V]  ~] = -i \, \mathcal{T}_{V} \frak{h}^{(\text{TT})-}_{\mu \nu},~~[~ {\Psi}^{(\text{TT})-}_{\mu } ,Q^{conf}[V]  ~] = -i \, \mathbb{T}_{V} {\Psi}^{(\text{TT})-}_{\mu}.
\end{align}
{Finally}, we have denoted the  hermitian $u(1)$  quantum charges as  $Q^{u(1)}$. These also consist of two independent $u(1)$ charges $Q^{u(1)} = Q^{u(1)-} + Q^{u(1)+}$, acting on negative-helicity and positive-helicity states respectively. For real constant parameters $\alpha$, they  act on our quantum fields as
\begin{align}
[~ \frak{h}^{(\text{TT})-}_{\mu \nu} ,Q^{u(1)}[\alpha]  ~] =  -i\, \delta^{phase}_{\alpha}\frak{h}^{(\text{TT})-}_{\mu \nu},~~~[~ {\Psi}^{(\text{TT})-}_{\mu } ,Q^{u(1)}[\alpha]  ~] = -i\, \delta^{phase}_{\alpha}{\Psi}^{(\text{TT})-}_{\mu},
\end{align}
where
\begin{align}\label{def:phase variations}
  \delta^{phase}_{\alpha}\frak{h}^{(\text{TT})-}_{\mu \nu} = i\,  \alpha \, \frak{h}^{(\text{TT})-}_{\mu \nu},~~~   \delta^{phase}_{\alpha}{\Psi}^{(\text{TT})-}_{\mu} = \frac{5i}{2} \, \alpha \, {\Psi}^{(\text{TT})-}_{\mu}.
\end{align}

{We similarly find} the following anti-commutator of quantum SUSY Noether charges: 
\begin{align}\label{SUSY algebra in terms of quant charges (e2,e1)}
  & \left\{ Q^{susy}{[\epsilon^{(\sigma;q)}]}, Q^{susy}{[\epsilon^{(\sigma' ; q')}]}^{\dagger} \right\} \nonumber \\
  = & - {i}\, \sum_{p= \pm} Q^{dSp}[Re(\xi^{((\sigma' ; q'),(\sigma ; q))}_{\mathbb{C}})] + {i}  \, \sum_{p= \pm}Q^{conf\,p}[Re(V^{((\sigma' ; q'),(\sigma ; q))}_{\mathbb{C}})]\nonumber\\
  &-{ i}\,\sum_{p= \pm}  Q^{u(1)p} \left[Re \left(\frac{1}{4}\overline{\epsilon}^{(\sigma' ; q')}  \gamma^{5}  \epsilon^{(\sigma;q)}\right)\right] \nonumber\\
  &+\sum_{p = \pm} p \,Q^{dSp}[Im(\xi^{((\sigma' ; q'),(\sigma ; q))}_{\mathbb{C}})]- \sum_{p= \pm} p \, Q^{conf\, p}[Im(V^{((\sigma' ; q'),(\sigma ; q))}_{\mathbb{C}}] \nonumber\\
  &+ \sum_{p = \pm }p Q^{u(1)p} \left[Im \left(\frac{1}{4}\overline{\epsilon}^{(\sigma' ; q')}  \gamma^{5}  \epsilon^{(\sigma;q)}\right)  \right].
\end{align}
 Note that the even quantum charges that depend on the imaginary parts of the Killing spinor bilinears are multiplied by factors of $p  = \pm$, as was already evident from the traced anti-commutator in  eq.~(\ref{traced anti-comtr final result}). Given a real Killing vector $\xi^{\mu}$, a real genuine conformal Killing vector $V^{\mu}$, and a real constant parameter $\alpha$, these quantum charges generate  transformations parametrised by the `imaginary counterparts' of $\xi^{\mu}, V^{\mu}$ and $\alpha$, as:
\begin{align}\label{even quant dS generators final imaginary prmtr}
&[~ \frak{h}^{(\text{TT})-}_{\mu \nu} ,\sum_{p = \pm} (-p) \,Q^{dSp}[\xi]  ~] = - \, \pounds_{i\, \xi} \frak{h}^{(\text{TT})-}_{\mu \nu},\nonumber \\
&[~ {\Psi}^{(\text{TT})-}_{\mu } ,\sum_{p = \pm} (-p) \,Q^{dSp}[\xi]  ~] = - \, \mathbb{L}_{i\, \xi} {\Psi}^{(\text{TT})-}_{\mu},
\end{align}
\begin{align}\label{even quant conf-like generators final imaginary prmtr}
&[~ \frak{h}^{(\text{TT})-}_{\mu \nu} ,\sum_{p = \pm} (-p) \,Q^{conf \, p}[V] ~] = - \, \mathcal{T}_{i \, V} \frak{h}^{(\text{TT})-}_{\mu \nu},  \nonumber \\
&[~ {\Psi}^{(\text{TT})-}_{\mu } ,  \sum_{p = \pm} (-p) \,Q^{conf \, p}[V]~] = - \, \mathbb{T}_{i \, V} {\Psi}^{(\text{TT})-}_{\mu},
\end{align}
and,
\begin{align}
&[~ \frak{h}^{(\text{TT})-}_{\mu \nu} ,\sum_{p = \pm} (-p) \,Q^{u(1) \, p}[\alpha] ~] = - \, \delta^{phase}_{i \, \alpha} \frak{h}^{(\text{TT})-}_{\mu \nu},  \nonumber \\
&[~ {\Psi}^{(\text{TT})-}_{\mu } ,  \sum_{p = \pm} (-p) \,Q^{u(1) \, p}[\alpha]~] = - \, \delta^{phase}_{i \, \alpha} {\Psi}^{(\text{TT})-}_{\mu},
\end{align}
where  $\delta^{phase}_{i \, \alpha}$ describes infinitesimal scale transformations [compare with the real-parameter case in eq.~(\ref{def:phase variations})]. Recalling that the quantum SUSY Noether charges can be expressed as a sum of two independent charges,
$Q^{susy}[\epsilon^{(\sigma;q)}] =Q^{susy-}[\epsilon^{(\sigma;q)}] +Q^{susy+}[\epsilon^{(\sigma;q)}] $,  generating separately SUSY UIRs with negative and positive helicity, respectively, we may re-express the anti-commutator (\ref{SUSY algebra in terms of quant charges (e2,e1)}) as 
\begin{align}
& \left\{ Q^{susy\,-}{[\epsilon^{(\sigma;q)}]}, Q^{susy\,-}{[\epsilon^{(\sigma' ; q')}]}^{\dagger} \right\} \nonumber \\
  = & - {i}\, Q^{dS-}[Re(\xi^{((\sigma' ; q'),(\sigma ; q))}_{\mathbb{C}})] + {i}  \, Q^{conf\,-}[Re(V^{((\sigma' ; q'),(\sigma ; q))}_{\mathbb{C}})]\nonumber\\
  &-{ i}\,  Q^{u(1)-} \left[Re \left(\frac{1}{4}\overline{\epsilon}^{(\sigma' ; q')}  \gamma^{5}  \epsilon^{(\sigma;q)}\right)\right] \nonumber\\
  &- \Bigg( \,Q^{dS-}[Im(\xi^{((\sigma' ; q'),(\sigma ; q))}_{\mathbb{C}})] \, - \, Q^{conf\, -}[Im(\,V^{((\sigma' ; q'),(\sigma ; q))}_{\mathbb{C}}\,)]  \nonumber\\
  &~~~~~~~~~~~+  Q^{u(1)-} \left[Im \left(\frac{1}{4}\overline{\epsilon}^{(\sigma' ; q')}  \gamma^{5}  \epsilon^{(\sigma;q)}\right)  \right] \Bigg),
\end{align}
and
\begin{align}
& \left\{Q^{susy\,+}{[\epsilon^{(\sigma' ; q')}]}^{\dagger} ,Q^{susy\,+}{[\epsilon^{(\sigma;q)}]}  \right\} \nonumber \\
  = &  {i}\, Q^{dS+}[Re(\xi^{((\sigma ; q),(\sigma '; q'))}_{\mathbb{C}})] - {i}  \, Q^{conf\,+}[Re(V^{((\sigma ; q),(\sigma' ; q'))}_{\mathbb{C}})]\nonumber\\
  &+{ i}\,  Q^{u(1)+} \left[Re \left(\frac{1}{4}\overline{\epsilon}^{(\sigma ; q)}  \gamma^{5}  \epsilon^{(\sigma';q')}\right)\right] \nonumber\\
  &+  \,Q^{dS+}[Im(\,\xi^{((\sigma ; q),(\sigma' ; q'))}_{\mathbb{C}}\,)] \, - \, Q^{conf\, +}[Im(\,V^{((\sigma ; q),(\sigma '; q'))}_{\mathbb{C}}\,)]  \nonumber\\
  &~~~~~~~~~~~+  Q^{u(1)+} \left[Im \left(\frac{1}{4}\overline{\epsilon}^{(\sigma ; q)}  \gamma^{5}  \epsilon^{(\sigma';q')}\right)  \right] .
\end{align}
In the {second equation,}  we have used the fact that the real parts of Killing spinor bilinears change sign under the exchange $(\sigma;q) \leftrightarrow (\sigma' ;q')$, while their imaginary parts remain the same - see eq.~(\ref{KS bilinears complex collecvly}).
We thus have two independent sets of generators that form  superalgebras separately. However, the superalgebra formed by the generators 
$$Q^{susy-}, Q^{susy- \, \dagger},  Q^{dS-}, Q^{conf-}, Q^{u(1)-}$$
is isomorphic to the superalgebra formed by the generators 
$$-Q^{susy+\dagger}, -Q^{susy+ },- Q^{dS+}, -Q^{conf+}, -Q^{u(1)+}.$$ 
In particular, the role of $Q^{susy-}$ is played by $-Q^{susy+\,\dagger}$. Moreover, the even generators $- Q^{dS+}$ and $-Q^{conf+}$ generate the algebra $so(2,4) \cong so(4,2)$, which is isomorphic to the algebra generated by
$Q^{dS-}$ and $Q^{conf-}$.

%%%%%%%%%%%%%%%%%%%%%%%%%%%%%%%%%%%%%%%%%%%%%%%%%%%%%%%%%%%%%%%%%%%%%%%%%%%%%%%%%%%%%%%%%%%%%%%%%%%%%%%%%%%%%%%%%%%%%%%%%%%%%%%%%%%%%%%%%%%%%%%%%%%%%%%%%%%%%%%%%%%%%%%%%%%%%%%%%%%%%%%%%%%%%%%%%%%%%%%%%
 \section{Discussions and Open Questions}\label{Discussions}
In this paper, we showed that the free supersymmetric theory of the chiral graviton and chiral gravitino fields on fixed $dS_{4}$ is unitary.
 This free unitary theory  cannot become interacting while preserving SUSY  in  a way that makes the spin-2 sector the true graviton sector of General Relativity, as the three-graviton coupling cannot be $u(1)$-invariant. Nevertheless, it remains worthwhile to investigate whether a non-linear version of the theory exists. If such a supergravity-like theory {were to exist,} SUSY would have to be locally realised. 

As a step toward exploring possible consistent interactions, it would be mathematically interesting, and perhaps natural, to reformulate the free chiral graviton-chiral gravitino theory in terms of spin-tensors  belonging to ``unbalanced'' representations of the Lorentz group, as in Ref.~\cite{Zhenya}. Then, the question of finding possible interactions can be investigated using, for example,  methods based on the presymplectic BV-AKSZ formulation \cite{Maxim1, Maxim2, Maxim3}.  {Interestingly,} a framework for the study of consistent interactions of local gauge theories in this formulation has been recently proposed in \cite{Maxim1}, and it simplifies significantly the analysis of consistent interactions. 

We note in passing that if an interacting theory involving our {supermultiplet
of chiral graviton and gravitino}  {were to exist,} it would require the gauging of global symmetries and might also require the inclusion of additional fields, {as}  suggested by recent work \cite{Thomee}, where consistent interactions were studied for a real (non-unitary) partially massless graviton and two Majorana gravitini on $AdS_{4}$. This field content forms the basis of what one would call  `linearised partially massless supergravity around $AdS_{4}$' \cite{Thomee, Zinoviev}. The point of resemblance with our theory lies in the fact that in \cite{Thomee}, it was found that the global symmetries of linearised  partially massless supergravity around $AdS_{4}$ include conformal-like symmetries for the gravitini, similar to those in our equation (\ref{conf-like gravitino TT}). However, it was also shown that there are obstructions to the Jacobi identity of the gauge algebra,  {i.e.}\ the global symmetries cannot be gauged, {unless the field content is modified.} Interestingly, it was suggested that by adding extra fields to the theory, so that the field content matches that of $N = 1$ pure conformal supergravity around $AdS_{4}$, the global algebra (including the conformal-like symmetries) can be gauged,  {and} consistent interactions {might be constructed}. 
 {We also speculate} that a non-linear version of the theory {presented in this paper} could be related to a complex, chiral version of conformal supergravity admitting $dS_{4}$ solutions.

Another interesting future direction is to investigate possible relations between our linear supersymmetric theory and `chiral Supergravity', as discussed in 
Refs.~\cite{Tsuda1,Tsuda2,Jacobson1,Jacobson2}. 

{Finally}, we note that it is likely that an analogue of our chiral {graviton-}gravitino supersymmetric theory exists on $dS_{2}$. In such a two-dimensional theory the question of consistent interactions would be easier to tackle.  In particular, the $\Delta=2$ and $\Delta= 3/2$ discrete series UIRs of $so(2,1)$, corresponding to a shift-symmetric `tachyonic' scalar \cite{Dio_charm} and a shift-symmetric imaginary-mass spinor \cite{Letsios_Alan}, respectively, on $dS_{2}$,  can be viewed as the two-dimensional analogues of the graviton and gravitino, respectively \cite{Anninos_SUGRA, Letsios_Alan}. Each of these two fields on $dS_{2}$ corresponds to a direct sum of two $so(2,1)$ UIRs with opposite `chirality', akin to the four-dimensional case. 
Moreover, the $\Delta=2$ discrete series scalar field on $dS_{2}$ (as well as the ones with $\Delta >2$) was recently shown to enjoy a hidden global conformal symmetry \cite{Hinter} (akin to the conformal-like symmetry for the four-dimensional graviton that we discussed in {this paper}). We expect that the fermionic counterparts \cite{Letsios_Alan} of the discrete series scalar fields on $dS_{2}$  will also enjoy such a conformal symmetry. Then, it would be interesting to investigate whether one can construct a unitary supersymmetric theory on $dS_{2}$ using a chiral $\Delta=2$ scalar field and a chiral $\Delta=3/2$ spinor field. If this theory resembles the four-dimensional theory presented in this paper, then the commutator between two SUSY transformations will close on the hidden conformal symmetries. We leave the investigation of this model for future work.

\

 %%%%%%%%%%%

 %%%%%%%%%%%%%%%%%%%%%%%%%%%%%%%%%%%%%%%%%%%%%%%%%%%%%%%%%%%%%%%%%%%%%%%%%%%%%%%%%%%
 
%%%%%%%%%%%%%%%%%%%%%%%%%%%%%%%%%%%%%%%%%%%%%%%%%%%%%%%%%%%%%%%%%%%%%%%%%%%%%%%%%%%%%%%%%%%%%%%%%%%%%%%%%%%%%%%%%%%%%%%%%%%%%%%%%%%%%% 

 %%%%%%%%%%%%%%%%%%%%%%%%%%%%%%%%%%%%%%%%%%%%%%%%%%%%%%%%%%%%%%%%%%%%%%%%%%%%%%%%%%%%%%%%%%%%%%%%%%%%%%%%%%%%%%%%%%%%%%%%%%%%%%%%%%%%%%%%%%%%%%%%%%%%%%%%%%%%%%%%%%%%%%%%%%%%%%%%%%%%%%%%%%%%%%%%%%%%%%%%%%%%%%%%%%%%%%%%%%%%%%%%%%%%%%%%%%%%%%%%%%%%%%%%%

.

\acknowledgments
We  would like to thank Dionysios Anninos for useful discussions and comments. We would also like to thank  Nicolas Boulanger, Evgeny Skvortsov, Maxim Grigoriev, Sylvain Thomée, Guillermo Silva, Mati Sempé, Charis Anastopoulos, and David Andriot for useful discussions. The work of V. A. L.  was supported by the ULYSSE Incentive Grant for Mobility in Scientific Research [MISU] F.6003.24, F.R.S.-FNRS, Belgium.  {In the early stage of this work, V. A. L.  was supported by a studentship from the Department of Mathematics at the University of York and a fellowship from the Eleni Gagon Survivor's Trust for research at the Department of Mathematics at King's College London.
%%%%%%%%%%%%%%%%%%%%%%%%%%%%%%%%%%%%%%%%%%%%%%%%%%%%%%%%%%%%%%%%%%%%%%%%%%%%%%%%%%%%%%%%%%%%%%%%%%%%%%%%%%%%%%%%%%%%%%%%%%%%%%%%%%%%%%%%%%%%%%%%%%%%%%%%%%%%%%%%%%%%%%%%%%%%%%%%%%%%%%%%%%%%%%%%%%%%%%%%%%%%%%%%%%%%%%%%%%%%%%%%%%%%%%%%%%%%%%%%%%%%%%%%%%%%%%%%%%%%%%%%%%%%%%%%%%%%%%%%%%%%%%%%%%%%%%%%%%%%%%%%%%%%
\appendix

\section{Classification of the UIRs of the dS algebra} \label{App_Classification_UIRs D=4}
The dS algebra $so(4,1)$ has 10 generators $J_{AB} = - J_{BA}$, with $A,B \in \{0,1,2,3,4 \}$. These satisfy the commutation relations:
\begin{align}
    [J_{AB},    J_{CD}] = \left(\eta_{BC} J_{AD} + \eta_{AD}  J_{BC}\right) - (A\leftrightarrow B),
\end{align}
where $\eta_{AB} = diag(-1,1,1,1,1)$. 
In the case of unitary representations, each of the generators $J_{AB}$ must be realised as an anti-hermitian operator with respect to a positive-definite scalar product.

Let us review the classification of the $so(4,1)$ UIRs under the decomposition $so(4,1) \supset$ $so(4)$~\cite{Schwarz, Ottoson}. 
An irreducible representation of $so(4)$ appears with multiplicity one in a UIR of $so(4,1)$ or it does not appear at all~\cite{Dixmier}. 
An irreducible representation of $so(4)$ is specified by the highest weight~\cite{barut_group,Dobrev:1977qv, Homma}
\begin{equation}\label{define_highest_weight_orthogonal D=4}
    \vec{f}=(f_{1},f_{2}),
\end{equation}
where
\begin{align}
 \label{highest_weight_so(4)}   &f_{1}  \geq |f_{2}|.
\end{align}
The numbers $f_{1}$ and $f_{2}$ are both integers or half-odd-integers, and $f_{2}$ can be negative. 

%%%%%%%%%%%%%%%%%%%%%%%%%%%%%%%%%%%%%%%%%%%%%%%%%%%%%%%%%%%%%%%%%%%%%%%%%%%%%%%%%%%%%%%%%%%%%%%%%%%%%%%%%%%%%

\noindent \textbf{UIRs of $\bm{so(4,1)}$.}  A UIR of $so(4,1)$ is specified by two numbers, the scaling dimension $\Delta$ and the spin $s$, denoted collectively as $\vec{\mathcal{F}}=(\Delta,s)$. The number $s \geq 0$ is an integer or half-odd integer. For the $so(4)$ representations $\vec{f}=(f_{1}, f_{2})$ contained in the UIR $\vec{\mathcal{F}}=(\Delta, s)$ we have:
\begin{align} \label{branching_rules_spin(2p,1)->spin(2p)}
  f_{1}\geq s\geq |f_{2}| . 
\end{align}
{The representation-theoretic labels in Refs.~\cite{Letsios_announce, Yale_Thesis} are related to the labels of the present paper as: $\Delta = F_{0} +3$ and $s=F_{1} $.}
The UIRs of $so(4,1)$ are listed below~\cite{Schwarz, Ottoson}:
\begin{itemize}
    \item \textbf{Principal Series} $\bm{D}_{\textbf{prin}}\bm{(\,\vec{\mathcal{F}}\,)}$\textbf{:}
    \begin{align}\label{Principal_UIR_D=4}
        \Delta=\frac{3}{2}+iy, \hspace{5mm} (y>0).
    \end{align}
     $s$ is an integer or half-odd integer.

    \item \textbf{Complementary Series} $\bm{D}_{\textbf{comp}}\bm{(\,\vec{\mathcal{F}}\,):}$ 
    \begin{align}\label{Compl_UIR_D=4}
     \frac{3}{2}\leq \Delta<3-\tilde{n}, \hspace{5mm}\tilde{n} \in\{0,1\}.   
    \end{align}
     If $ \tilde{n}=0$, then $s=0$, and for the $so(4)$ content we have $f_{2}=0$. If $\tilde{n} = 1$, then $s$ is a positive integer.
    
     \item \textbf{Exceptional Series} $\bm{D}_{\textbf{ex}}\bm{(\,\vec{\mathcal{F}}\,):}$ 
     \begin{align}\label{Exceptional_UIR_D=4}
        \Delta=2. 
        \end{align}
    $s$ is a positive integer and $f_{2}=0$.

 \item \textbf{Discrete Series} $\bm{D}^{\bm{\pm}}\bm{(\,\vec{\mathcal{F}}\,):}$ 
$\Delta$ is real.  The representation-theoretic labels $\Delta$ and $s$ are both integers or half-odd integers. There are two different cases of discrete series UIRs depending on the $so(4)$ content:
\begin{align}\label{condition_for_discrete_series_+ D=4}
   s \geq f_{2}  \geq \Delta-1 \geq \frac{1}{2}\hspace{9mm}\text{for}~D^{+}(\,\vec{\mathcal{F}}\,),
\end{align}
\begin{align}\label{condition_for_discrete_series_- D=4}
  -s \leq f_{2} \leq -\Delta+1 \leq -\frac{1}{2}\hspace{6mm}\text{for}~D^{-}(\,\vec{\mathcal{F}}\,).
  \end{align}
  Form eq.~(\ref{condition_for_discrete_series_+ D=4}), it is clear that the $so(4)$ content of $D^{+}$ UIRs corresponds to $so(4)$ irreps with positive last component, $f_{2}$, of the highest weight~(\ref{highest_weight_so(4)}). Similarly, according to eq.~(\ref{condition_for_discrete_series_- D=4}), only $so(4)$ irreps with negative $f_{2}$ are contained in $D^{-}$ UIRs.
\end{itemize}
The graviton on $dS_{4}$ corresponds to $\Delta = 3$ and $s=2$. In particular, the positive frequency modes of the graviton on $dS_{4}$ form the direct sum of discrete series UIRs $D^{-}(3,2) \bigoplus D^{+}(3,2)$ \cite{Yale_Thesis, HiguchiLinearised}. The gravitino (i.e.\ strictly massless spin-3/2 field) on $dS_{4}$ corresponds to $\Delta = 5/2$ and $s=3/2$. The positive-frequency modes of the gravitino on $dS_{4}$ form the direct sum of discrete series UIRs $D^{-}(5/2,3/2) \bigoplus D^{+}(5/2,3/2)$~\cite{Letsios_announce, Letsios_announce_II}. In general, the positive-frequency modes for any strictly massless boson or fermion of any spin $ s\geq 1/2$ on $dS_{4}$ correspond to the direct sum $D^{-}(s+1,s) \bigoplus D^{+}(s+1,s)$~\cite{Yale_Thesis, Letsios_announce, Letsios_conformal-like}. For further discussions on representation-theoretic aspects of fields on dS spacetime see Refs.~\cite{Yale_Thesis, STSHS, Letsios_announce, Letsios_announce_II, Letsios_conformal-like, Sengor, Gazeau, Sun, Mixed_Symmetry_dS, Hinter, Letsios_Alan, Prof_Alan, Letsios_prof, Dio_charm, Penedones, Kamran}. 
%%%%%%%%%%%%%%%%%%%%%%%%%%%%%%%%%%%%%%%%%%%%%%%%%%%%%%%%

The quadratic Casimir of $so(4,1)$ is defined as
\begin{align}
   C_{2} \equiv \sum_{A = 1}^{4} \left( J_{0A}  \right)^{2} - \frac{1}{2} \delta^{IK} \delta^{JL} J_{I J} J_{K L}\hspace{8mm} (I,J,K,L \in \{ 1,2,3,4 \} ).
\end{align}
For a $so(4,1)$ UIR labelled by $\vec{\mathcal{F}}=(\Delta,s)$ the quadratic Casimir has the (real) eigenvalue:
\begin{align}\label{Casimir_Spin(4,1)_eigenvalue}
    c_{2}(\vec{\mathcal{F}})=(\Delta-3)\,\Delta+s(s+1).
\end{align}

%%%%%%%%%%%%%%%%%%%%%

%%%%%%%%%%%%%%%%%%%%%%%%%%%%%%%%%%%%%%%%%%%%%%%%%%%%%%%%%%%%%%%%%%%%%%%%%
\section{Global dS geometry (Christoffel symbols, spin connection and all that)} \label{Appendix_global dS details}
In global coordinates (\ref{dS_metric}), the non-zero Christoffel symbols are 
\begin{align}\label{Christoffels_dS}
    &\Gamma^{t}_{\hspace{0.2mm}\tilde{\mu} \tilde{\nu}}=\cosh{t} \sinh{t} \hspace{1mm}\tilde{g}_{\tilde{\mu} \tilde{\nu}}, \hspace{2mm} \Gamma^{\tilde{\mu}}_{\hspace{0.2mm}\tilde{\nu} t} =\tanh{t}  \hspace{1mm}\tilde{g}^{\tilde{\mu}}_{\tilde{\nu}}, \nonumber \\ 
& \Gamma^{\tilde{\kappa}}_{\hspace{0.2mm}\tilde{\mu} \tilde{\nu}}=\tilde{\Gamma}^{\tilde{\kappa}}_{\hspace{0.2mm}\tilde{\mu} \tilde{\nu}},~~~~~~~~~~~\tilde{\mu},\tilde{\nu},\tilde{\kappa}\in \{ \theta_{1}, \theta_{2} ,  \theta_{3}  \},
\end{align}
where $\tilde{g}_{\tilde{\mu} \tilde{\nu} }$ and $\tilde{\Gamma}^{\tilde{\kappa}}_{\hspace{0.2mm}\tilde{\mu} \tilde{\nu}}$ are the metric tensor and the Christoffel symbols, respectively, on $S^{3}$.

 We work with the following representation of gamma matrices:
\begin{equation}\label{gammas rep}
 \gamma^{0}=i \begin{pmatrix}  
   0 & \bm{1} \\
   \bm{1} & 0
    \end{pmatrix} , \hspace{5mm}
    \gamma^{j}=\begin{pmatrix}  
   0 & i\widetilde{\gamma}^{j} \\
   -i\widetilde{ \gamma}^{j} & 0
    \end{pmatrix} ,
     \end{equation} 
($ j=1,2,3$) where $\bm{1}$ is the 2-dimensional spinorial identity matrix. The timelike gamma matrix is anti-hermitian, while the spacelike ones are hermitian. The lower-dimensional gamma matrices, $\widetilde{\gamma}^{j}$, satisfy the Euclidean Clifford algebra in 3 dimensions:
\begin{equation} \label{Euclidean_3D gammas}
    \{ \widetilde{\gamma}^{j}, \widetilde{\gamma}^{k}\} = 2 \delta^{jk} \bm{1}, \hspace{7mm}j,k=1,2,3.
\end{equation}
As in Refs.~\cite{Camporesi, Letsios_announce, Letsios_announce_II},  the representation of the lower-dimensional gamma matrices we use is:
\begin{align}
 \widetilde{\gamma}^{1}=\begin{pmatrix}  
   0 & i \\
   -i & 0
    \end{pmatrix}, ~ ~~  \widetilde{\gamma}^{2}=\begin{pmatrix}  
   0 & 1 \\
   1 & 0
    \end{pmatrix}, ~~~ \widetilde{\gamma}^{3}=\begin{pmatrix}  
   1 & 0 \\
   0 & -1
    \end{pmatrix}.
\end{align}
 In our representation for the four-dimensional gamma matrices, the fifth gamma matrix~(\ref{def_gamma5}) is  given by
\begin{align} \label{gamma5 chiral rep}
    \gamma^{5}=\begin{pmatrix}
    \bm{1} & 0\\
    0      & -\bm{1}
    \end{pmatrix},
\end{align}
and we note that (\ref{def_gamma5}) can be re-written as $\varepsilon_{\mu \nu \rho \sigma}  = i   \gamma^{5}  \gamma_{\mu \nu \rho \sigma}$. 
Also, under hermitian conjugation we have:
$ (\gamma^{\mu})^{\dagger} = \gamma^{0}  \gamma^{\mu}   \gamma^{0}$, $(\gamma^{\rho \sigma})^{\dagger} = \gamma^{0}   \gamma^{\rho \sigma}   \gamma^{0}$ and $(\gamma^{\mu \rho \sigma})^{\dagger} =- \gamma^{0}   \gamma^{\mu\rho \sigma}   \gamma^{0}.$
Note also the following useful properties \cite{Freedman}:
\begin{align} \label{properties of epsilon}
    &\varepsilon_{\alpha \beta \rho \gamma} \varepsilon^{\alpha \beta \mu \nu}=-4 \delta_{\rho}^{[\mu} \delta_{\gamma}^{\nu]}, \nonumber \\
    & \varepsilon_{\alpha \eta \rho \gamma}\varepsilon^{\alpha \kappa \mu \nu}=(-3!)\delta_{\eta}^{[\kappa} \delta_{\rho}^{\mu} \delta_{\gamma}^{\nu]}, \nonumber \\
    &\varepsilon_{\alpha \beta \sigma \delta}\varepsilon^{\kappa \lambda \mu \nu}=(-4!)\delta_{\alpha}^{[\kappa}\delta_{\beta}^{\lambda}\delta_{\sigma}^{\mu}\delta_{\delta}^{\nu]}.
\end{align}

For the vierbein fields on global $dS_{4}$, we choose the expressions:
\begin{equation}\label{vielbeins}
    e^{t}{\hspace{0.2mm}}_{0}=1, \hspace{5mm}  e^{\tilde{\mu}}{\hspace{0.2mm}}_{i}=\frac{1}{\cosh{t}} \tilde{e}^{\tilde{\mu}}{\hspace{0.2mm}}_{i} , \hspace{7mm}i=1,2,3,
\end{equation}
 where $\tilde{e}^{\tilde{\mu}}{\hspace{0.2mm}}_{i}$ are the dreibein fields on $S^{3}$. The non-zero components of the dS spin connection are given by
\begin{equation}\label{spin_connection_dS}
   \omega_{ijk} = \frac{  \tilde{\omega}_{ijk}}{ \cosh{t}  } , \hspace{6mm}  \omega_{i0k}= -\omega_{ik0}=  -\tanh{t}\hspace{1mm} \delta_{ik},\hspace{5mm}i,j,k \in \{  1,2,3\},
        \end{equation} 
where $\tilde{\omega}_{ijk}$ is the spin connection on $S^{3}$.

%%%%%%%%%%%%%%%%%%%%%%%%%%%%%%%%%%%%%%%%%%%%%%%%%%%%%%%%%%%%%%%%%5

\section{Transverse, \texorpdfstring{$\tilde{\gamma}$}{gamma}-traceless delta function (\ref{def: TT delta function 3/2}) and locality of the equal-time anti-commutator (\ref{local eq-time anti-com 3/2 special})}  
In this Appendix, we will explicitly demonstrate the locality of the equal-time anti-commutator (\ref{local eq-time anti-com 3/2 special}). To achieve this, we will first show that the transverse and gamma-traceless delta function {on $S^3$} which appears in the anti-commutator (\ref{local eq-time anti-com 3/2 special}), and is defined in (\ref{def: TT delta function 3/2}), can be re-expressed as
\label{Appenix_Locality and TT-delta function}
     \begin{align} \label{TT delta function 3/2 final}
         \Delta^{TT}_{\tilde{\mu}  \tilde{\nu}'}(\bm{\theta_{3}},\bm{\theta_{3}}') 
         = & \left(\tilde{g}_{\tilde{\mu} \tilde{\nu}'} \mathbb{U}(\bm{\theta_{3}}, \bm{\theta_{3}'}) -\frac{1}{3} \tilde{\gamma}_{\tilde{\mu} }\mathbb{U}(\bm{\theta_{3}}, \bm{\theta_{3}'})  \tilde{\gamma}_{\tilde{\nu}'} \right)\frac{\delta(\bm{\theta_{3} - \bm{\theta_{3}}'})}{\sqrt{\tilde{g}}} \nonumber \\
         &+ \frac{3}{2} \tilde{\nabla}^{T}_{\tilde{\mu}} \left(  ~\frac{1}{\tilde{\slashed{\nabla}}^{2} +9/4} \sum_{\sigma \in \{ +,-  \}}\sum_{n=1}^{\infty} \sum_{l,q}\chi_{\sigma}^{(n;l;q)}(\bm{\theta_{3}})\otimes\chi_{\sigma}^{(n;l;q)}(\bm{\theta_{3}}')^{\dagger} \right)\overset{\leftarrow}{\tilde{\nabla}^{T}}_{\tilde{\nu}'},
     \end{align}
  where $\mathbb{U}(\bm{\theta_{3}}, \bm{\theta_{3}'})$ is the spinor parallel propagator on $S^{3}$ with $\mathbb{U}(\bm{\theta_{3}}, \bm{\theta_{3}})= \bm{1}$, and  $\tilde{g}_{\tilde{\mu}    \tilde{\nu}'}$, $\tilde{g}$,   $\tilde{\gamma}_{\tilde{\mu}}$ are the bi-vector of parallel transport, determinant of the metric and gamma matrices, respectively, on $S^{3}$. The superscript ${T}$ on the covariant derivatives denotes their gamma-traceless part: 
 \begin{align}\label{TT cov deriv on S^3}
      \tilde{\nabla}^{T}_{\tilde{\mu}}   =\tilde{\nabla}_{\tilde{\mu}} - \frac{1}{3}  \tilde{\gamma}_{\tilde{\mu}} \tilde{\gamma}^{\tilde{\alpha}}\tilde{{\nabla}}_{\tilde{\alpha}}~~\text{and}~~\overset{\leftarrow}{\tilde{\nabla}^{T}}_{\tilde{\nu}'}   = \overset{\leftarrow}{\tilde{\nabla}}_{\tilde{\nu}'} -\frac{1}{3}\overset{\leftarrow}{\tilde{\nabla}}_{\tilde{\alpha}'} \tilde{\gamma}^{\tilde{\alpha}'}     \tilde{\gamma}_{\nu'}.
 \end{align}
     The spinors $\chi_{\pm}^{(n;l;q)}$ in (\ref{TT delta function 3/2 final}) are the spinor spherical harmonics on the unit $S^{3}$ which are eigenfunctions of the Dirac operator, $\tilde{\slashed{\nabla}} = \tilde{\gamma}^{\tilde{\alpha}}  \tilde{\nabla}_{\tilde{\alpha}}$,  \cite{Camporesi}
     \begin{align} \label{eigenspinors on S^3}
   & \tilde{\slashed{\nabla}}\chi_{\pm}^{(n;l;q)}(\bm{\theta_{3}})= \pm i \left(n+\frac{3}{2}\right)  \chi_{\pm}^{(n;l;q)}(\bm{\theta_{3}}) ,~~~n \in \{0,1,2,... \},
\end{align}
where the quantum numbers $n, l ,q$ correspond to the chain of subalgebras $so(4) \supset$ $so(3)  \supset$ $so(2)$ with $n +\tfrac{1}{2} \geq l+\tfrac{1}{2} \geq | q + \tfrac{1}{2} | \geq \tfrac{1}{2}$. The spinor spherical harmonics are normalised on $S^{3}$ as
\begin{align}
   \int_{S^{3}} d\bm{\theta_{3}} \sqrt{\tilde{g}}~ \chi_{\sigma}^{(n;l;q)}(\bm{\theta_{3}})^{\dagger}\chi_{\sigma'}^{(n';l';q')}(\bm{\theta_{3}}) = \delta_{\sigma \sigma'}\delta_{nn'} \delta_{l l '}\delta_{qq'}.
\end{align}
They also satisfy the following completeness relation:
\begin{align} \label{completeness spinors S^3}
\sum_{\sigma \in \{ +, - \}} \sum_{n=0}^{\infty}   \sum_{l,q}   \chi_{\sigma}^{(n;l;q)}(\bm{\theta_{3}}) 
 \otimes \chi_{\sigma}^{(n;l;q)}(\bm{\theta_{3}'})^{\dagger} =\frac{\delta(\bm{\theta_{3} - \bm{\theta_{3}}'})}{\sqrt{\tilde{g}}} \mathbb{U}(\bm{\theta_{3}}, \bm{\theta_{3}'}).
\end{align}
 For later convenience, some comments are in order:\\
 $\bullet$ Although the value $n=0$ is allowed in the spectrum of the Dirac operator in (\ref{eigenspinors on S^3}), this value is omitted from the sum in (\ref{TT delta function 3/2 final}) as it renders the denominator ill-defined [this also becomes clear in our proof of (\ref{TT delta function 3/2 final}) below].  \\
 $\bullet$ For $n=0$, {for which} the allowed values for the rest of the angular momentum numbers are $l = 0$ and $q=-1,0$, the spinor spherical harmonics (\ref{eigenspinors on S^3}) coincide with the Killing spinors on $S^{3}$, satisfying:
\begin{align}\label{Killing spinor eqn S^3}
    \tilde{\nabla}_{\tilde{\mu}}\  \chi_{\pm}^{(0;0;q)} =\pm \frac{i}{2}  \tilde{\gamma}_{\tilde{\mu}}  \chi_{\pm}^{(0;0;q)}.
\end{align} 
 It is clear that this equation is identical with $\tilde{\nabla}^{T}_{\tilde{\mu}} \chi_{\pm}^{(0;0;q)}=0$ [see eq.~(\ref{TT cov deriv on S^3})]. In the main text, the Killing spinors $\chi_{\pm}^{(0;0;q)}$
 are denoted as $\tilde{\epsilon}_{\pm,q}$ - see eq.~(\ref{dS Killing spinors explicit}).
\\
$\bullet$ The commutator of covariant derivatives acting on spinors on $S^{3}$ is 
\begin{align} \label{commutator cov der spinors S^3}
        [\tilde{\nabla}_{\tilde{\mu}}, \tilde{\nabla}_{\tilde{\nu}}]=\frac{1}{4}\tilde{R}_{\tilde{\mu} \tilde{\nu} \tilde{\kappa} \tilde{\lambda}} \tilde{\gamma}^{\kappa}\tilde{\gamma}^{\tilde{\lambda}},
\end{align}
where the Riemann tensor of the unit $S^{3}$ is
\begin{align}\label{Riemann_tens S^3}
    \tilde{R}_{\tilde{\mu} \tilde{\nu} \tilde{\kappa} \tilde{\lambda}}=\tilde{g}_{\tilde{\mu} \tilde{\kappa}} \tilde{g}_{\tilde{\nu} \tilde{\lambda}}-\tilde{g}_{\tilde{\nu} \tilde{\kappa}} \tilde{g}_{\tilde{\mu} \tilde{\lambda}}.
\end{align}
Also, when acting on spinors on $S^{3}$, the squared Dirac operator is related to the Laplace-Beltrami operator as
\begin{align}\label{Dirac^2 and LB on S^3}
   \tilde{g}^{\tilde{\mu}  \tilde{\nu}} \tilde{\nabla}_{\tilde{\mu}}  \tilde{\nabla}_{\tilde{\nu}} =  \tilde{\slashed{\nabla}}^{2} + \frac{\tilde{R}}{4},
\end{align}
where the Ricci scalar is $\tilde{R} = 6 $.
%%%%%%%%%%%%%%%%%%%%%%%%%%%%%%%%%%%%%%%%%%%%%%%%%%%%%%%%%%%%%%%%%%%%%%
Let us now start proving eq.~(\ref{TT delta function 3/2 final}).

 $$ \textbf{Proof of (\ref{TT delta function 3/2 final}):}$$
 To prove (\ref{TT delta function 3/2 final}), we need to make use of the completeness of the vector-spinor eigenfunctions of the Dirac operator, also known as vector-spinor spherical harmonics, on $S^{3}$. There are two kinds of vector-spinor spherical harmonics on $S^{3}$ \cite{Chen}: the transverse-traceless harmonics (\ref{vector-spinor+-eigen_S3}) and the longitudinal ones.  We denote the latter as $\tilde{\lambda}^{(\text{P};\pm n; {l};q)}_{ \tilde{\mu}}(\bm{\theta_{3}})$ and $\tilde{\lambda}^{(\text{M};\pm n; {l};q)}_{ \tilde{\mu}}(\bm{\theta_{3}}) $. To show that $\Delta^{TT}_{\tilde{\mu}  \tilde{\nu}'}(\bm{\theta_{3}},\bm{\theta_{3}}')$ [eq.~(\ref{def: TT delta function 3/2})] is given by (\ref{TT delta function 3/2 final}), we need  to exploit the fact that the TT and longitudinal vector-spinor spherical harmonics form a complete set on $S^{3}$. The corresponding completeness relation is
\begin{align}\label{completeness vec-spinors S^3 step1}
 \frac{\delta(\bm{\theta_{3} - \bm{\theta_{3}}'})}{\sqrt{\tilde{g}}} \tilde{g}_{\tilde{\mu}  \tilde{\nu}'} \mathbb{U}(\bm{\theta_{3}}, \bm{\theta_{3}'})  =&  ~\Delta^{TT}_{\tilde{\mu}  \tilde{\nu}'}(\bm{\theta_{3}},\bm{\theta_{3}}')   + \sum_{\sigma \in \{+ , -  \}}\sum_{n , l , q} \tilde{\lambda}_{\tilde{\mu}}^{(\text{P}; \sigma n; l;q)} (\bm{\theta_{3}})  \otimes  \tilde{\lambda}_{\tilde{\nu}'}^{(\text{P}; \sigma n; l;q)} (\bm{\theta_{3}} ')^{\dagger} \nonumber \\
 &~+ \sum_{\sigma \in \{+ , -  \}}\sum_{n , l , q} \tilde{\lambda}_{\tilde{\mu}}^{(\text{M}; \sigma n; l;q)} (\bm{\theta_{3}})  \otimes  \tilde{\lambda}_{\tilde{\nu}'}^{(\text{M}; \sigma n; l;q)} (\bm{\theta_{3}} ')^{\dagger},
\end{align}
where $\Delta^{TT}_{\tilde{\mu}  \tilde{\nu}'}(\bm{\theta_{3}},\bm{\theta_{3}}')$ is the sum over the transverse harmonics [see (\ref{def: TT delta function 3/2})], while the rest of the sums in (\ref{completeness vec-spinors S^3 step1}) concern the longitudinal harmonics with all the allowed values of the quantum numbers $n, l,q$ (these allowed values are discussed below).

To proceed, we need more information concerning the longitudinal harmonics.
The longitudinal vector-spinor harmonics $\tilde{\lambda}^{(\text{P};\pm n; {l};q)}_{ \tilde{\mu}}(\bm{\theta_{3}}),\tilde{\lambda}^{(\text{M};\pm n; {l};q)}_{ \tilde{\mu}}(\bm{\theta_{3}}) $ satisfy \cite{Chen}
\begin{align} 
    \tilde{\slashed{\nabla}} \tilde{\lambda}^{(\text{P};\pm n; {l};q)}_{ \tilde{\mu}}(\bm{\theta_{3}}) = + i\, \sqrt{(n+{3}/{2})^{2}-2}~ \tilde{\lambda}^{(\text{P};\pm n; {l};q)}_{ \tilde{\mu}}(\bm{\theta_{3}})
\end{align}
and
\begin{align}
    \tilde{\slashed{\nabla}} \tilde{\lambda}^{(\text{M};\pm n; {l};q)}_{ \tilde{\mu}}(\bm{\theta_{3}}) = - i\, \sqrt{(n+{3}/{2})^{2}-2}~ \tilde{\lambda}^{(\text{M};\pm n; {l};q)}_{ \tilde{\mu}}(\bm{\theta_{3}}),
\end{align}
and they are expressed in terms of the  
spinor harmonics (\ref{eigenspinors on S^3}) as \cite{Chen}:
\begin{align}\label{def: long vec-spin P}
 \tilde{\lambda}^{(\text{P};\pm n; {l};q)}_{ \tilde{\mu}}(\bm{\theta_{3}}) =\frac{c^{(\text{P};\pm n)}}{\sqrt{2}} \left(  \tilde{\nabla}_{\tilde{\mu}}+\frac{i}{2} \left\{ \mp (n+3/2)  + \sqrt{(n+{3}/{2})^{2}-2}~ \right \} \tilde{\gamma}_{\tilde{\mu}}\right )   \chi_{\pm}^{(n;l;q)}(\bm{\theta_{3}}), 
\end{align}
\begin{align}\label{def: long vec-spin M}
 \tilde{\lambda}^{(\text{M};\pm n; {l};q)}_{ \tilde{\mu}}(\bm{\theta_{3}}) =\frac{c^{(\text{M};\pm n)}}{\sqrt{2}} \left(  \tilde{\nabla}_{\tilde{\mu}}+\frac{i}{2} \left\{ \mp (n+3/2)  - \sqrt{(n+{3}/{2})^{2}-2}~ \right \} \tilde{\gamma}_{\tilde{\mu}}\right )   \chi_{\pm}^{(n;l;q)}(\bm{\theta_{3}}). 
\end{align}
The normalisation factors $c^{(\text{P};\pm n)}$ and $c^{(\text{M};\pm n)}$ were not introduced in Ref.~\cite{Chen}. We introduce them here such that the longitudinal vector-spinor harmonics satisfy
\begin{align}
   \int_{S^{3}} d\bm{\theta_{3}} \sqrt{\tilde{g}}~  \tilde{g}^{\tilde{\mu}  \tilde{\nu}}~\tilde{\lambda}^{(\text{S};\sigma n; {l};q)}_{ \tilde{\mu}}(\bm{\theta_{3}})^{\dagger} \tilde{\lambda}^{(\text{S'};\sigma' n'; {l'};q')}_{ \tilde{\nu}}(\bm{\theta_{3}}) = \delta_{\text{SS}'} \delta_{\sigma \sigma'} \delta_{nn'} \delta_{l l '}\delta_{qq'},
\end{align}
where   $\text{S, S}'\in \{ \text{P, M} \}$ and $\sigma \in \{ +,-  \}$. It is straightforward to find that:
\begin{align}
 \left|  \frac{c^{(\text{P};\sigma n)}}{\sqrt{2}} \right|^{2} = \Bigg(  \frac{3}{2} \left( (n+3/2)^{2}-2 \right) -\sigma \frac{1}{2} (n+3/2)\sqrt{(n+3/2)^{2}-2 }  \Bigg)^{-1},
\end{align}
where for $\sigma = -$ we have that all the values of $n \geq 0$ are allowed, while for $\sigma = +$ we have $ n \geq 1$  because for $n=0$ the harmonic $\tilde{\lambda}^{(\text{P};+ n; {l};q)}_{ \tilde{\mu}}$ is identically zero (and, thus, its normalisation factor is not defined) as the spinor harmonic in (\ref{def: long vec-spin P}) is a Killing spinor (for $n=0$) and the differential operator acting on it takes the form of the operator in the Killing spinor equation (\ref{Killing spinor eqn S^3}).
Similarly, we  find 
\begin{align}
 \left|  \frac{c^{(\text{M};\sigma n)}}{\sqrt{2}} \right|^{2} = \Bigg(  \frac{3}{2} \left( (n+3/2)^{2}-2 \right) +\sigma \frac{1}{2} (n+3/2)\sqrt{(n+3/2)^{2}-2 }  \Bigg)^{-1},
\end{align}
where, now, for $\sigma = +$ we have $n \geq 0$, while for $\sigma = -$ we have $ n \geq 1$ {for} the same {reason} as in the case of $c^{(\text{P};+ n)}$ above. 

Now that we know the allowed values of the quantum number $n$, let us re-write (\ref{completeness vec-spinors S^3 step1}) as
\begin{align}\label{completeness vec-spinors S^3 step2}
\Delta^{TT}_{\tilde{\mu}  \tilde{\nu}'}(\bm{\theta_{3}},\bm{\theta_{3}}')    =&  ~\frac{\delta(\bm{\theta_{3} - \bm{\theta_{3}}'})}{\sqrt{\tilde{g}}} \tilde{g}_{\tilde{\mu}  \tilde{\nu}'} \mathbb{U}(\bm{\theta_{3}}, \bm{\theta_{3}'})  - \sum_{n=1}^{\infty}\sum_{ l , q} \tilde{\lambda}_{\tilde{\mu}}^{(\text{P}; + n; l;q)} (\bm{\theta_{3}})  \otimes  \tilde{\lambda}_{\tilde{\nu}'}^{(\text{P}; + n; l;q)} (\bm{\theta_{3}} ')^{\dagger} \nonumber \\
& - \sum_{n=0}^{\infty}\sum_{ l , q} \tilde{\lambda}_{\tilde{\mu}}^{(\text{P}; - n; l;q)} (\bm{\theta_{3}})  \otimes  \tilde{\lambda}_{\tilde{\nu}'}^{(\text{P}; - n; l;q)} (\bm{\theta_{3}} ')^{\dagger} \nonumber \\
 &~- \sum_{n=1}^{\infty}  \sum_{l , q} \tilde{\lambda}_{\tilde{\mu}}^{(\text{M}; - n; l;q)} (\bm{\theta_{3}})  \otimes  \tilde{\lambda}_{\tilde{\nu}'}^{(\text{M}; - n; l;q)} (\bm{\theta_{3}} ')^{\dagger} \nonumber \\
 &- \sum_{n=0}^{\infty}  \sum_{l , q} \tilde{\lambda}_{\tilde{\mu}}^{(\text{M}; + n; l;q)} (\bm{\theta_{3}})  \otimes  \tilde{\lambda}_{\tilde{\nu}'}^{(\text{M}; + n; l;q)} (\bm{\theta_{3}} ')^{\dagger} .
\end{align}
Substituting (\ref{def: long vec-spin P}) and (\ref{def: long vec-spin M}) into (\ref{completeness vec-spinors S^3 step2}), and after a long but straightforward calculation, we find that all the $n=0$ terms cancel among themselves with the help of (\ref{Killing spinor eqn S^3}), while the remaining terms ($ n=1,2,...$) can be written in the simpler form:
   \begin{align} \label{completeness vec-spinors S^3 step3}
         \Delta^{TT}_{\tilde{\mu}  \tilde{\nu}'}(\bm{\theta_{3}},\bm{\theta_{3}}') =\left(\tilde{g}_{\tilde{\mu} \tilde{\nu}'}\mathbb{U}(\bm{\theta_{3}}, \bm{\theta_{3}'})-\frac{1}{3} \tilde{\gamma}_{\tilde{\mu} } \mathbb{U}(\bm{\theta_{3}}, \bm{\theta_{3}'}) \tilde{\gamma}_{\tilde{\nu}'}\right) \frac{\delta(\bm{\theta_{3} - \bm{\theta_{3}}'})}{\sqrt{\tilde{g}}} +Y_{\tilde{\mu}  \tilde{\nu}'}(\bm{\theta_{3}}, \bm{\theta_{3}}') ,
         \end{align}
         where we have used (\ref{completeness spinors S^3}), and we have also defined
         \begin{align}
        Y_{\tilde{\mu}  \tilde{\nu}'}&(\bm{\theta_{3}}, \bm{\theta_{3}}') \nonumber \\
       =   \frac{3}{2}& \sum_{\sigma \in \{ +,-  \}}\sum_{n=1}^{\infty} \sum_{l,q} \frac{1}{-(n+3/2)^{2} +9/4} \Bigg \{ \left(\tilde{\nabla}_{\tilde{\mu}}\chi_{\sigma}^{(n;l;q)}(\bm{\theta_{3}})-\frac{i \sigma (n+3/2)}{3} \tilde{\gamma}_{\tilde{\mu}}\chi_{\sigma}^{(n;l;q)}(\bm{\theta_{3}}) \right) ~  \nonumber \\
       & \otimes \Bigg(\tilde{\nabla}_{\tilde{\nu}'}\chi_{\sigma}^{(n;l;q)}(\bm{\theta_{3}}')^{\dagger}+\frac{i \sigma (n+3/2)}{3} \chi_{\sigma}^{(n;l;q)}(\bm{\theta_{3}}')^{\dagger}\,\tilde{\gamma}_{\tilde{\nu}'} \Bigg) \Bigg\}.
     \end{align}
Then, using (\ref{eigenspinors on S^3}), it is easy to show that $ \Delta^{TT}_{\tilde{\mu}  \tilde{\nu}'}(\bm{\theta_{3}},\bm{\theta_{3}}')$ in (\ref{completeness vec-spinors S^3 step3}) is equal to the desired expression (\ref{TT delta function 3/2 final}). {\hfill $\Box$}

\medskip

%%%%%%%%%%%%%%%%%%%%5
Now let us use eq.~(\ref{TT delta function 3/2 final}) to show that the equal-time anti-commutator (\ref{local eq-time anti-com 3/2 special}) is local (i.e.\ vanishes for ${\bm{\theta_{3}}} \neq \bm{\theta_{3}}'$)  {despite} that $\Delta^{TT}_{\tilde{\mu}  \tilde{\nu}'}$ is non-local due to the appearance of $\left(\tilde{\slashed{\nabla}}^{2} + 9/4 \right)^{-1}$ {in (\ref{TT delta function 3/2 final}).} {It is clear that} the locality of the anti-commutator (\ref{local eq-time anti-com 3/2 special}) 
 {reduces to} the locality of the following quantity:
\begin{align*}
    &\left( \tilde{\slashed{\nabla}}^{2}  + \frac{1}{4}\right)  {\Delta^{TT}_{\tilde{\mu}  \tilde{\nu}'}(\bm{\theta_{3}},\bm{\theta_{3}}')}\\
   = &  \left( \tilde{\slashed{\nabla}}^{2}  + \frac{1}{4}\right)\left(\tilde{g}_{\tilde{\mu} \tilde{\nu}'}\mathbb{U}(\bm{\theta_{3}}, \bm{\theta_{3}'})-\frac{1}{3} \tilde{\gamma}_{\tilde{\mu} }  \mathbb{U}(\bm{\theta_{3}}, \bm{\theta_{3}'})\tilde{\gamma}_{\tilde{\nu}'}\right)\frac{\delta(\bm{\theta_{3} - \bm{\theta_{3}}'})}{\sqrt{\tilde{g}}}   \\
         &+ \frac{3}{2} \sum_{\sigma \in \{ +,-  \}}\sum_{n=1}^{\infty} \sum_{l,q} \left( \tilde{\slashed{\nabla}}^{2}  + \frac{1}{4}\right)\tilde{\nabla}^{T}_{\tilde{\mu}} \left(  ~\frac{1}{\tilde{\slashed{\nabla}}^{2} +9/4} \chi_{\sigma}^{(n;l;q)}(\bm{\theta_{3}})\otimes\chi_{\sigma}^{(n;l;q)}(\bm{\theta_{3}}')^{\dagger} \right)\overset{\leftarrow}{\tilde{\nabla}^{T}}_{\tilde{\nu}'}.
\end{align*}
It is straightforward to commutate the two differential operators $ \tilde{\slashed{\nabla}}^{2} +1/4$ and $\tilde{\nabla}^{T}_{\tilde{\mu}}$ using (\ref{commutator cov der spinors S^3}), as
$$   \left( \tilde{\slashed{\nabla}}^{2}  + \frac{1}{4}\right) \tilde{\nabla}^{T}_{\tilde{\mu}}=\left( \tilde{\slashed{\nabla}}^{2}  + \frac{1}{4}\right) \tilde{\nabla}_{\tilde{\mu}} -\frac{1}{3}\tilde{\gamma}_{\tilde{\mu}} \tilde{\slashed{\nabla}}\left( \tilde{\slashed{\nabla}}^{2}  + \frac{1}{4}\right)  = \tilde{\nabla}^{T}_{\tilde{\mu}} \left( \tilde{\slashed{\nabla}}^{2}  + \frac{9}{4}\right),$$
where we have used that gamma matrices commute with the squared Dirac operator because of (\ref{Dirac^2 and LB on S^3}). It is now clear that there is no non-local term in  $\left( \tilde{\slashed{\nabla}}^{2}  + \frac{1}{4}\right)  \Delta^{TT}_{\tilde{\mu}  \tilde{\nu}'}$, as:
\begin{align*}
    \left( \tilde{\slashed{\nabla}}^{2}  + \frac{1}{4}\right)  {\Delta^{TT}_{\tilde{\mu}  \tilde{\nu}'}(\bm{\theta_{3}},\bm{\theta_{3}}')} = & \left( \tilde{\slashed{\nabla}}^{2}  + \frac{1}{4}\right)\left(\tilde{g}_{\tilde{\mu} \tilde{\nu}'}\mathbb{U}(\bm{\theta_{3}}, \bm{\theta_{3}'})-\frac{1}{3} \tilde{\gamma}_{\tilde{\mu} }\mathbb{U}(\bm{\theta_{3}}, \bm{\theta_{3}'})  \tilde{\gamma}_{\tilde{\nu}'}\right) \frac{\delta(\bm{\theta_{3} - \bm{\theta_{3}}'})}{\sqrt{\tilde{g}}}   \\
         &+ \frac{3}{2} \sum_{\sigma \in \{ +,-  \}}\sum_{n=1}^{\infty} \sum_{l,q} \tilde{\nabla}^{T}_{\tilde{\mu}} \left(\chi_{\sigma}^{(n;l;q)}(\bm{\theta_{3}})\otimes\chi_{\sigma}^{(n;l;q)}(\bm{\theta_{3}}')^{\dagger} \right)\overset{\leftarrow}{\tilde{\nabla}^{T}}_{\tilde{\nu}'}.
\end{align*}
We can now include the value $n=0$ in the summation as it gives zero contribution because of the Killing spinor equation (\ref{Killing spinor eqn S^3}). Finally, using the completeness of the spinor spherical harmonics (\ref{completeness spinors S^3}), we arrive at the local expression
\begin{align} \label{(diff op) on TT delta function 3/2 final}
    \left( \tilde{\slashed{\nabla}}^{2}  + \frac{1}{4}\right)  {\Delta^{TT}_{\tilde{\mu}  \tilde{\nu}'}(\bm{\theta_{3}},\bm{\theta_{3}}')} = & \left( \tilde{\slashed{\nabla}}^{2}  + \frac{1}{4}\right)\left(\tilde{g}_{\tilde{\mu} \tilde{\nu}'}\mathbb{U}(\bm{\theta_{3}}, \bm{\theta_{3}'})-\frac{1}{3} \tilde{\gamma}_{\tilde{\mu} } \mathbb{U}(\bm{\theta_{3}}, \bm{\theta_{3}'}) \tilde{\gamma}_{\tilde{\nu}'}\right) \frac{\delta(\bm{\theta_{3} - \bm{\theta_{3}}'})}{\sqrt{\tilde{g}}}  \nonumber \\
         &+ \frac{3}{2} \tilde{\nabla}^{T}_{\tilde{\mu}}\left( \frac{\delta(\bm{\theta_{3} - \bm{\theta_{3}}'})}{\sqrt{\tilde{g}}}~\mathbb{U}(\bm{\theta_{3}}, \bm{\theta_{3}'}) \right) \overset{\leftarrow}{\tilde{\nabla}^{T}}_{\tilde{\nu}'}\, .
\end{align}
This shows that the equal-time anti-commutator (\ref{local eq-time anti-com 3/2 special}) is local.

%%%%%%%%%%%%%%%%%%%%%%%%%%%%%%%%%%%%%%%%%%%%%%%%%%%%%%%%%%%%%%%%%%%%%%%%%%%%%%%%%%%%%%%%%%%%%%%%%%%%%%%%%%%%%%%%%%%%%%%%%%%%%%%%%%%%%%%%%%%%%%%%%%%%%%%%%%%%%%%%%%%555
\section{Useful expressions concerning the conformal-like symmetry of the graviton}\label{Appendix_conf-like symmetry graviton}

 Let $B_{\mu \nu}$ be any (complex or real) symmetric spin-2 tensor field on $dS_{4}$. Its conformal-like transformation is 
{defined in} (\ref{conf-like graviton diff operator compact}) as 
\begin{align}
   T_{V}B_{\mu \nu}= V^{\rho}  \varepsilon_{\rho \sigma \lambda (\mu}  \nabla^{\sigma}B^{\lambda}_{~ \nu)}.
\end{align}
Recall that  $V^{\mu}$ is any genuine conformal Killing vector (\ref{V=nabla phi}).
 One can straightforwardly prove the following:
\begin{align}\label{useful quantities conf-like graviton}
    & g^{\mu \nu}~ T_{V}B_{\mu \nu} = 0, \nonumber \\
    & \nabla^{\alpha}\, T_{V}B_{\alpha \nu} = \frac{1}{2} V^{\rho} \varepsilon_{\rho  \sigma \lambda   \nu} \nabla^{\sigma}\nabla^{\alpha}B_{\alpha}^{~\lambda} , \nonumber \\
    & \nabla_{(\mu}\nabla^{\alpha}\, T_{V}B_{\nu)\alpha} = \frac{1}{2} V^{\rho} \varepsilon_{\rho \sigma \lambda (\nu} \nabla^{\sigma}\nabla_{\mu)}\nabla^{\alpha}B_{\alpha}^{~\lambda}, \nonumber \\
     & \nabla^{\nu} \nabla^{\alpha}\, T_{V}B_{\nu\alpha} = 0, \nonumber \\
     & \Box T_{V}B_{\mu \nu} = V^{\rho}  \varepsilon_{\rho \sigma \lambda (\mu}  \nabla^{\sigma}~ \Box B^{\lambda}_{~ \nu)}. 
\end{align}
These expressions can be used to prove that $T_{V}$ is a symmetry of the full linearised Einstein equations (\ref{EOM_graviton general}), as well as of the ones in the TT gauge (\ref{EOM graviton TT}). Moreover, one can similarly show that both $T_{V}$ and $\mathcal{T}_{V} = i T_{V}$ are symmetries of the non-gauge-fixed complex linearised Einstein equations (\ref{EOM_ cmplx graviton general}), as well as of the complex graviton equations in the TT gauge (\ref{EOM cmplx graviton TT}).
%%%%%%%%%%%%%%%%%%%%%%%%%%%%%%%%%%%%%%%%%%%%%%%%%%%%%%%%%%%%%%%%%%%%%%%%%%%%%%%%%

\section{Some properties of the field strengths} \label{Append_field strengths}
Let us recall some properties of the field strengths for the complex graviton 
[eq.~(\ref{def:graviton_field-strength})] and complex gravitino 
[eq.~(\ref{def:gravitino_field-strength})]. Without making use of the equations of motion, it is easy to show that
\begin{align}
     & \nabla_{[\kappa} U_{\alpha \beta] \mu \nu}=0 , ~~U_{[\alpha \beta \mu] \nu}=0,
\end{align}
and \cite{Letsios_conformal-like}
\begin{align}
     \left(\nabla_{[\kappa}   + \frac{i}{2}\gamma_{[\kappa}   \right) F_{\alpha \beta] } =0.
\end{align}
If the complex graviton satisfies the field equations, then the complex linearised Weyl tensor satisfies
\begin{align}
&    g^{\alpha \mu} U_{\alpha \beta \mu \nu} =0,~~   g^{\beta \nu} U_{\alpha \beta \mu \nu} =0\\
    &\nabla^{\alpha}  U_{\alpha \beta \mu \nu}=0 .
\end{align}
The dual, $\widetilde{U}_{\alpha   \beta   \mu \nu} = \frac{1}{2}\varepsilon_{\alpha \beta}^{~~~~\kappa \lambda}   U_{\kappa   \lambda \mu \nu} $, can be also expressed {(using the equations of motion) as}
\begin{align}\label{left dual=right dual Weyl}
  \widetilde{U}_{\alpha   \beta   \mu \nu}  =  \frac{1}{2} U_{\alpha \beta \kappa   \lambda}~\varepsilon_{~~~\mu \nu}^{\kappa \lambda}.
\end{align}
Equation (\ref{left dual=right dual Weyl}) can be proved as follows. Let us denote the tensor on the right-hand side of eq.~(\ref{left dual=right dual Weyl}) as 
$P_{\alpha \beta \mu \nu} = \frac{1}{2}\varepsilon_{\mu \nu}^{~~~\kappa \lambda} U_{\alpha \beta \kappa   \lambda}$. Contracting $P_{\alpha \beta \mu \nu}$ with  $\frac{1}{2}\varepsilon_{\rho   \sigma}^{~~~\alpha   \beta}$, and using 
eq.~(\ref{properties of epsilon}) and the equations of motion, one finds
$$   \widetilde{P}_{\rho \sigma \mu \nu}  = - U_{\rho    \sigma  \mu \nu}.$$
Then, contracting this equation with $\frac{1}{2}\varepsilon_{\alpha   \beta}^{~~~\rho  \sigma}$, and using (\ref{properties of epsilon}), we have
$$  \frac{1}{2}\varepsilon_{\alpha   \beta}^{~~~\rho  \sigma} \widetilde{P}_{ \rho \sigma \mu \nu} = -P_{\alpha \beta \mu \nu} = - \widetilde{U}_{\alpha  \beta \mu \nu} ,  $$
 {thus proving} eq.~(\ref{left dual=right dual Weyl}).

If the complex gravitino satisfies the  equations of motion, then its  field strength satisfies \cite{Letsios_conformal-like}:
\begin{align}
    &\nabla^{\alpha}  F_{\alpha  \nu}=  \gamma^{\alpha}  F_{\alpha  \nu} =0, \\
  & \nabla_{[\kappa} F_{\alpha \beta] }=0 ,\\
  &\gamma_{[\kappa} F_{\alpha \beta] } =0,  \label{gamma bianchi identity} \\
  & \slashed{\nabla} F_{\mu \nu}=0, \\
  &  \gamma_{\mu\nu \alpha \beta} F^{\alpha \beta }= - 2 F_{\mu \nu}, ~~ \text{and thus,}~~\frac{1}{2}\varepsilon_{\mu\nu \alpha \beta} F^{\alpha \beta }=- i \gamma^{5} F_{\mu \nu}. \label{duality properties3/2 fieldstrngth}
\end{align}
%%%%%%%%%%%%%%%%%%%%%%%%%%%%%%%%%%%%%%%%%%%%%%%%%%%%%%%%%%%%
\subsection{Deriving the SUSY transformation (\ref{SUSY_transf of spin2 fieldstrength FINAL}) of the spin-2 field strength from the initial SUSY transformation (\ref{SUSY_transf of spin2 fieldstrength})}
To derive eq.~(\ref{SUSY_transf of spin2 fieldstrength FINAL}) from 
eq.~(\ref{SUSY_transf of spin2 fieldstrength}) we have to make use of 
eq.~(\ref{simplifier of Weyl SUSY transf}). Let us now prove 
eq.~(\ref{simplifier of Weyl SUSY transf}). We start by considering the following quantity:
\begin{align}\label{step 1 simplifier proof}
   2~\varepsilon_{\rho \sigma \alpha \beta}~  \gamma_{[\mu}^{\hspace{2mm}\alpha}F^{\beta}_{\hspace{2mm}\nu]} =~ i \gamma^{5}\gamma_{\rho \sigma \alpha \beta}~  \left(\gamma_{\mu}^{\hspace{2mm}\alpha}F^{\beta}_{\hspace{2mm}\nu}- \gamma_{\nu}^{\hspace{2mm}\alpha}F^{\beta}_{\hspace{2mm}\mu}  \right),  
\end{align}
where, on the right-hand side, we have used (\ref{gamma_abcd-e_abcd gamma5}) and we have expanded the anti-symmetrisation of the indices $\mu$ and $\nu$.
Using $\gamma_{\rho \sigma \alpha \beta}~\gamma_{\mu}^{\hspace{2mm}\alpha}=g_{\rho \mu}  \gamma_{\sigma \beta} -g_{\sigma   \mu}  \gamma_{\rho  \beta}   +g_{\beta \mu}     \gamma_{\rho  \sigma}$, as well as the fact that the spin-3/2 field strength is gamma traceless on-shell, and thus $\gamma_{\rho \beta} F^{\beta}_{~~\nu}= - F_{\rho \nu}$, eq.~(\ref{step 1 simplifier proof}) gives
\begin{align}\label{step 2 simplifier proof}
   2~\varepsilon_{\rho \sigma \alpha \beta}~  \gamma_{[\mu}^{\hspace{2mm}\alpha}F^{\beta}_{\hspace{2mm}\nu]} =~ i \gamma^{5}\left(2 g_{\rho [\mu} F_{\nu]  \sigma}- 2 g_{\sigma [\mu} F_{\nu]  \rho} +2 \gamma_{\rho  \sigma}F_{\mu \nu}  \right).  
\end{align}
Then, the  first two terms on the right-hand side of eq.~(\ref{step 2 simplifier proof}) can be re-expressed as
$$ 2 g_{\rho [\mu} F_{\nu]  \sigma}- 2 g_{\sigma [\mu} F_{\nu]  \rho} =\gamma_{\mu  \nu}F_{\rho \sigma}   -\gamma_{\rho  \sigma}F_{\mu \nu}.  $$
This can be straightforwardly proved by {using} $\gamma_{\kappa \lambda}= {(\gamma_{\kappa} \gamma_{\lambda} - \gamma_{\lambda} \gamma_{\kappa})/2} $ on the right-hand side, and then making use of  the on-shell property (\ref{gamma bianchi identity}). Thus, eq.~(\ref{step 2 simplifier proof}) gives
\begin{align}\label{step 3 simplifier proof}
   {2}~\varepsilon_{\rho \sigma {\gamma\delta}}~  \gamma_{[\mu}^{\hspace{2mm}{\gamma}}F^{{\delta}}_{\hspace{2mm}\nu]} =~ {i} \gamma^{5}\left(\gamma_{\mu  \nu}F_{\rho \sigma} + \gamma_{\rho  \sigma}F_{\mu \nu}  \right).
\end{align}
Contracting both sides of eq.~(\ref{step 3 simplifier proof})  with ${\varepsilon^{\alpha \beta \rho \sigma}}$
{and dividing the result by $2$}, we have
\begin{align*}
   -2\,  \gamma_{[\mu}^{\hspace{2mm}[\alpha}F^{\beta]}_{~~~\nu]} =~ \frac{i}{2} \gamma^{5}\left(\gamma_{\mu  \nu} \widetilde{F}^{\alpha \beta} + \frac{1}{2}\varepsilon^{\alpha \beta \rho \sigma}\gamma_{\rho  \sigma}F_{\mu \nu}  \right).
\end{align*}
Then, using the on-shell property (\ref{duality properties3/2 fieldstrngth}), as well as $\varepsilon^{\alpha \beta \rho \sigma}\gamma_{\rho  \sigma}/2 = -i \gamma^{5} \gamma^{\alpha \beta} $, we find
\begin{align}
   -2\,  \gamma_{[\mu}^{\hspace{2mm}[\alpha}F^{\beta]}_{~~~\nu]} =~ \frac{1}{2} \left(\gamma_{\mu  \nu} {F}^{\alpha \beta} + \gamma^{\alpha \beta}F_{\mu \nu}  \right),
\end{align}
{proving eq.~(\ref{simplifier of Weyl SUSY transf}).}
}
%%%%%%%%%%%%%%%%%%%%%%%%%%%%%%%%%%%%%%%%%%%%%%%%%%%%%%%%%%%%%%%%%%%%%%%%%%%%55

%%%%%%%%%%%%%%%%%%%%%%%%%%%%%%%%%%%%%%%%%%%%%%%%%%%%%%%%%%%%%%%%%%%%%%%%%%%%%%%%%%%%%%%%%%%%%%%%%%%%%%%%%%%%%%%%%%%%

%\end{comment}
\providecommand{\noopsort}[1]{}\providecommand{\singleletter}[1]{#1}%

\end{document}